%% file: arXiv_main.tex
\documentclass[conference]{IEEEtran}

\RequirePackage[
  datamodel=acmdatamodel,
  style=acmnumeric,
  ]{biblatex}

\usepackage{graphicx} %
\usepackage{hyperref}
\usepackage{enumitem}
\usepackage{amssymb}   %
\usepackage{amsmath} 
\usepackage{pifont}
\usepackage{diagbox}
\usepackage{multirow}
\usepackage{mathtools}
\usepackage{tikz}
\usetikzlibrary{shapes.geometric, arrows.meta, bending, positioning, fit}
\tikzstyle{arrow} = [thick,->,>=stealth]
\usepackage{subcaption}
\usepackage{tabularx}

\usepackage{xspace}
\usepackage{ifthen}
\usepackage{placeins}

\usepackage{pgfplots}
\usepgfplotslibrary{polar}
\pgfplotsset{compat=1.18}

\DeclareMathOperator*{\argmin}{arg\,min}

\newboolean{showcomments}
\setboolean{showcomments}{true} 

\newcommand{\subheading}[1]{\vspace{0.5 em}\noindent \textbf{#1}}

\newcommand*\tageq{\refstepcounter{equation}\tag{\theequation}}

\newcommand{\mta}{\ensuremath{\mathtt{MTA}}\xspace}
\newcommand{\mtaafter}{\ensuremath{\mathtt{MTA^*}}\xspace}
\newcommand{\bda}{\ensuremath{\mathtt{BDA}}\xspace}
\newcommand{\bdaafter}{\ensuremath{\mathtt{BDA^*}}\xspace}
\newcommand{\lifespan}{\ensuremath{\mathtt{Span}_{50}}\xspace}
\newcommand{\nsga}{{{Stochastic}}\xspace}

\begin{document}

\title{On Hyperparameters and Backdoor-Resistance in Horizontal Federated Learning}

\author{
\IEEEauthorblockN{Simon Lachnit}
\IEEEauthorblockA{Ruhr University Bochum \\
Bochum, Germany \\
simon.lachnit@rub.de}
\and
\IEEEauthorblockN{Ghassan Karame}
\IEEEauthorblockA{Ruhr University Bochum \\
Bochum, Germany \\
ghassan@karame.org}
}

\IEEEtitleabstractindextext{
\begin{abstract}

Horizontal Federated Learning (HFL) is particularly vulnerable to backdoor attacks as adversaries can easily manipulate both the training data and processes to execute sophisticated attacks.
In this work, we study the impact of training hyperparameters on the effectiveness of backdoor attacks and defenses in HFL. More specifically, we show both analytically and by means of measurements that the choice of hyperparameters by benign clients does not only influence model accuracy \textit{but also significantly impacts backdoor attack success}. This stands in sharp contrast with the multitude of contributions in the area of HFL security, which often rely on custom ad-hoc hyperparameter choices for benign clients---leading to more pronounced backdoor attack strength and diminished impact of defenses. 
Our results indicate that properly tuning benign clients' hyperparameters---such as learning rate, batch size, and number of local epochs---can significantly curb the effectiveness of backdoor attacks, \emph{regardless of the malicious clients’ settings}. We support this claim with an extensive robustness evaluation of state-of-the-art attack-defense combinations, showing that carefully chosen hyperparameters yield across-the-board improvements in robustness without sacrificing main task accuracy. For example, we show that the 50\%-lifespan of the strong A3FL attack can be reduced by $98.6\%$, respectively---all without using any defense and while incurring only a $2.9$ percentage points drop in clean task accuracy. %

\end{abstract}}

\maketitle

\IEEEdisplaynontitleabstractindextext

\input{section_01}

\input{section_02}

\input{section_03}

\input{section_04}

\input{section_05}

\input{section_06}

\input{section_07}

\section*{Acknowledgments}

This work has been co-funded by the Deutsche Forschungsgemeinschaft (DFG, German Research Foundation) under Germany’s Excellence Strategy- EXC 2092 CASA- 390781972, by the German Federal Ministry of Education and Research (BMBF) through the project TRAIN (01IS23027A). 
Calculations for this publication were performed on the HPC cluster Elysium of the Ruhr University Bochum, subsidised by the DFG (INST 213/1055-1).

\printbibliography

\appendices

\input{appendix}

\end{document}

%% file: section_01.tex
\section{Introduction}

Horizontal Federated Learning (HFL), initially proposed by Google in 2017~\cite{mcmahan_communication-efficient_2017}, is a distributed Machine Learning (ML) algorithm that enables multiple participants to train a global model on their local training data without explicitly exchanging the data. This is achieved through a central server that coordinates the training process in multiple rounds, aggregating locally updated models from all participants after each round and then broadcasting the updated global model back to them.
A popular example is Google's smart keyboard application \verb+Gboard+, which uses HFL to improve the next-word predictions~\cite{hard_federated_2019}.
Other examples include the personalization of Apple's voice assistant \verb+Siri+~\cite{paulik_federated_2021} and their keyboard application \verb+QuickType+~\cite{apple_designing_2019}.

Despite its advantages, HFL, however, comes with a severe security limitation: given its lack of authorization mechanisms, clients cannot be easily made accountable for their contributions to the global model---much like in open-access open-source systems~\cite{noauthor_wikipedia_2025}. This lack of accountability allows malicious clients to submit manipulated updates and tamper with the training process.
A prominent example is backdoor attacks, which enable an adversary to assign a wrong classification to all inputs with a certain (adversary-chosen) feature while ensuring that the classifier behaves normally on all other inputs. 
For instance, attacks like A3FL~\cite{zhang_a3fl_2023} and IBA~\cite{nguyen_iba_2023} adaptively optimize the trigger pattern such that it becomes more challenging for the benign clients to unlearn the backdoor behavior.
Other works, like DarkFed~\cite{DBLP:conf/ijcai/LiWNHXZW24}, show that it is possible to backdoor the global model even without access to task-specific data. This is achieved by training on a shadow dataset and imitating the behavior of the global model. 
Popular approaches to limit the influence of malicious clients on the global model consist of constraining the size of the model updates~\cite{sun_can_2019} or relying on some form of anomaly detection~\cite{fung_limitations_2020, mhamdi_hidden_2018}.

While current research has significantly improved the security of HFL deployments, we note that existing backdoor attacks and defenses against HFL rely on an ad-hoc choice of hyperparameters for benign clients, often prioritizing the tuning of the parameters of malicious clients. For instance, a quick glimpse over 15 top-tier papers published in the last five years (see Table~\ref{tab:hyper_params}) shows that existing proposals exhibit (i) a wide range of choices for the batch size, local epochs, momentum, and learning rates (sometimes even varying during the attack window) for malicious clients, \emph{coupled with} (ii) less-diverse ad-hoc choices for benign clients. 
Given that the interplay of these hyperparameters directly impacts model stability and convergence\footnote{For instance, a high learning rate can cause instability, while a low one may hinder convergence~\cite{DBLP:conf/nips/HeLT19, DBLP:conf/iclr/CohenKLKT21, DBLP:conf/icml/AndriushchenkoV23}. Similarly, smaller batches add gradient noise that may help generalize but slow down convergence~\cite{DBLP:conf/nips/HeLT19}.}~\cite{li_convergence_2020,DBLP:conf/icml/AndriushchenkoV23,DBLP:conf/aaai/LiuZFYXM024, DBLP:conf/icml/FarhadkhaniGGPS22}, 
\emph{this naturally raises the question of whether properly tuning the hyperparameters of benign clients in HFL also has a significant impact on backdoor resistance.} Namely, a well-generalized model learns meaningful patterns from the data, making it less likely to be misled by maliciously injected triggers. In contrast, an overfitted model memorizes training data rather than learning generalizable features, making it more vulnerable to backdoor attacks that exploit this memorization to embed hidden behaviors. Moreover, 
overfitting can undermine the effectiveness of existing backdoor defenses---making attacks ``appear more successful than they truly are''.  %
Importantly, the benign clients' hyperparameters are not controlled by malicious nodes and could therefore be optimized to proactively reduce the impact of backdoor attacks.

In this paper, we address this gap and analyze, for the first time, the impact of the choice of hyperparameters, such as the learning rate, the batch size, the number of local epochs, and the weight decay, on the backdoor robustness of HFL. 
More specifically, we show, both analytically and empirically by means of measurements, that a proper choice of hyperparameters among honest learners in HFL can drastically improve the aggregated model's robustness against state-of-the-art backdoor attacks---irrespective of the choice of hyperparameters assumed by malicious clients. %
Compared to most existing defensive approaches, which react to backdoor attacks by, e.g., filtering malicious updates, we aim to \emph{proactively} (as opposed to reactively) improve security against backdoor attacks.
For example, we show that there exists a set of Pareto-near-optimal hyperparameter choices that achieve across-the-board improvements in accuracy-robustness tradeoffs against state-of-the-art backdoor attacks that could otherwise not be realized by existing defenses. Our analysis further reveals that some attacks are weaker than they initially appear, while certain defenses, such as FoolsGold~\cite{fung_limitations_2020} and Bulyan~\cite{mhamdi_hidden_2018}, are more effective than commonly reported.

In summary, we make the following contributions in this work:
\begin{description}[leftmargin=0.5 cm]
    \item[Gaps in existing HFL evaluations: ] 
    We observe that existing attacks and defenses in HFL rely on an ad-hoc selection of hyperparameters. 
    A thorough study of 15 top-tier papers published in the past five years reveals that current proposals employ a diverse range of (i) learning rate values (sometimes varying within the attack window), (ii) batch sizes and local epoch settings, and (iii) momentum and weight decay values. 
    We note that while the hyperparameters of malicious clients are not consistent across attacks, the hyperparameters used for benign clients seem to be ad-hoc and less diverse (Section~\ref{sec:motivation}).
    \item[Impact of the hyperparameters: ] We analyze both analytically and empirically the impact of the learning rate, batch size, number of local epochs, momentum, and weight decay on the security of HFL against state-of-the-art attacks, such as A3FL \cite{zhang_a3fl_2023}, Chameleon \cite{dai_chameleon_2023}, DarkFed \cite{DBLP:conf/ijcai/LiWNHXZW24}, and FCBA \cite{DBLP:conf/aaai/LiuZFYXM024}. 
    Our results demonstrate that selecting appropriate hyperparameters for benign clients can effectively limit attack success, regardless of the attacker's hyperparameter choices (Section~\ref{sec:lr_effect} and Section~\ref{sec:other_params}). We additionally show that our findings hold across diverse attacks, multiple datasets, and model architectures.
    \item[Adaptive Adversary: ] We additionally explore adaptive strategies for an adversary to increase backdoor accuracy by fine-tuning malicious clients' hyperparameters in response to benign hyperparameter choices. We, however, demonstrate that a careful selection of benign parameters can severely cap the impact of such adaptive strategies. 
    \item[Recommended benign configurations: ] By leveraging our findings, we conduct an extensive robustness evaluation of carefully chosen benign hyperparameters with state-of-the-art backdoor attacks and find across-the-board improvements in robustness against all considered attacks. 
    For instance, our experiments on A3FL~\cite{zhang_a3fl_2023}, one of the most powerful backdoor attacks, show that properly tuning benign hyperparameters decreases backdoor accuracy and 50\%-Lifespan by up to $20.9$ percentage points and $98.6\%$, with a minimal impact on clean task accuracy (of $2.9$ percentage points) and without relying on any defense.
    When combined with state-of-the-art defenses, our recommended benign hyperparameters further reduce backdoor accuracy, for example, for the Chameleon attack under the FoolsGold~\cite{fung_limitations_2020} defense, by up to $59.9$ percentage points (Section~\ref{sec:pareto}).
    \item[Open science: ] All the code and data required to reproduce our results in the paper are publicly available at \url{https://github.com/RUB-InfSec/federated_learning_hyperparams}.
\end{description}

%% file: section_02.tex
\section{Background}

Machine Learning (ML) is a broad term encompassing various methods for learning desired functionalities from raw data~\cite{goodfellow_deep_2016}, often by training Artificial Neural Networks (ANNs) via Stochastic Gradient Descent (SGD). 
SGD updates the model parameters iteratively to minimize a loss function $\ell(\cdot)$ on data batches $(X, Y) \subset (\mathcal{X}, \mathcal{Y})$ with batch size $B = |X|$: $\theta^{t+1} = \theta^t - \eta \cdot \nabla_{\theta^t} \sum_{(x, y) \in (X, Y)} \ell \left(\theta^t, x, y\right)$, where $\eta$ is the \textit{learning rate} that controls the update step size. 

To accelerate convergence, SGD can be enhanced with momentum~\cite{DBLP:conf/icml/SutskeverMDH13}, which maintains an exponentially decaying average of past gradients and incorporates it into the update step to speed up optimization in regions of consistent gradient directions. 
Another common extension is weight decay~\cite{DBLP:conf/nips/KroghH91}, which adds a penalty term on the $\ell_2$-norm of the model parameters. This acts as a regularization technique, promoting more robust feature learning.

\subsection{Horizontal Federated Learning}

In HFL~\cite{mcmahan_communication-efficient_2017, shejwalkar_back_2022, cao_fedrecover_2023, fereidooni_freqfed_2024, rieger_crowdguard_2024}, a central \emph{server} $S$ coordinates training among $N$ \emph{clients} $C = {C_1, C_2, \dots, C_N}$, each with a local dataset $\mathcal{D}_{C_k}$.
Each client $C_k$ has a loss function $F_k(\theta) = \frac{1}{n_k} \sum_{(x,y) \in \mathcal{D}_{C_k}}{\ell(\theta, x, y)}$.
The size of the local dataset $\mathcal{D}_{C_i}$ is denoted by $n_i$.
This gives the following distributed optimization: \[\theta^* = \argmin_\theta \left\{ F(\theta) = \sum_{k \in C}{\frac{n_k}{\sum_{c \in C}{n_c}}F_k(\theta)} \right\}\]

Prior to the start of training, during the \textit{setup phase}, the central server $\mathcal{S}$ selects an effective set of hyperparameters for local ML training and distributes them to all clients.
The training in HFL then proceeds in rounds: the server sends the current global model $\theta^t$ to a subset of $M$ clients $C^t \subset C$ ($|C^t| = M$). Each client $c \in C^t$ updates its local model and trains it on $\mathcal{D}_c$ for $E$ epochs, then computes its \emph{model update} as $\Delta_c^t = \theta^t_c - \theta^t$.
The server aggregates these updates using an \emph{aggregation rule} $f_{agg}(\cdot)$ to update the global model: $\theta^{t+1} = f_{agg}\left(\left\{\Delta^t_c \mid \forall c \in C^t\right\}\right)$. 

The most prominent aggregation rule used in current academic and practical applications is Federated Averaging (FedAvg)~\cite{daly_federated_2024}, where updates are weighted by dataset sizes~\cite{mcmahan_communication-efficient_2017}:
$\theta^{t+1} = \theta^{t} + \sum_{k \in C^t}{\frac{n_k}{\sum_{c \in \mathcal{C}^t}{n_c}}\Delta_k^t}$.
In HFL, clients typically have disjoint local datasets.
While centralized ML assumes IID-sampled training and test data~\cite{sherman_identification_2018}, HFL typically operates with non-IID client data~\cite{mcmahan_communication-efficient_2017}.
Most often, samples are distributed across clients using a Dirichlet distribution~\cite{wang_attack_2020, yoo_backdoor_2022, zhang_a3fl_2023, bagdasaryan_how_2020}.

\subsection{Backdoor Attacks against HFL}

Poisoning attacks on ML models during training are a well-known threat~\cite{gu_badnets_2017, liu_trojaning_2018, wang_neural_2019}, extensively studied in traditional ML.
However, the threat model of HFL exhibits a key difference compared to traditional ML: 
In HFL, malicious clients, so-called \emph{model poisoning attackers}~\cite{bagdasaryan_how_2020}, can influence not only the training data but the (local) training procedure and the uploaded model update, respectively.
Therefore, a common assumption in HFL is that the attacker can compromise a set of clients completely, i.e., access and control their training procedure, training data, and in- and outgoing communication~\cite{shejwalkar_back_2022}.

A \emph{backdoor attack} is a particularly dangerous type of poisoning attack in which an attacker modifies the model to misbehave under specific conditions, i.e., when a so-called \emph{trigger} is present in the input while remaining benign on all other inputs~\cite{gu_badnets_2017}.
This trigger can either be artificially added to the input, e.g., a white rectangle in one corner of an image, or be an inherent input property that can be expressed semantically, e.g., all images of green cars~\cite{bagdasaryan_how_2020}.
The former is called an \textit{artificial trigger} backdoor, while the latter is a \textit{semantic} backdoor.
Recent approaches like, e.g., A3FL~\cite{zhang_a3fl_2023} and IBA~\cite{nguyen_iba_2023}, extend artificial trigger attacks further by considering adapting the trigger as well to achieve the maximum attack effect.

Several metrics quantify backdoor attack success in HFL:

\begin{description}[leftmargin=*]
    \item[Main Task Accuracy:] The \textit{Main Task Accuracy (MTA)}, also referred to as the \textit{clean accuracy}, captures the performance of the poisoned model on clean data~\cite{rieger_crowdguard_2024}.
    It is measured on a hold-out test dataset $(X_{test} \subset \mathcal{X}, Y_{test} \subset \mathcal{Y})$ and defined as: 
    $$MTA_{X_{test}, Y_{test}}(f) = \frac{\lvert \{ (x, y) \in (X_{test}, Y_{test}): f(x) = y \} \rvert}{\lvert X_{test} \rvert}$$
    \item[Backdoor Accuracy:] The \textit{Backdoor Accuracy (BDA)} captures how well the model misclassifies inputs $x \in \mathcal{X}^*$, with $\mathcal{X}^* \subset \mathcal{X}$ being the subset of the input space that contains samples with the backdoor trigger, as the target class $y_t$~\cite{rieger_crowdguard_2024}.
    It is measured on a backdoored dataset $X^* \subset \mathcal{X}^*$:
    $$BDA_{X^*,y_{t}}(f) = \frac{\lvert \{ x \in X^*: f(x) = y_{t} \} \rvert}{\lvert X^* \rvert}$$
    \item[Lifespan:] In HFL, attackers might be excluded from training at some point~\cite{zhang_neurotoxin_2022}.
    Catastrophic forgetting~\cite{kirkpatrick_overcoming_2017} causes the backdoor effect to fade upon benign re-training.
    The $\gamma$-lifespan metric~\cite{zhang_neurotoxin_2022, dai_chameleon_2023, yoo_backdoor_2022, zhang_a3fl_2023} quantifies this: let $\gamma \in [0, 1]$ denote a threshold accuracy, $t_0$ the round the attack ends, and $f_{\theta^t}$ the global model at round $t$. The lifespan is:
    $$Span_\gamma = \max\{t \mid BA_{X^*, y_t}(f_{\theta^t}) > \gamma\} - t_0$$
\end{description}

When reporting the average \text{MTA} and \text{BDA}, we distinguish between two important phases: (1) the attack phase, when the adversary is active, and (2) the post-attack phase. Both phases are crucial for assessing the overall effectiveness of the backdoor attack. We denote the averages during and after the attack by \mta / \bda and \mtaafter / \bdaafter, respectively.
We summarize the various notations that we will use throughout this paper in Table~\ref{tab:notation}. 

\begin{table}
    \centering
    \footnotesize
    \scalebox{0.95}{\begin{tabular}{|ll|}
        \hline
        $\eta_b / \eta_m$: & benign / malicious learning rate \\
        $\beta$: & relative malicious learning rate $\left(\frac{\eta_m}{\eta_b}\right)$ \\
        $\mu_b / \mu_m$: & benign / malicious momentum \\
        $\lambda_b / \lambda_m$: & benign / malicious weight decay\\
        $E_b / E_m$: & benign / malicious number of local training epochs\\
        $B_b / B_m$: & benign / malicious batch size\\
        $N$: & number of clients \\
        $M$: & number of clients selected per round\\
        $\theta^t_k$: & model parameters of client $k$'s local model in global round $t$ \\
        \mta / \bda: & average main task / backdoor accuracy during the attack\\
        \mtaafter / \bdaafter: & average main task / backdoor accuracy after the attack\\
        \lifespan & $50\%$-lifespan of the backdoor \\
        \hline
    \end{tabular}}
    \caption{Summary of notations used throughout this work. A subscript $_b$ refers to the benign clients' choice, while subscript $_m$ refers to its malicious version.}
    \label{tab:notation}
\end{table}

\subsubsection{Backdoor Attacks}
\label{sec:background_attacks}
In this work, we consider the following backdoor attacks\footnote{Please refer to Section~\ref{sec:methodology} for a justification of the selection.}:

\begin{description}[leftmargin=0.2 cm]
    \item[A3FL] \label{par:a3fl} A3FL~\cite{zhang_a3fl_2023} is a trigger-optimization backdoor attack whose core idea is to predict the movement of the global model in each round and account for it by tweaking the trigger accordingly.
    Assuming explicit trigger unlearning as the "{}worst-case"{} server-side defense, A3FL optimizes the trigger to bypass this defense.
    \item[Chameleon] Chameleon~\cite{dai_chameleon_2023} extends backdoor lifespan by making poisoned samples' latent space representations more similar to samples belonging to the target class and more dissimilar to samples belonging to the source class.
    It splits the model into a representation encoder (trained with contrastive learning) and a classifier (trained with standard poisonous training).
    \item[DarkFed] The DarkFed attack~\cite{DBLP:conf/ijcai/LiWNHXZW24} assumes malicious clients lack task-specific data. 
    Instead, it uses synthetic data to implant backdoors based on two observations: (1) similar logits between the local backdoored model and the global model on synthetic data imply similar MTA on real data, and (2) a synthetic dataset with triggers can establish the desired trigger-target class relationship.
    \item[FCBA] FCBA~\cite{DBLP:conf/aaai/LiuZFYXM024} extends the Distributed Backdoor Attack~\cite{xie_dba_2020}, which splits the backdoor trigger across malicious clients to increase the susceptibility to the combined trigger. 
    The FCBA attack advances this concept by additionally training on sub-trigger combinations, boosting attack persistence.
\end{description}

\subsubsection{Defenses}
\label{sub:defenses}

Defenses against backdoor attacks in HFL are mostly realized as specific instantiations or extensions of the server-side aggregation rule.
In this paper, we consider four defenses, namely Krum~\cite{blanchard_machine_2017}, Multi-Krum~\cite{blanchard_machine_2017}, Bulyan~\cite{mhamdi_hidden_2018}, and FoolsGold~\cite{fung_limitations_2020}.
For a justification of the selection of these specific defenses and a short explanation of each, please refer to Appendix~\ref{app:defense_selection}.

%% file: section_03.tex
\section{Motivation \& System Model}
\label{sec:motivation}

While recent research on backdoor attacks and defenses in HFL has substantially enhanced the security of HFL systems, our empirical analysis of 15 top-tier papers published in the last five years (cf. Section~\ref{sec:methodology}) 
shows that existing attacks and defenses in HFL often rely on an ad-hoc choice of hyperparameters---in addition to notable differences in their system and threat models.

\addtocounter{footnote}{+1}
\setlength\tabcolsep{2.9pt} %
\begin{table*}
    \centering
    \scriptsize
    \scalebox{0.88}{
        \begin{tabular}{|c|cc|cc|cc|cc|cc|cc|cc|}
            \hline
            \multirow{2}{*}{} & \multicolumn{2}{c|}{\textbf{System Model}} & \multicolumn{2}{c|}{\textbf{Threat Model}} & \multicolumn{10}{c|}{\textbf{Hyperparameters}}\\
            \cline{2-15}
            & $N$ & $M$ & atk. participation & $a_e - a_s$ & $\eta_b$ & $\eta_m$ & $\mu_b$ & $\mu_m$ & $\lambda_b$ & $\lambda_m$ & $B_b$ & $B_m$ & $E_b$ & $E_m$ \\ 
            \hline
            Optim~\cite{DBLP:conf/cikm/Yang0NHW24} & N.A. & N.A. & pool $(5 - 30\%)$ & all rounds & $0.1$ & $0.1$ & N.A. & N.A. & N.A. & N.A. & N.A. & N.A. & N.A. & N.A. \\ 
            FCBA~\cite{DBLP:conf/aaai/LiuZFYXM024} & $100$ & $10$ & freq. $\left(\frac{1}{2}\right)$ & 27 & $0.1$ & $0.05$ & $0.9$ & $0.9$ & $0.0005$ & $0.0005$ & $64$ & $64$ & $2$ & $6$ \\        
            DarkFed~\cite{DBLP:conf/ijcai/LiWNHXZW24} & $20$ & $20$ & pool $(20\%)$ & $100$ & $0.1$ & $0.005$ & $0.9$ & $0$ & $0.0005$ & $0$ & $64$ & $64$ & $2$ & $2$ \\ 
            A3FL~\cite{zhang_a3fl_2023} & 100 & 10 & pool $(5\%)$ & 100 & $\begin{cases}\frac{0.01t-15}{1499} & t \le 1500 \\ \frac{-0.02t + 60}{1500} & t \in [1501, a_s] \\ 0.002 & t > a_s\end{cases}$ & 0.002 & 0.9 & 0.9 & 0.0005 & 0.0005 & 64 & 64 & 2 & 2 \\
            Chameleon~\cite{dai_chameleon_2023} & 100 & 10 & freq. $(1)$ & 230 & $\begin{cases} \frac{0.199t-0.3}{499} & t \le 500 \\ \max \left(\frac{-0.2t + 400}{1500}, 0.0001 \right) & t \in [501, a_e] \\ 0.005 & t > a_e \end{cases}$ & 0.01 & 0.9 & 0.9 & 0.0005 & 0.005 & 64 & 64 & 2 & 3 \\
            3DFed~\cite{li_3dfed_2023} & 100 & 100 & pool $(30\%)$ & 30 & 0.1 & 0.1 & 0.9 & 0.9 & 0.0005 & 0.0005 & 64 & 64 & 2 & 15 \\
            IBA~\cite{nguyen_iba_2023} & 200 & 10 & freq. $\left(\frac{1}{10}\right)$ & all rounds & $0.02 \cdot 0.998^t$ & $0.02 \cdot 0.998^t$ & 0.9 & 0.9 & 0.0001 & 0.0001 & 32 & 32 & 2 & 2 \\
            CerP~\cite{lyu_poisoning_2023}\footnotemark & $100$ & $20$ & pool $(4\%)$ & $330$ & $0.01$ & $0.005$ & $0.9$ & $0.9$ & $0.0005$ & $0.005$ & $64$ & $64$ & $2$ & $1$ \\
            F3BA~\cite{fang_vulnerability_2023} & $20$ & $10$ & pool $(20\%)$ & all rounds & $0.001$ & $0.001$ & $0.9$ & $0.9$ & $0.0001$ & $0.0001$ & $64$ & $64$ & $2$ & $2$ \\ 
            Neurotoxin~\cite{zhang_neurotoxin_2022} & $1000$ & $10$ & freq. $(1)$ & $250$ & $\begin{cases} \frac{0.199t-0.3}{499} & t \le 500 \\ \max \left(\frac{-0.2t + 400}{1500}, 0.0001 \right) & t > 501\end{cases}$ & $0.02$ & $0.9$ & $0.9$ & $0.0005$ & $0.005$ & $64$ & $64$ & $2$ & $10$ \\
            Tails~\cite{wang_attack_2020} & $200$ & $10$ & freq. $\left(\frac{1}{10}\right)$ & all rounds & $0.02 \cdot 0.998^t$ & $0.02 \cdot 0.998^t$ & $0.9$ & $0.9$ & $0.0001$ & $0.0001$ & $32$ & $32$ & $2$ & $2$ \\ 
            Model Replacement~\cite{bagdasaryan_how_2020} & $100$ & $10$ & pool $(1\%)$ & $1$ & $0.1$ & $0.05$ & $0.9$ & $0.9$ & $0.0005$ & $0.005$ & $64$ & $64$ & $2$ & $15$ \\ 
            DBA~\cite{xie_dba_2020} & $100$ & $10$ & freq. $(1)$ & $70$ & $0.1$ & $0.05$ & $0.9$ & $0.9$ & $0.0005$ & $0.005$ & $64$ & $64$ & $2$ & $6$ \\ 
            Model Poisoning~\cite{bhagoji_analyzing_2019}\footnotemark & $10$ & $10$ & freq. $(1)$ & all rounds & $0.001$ & $0.001$ & $0$ & $0$ & $0$ & $0$ & $100$ & $100$ & $5$ & $5$ \\ 
            LIE~\cite{baruch_little_2019} & $51$ & $51$ & pool $(24\%)$ & all rounds & $0.1$ & $0.1$ & $0.9$ & $0.9$ & $0.0005$ & $0.0001$ & $128$ & $128$ & $1$ & $5$\\
            \hline
        \end{tabular}
    }
    \caption{Hyperparameters used in the empirical evaluation of SoTA backdoor attacks on HFL. See Table~\ref{tab:notation} for notation. Attacks use either fixed-pool or fixed-frequency \textit{participation} for an attack window comprising global rounds $t \in [a_s, a_e]$. If the values for a parameter differed between the paper and the source code, we reported the value from the source code. For parameters varying for different datasets, we always reported those for CIFAR-10, the most frequently used dataset in all considered attacks. N.A. refers to those cases where we could not acquire information about the parameters from the source code.}
    \label{tab:hyper_params}
\end{table*}
\setlength\tabcolsep{6pt} %
\addtocounter{footnote}{-3}

\subsection{Ad-hoc Choices} 
Our results (cf. Table~\ref{tab:hyper_params}) suggest that there appears to be a lack of consensus within the community regarding system and threat models, as different models make varying assumptions. 
For example, some contributions like~\cite{baruch_little_2019, bagdasaryan_how_2020, zhang_a3fl_2023} assume a fixed-pool attacker participation scheme, while others, e.g., \cite{nguyen_iba_2023, xie_dba_2020, wang_attack_2020}, employ a fixed-frequency scheme. %

Moreover, there seems to be no clear consensus on the hyperparameters for benign client training. However, some custom choices, such as the benign momentum $\mu_b = 0.9$ and the benign number of local epochs $E_b=2$, appear common. 
Other parameters, like the benign weight decay $\lambda_b$, the benign batch size $B_b$, and the benign learning rate $\eta_b$, exhibit more significant variations.

More importantly, we note that malicious clients significantly exhibit higher variance in hyperparameters than benign clients, suggesting that a careful selection of hyperparameters may have a significant influence on the attack's success. For instance, the malicious learning rate $E_m$ ranges from $1$ to $15$ while its benign counterpart $E_b$ only ranges from $1$ to $5$, with $2$ being the default choice in almost all cases.
This highly suggests that optimizing these parameters probably substantially impacts the attack's effectiveness. 
Note that \cite{lyu_poisoning_2023} studied the effect of $E_m$ and found that increasing $E_m$ increases attack success up to a certain point, after which it decreases due to too much deviation between the local and global model.
\emph{Naturally, this raises the question of whether their benign counterparts also have a significant impact on backdoor resistance.} These parameters are not under the control of the adversary or malicious nodes and could, ideally, be optimized to reduce the impact of backdoor attacks.

Such ad-hoc choices are particularly surprising given the extensive research on hyperparameter impact on the accuracy and robustness of ML models in traditional ML~\cite{DBLP:conf/nips/DAngeloAVF24,DBLP:conf/icml/Croce020a,DBLP:conf/icml/AndriushchenkoV23}.
For instance, smaller batch sizes and higher learning rates are known to improve generalization~\cite{DBLP:conf/nips/HeLT19, DBLP:conf/icml/AndriushchenkoV23}, a property often linked to increased robustness~\cite{DBLP:conf/cvpr/Stutz0S19}.
This raises the question of why hyperparameter choices in backdoor-related research are not subject to similar scrutiny---a gap that we address in this work.

\begin{figure}
    \centering
    \subfloat[$\eta_b$ schedule \label{fig:a3fl_lr_schedule}]{
        \centering
        \includegraphics[width=0.47\linewidth]{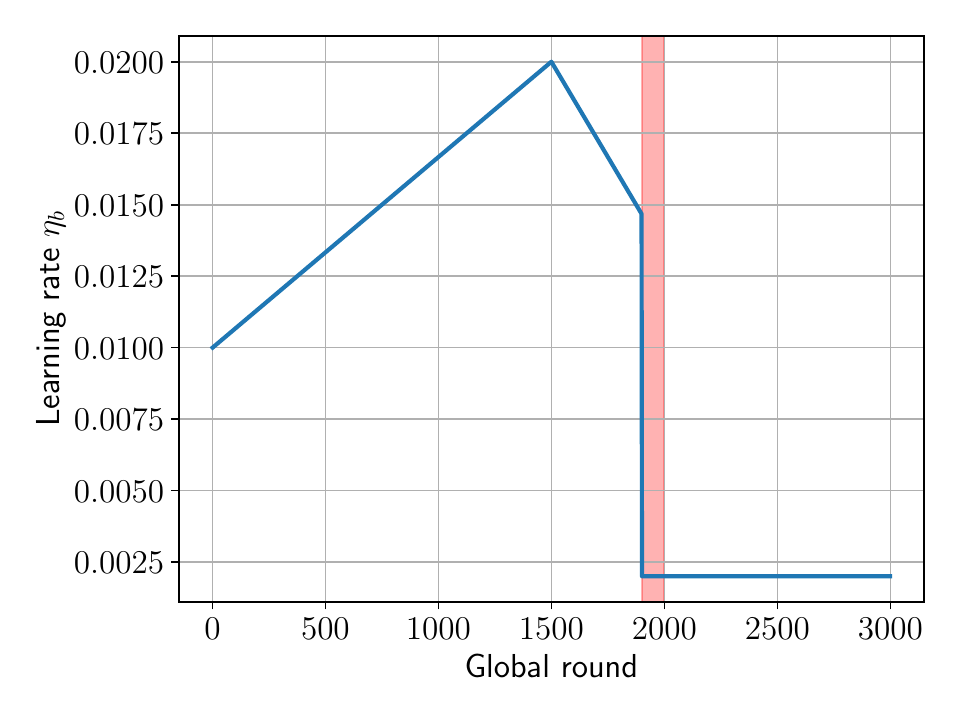}
    }
    \hfill
    \subfloat[\bda for different $\eta_b$ schedules \label{fig:a3fl_original_vs_static_lr}]{
        \centering
        \includegraphics[width=0.47\linewidth]{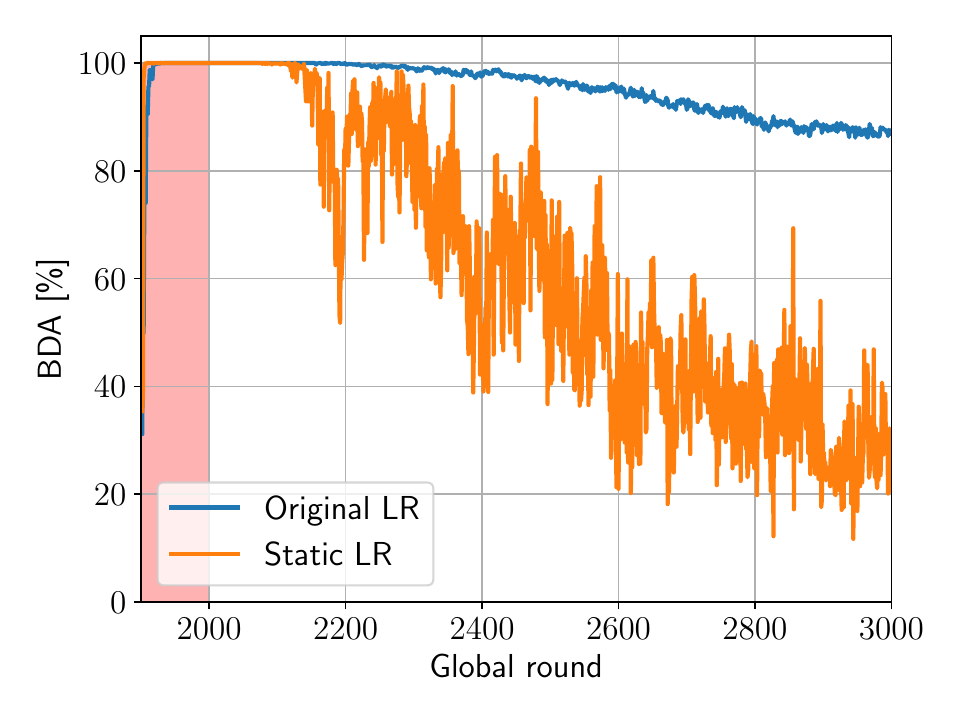}
    }
    \caption{The schedule of A3FL~\cite{zhang_a3fl_2023}'s benign learning rate $\eta_b$. The red area marks the attack window; when the attack starts, the benign learning rate suddenly drops.}
\end{figure}

\subheading{Example.}
Take as an example the A3FL attack~\cite{zhang_a3fl_2023} (cf. Section~\ref{par:a3fl}). 
Figure~\ref{fig:a3fl_lr_schedule} shows the learning rate schedule for the benign clients.
Notice that \textit{the learning rate of the benign clients is reduced significantly at the start of the attack window}.

To better understand the impact of this unexpected learning rate adjustment, we replaced the \emph{benign} learning rate schedule with a constant $\eta=0.02$ and re-run the attack\footnote{Here, we kept all other parameters and hyperparameters unchanged from their original values in~\cite{zhang_a3fl_2023}.}.
Our results (cf. Figure~\ref{fig:a3fl_original_vs_static_lr}) show a significant decrease in the achieved \bdaafter of $\sim 42.12\%$ and in the \lifespan by up to $78.95\%$ 
while reducing \mta by up to $0.7\%$.
This showcases that the choice of the benign clients' learning rate not only influences model training and convergence, as commonly assumed, but also drastically impacts the success of potential training-time attacks.

\subheading{Research question.}
Motivated by these observations, we set forth in this paper to answer the following research question: \emph{Is there a choice of hyperparameters for benign clients that makes it easier to defend against backdoor attacks in FL? Conversely, are there specific parameters that facilitate attacks on FL?}

\subsection{System \& Threat Model}
\label{sec:general_system_threat_model}
Before addressing our research question, we outline the system and threat models considered in this work. As summarized in Table~\ref{tab:hyper_params}, prior studies explore various dimensions for modeling the HFL setup and adversary capabilities.

\begin{description}[leftmargin=0.5 cm]
    \item[Client participation: ] Out of $N$ total clients, a fixed subset of $M$ may be selected each round, or clients may be sampled independently with some individual probability $p_{train}$~\cite{DBLP:conf/infocom/LuoXWHT22}. 
    \item[Dataset partitioning: ] Datasets are partitioned IID or non-IID across clients. The latter is typically achieved using label-wise Dirichlet sampling~\cite{zhang_neurotoxin_2022, lyu_poisoning_2023, zhang_a3fl_2023, bagdasaryan_how_2020}.
    \item[Aggregation rule: ] The coordinating server $\mathcal{S}$ aggregates client updates using schemes such as FedAvg~\cite{mcmahan_communication-efficient_2017} or more robust alternatives.

\stepcounter{footnote}\footnotetext{Parameters given for CIFAR-100 since CIFAR-10 was not implemented.}
\stepcounter{footnote}\footnotetext{Parameters given for FashionMNIST since CIFAR-10 was not implemented.}

    \item[Attacker participation: ] Either a fixed-pool (a static $\alpha$ fraction of clients is compromised) or fixed-frequency (one client corrupted every $n$-th round) participation model is assumed. 
    \item[Attack window: ] Attacks may begin at training start or later (e.g., near convergence). To simulate finite attacker presence, a limited attack window is typically enforced.
    \item[Collusion: ] Malicious clients may act independently, in coordination (swarm), or as puppets forwarding crafted updates~\cite{fung_limitations_2020}.
    \item[Malicious clients' knowledge: ] Attackers are unaware of benign clients' local data and updates, but do know the broadcasted benign hyperparameters.
    \item[Server: ] For backdoor attacks, the server is assumed to be benign.
    \item[Backdoor Trigger: ] Triggers can be artificial~\cite{li_3dfed_2023}, semantic~\cite{bagdasaryan_how_2020}, optimized~\cite{zhang_a3fl_2023}, or input-specific~\cite{nguyen_iba_2023}.
\end{description}

In this work, we assume the following system and threat models. 
Following prior work~\cite{fung_limitations_2020, zhang_a3fl_2023, bagdasaryan_how_2020, fang_vulnerability_2023}, we assume a benign server and restrict our focus to backdoor attacks launched by malicious clients. 
The adversary operates within a fixed window, using a fixed-pool pattern and a static corruption model: a static $\alpha$ fraction of clients is compromised at the outset and behaves arbitrarily (Byzantine) during the attack phase. 
This captures an unrestricted choice of trigger by the adversary.

We assume the coordinator performs pre-deployment testing to select benign ML hyperparameters and may apply robust aggregation rules (cf. Section~\ref{sub:defenses}) to defend against backdoor attacks. Ideally, the coordinators would like  the \mta of the defended model ($\mta_\text{def}$) to converge within a threshold $\epsilon_{def}$ to the achievable \mta without defenses ($\mta_\text{ideal}$) to ensure high model utility in spite of attacks:
\begin{align}
\min{\left\{\bda \mid \mta_\text{ideal} - \mta_\text{def} \leq \epsilon_{def}\right\}}  \tageq \label{eq:mta_constraint}
\end{align}

Crucially, the defender---comprising honest clients and the coor\-dinator---has no knowledge of the presence, identity, hyperparameters, or strategies of malicious clients.

We assume that malicious clients have no access to the local datasets or model updates of benign clients, nor any knowledge of the server-side aggregation rule or configuration. However, they do observe the benign clients’ training hyperparameters, as these are selected and broadcast by the central server. This allows adversaries to adapt their behavior based on the chosen benign hyperparameters. We further assume that the adversary can do so in an informed manner---that is, they can estimate the impact of their own hyperparameter choices using techniques such as predictive betting or probing to guide their adaptations in response to those of the benign clients.
A backdoor adversary pursues a two-fold goal: it aims to maximize the global model’s \bda while ensuring that the resulting drop in \mta between the clean ($\mta_\text{clean}$) and the poisoned model ($\mta_\text{poison}$) remains below a predefined threshold $\epsilon_{adv}$, hereby preserving the model’s overall utility.
Keeping the \mta high enough ensures that the adversary can avoid easy detection (a low MTA might cause the coordinating entity to raise an alarm).
This two-fold goal can be expressed as: 
\begin{align}
\max{\left\{\bda \mid \mta_\text{clean} - \mta_\text{poison} \leq \epsilon_{adv}\right\}} 
\tageq \label{eq:bda_adv}
\end{align}

Notice that defenders face stricter constraints than attackers: benign clients must fix their hyperparameters in advance, while adversaries can adapt theirs dynamically to suit the chosen attack strategy. 
We also note that different (honest) clients are unlikely to use different hyperparameters, since such heterogeneity typically leads to incoherent updates and unstable training dynamics~\cite{mcmahan_communication-efficient_2017}.
To empirically validate this, we conducted experiments in which each client randomly selected values for $\eta_b$, $\mu_b$, $\lambda_b$, $E_b$, and $B_b$ prior to the start of HFL training. 
We evaluated this setup under the A3FL~\cite{zhang_a3fl_2023} and Chameleon~\cite{dai_chameleon_2023} attacks, using three aggregation rules: plain FedAvg~\cite{mcmahan_communication-efficient_2017}, Bulyan~\cite{mhamdi_hidden_2018}, and FoolsGold~\cite{fung_limitations_2020}. 
As shown in Figure~\ref{fig:diversifying_hyperparams} in Appendix~\ref{app:diversifying_hyperparams}, model convergence failed entirely in four out of six cases, with \mta remaining below 25\%. 

\subsection{Methodology}
\label{sec:methodology}
To answer the aforementioned research question, we first analytically evaluate the impact of each hyperparameter on backdoor robustness. We then validate our analysis via extensive measurements using state-of-the-art backdoor attacks.

To ensure an unbiased and comprehensive selection of attacks / defenses used in our empirical analysis, we conducted an extensive literature survey of papers about backdoor attacks and defenses in FL.
To this end, we crawled \url{dblp.org} for all papers that match the query "{}federated (backdoor | attack | defense | poisoning)"{}.
This yielded an initial list of 182 papers that were published at A/A* conferences according to the CORE rankings\footnote{\url{https://portal.core.edu.au/conf-ranks/}} between 2019 and 2024. 
From this list, we manually filtered those that propose \textit{new backdoor attacks} in \textit{horizontal} federated learning using a \textit{synchronous} communication model, resulting in 31 papers.
Since proper evaluation requires access to official implementations, we excluded papers without publicly available source code, leaving us with 20 attacks.
From there, we only kept those attacks that target classification tasks, which left us with a total of 15 attacks (cf. Figure~\ref{fig:related_work_chain} in Appendix~\ref{app:RW_chain}).
This means that, for example, we excluded poisoning attacks against federated recommendation systems~\cite{DBLP:conf/icde/RongYZYCH22, DBLP:conf/ijcai/RongHC22, DBLP:conf/wsdm/ZhangYCHNC22} since they optimize for metrics different to the backdoor accuracy, e.g., the exposure rate, and do not necessarily use neural networks~\cite{DBLP:conf/icde/RongYZYCH22}.
Based on this selection, we evaluate the four most recent attacks, namely A3FL~\cite{zhang_a3fl_2023}, Chameleon~\cite{dai_chameleon_2023}, DarkFed~\cite{DBLP:conf/ijcai/LiWNHXZW24}, and FCBA~\cite{DBLP:conf/aaai/LiuZFYXM024}.
Further details on selecting this subset and the individual attacks are in Appendix~\ref{app:RW_chain} and Section~\ref{sec:background_attacks}, respectively.

\begin{figure}
    \centering
    \includegraphics[width=0.85\linewidth]{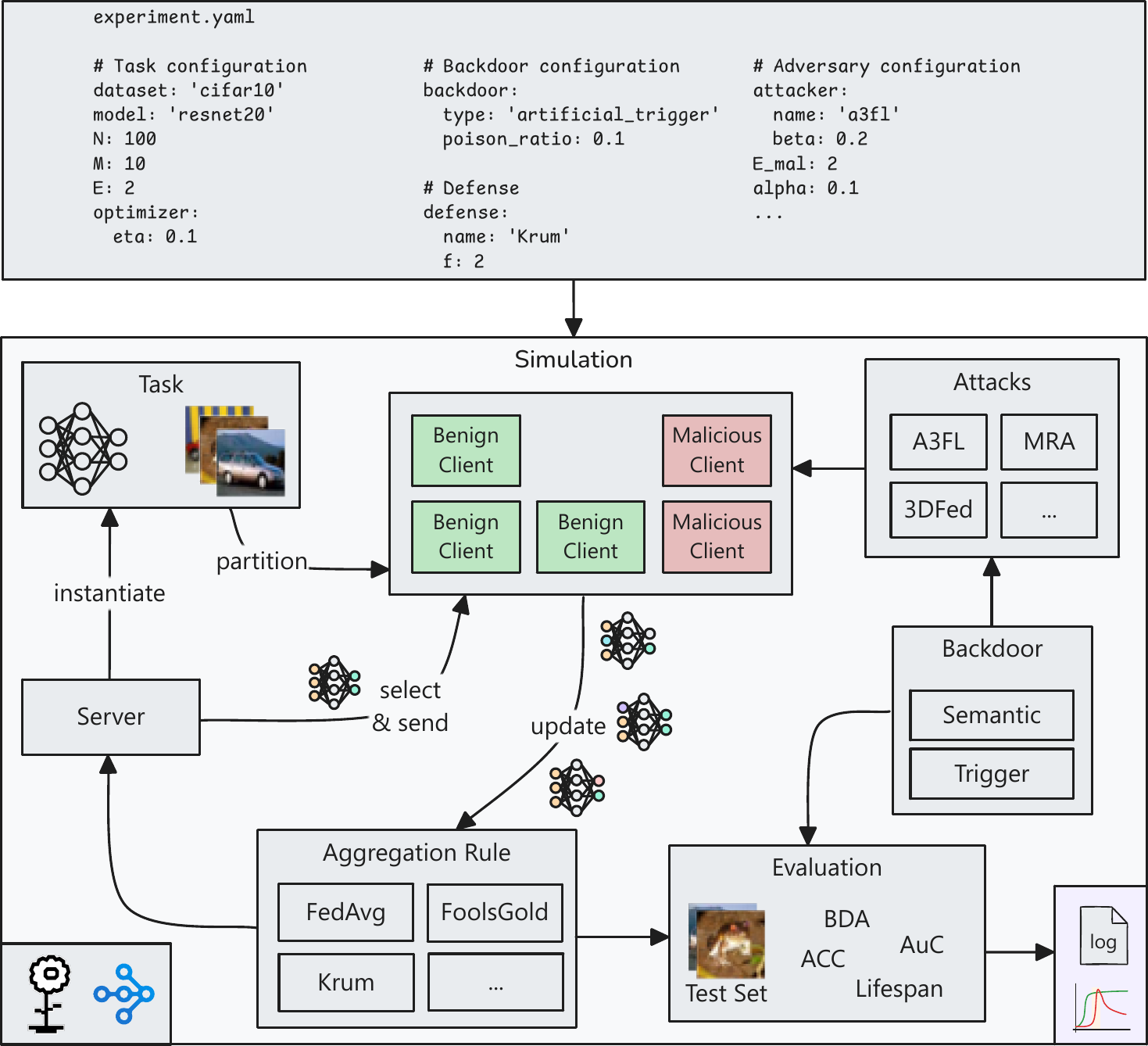}
    \caption{Overview of our empirical evaluation.}
    \label{fig:beaf_schematics}
\end{figure}

To compare attack success under different hyperparameters, we migrated all selected attacks into an evaluation framework built on Flower~\cite{beutel_flower_2022}, chosen for its flexibility, mature FL engine, and multi-GPU support via Ray~\cite{moritz_ray_2018}.
Figure~\ref{fig:beaf_schematics} overviews our evaluation setup.
For reproducibility, we fixed all PRNG seeds and enabled deterministic PyTorch operations.
We evaluated all attacks with PyTorch version 2.2.2 and Python 3.10 on a computing cluster providing compute nodes equipped with two AMD EPYC 9254, three NVIDIA A30 GPUs, and 384 GB RAM, and bigger nodes equipped with two AMD EPYC 9454, eight NVIDIA H100 GPUs, and 1152GB RAM.

\subsection{Comparison to related work} %
Shejwalkar et al.~\cite{shejwalkar_back_2022} examine how system-level parameters---such as the number of clients---influence the success of poisoning attacks. This analysis does not consider backdoor attacks, and does not explore the effect of ML training hyperparameters on robustness.

Bagdasaryan and Shmatikov~\cite{DBLP:conf/uss/BagdasaryanS21}, on the other hand, propose an automated hyperparameter-tuning framework that helps practitioners identify more robust configurations by estimating how much of the dataset must be poisoned for a simple attacker to succeed.
The key distinction---and the reason why the automated centralized-ML tuning framework of~\cite{DBLP:conf/uss/BagdasaryanS21} is not directly applicable to our analysis---is that it automatically identifies configurations that perform well under a non-adaptive adversary that uses the same exact hyperparameters as the benign clients, and cannot react by also tuning its hyperparameters in response to the chosen benign one. %
In contrast, we explicitly consider adaptive malicious clients that tailor their strategy in response to the benign configuration chosen by the benign clients (which is inherent to FL). 
Unfortunately, this kind of adaptive behavior falls outside the scope of what~\cite{DBLP:conf/uss/BagdasaryanS21} can analyze.

%% file: section_04.tex
\section{Impact of Learning Rate}
\label{sec:lr_effect}

In what follows, we analyze the impact of the benign and malicious learning rates on the success of backdoor attacks in FL. 

\subsection{Analysis}
To this end, we model an HFL system with an $\alpha \in [0, 0.5]$ portion of all clients being malicious and the remaining $(1-\alpha)$ fraction being benign, analytically.

Let $\eta_{b}$ denote the learning rate of the benign clients and $\eta_m = \beta \cdot \eta_b$ with $\beta > 0$ refer to the learning rate of the malicious clients.
To simplify the analysis, we model the system as comprising, instead of individual clients, two distinct groups---the benign and malicious clients---each optimizing their respective loss function $F_b$ and $F_m$ with corresponding optima $\theta_b^*$ and $\theta^*_m$.
This abstraction ensures that any set of malicious or benign clients, each with individual per-client loss functions and optima, can be represented using a single loss function encapsulating the collective optimization objective of the group and incorporating each loss function.

We model the evolution of the malicious loss---directly corresponding to the backdoor accuracy---in an HFL setting where the central server employs the FedAvg aggregation rule and clients perform gradient descent, i.e., SGD with full batch size, for a single local epoch using the following recursive equation:
\begin{align*}
F_{m}\left( \theta^t \right)  =& F_{m}\left( \theta^{t-1} - (1-\alpha) \Delta_{b}^t - \alpha \Delta_{m}^t \right) \tageq \label{eq:lr}\\
\text{where } \Delta_{b}^t = & \eta_b \nabla F_{b}\left( \theta^{t-1} \right) \text{ and } \Delta_{m}^t = \beta \eta_b \nabla F_{m}\left( \theta^{t-1} \right)
\end{align*}

Given Equation~\ref{eq:lr}, it is evident that the global model’s update step can be computed as a weighted combination of malicious and benign updates. To illustrate this, consider a simplified scenario where $\nabla F_b\left(\theta^{t-1}\right)$ and $\nabla F_m\left(\theta^{t-1}\right)$ are vectors pointing in opposite directions. In this case, the resulting update vector will align with the direction of the malicious update if $\alpha \beta \eta_b > (1-\alpha) \eta_b$.
In practice, the update vectors from benign and malicious clients are expected to conflict with each other in at least some dimensions~\cite{ozdayi_defending_2021}. 
The Model Replacement attack~\cite{bagdasaryan_how_2020} supports this intuition; here, an attacker can replace the global model with a malicious one in a single update step by artificially increasing their local update (and subtracting an approximation of the benign updates), which mimics increasing the local learning rate.

\subheading{Modeling $F_m$ and $F_b$.}
Inspired by Andriushchenko et\,al.~\cite{DBLP:conf/icml/AndriushchenkoV23}, we use simple diagonal linear networks with a single hidden layer to model the unknown loss functions $F_b$ and $F_m$.
For the main task, we define a straightforward classification problem for two-dimensional inputs $x = (x_1, x_2) \in [-1, 1] \times [-1, 1]$, where the labels are assigned as $y = \begin{cases}0 & x_1 \ge x_2 \\ 1 & x_2 < x_1\end{cases}$.
Additionally, we introduce a backdoor task designed to misclassify all $(x_1, x_2) \in [0, 1] \times [0, x_1]$ as class 1.
This gives us a concrete and realistic instantiation of $F_b$ and $F_m$, enabling us to evaluate better the impact of $\eta_b$ and $\beta$. 
Here, we instantiate all other hyperparameters except for the learning rates to default values, i.e., $E_b = E_m = 2, B_b = B_m = 64, \mu_b = \mu_m = 0.9, \lambda_b = \lambda_m = 0.0005$, which we derive from Table~\ref{tab:hyper_params} by selecting the most frequently used benign values.

\begin{figure}
    \centering
    \subfloat[Learning rate \label{fig:lr_effect_numerical}]{
        \centering
        \includegraphics[width=0.47\linewidth]{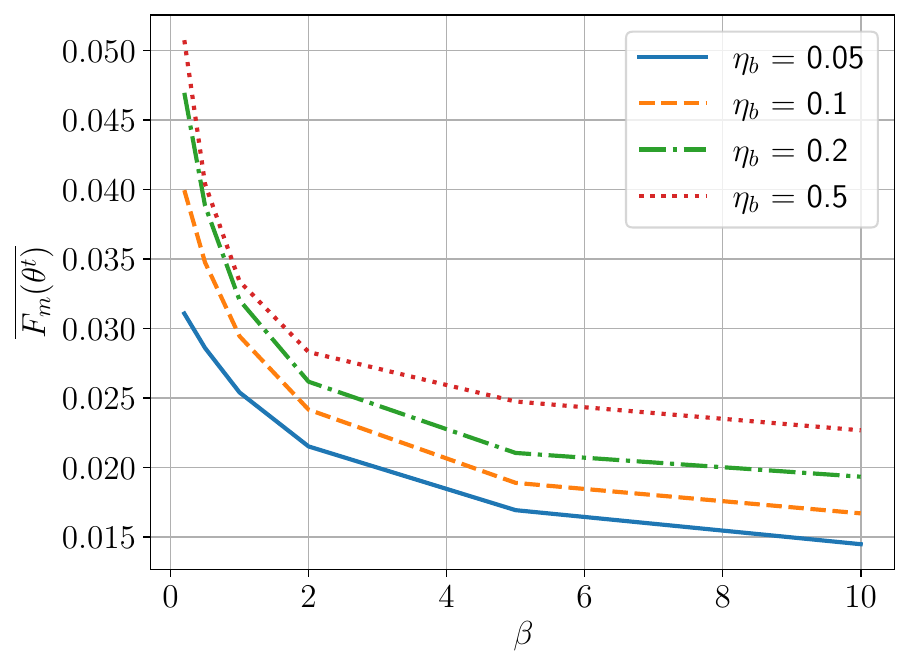}
    }
    \hfill
    \subfloat[Momentum \label{fig:mu_effect_numerical}]{
        \centering 
        \includegraphics[width=0.47\linewidth]{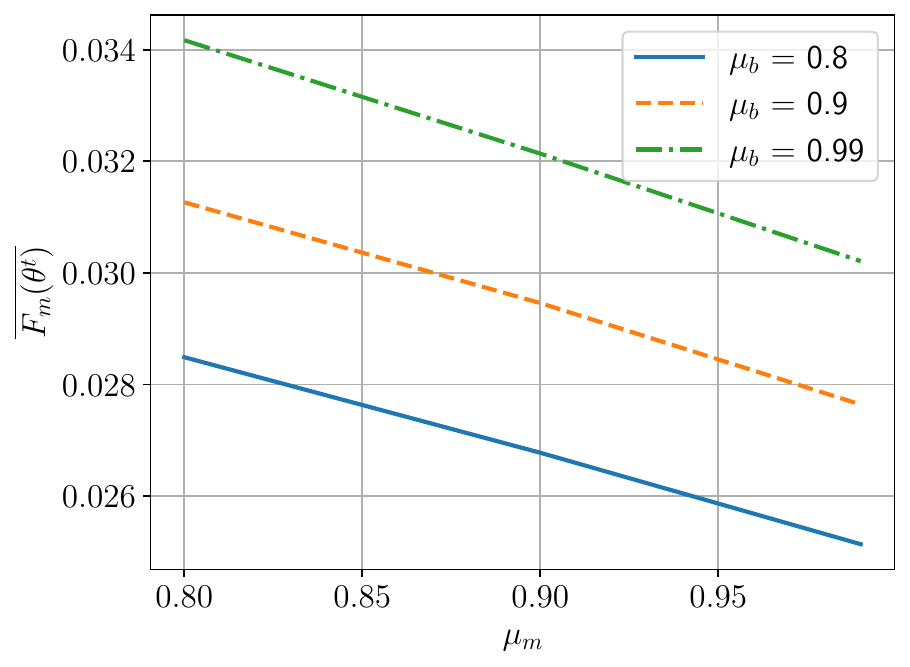}
    }
    \caption{Average malicious loss for various benign and malicious values for the learning rate and momentum over 200 rounds. Lower loss values correspond to higher \bda values.}
\end{figure}

In Figure~\ref{fig:lr_effect_numerical}, we evaluate the resulting loss difference averaged over 200 rounds $\overline{F_m(\theta^t)}- F_m(\theta_m^*) \sim \overline{F_m(\theta^t)}$ .
Our analytical results clearly show that higher benign learning rates $\eta_b$ lead to increased malicious loss, indicating a lower attack success. 
There are two main reasons for this: (i) since the learning rate directly influences the magnitude of model updates, using a higher benign learning rate ensures that updates from benign clients have a greater impact on the global model, and (ii) higher learning rates act as a regularizer~\cite{DBLP:conf/icml/AndriushchenkoV23}, helping the model generalize better and reducing its vulnerability to backdoor injections. 
Further, we observe that the loss exhibits an asymptotic behavior with respect to $\beta$, suggesting that an appropriate choice of $\eta_b$ can lower-bound the malicious loss, irrespective of the malicious learning rate $\eta_m = \beta \cdot \eta_b$.

\subsection{Empirical Validation}

We now evaluate the impact of $\eta_b$ using more complex instantiations of $F_b$ and $F_m$ derived from our selection of state-of-the-art backdoor attacks in Section~\ref{sec:methodology}. 

\subheading{Setup.} 
\label{sec:system_model}
Unless otherwise specified, and in line with prior work (cf. Table~\ref{tab:hyper_params}), we assume a total of $N=100$ clients, with a fixed number of $M=10$ clients chosen in every round.
As shown in Equation~\ref{eq:lr}, the specific values of $M$ and $N$ have no direct impact on the malicious loss. 
However, it is essential to avoid setting $N$ too high, as splitting the data across too many clients can result in overly weak local learners, increasing the risk of overfitting.
We assume that an $\alpha=10\%$ fraction of all clients is malicious.

Following standard FL research practices, we also partition the training dataset in a non-IID manner using label-wise Dirichlet sampling with $0.9$ as hyperparameter~\cite{zhang_neurotoxin_2022, lyu_poisoning_2023, zhang_a3fl_2023, bagdasaryan_how_2020}.
As our running example that we will use throughout this paper, we train a ResNet20~\cite{he_deep_2016} model on CIFAR-10~\cite{krizhevsky_learning_2009} using the widely adopted FedAvg aggregation rule~\cite{mcmahan_communication-efficient_2017}---a typical setup used in FL backdoor research~\cite{fang_vulnerability_2023, zhang_a3fl_2023, DBLP:conf/cikm/Yang0NHW24}.

Since Li et al.~\cite{li_convergence_2020} showed that a decaying learning rate is necessary for FedAvg to converge, we apply a schedule that decays $\eta_b$ by a factor of $\gamma=0.999$ every round.
This choice of $\gamma$ satisfies the criterion in~\cite{li_convergence_2020}, namely: $\frac{1}{2} n_t \le n_{t+E} \le n_t \iff \gamma \in \left[2^{-1/E}, 1\right]$ for all tested values of $E$.
However, we also conduct additional experiments using a constant learning rate to verify that this aspect does not bias our analysis.
The global model trained on these parameters reaches a clean accuracy on the hold-out test set of 86.06\% after 1000 rounds of benign training. Consistent with state-of-the-art backdoor attacks~\cite{zhang_a3fl_2023, dai_chameleon_2023, li_3dfed_2023, bagdasaryan_how_2020, xie_dba_2020}, the attack window begins near convergence (round 1000) and lasts for 200 rounds---the longest among evaluated attacks. 
To assess backdoor persistence, training continues for an additional 4000 rounds.

Our experiments focus on evaluating the impact of the learning rate $\eta$\footnote{We evaluate the impact of hyperparameters like momentum $\mu$, weight decay $\lambda$, local epochs $E$, and batch size $B$ in Section~\ref{sec:other_params}.}, while fixing the other hyperparameters to default values, derived from commonly used values in Table~\ref{tab:hyper_params}, i.e., $\eta_b = 0.1, \mu_b=0.9$, $\lambda_b=0.0005$, $B_b=64$, and $E_b=2$. 
In Section~\ref{sec:pareto}, we examine how the interplay of these hyperparameters influences the effectiveness of backdoor attacks.

\subheading{Evaluation results.}
To confirm our analytical analysis using more complex instantiations for $F_m$ and $F_b$, we now empirically measure the impact of the learning rate on the success of all previously selected attacks, namely the A3FL~\cite{zhang_a3fl_2023} attack, the Chameleon~\cite{dai_chameleon_2023} attack, the DarkFed~\cite{DBLP:conf/ijcai/LiWNHXZW24}, and the FCBA~\cite{DBLP:conf/aaai/LiuZFYXM024} attack. 
Here, we left the hyperparameters that we do not ablate at their default values for all attacks, either extracted from the corresponding paper or the respective source code.
Note that for the FCBA~\cite{DBLP:conf/aaai/LiuZFYXM024} and DarkFed~\cite{DBLP:conf/ijcai/LiWNHXZW24} attack, we had to apply slight modifications to make them compatible with our system and threat model.
We discuss these changes in Appendix~\ref{app:changed_attacks} and show that they do not influence the attack's success.

\begin{figure}[tb]
    \centering
    \subfloat[A3FL~\cite{zhang_a3fl_2023} \label{fig:lr_effect_a3fl_bda_during}]{
        \centering
        \includegraphics[width=0.47\linewidth]{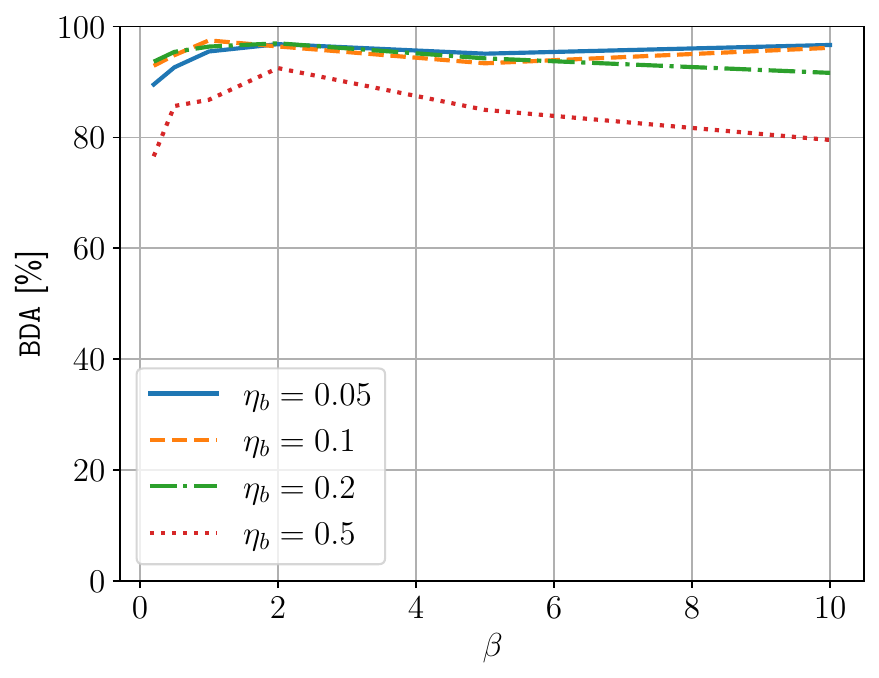}
    }
    \hfill
    \subfloat[Chameleon~\cite{dai_chameleon_2023} \label{fig:lr_effect_chameleon_bda_during}]{
        \centering
        \includegraphics[width=0.47\linewidth]{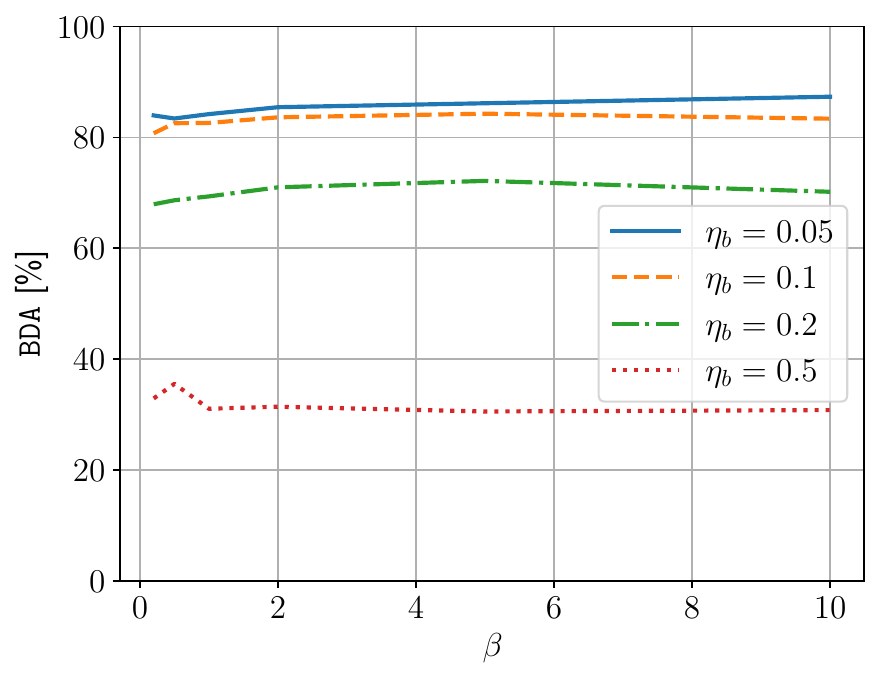}
    }

    \vspace{0.6em}
    
    \subfloat[DarkFed~\cite{DBLP:conf/ijcai/LiWNHXZW24} \label{fig:lr_effect_darkfed_bda_during}]{
        \centering
        \includegraphics[width=0.47\linewidth]{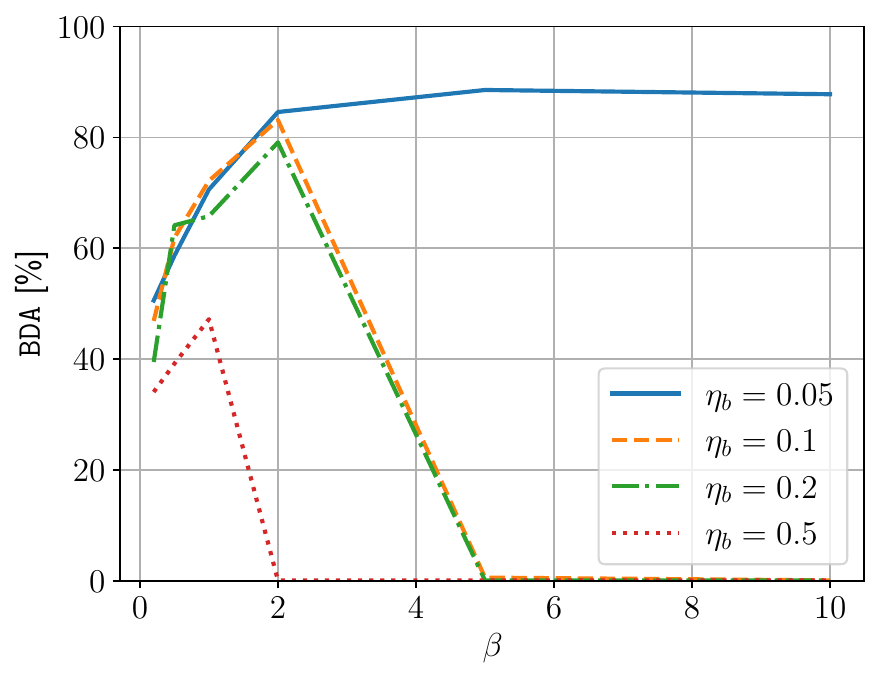}
    }
    \hfill
    \subfloat[FCBA~\cite{DBLP:conf/aaai/LiuZFYXM024} \label{fig:lr_effect_optim_bda_during}]{
        \centering
        \includegraphics[width=0.47\linewidth]{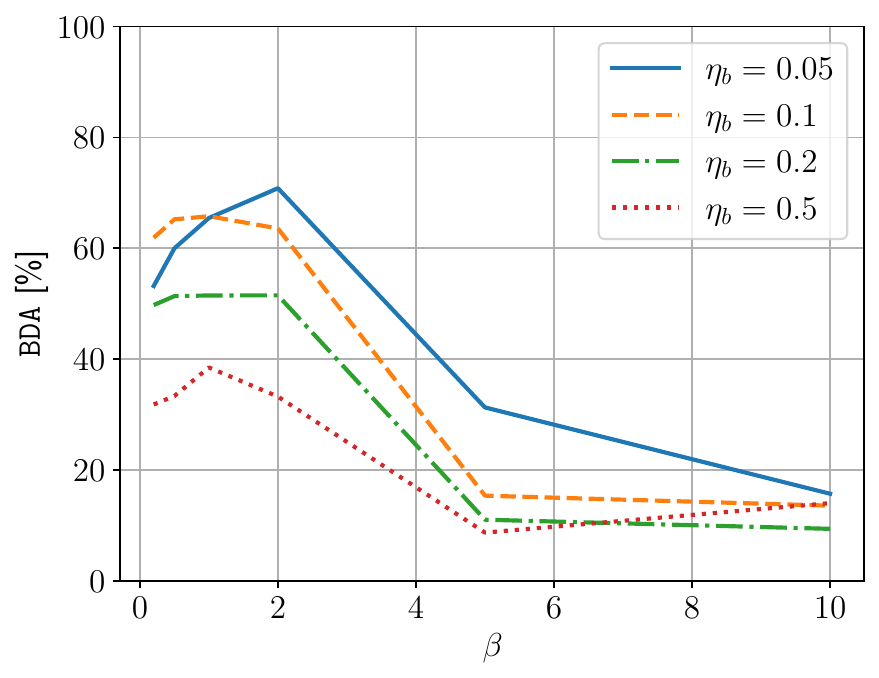}
    }
    \caption{Impact of $\eta_b$ and $\beta$ on the \bda of SoTA attacks.}
    \label{fig:lr_effect_bda_during}
\end{figure}

We evaluate the \bda in Figure~\ref{fig:lr_effect_bda_during}.
Note that the \bda is inversely proportional to the malicious loss in Figure~\ref{fig:lr_effect_numerical}.
For all four attacks, we see a strong effect of $\eta_b$ on the attack's success, effectively limiting the maximum achievable \bda values.
We note that increasing $\eta_b$ from $0.05$ to $0.5$ reduces the maximum \bda by a large amount of $4.32$ percentage points, $51.79$ percentage points, $41.34$ percentage points, and $32.33$ percentage points for the A3FL, Chameleon, DarkFed, and FCBA attack. 
Our results for (i) the IBA attack~\cite{nguyen_iba_2023} (Appendix~\ref{app:IBA}) and (ii) a straightforward label-flipping poisoning scheme inspired by~\cite{gu_badnets_2017}, which we dub ``baseline'' in the sequel (Appendix~\ref{app:Baseline}), are also consistent with these findings. 

Interestingly, the malicious learning rate $\beta \cdot \eta_b$ for the Chameleon attack seems to have considerably less impact than for the other attacks.
We attribute this discrepancy to the fact that the Chameleon attack uses supervised contrastive loss---instead of the standard cross-entropy loss used in other attacks~\cite{dai_chameleon_2023}. 
In the case of DarkFed, we observe a sudden decrease in \bda when $\beta$ is chosen too high.
A closer investigation of the corresponding experimental results indicates that high malicious learning rates, in combination with the training on a shadow dataset, cause too much divergence, leading to exploding gradients that disrupt further training.

In Appendix~\ref{app:lr_effect}, we also include the results measured in terms of the \bdaafter and the \lifespan in Figures~\ref{fig:lr_effect_bda_after}, and \ref{fig:lr_effect_lifespan}.
When evaluating the \bdaafter for the A3FL attack, we see an even more pronounced effect with a decrease of $27.61$ percentage points
In the case of Chameleon, DarkFed, and FCBA attacks, \bdaafter did not exceed $18\%$ for any configuration. 
In terms of \lifespan, we observed even higher reduction by $87.26\%$, $100\%$, $97.26\%$, and $100\%$ for the A3FL, Chameleon, DarkFed, and FCBA attacks, respectively.
These results confirm that increasing $\eta_b$ significantly simplifies backdoor unlearning.

All in all, our empirical findings confirm that our previous analysis remains valid: increasing the \textit{benign} learning rate reduces attack success and caps the maximum achievable success---regardless of the malicious learning rate.
The choice of $\eta_b$, though, is constrained by the primary task itself since learning rate tuning is essential for achieving optimal main task accuracy.
When examining the \mta, shown in Figure~\ref{fig:lr_effect_mta_during} in Appendix~\ref{app:lr_effect}, we find that a choice of $\eta_b=0.2$ achieves comparable \mta to $\eta_b = 0.05$, while significantly improving the backdoor robustness.
In the case of DarkFed, the \mta drops to $10\%$ (for a dataset comprising ten classes, this is equivalent to random guessing) when $\beta$ is too high and $\eta_b \in {0.1, 0.2, 0.5}$ due to exploding gradients caused by an overly aggressive malicious learning rate.
However, in these settings, the \bda also drops drastically. We observe similar behavior for the FCBA attack as well, which we attribute to its use of model update scaling that, in some circumstances, leads to complete model divergence.
Therefore, we argue that no rational adversary would choose such high learning rates, as this would lead to a completely ineffective attack.

\begin{figure}[tb]
    \centering
    \subfloat[MobileNetV2]{
        \centering
        \includegraphics[width=0.47\linewidth]{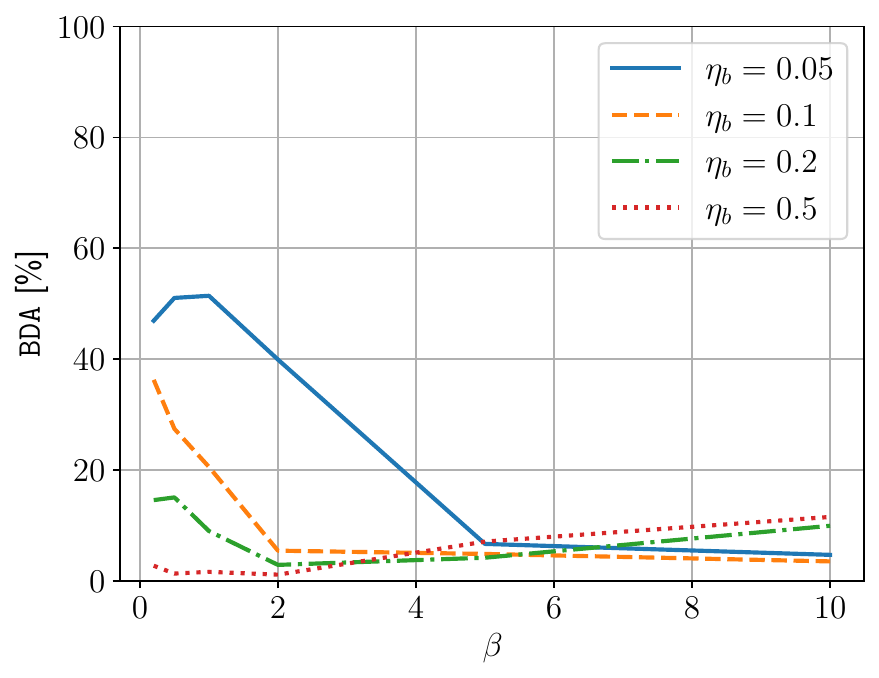}
    }
    \hfill
    \subfloat[Tiny-ImageNet]{
        \centering
        \includegraphics[width=0.47\linewidth]{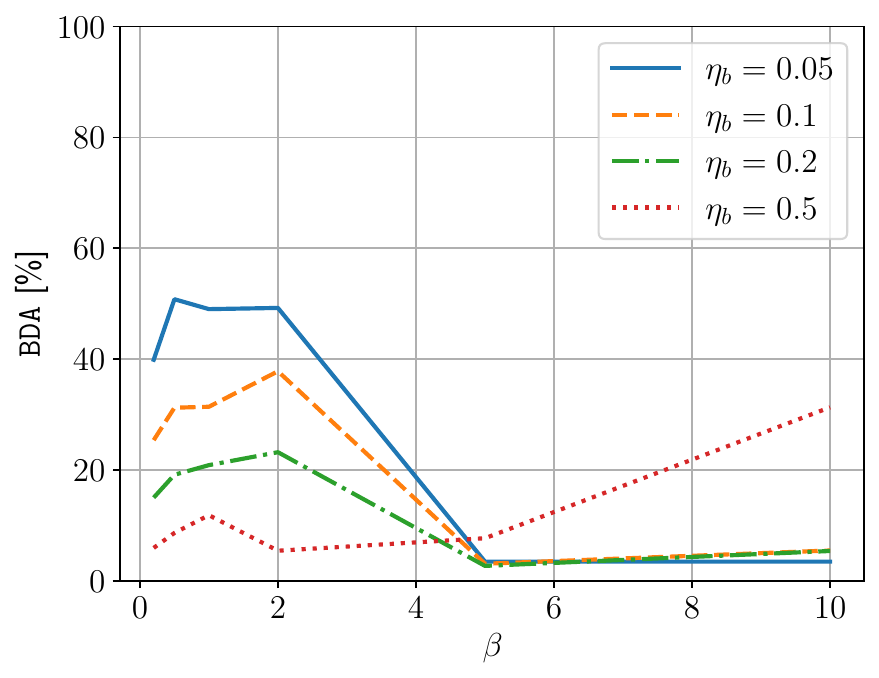}
    }

    \vspace{0.6em}
    
    \subfloat[VGG17]{
        \centering
        \includegraphics[width=0.47\linewidth]{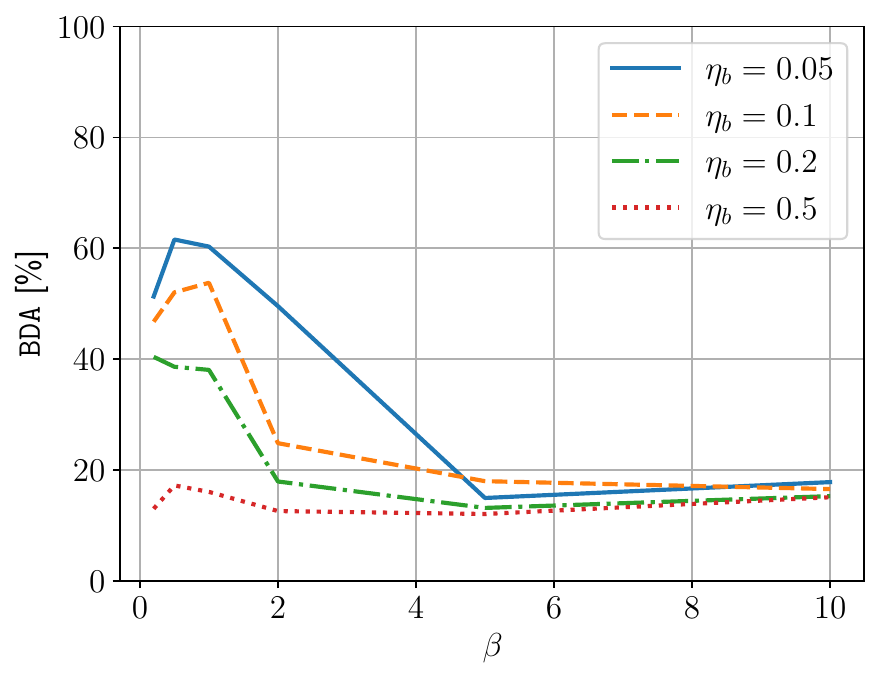}
    }
    \hfill
    \subfloat[Uniform Schedule \label{fig:diverse_architectures_uniform_schedule}]{
        \centering
        \includegraphics[width=0.47\linewidth]{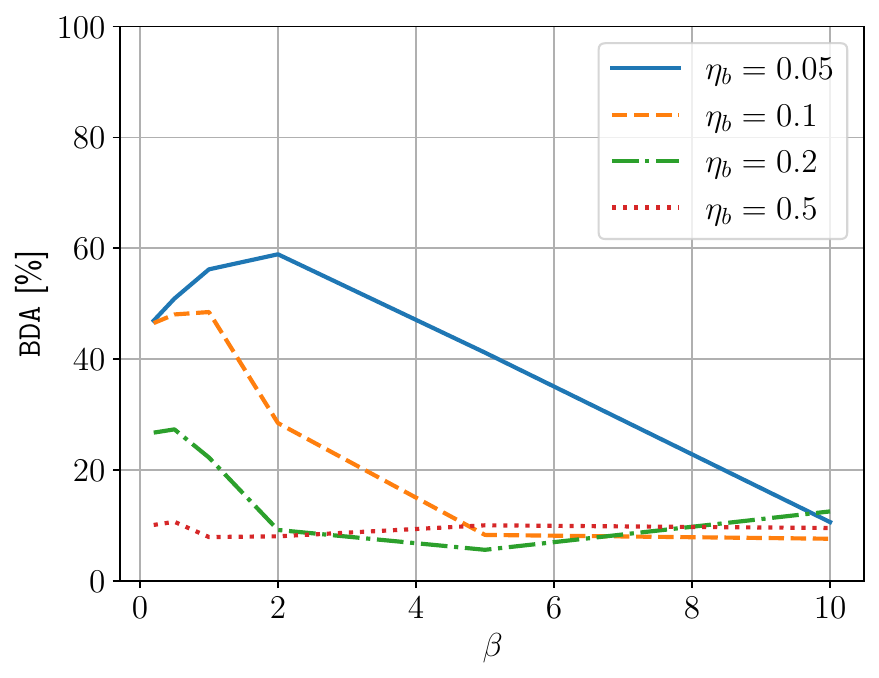}
    }
    
    \caption{Impact of $\eta_b$ and $\beta$ on the \bda of the FCBA attacks when using MobileNetV2 or VGG17 as the model, when using the dataset Tiny-ImageNet, or when using a uniform learning rate schedule.}
    \label{fig:diverse_architectures}
\end{figure}

To verify that our findings are not specific to a particular model or dataset, we repeated our analysis using (i) common architectures used in HFL, such as MobileNetV2~\cite{DBLP:conf/cvpr/SandlerHZZC18} and VGG17~\cite{DBLP:journals/corr/SimonyanZ14a} as adapted in~\cite{nguyen_iba_2023}, and (ii) the Tiny-ImageNet dataset~\cite{tinyimagenet}, also popularly used in HFL (see Figure~\ref{fig:diverse_architectures} for our results).
While these changes affect absolute \bda values (e.g., MobileNetV2 yields higher \bda for low $\beta$ than ResNet20), the overall trends hold for all considered architectures. 
In particular, for $\beta \leq 5$, higher benign learning rates $\eta_b$ consistently reduce \bda across the considered architectures. 
Furthermore, it is evident that the more complex VGG17 architecture---with approximately $74$ times the number of parameters than ResNet20---has greater capacity to learn both the primary and backdoor tasks simultaneously. This results in a significantly higher maximum \lifespan for $\eta_b = 0.05$. However, this advantage diminishes further as $\eta_b$ increases.
Due to the lack of space, we include our full results in Appendix~\ref{app:lr_imagenet} and Appendix~\ref{app:lr_mobilenet}. 

We also experimented with a uniform learning rate schedule (cf. Figure~\ref{fig:diverse_architectures_uniform_schedule} and Appendix~\ref{app:lr_uniform}), which confirmed the same trends with only minor differences in \mta degradation, consistent with~\cite{li_convergence_2020}.

%% file: section_05.tex
\section{Impact of Other Hyperparameters}
\label{sec:other_params}

Building on the approach from Section~\ref{sec:lr_effect}, we find that momentum ($\mu$) exhibits a similar, though less pronounced, influence compared to the learning rate. Additionally, our results reveal a strong interdependence between the number of local epochs ($E$) and the batch size ($B$), with both playing a critical role in determining the success of the attack. Finally, we observe that the weight decay parameter ($\lambda$) has a particularly notable effect on \bda and \lifespan.

\subsection{Analysis}

\subheading{Impact of Momentum.}
We now incorporate in Equation~\ref{eq:lr} the notion of multiple local steps, whose gradients are accumulated in a decaying average depending on the benign and malicious momentum factors $\mu_b$ and $\mu_m$, respectively:

\begin{align*}
    F_{m}\left( \theta^t \right)  =& F_{m}\left( \theta^{t-1} - (1-\alpha) \Delta_{b}^t - \alpha \Delta_{m}^t \right) \tageq \label{eq:momentum}\\
    \Delta_b^t = & \theta_b^{t-1,S_b} - \theta^{t-1,0}, \Delta_m^t = \theta_m^{t-1,S_m} - \theta^{t-1,0} \\
    \theta_b^{t,s}  = & \theta_b^{t,s-1} - \eta_b v_b^{t, s-1}, \theta_m^{t,s} = \theta_m^{t,s-1} - \eta_b \beta v_m^{t, s-1} \\
    \text{where } v_b^{t, s} = & \nabla F_b\left( \theta^{t,s} \right) + \begin{cases} 0 & s = 0 \\ \mu_b v_b^{t, s-1} & s > 0\end{cases}\\
    v_m^{t, s} = & \nabla F_m\left( \theta^{t,s} \right) + \begin{cases} 0 & s = 0 \\ \mu_m v_m^{t, s-1} & s > 0\end{cases}\\
    \theta_m^{t,0} = & \theta_b^{t,0} = \theta^{t}
\end{align*}

Here, $S_b$ and $S_m$ refer to a generic notion for the number of SGD steps that benign and malicious clients perform before uploading their model updates, respectively.
Given Equation~\ref{eq:momentum}, we observe that the magnitude of a single local update step depends on $\mu$ and the alignment between the decaying average of previous updates and the current gradient.
The stronger is the alignment, the larger is the update step.
In HFL, where each client has a relatively small local dataset~\cite{mcmahan_communication-efficient_2017}, update steps tend to be similar.
As a result, we expect increasing $\mu$ to have a similar effect to increasing $\eta$.
Thus, we expect a higher backdoor accuracy and an extended lifespan for lower $\mu_b$ and higher $\mu_m$.
However, if $\mu$ is too large, we anticipate a negative impact on the main task accuracy.
In the extreme case of $\mu=1$, there is no decay of past gradients, causing update steps to grow continuously. 
Eventually, their magnitude could reach levels comparable to an excessively high learning rate, leading to model divergence.
The same applies if a (malicious) client had a significantly larger dataset than other clients: 
Performing more local update steps amplifies the momentum’s impact and may lead to model divergence. 

Using the identical instantiations for $F_m$ and $F_b$ as in Section~\ref{sec:lr_effect}, i.e., by modeling $F_m$ and $F_b$ utilizing the loss of a small DLN on a simple main and backdoor task, we evaluate the resulting average loss distance in Figure~\ref{fig:mu_effect_numerical}.
We observe that increasing $\mu_b$ improves robustness against backdoor attacks.
However, unlike other parameters, we do not observe an asymptotic trend for $\mu_m$.
When comparing Figures~\ref{fig:mu_effect_numerical} and~\ref{fig:lr_effect_numerical}, the impact of momentum on the loss appears to be more localized and less pronounced compared to the effect of the learning rate. 
The momentum's effect spans only a narrow region of the learning rate's effect, suggesting that while both parameters shape the loss landscape in a similar manner, momentum does so within a more confined range and with a subtler impact.
We attribute this to the fact that reasonable values for momentum are limited to $[0, 1)$, whereas the learning rate can vary over a much broader scale, leading to more pronounced effects.

\subheading{Impact of Batch Size \& Local Epochs.}
Since both the batch size $B$ and the number of local epochs $E$ determine the number of SGD steps taken by a client before submitting its model update, we analyze these parameters jointly.
More specifically, we replace $S_b$ and $S_m$ and the definitions of $v_b^{t,s}$ and $v_m^{t,s}$ from Equation~\ref{eq:momentum} to account for the number of local epochs $E_b$ and $E_m$ and the used batch sizes $B_b$ and $B_m$, yielding:
\begin{align*}
    F_{m}\left( \theta^t \right)  =& F_{m}\left( \theta^{t-1} - (1-\alpha) \Delta_{b}^t - \alpha \Delta_{m}^t \right) \tageq \label{eq:local_epochs_batch_size}\\
    \Delta_b^t = & \theta_b^{t-1,E_b \cdot \left\lfloor \frac{|\mathcal D_b|}{B_b} \right\rfloor} - \theta^{t-1,0}\\
    \Delta_m^t =& \theta_m^{t-1,E_m \cdot \left\lfloor \frac{|\mathcal D_m|}{B_m} \right\rfloor} - \theta^{t-1,0} \\
    \theta_b^{t,s}  = & \theta_b^{t,s-1} - \eta_b v_b^{t, s-1}\\
    \theta_m^{t,s} = & \theta_m^{t,s-1} - \eta_b \beta v_m^{t, s-1} \\
    \text{where } v_b^{t, s} = & \nabla F_b\left( \theta^{t,s} , \mathcal{B} \subset_\$ \mathcal{D}_b \right) + \begin{cases} 0 & s = 0 \\ \mu_b v_b^{t, s-1} & s > 0\end{cases}\\
    v_m^{t, s} = & \nabla F_m\left( \theta^{t,s}, \mathcal{B} \subset_\$ \mathcal{D}_m \right) + \begin{cases} 0 & s = 0 \\ \mu_m v_m^{t, s-1} & s > 0\end{cases}\\
    \theta_m^{t,0} = & \theta_b^{t,0} = \theta^{t}
\end{align*}

\noindent Here, $\mathcal{D}_m$ and $\mathcal{D}_b$ represent the local datasets of malicious and benign clients, respectively, while $\mathcal{B} \subset_\$ \mathcal{D}$ denotes a randomly sampled batch from dataset $\mathcal{D}$ with $|\mathcal{B}| = B$.

Given Equation~\ref{eq:local_epochs_batch_size}, we see that performing more local update steps---either by increasing $E$ (the number of local epochs) or decreasing $B$ (the batch size)---amplifies the impact of the respective clients on the global model due to a more substantial overall change in the model. 
Additionally, we expect implicit regularization effects to arise from the noise introduced by the increased stochasticity of SGD when using smaller batch sizes~\cite{DBLP:conf/nips/HeLT19} or, conversely, from more local training epochs.
Given the interaction between benign and malicious clients, where benign clients in round $t+1$ can potentially mitigate the backdoor effect introduced in round $t$, the robustness of the injected backdoor behavior is a critical factor influencing its persistence.
Therefore, we believe that a backdoored model trained with higher regularization (i.e., using a smaller batch size) is more difficult to unlearn.

\begin{figure}
    \centering
    \subfloat[Number of local epochs $E$ \label{fig:E_effect_numerical}]{
        \centering
        \includegraphics[width=0.47\linewidth]{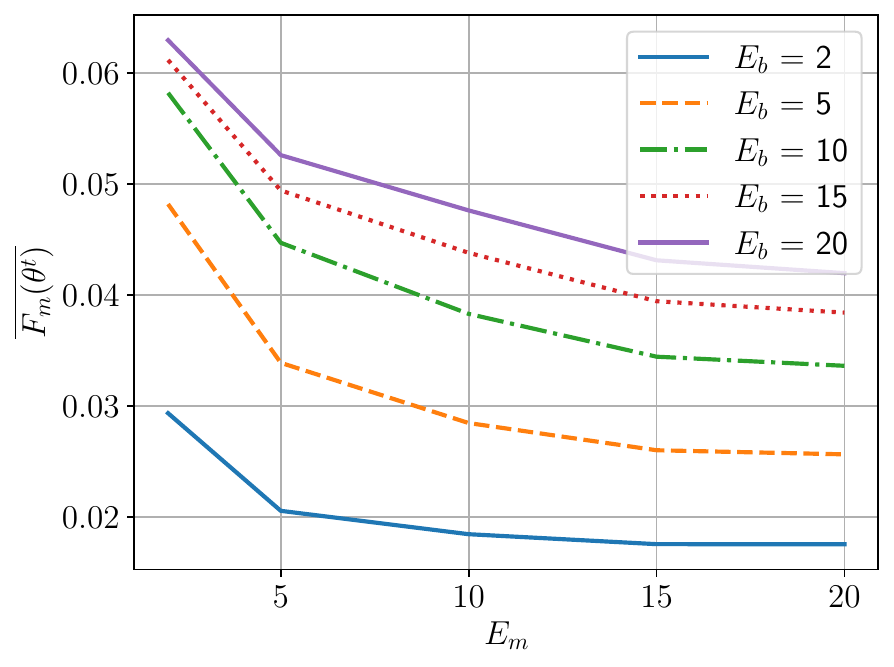}
    }
    \hfill
    \subfloat[Batch Size $B$ \label{fig:B_effect_numerical}]{
        \centering
        \includegraphics[width=0.47\linewidth]{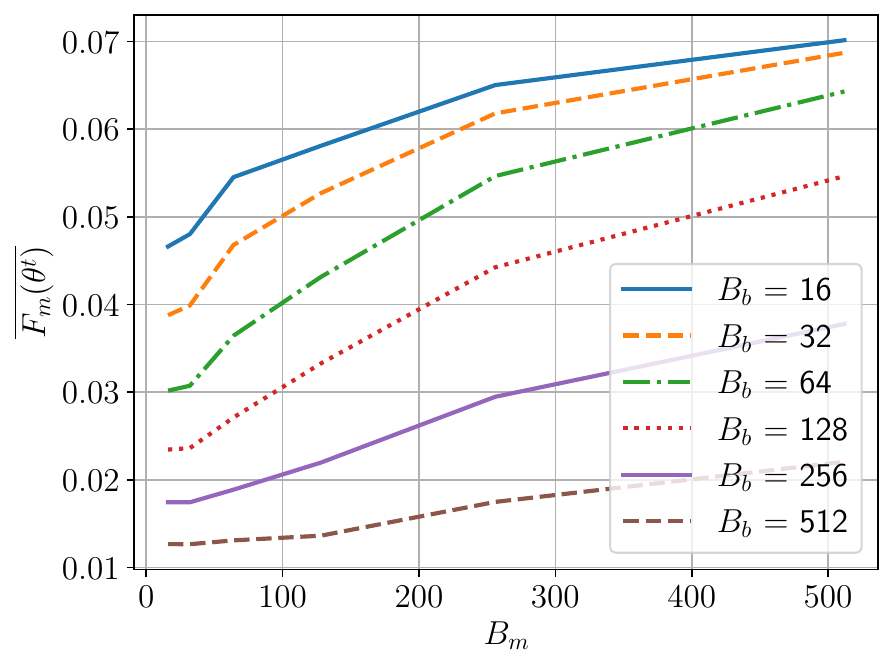}
    }
    \caption{Average malicious loss for various values of $E_b$ and $E_m$, and $B_b$ and $B_m$, over 200 rounds, respectively. Lower loss values correspond to higher \bda values.}
    \label{fig:B_E_effect_numerical}
\end{figure}

Our results in Figure~\ref{fig:B_E_effect_numerical} (where we model $F_m$ and $F_b$ utilizing the loss of a small DLN on a simple main and backdoor task) clearly show that an increase in benign update steps---either through a higher $E_b$ or a smaller $B_b$---leads to a higher malicious loss and, thereby, reduces attack success.
Moreover, we observe an asymptotic trend with respect to the malicious hyperparameters $E_m$ and $B_m$.
While increasing $E_m$ or decreasing $B_m$ can improve attack success, benign clients can ultimately upper-bound this effect, limiting the attack's effectiveness.

\subheading{Impact of Weight Decay.}
Last but not least, we extend Equation~\ref{eq:local_epochs_batch_size} as follows to analyze the impact of the weight decay:
\begin{align*}
    F_{m}\left( \theta^t \right)  =& F_{m}\left( \theta^{t-1} - (1-\alpha) \Delta_{b}^t - \alpha \Delta_{m}^t \right) \tageq \label{eq:weight_decay}\\
    \Delta_b^t = & \theta_b^{t-1,E_b \cdot \left\lfloor \frac{|\mathcal D_b|}{B_b} \right\rfloor} - \theta^{t-1,0}\\
    \Delta_m^t =& \theta_m^{t-1,E_m \cdot \left\lfloor \frac{|\mathcal D_m|}{B_m} \right\rfloor} - \theta^{t-1,0} \\
    \theta_b^{t,s}  = & \theta_b^{t,s-1} - \eta_b v_b^{t, s-1}\\
    \theta_m^{t,s} = & \theta_m^{t,s-1} - \eta_b \beta v_m^{t, s-1} \\
    \text{where } v_b^{t, s} =& \nabla F_b\left( \theta^{t,s} , \mathcal{B} \subset_\$ \mathcal{D}_b \right) + \lambda_b \theta^{t,s} \\
    &+ \begin{cases} 0 & s = 0 \\ \mu_b v_b^{t, s-1} & s > 0\end{cases}\\
    v_m^{t, s} =& \nabla F_m\left( \theta^{t,s}, \mathcal{B} \subset_\$ \mathcal{D}_m \right) + \lambda_m \theta^{t,s} \\
    &+ \begin{cases} 0 & s = 0 \\ \mu_m v_m^{t, s-1} & s > 0\end{cases}\\
    \theta_m^{t,0} = & \theta_b^{t,0} = \theta^{t}
\end{align*}
Here, we see that each model update step is modified to take a step in a direction that reduces the absolute size of each model parameter, with larger values of $\lambda$ enforcing stronger regularization. 
This constraint suppresses large parameter deviations, thereby impeding backdoor injection. In the next section, we confirm this analysis by means of measurements. 

\subsection{Empirical Validation}
We now empirically confirm our aforementioned analysis using the same backdoor attacks we considered and selected in Section~\ref{sec:methodology} as instantiations for $F_b$ and $F_m$.
To ease comparison, we employ the same system and threat model of Section~\ref{sec:system_model} and use a round-wise decay with $\gamma=0.999$ when setting the benign learning rate $\eta_b$.

\begin{figure}
    \centering
    \subfloat[A3FL~\cite{zhang_a3fl_2023} \label{fig:mu_effect_a3fl_bda_during}]{
        \centering
        \includegraphics[width=0.47\linewidth]{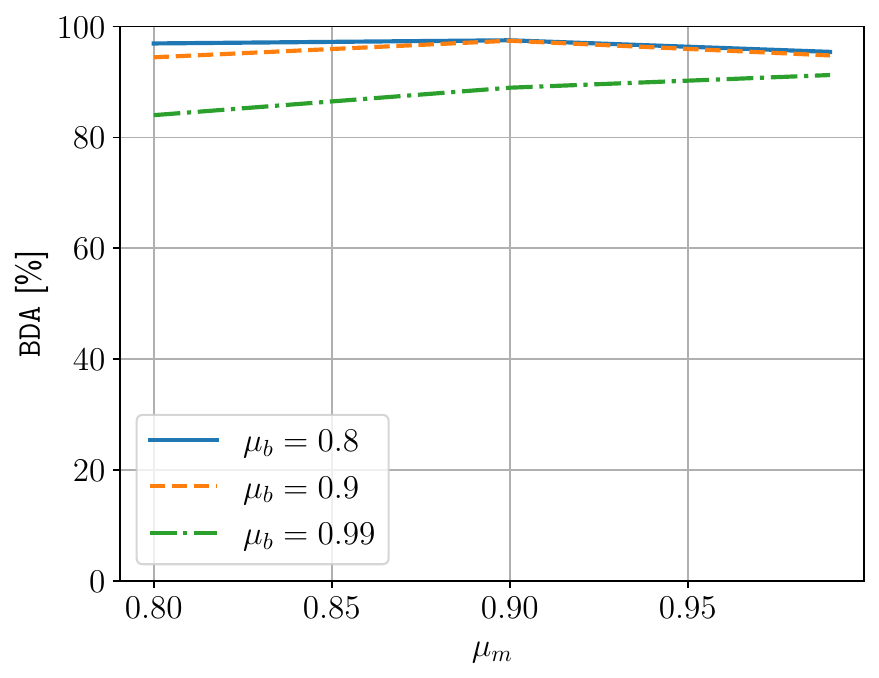}
    }
    \hfill
    \subfloat[Chameleon~\cite{dai_chameleon_2023} \label{fig:mu_effect_chameleon_bda_during}]{
        \centering
        \includegraphics[width=0.47\linewidth]{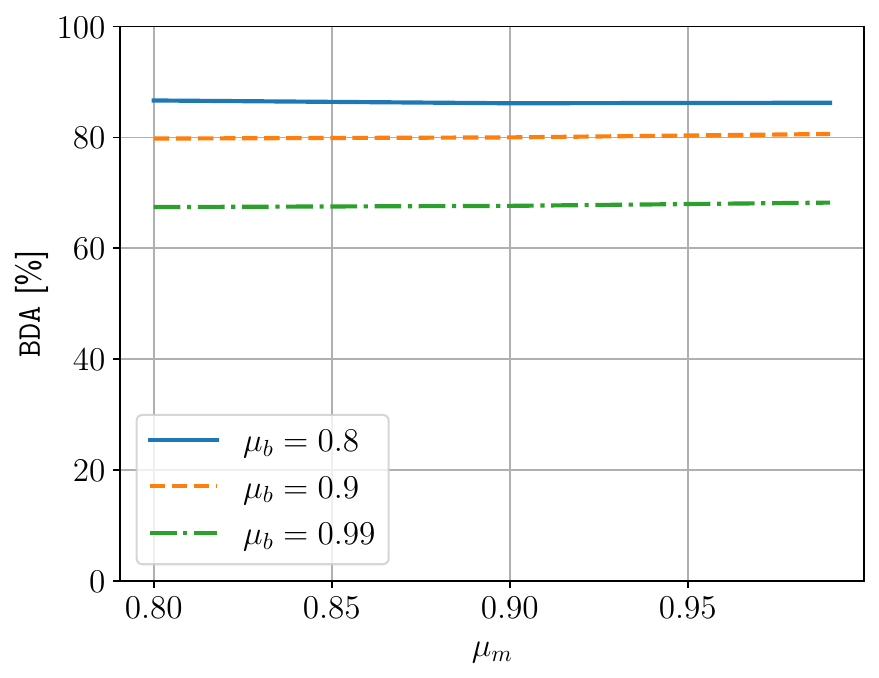}
    }
    
    \vspace{0.6em}
    
    \subfloat[DarkFed~\cite{DBLP:conf/ijcai/LiWNHXZW24} \label{fig:mu_effect_darkfed_bda_during}]{
        \centering
        \includegraphics[width=0.47\linewidth]{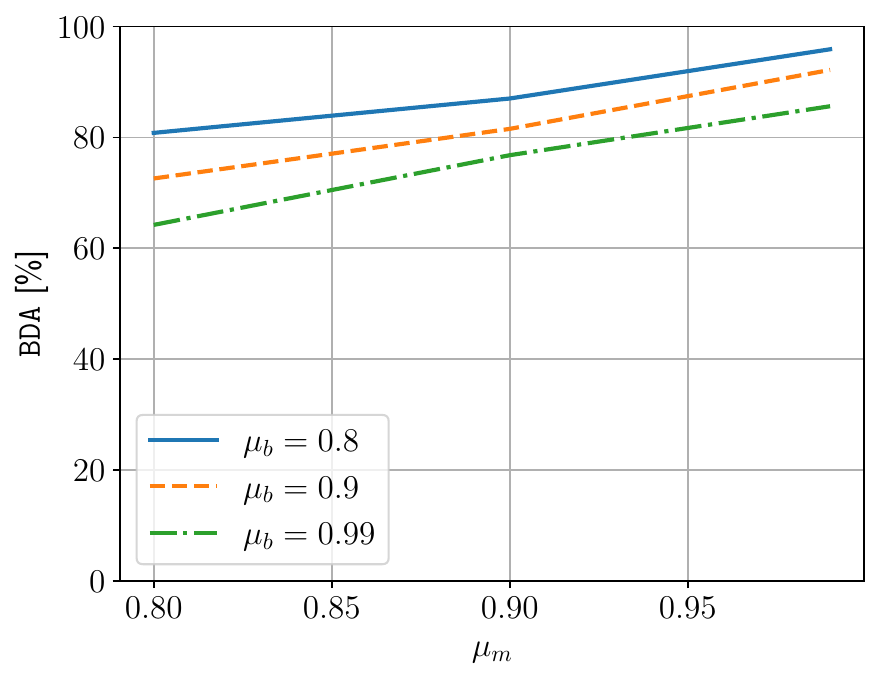}
    }
    \hfill
    \subfloat[FCBA~\cite{DBLP:conf/aaai/LiuZFYXM024} \label{fig:mu_effect_fcba_bda_during}]{
        \centering
        \includegraphics[width=0.47\linewidth]{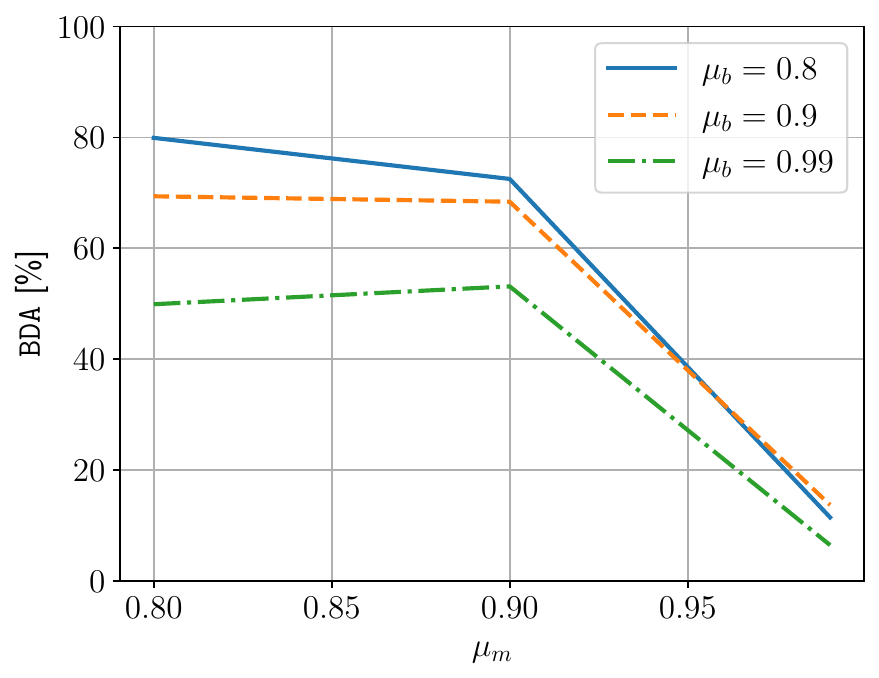}
    }

    \caption{Impact of $\mu_b$ and $\mu_m$ on the \bda of SoTA attacks.}
    \label{fig:mu_effect_bda_during}
\end{figure}

\subheading{Impact of Momentum.} 
\label{sec:mu_effect}
The impact of the benign and malicious momentum factors on the \bda is shown in Figure~\ref{fig:mu_effect_bda_during}. 
Overall, our empirical results confirm our analytical results:
Increasing $\mu_b$ from $0.8$ to $0.99$ reduces the maximum achievable \bda by $6.27$ percentage points, $18.42$ percentage points, $10.28$ percentage points and $26.79$ percentage points for the A3FL, Chameleon, DarkFed, and FCBA attacks, respectively.
As shown in Figure~\ref{fig:mu_effect_bda_after} in Appendix~\ref{app:mu_effect}, we observe the same trend for \bdaafter when increasing $\mu_b$ from $0.8$ to $0.99$, however, with smaller absolute decreases of $0.27$ percentage points, $2.29$ percentage points, $1.57$ percentage points, and $7.88$ percentage points for the A3FL, Chameleon, DarkFed, and FCBA attacks, respectively.
In contrast, when considering lifespan, the benign momentum has a substantial effect. For instance, increasing $\mu_b$ from 0.8 to 0.99 reduces the maximum achievable \lifespan by $46.85\%$, $72.73\%$, $62.67\%$, and $84.62\%$ for the A3FL, Chameleon, DarkFed, and FCBA attacks, respectively. 

The momentum, however, considerably impacts the \mta, as shown in Figure~\ref{fig:mu_effect_mta_during} in Appendix~\ref{app:mu_effect}. Namely, we see significant reductions of up to $24.8$ percentage points when increasing $\mu_b$ from $0.8$ to $0.99$, indicating that there exists a substantial trade-off between main task and backdoor accuracy for $\mu_b$. 
For additional comparison, we also include results for our ``baseline'' and the IBA~\cite{nguyen_iba_2023} attack in Appendix~\ref{app:Baseline} and \ref{app:IBA}, respectively.

\begin{figure}
    \centering
    \subfloat[A3FL~\cite{zhang_a3fl_2023} \label{fig:E_effect_a3fl_bda_during}]{
        \centering
        \includegraphics[width=0.47\linewidth]{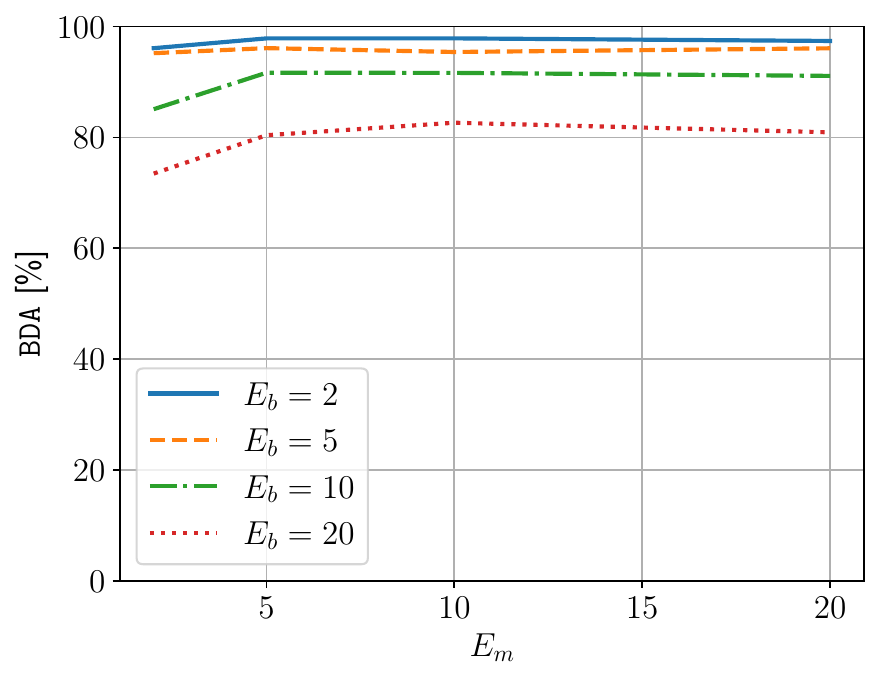}
    }
    \hfill
    \subfloat[Chameleon~\cite{dai_chameleon_2023} \label{fig:E_effect_chameleon_bda_during}]{
        \centering
        \includegraphics[width=0.47\linewidth]{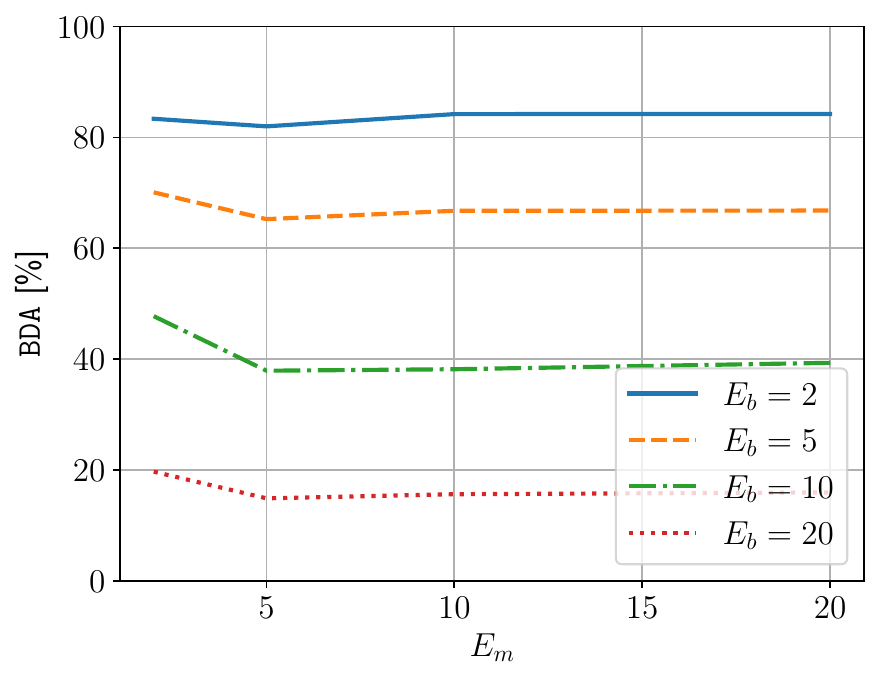}
    }

    \vspace{0.6em}
    
    \subfloat[DarkFed~\cite{DBLP:conf/ijcai/LiWNHXZW24} \label{fig:E_effect_darkfed_bda_during}]{
        \centering
        \includegraphics[width=0.47\linewidth]{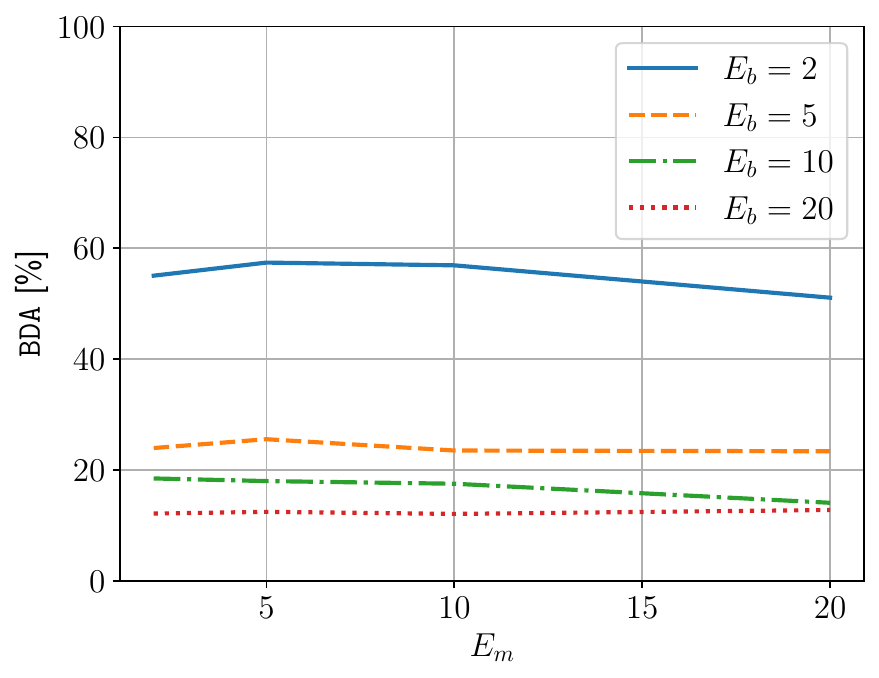}
    }
    \hfill
    \subfloat[FCBA~\cite{DBLP:conf/aaai/LiuZFYXM024} \label{fig:E_effect_fcba_bda_during}]{
        \centering
        \includegraphics[width=0.47\linewidth]{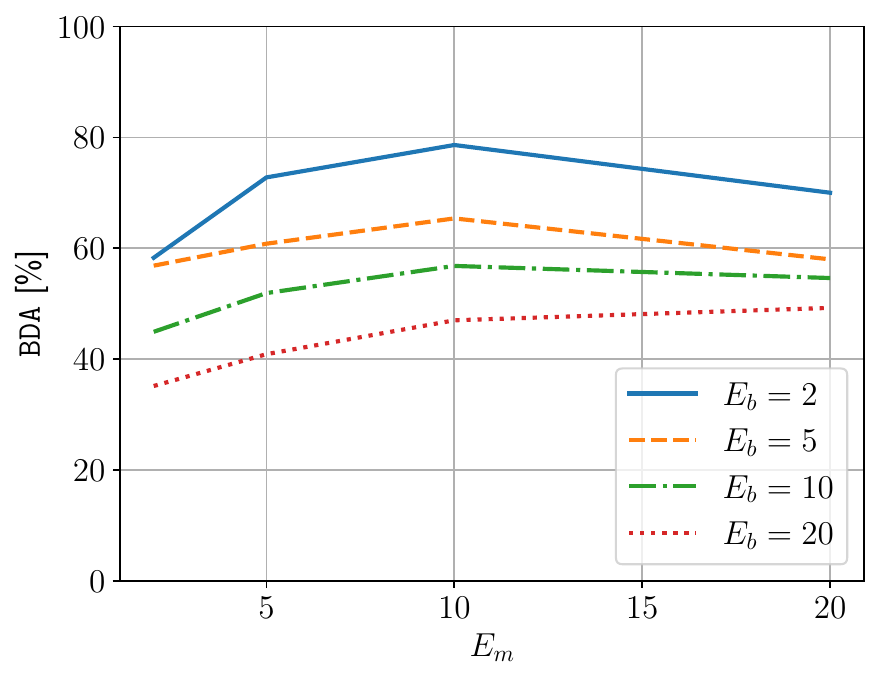}
    }

    \caption{Impact of $E_b$ and $E_m$ on the \bda of SoTA attacks.}
    \label{fig:E_effect_bda_during}
\end{figure}

\begin{figure}
    \centering
    \subfloat[A3FL~\cite{zhang_a3fl_2023} \label{fig:B_effect_a3fl_bda_during}]{
        \centering
        \includegraphics[width=0.47\linewidth]{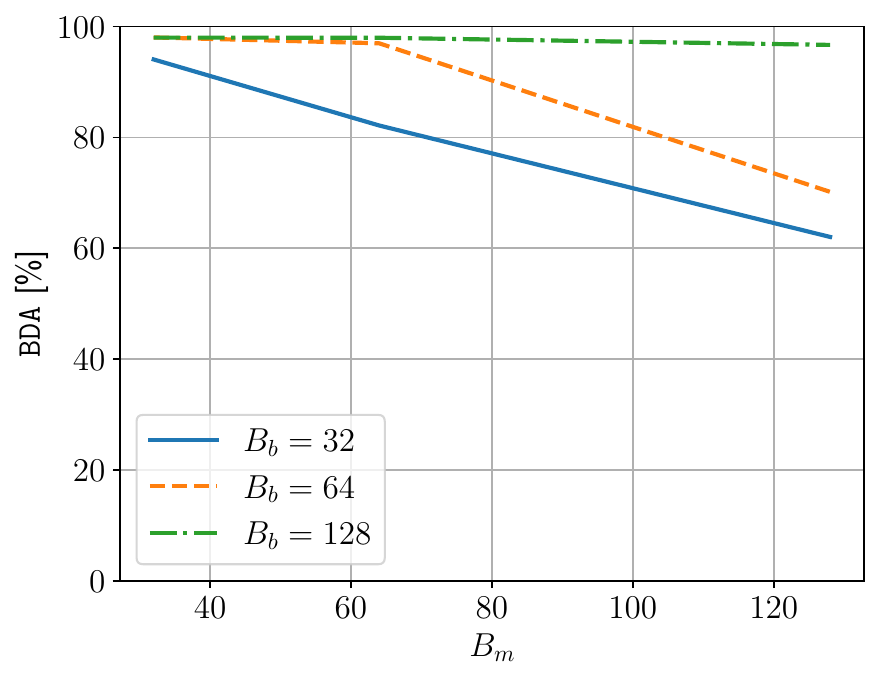}
    }
    \hfil
    \subfloat[Chameleon~\cite{dai_chameleon_2023} \label{fig:B_effect_chameleon_bda_during}]{
        \centering
        \includegraphics[width=0.47\linewidth]{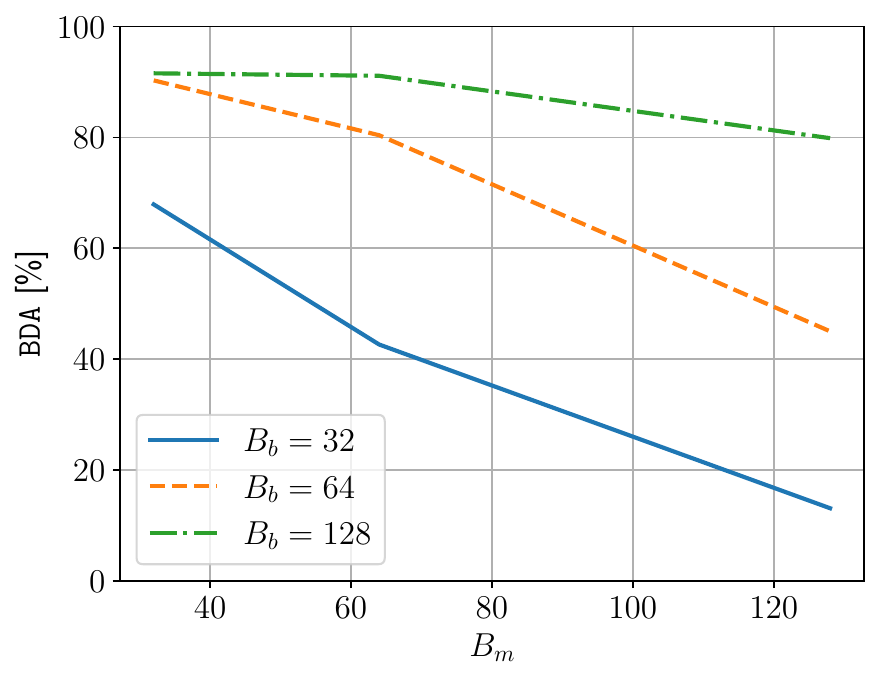}
    }

    \vspace{0.6em}
    
    \subfloat[DarkFed~\cite{DBLP:conf/ijcai/LiWNHXZW24} \label{fig:B_effect_darkfed_auc}]{
        \centering
        \includegraphics[width=0.47\linewidth]{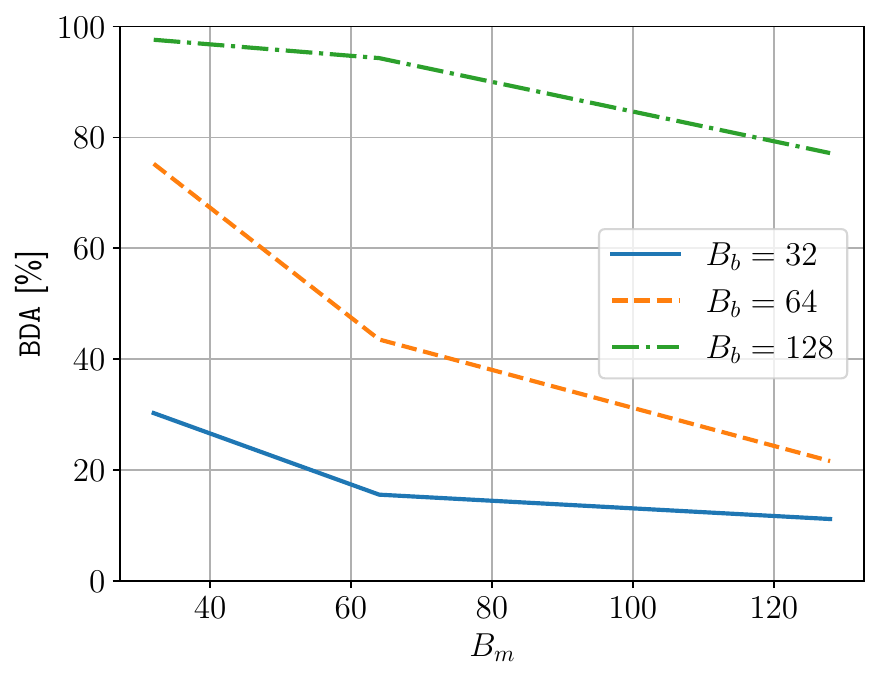}
    }
    \hfill
    \subfloat[FCBA~\cite{DBLP:conf/aaai/LiuZFYXM024} \label{fig:B_effect_fcba_bda_during}]{
        \centering
        \includegraphics[width=0.47\linewidth]{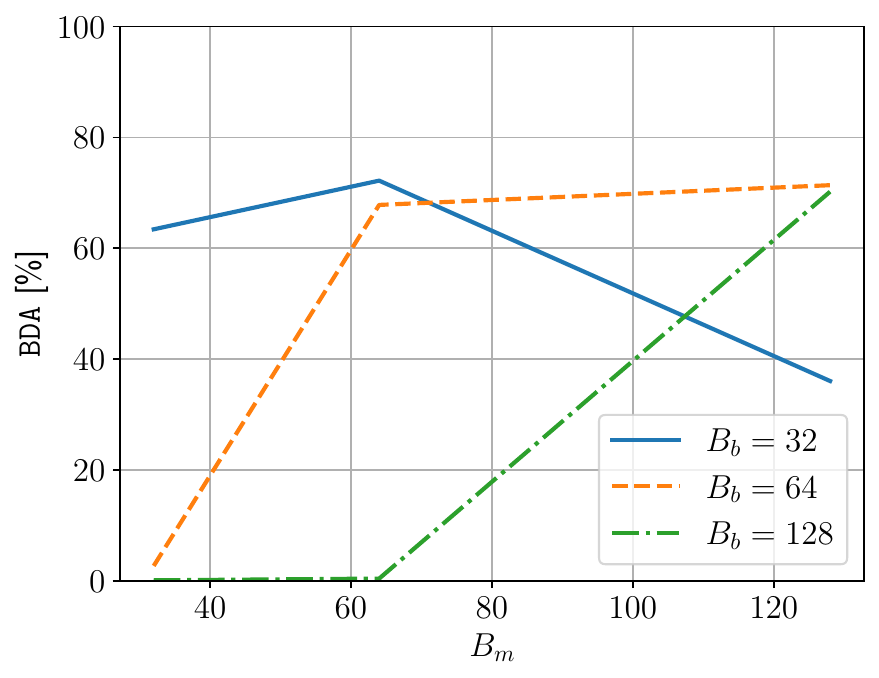}
    }

    \caption{Impact of $B_b$ and $B_m$ on the \bda of SoTA attacks.}
    \label{fig:B_effect_bda_during}
\end{figure}

\subheading{Impact of Batch Size \& Local Epochs.} \label{sec:E_B_effect}
The impact of the number of local epochs $E$ and the batch size $B$ on the \bda is shown in Figures~\ref{fig:E_effect_bda_during} and~\ref{fig:B_effect_bda_during}. 
Our empirical results confirm our analytical findings.
Namely, increasing $E_b$ or decreasing $B_b$ significantly reduces the \bda.
The maximum achievable \bda decreases by $13.02$ percentage points, $62.20$ percentage points, $35.49$ percentage points, and $27.28$ percentage points for the A3FL, Chameleon, DarkFed, and FCBA attack, respectively, when increasing $E_b$ from 2 to 20.
When decreasing the batch size from 128 to 32, on the other hand, the maximum achievable \bda for the A3FL, Chameleon, and DarkFed attacks decrease by $3.94$ percentage points, $23.65$ percentage points, and $67.33$ percentage points, respectively.
For the FCBA attack, however, we see a slightly different behavior:
for $(B_b, B_m) \in \{(64,32), (128,32), (128,32)\}$, the \bda drops to nearly $0\%$, while the \mta drops to $10\%$, i.e., random guessing across ten classes (see Figure~\ref{fig:B_effect_mta_during} in Appendix~\ref{app:E_B_effect}).
We attribute this to the use of model update scaling~\cite{bagdasaryan_how_2020}. 
When $B_m$ is small, attackers perform more updates on smaller batches, overfitting to their local data and diverging further from the global model. 
Upscaling such divergent updates increases the risk of destabilizing the global model, especially when $B_b$ is large and benign updates contribute less overall.

As shown in Figure~\ref{fig:E_effect_bda_after} and \ref{fig:B_effect_bda_after} in Appendix~\ref{app:E_B_effect}, increasing the number of local epochs from $2$ to $20$ leads to consistent drops in \bdaafter: $15.37$ percentage points (A3FL), $1.68$ percentage points (Chameleon), and $7.77$ percentage points (FCBA). 
For DarkFed, no decrease is observed, as the \bdaafter never exceeds 10\%---the baseline guessing probability—beyond what can be attributed to random fluctuations.
Similarly, reducing the benign batch size from $128$ to $32$ results in even stronger decreases: $44.69$ percentage points, $42.23$ percentage points, $4.07$ percentage points, and $14.40$ percentage points for the A3FL, Chameleon, FCBA, and DarkFed attacks.
This trend is even more pronounced in terms of \lifespan (Figures~\ref{fig:E_effect_lifespan} and \ref{fig:B_effect_lifespan} in Appendix~\ref{app:E_B_effect}), where all attacks witness reductions of over $70\%$ with lower batch size and over $91\%$ with number of local epochs.

Finally, measuring the \mta (Figures~\ref{fig:E_effect_mta_during} and \ref{fig:B_effect_mta_during} in Appendix~\ref{app:E_B_effect}) shows negligible differences for the A3FL, Chameleon, and DarkFed attacks: below $3$ percentage points for variations in $E$ and below $1$ percentage points for variations in $B$.
For the FCBA attack, the differences are more significant, namely up to $6.3$ percentage points when increasing $E_b$.
When decreasing $B_b$, on the other hand, the \mta does not decrease for the FCBA attack but instead increases by up to $71$ percentage points
This is due to the increased model divergence caused by scaling model updates, as explained before. 
For additional comparison, we also include results for our ``baseline'' and the IBA~\cite{nguyen_iba_2023} attack in Appendix~\ref{app:Baseline} and \ref{app:IBA}, respectively.

\subheading{Impact of Weight Decay.}
\label{sec:lambda_effect}
Figure~\ref{fig:lambda_effect_bda_during} in Appendix~\ref{app:mu_effect} shows the impact of the benign and malicious weight decay parameters on the \bda.
We find that weight decay has a relatively minor effect on both the \mta and \bda compared to other hyperparameters; \emph{it, however, has a substantial impact on \bdaafter and \lifespan}. 

Specifically, increasing $\lambda_b$ from $0.0001$ to $0.001$ leads to reductions in the maximum \bda by $1.78$ percentage points for A3FL, $4.72$ percentage points for Chameleon, $14.49$ percentage points for DarkFed, and $9.85$ percentage points for FCBA.
This effect is more pronounced in terms of \bdaafter and \lifespan (see Figures~\ref{fig:lambda_effect_bda_after} and \ref{fig:lambda_effect_lifespan} in Appendix~\ref{app:lambda_effect}), resulting in \bdaafter reductions of up to $35.09$ percentage points (A3FL), $3.47$ percentage points (Chameleon), $1.09$ percentage points (DarkFed), and $11.92$ percentage points (FCBA) and \lifespan reductions of $95.60\%$ (A3FL), $70\%$ (Chameleon), $100\%$ (DarkFed), and $61.54\%$ (FCBA). 
The relatively smaller drops in \bdaafter for Chameleon and DarkFed are due to their already low backdoor accuracy levels after the attack (below $15\%$). 
Similarly, increasing $\lambda_b$ from $0.0001$ to $0.001$ reduces the \mta (see Figure~\ref{fig:lambda_effect_mta_during} in Appendix~\ref{app:lambda_effect}) by no more than $2.4$ percentage points, except for the FCBA attack, where reductions of up to $6.2$ percentage points are observed---likely due to increased model divergence caused by scaling the updates.
For additional comparison, we also include results for our ``baseline'' and the IBA~\cite{nguyen_iba_2023} attack in Appendix~\ref{app:Baseline} and \ref{app:IBA}, respectively.

%% file: section_06.tex
\section{Interpretation of Findings}
\label{sec:pareto}

\begin{figure*}
    \centering
    \subfloat[A3FL~\cite{zhang_a3fl_2023} \label{fig:pareto_a3fl_accs_during}]{
        \centering
        \includegraphics[width=0.23\linewidth]{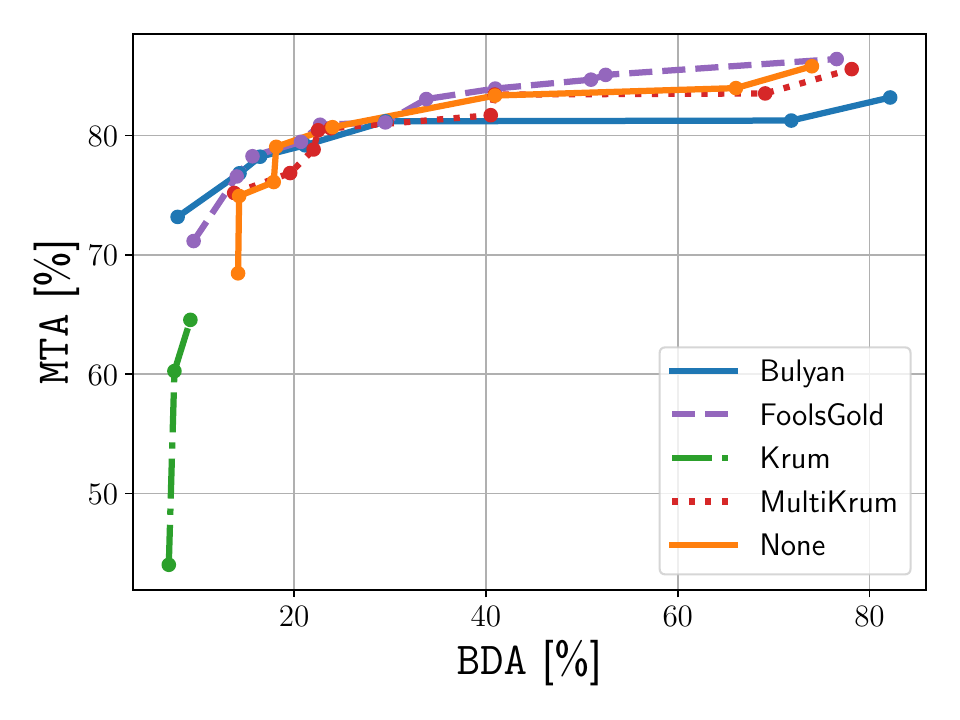}
    }
    \hfill
    \subfloat[Chameleon~\cite{dai_chameleon_2023} \label{fig:pareto_chameleon_accs_during}]{
        \centering
        \includegraphics[width=0.23\linewidth]{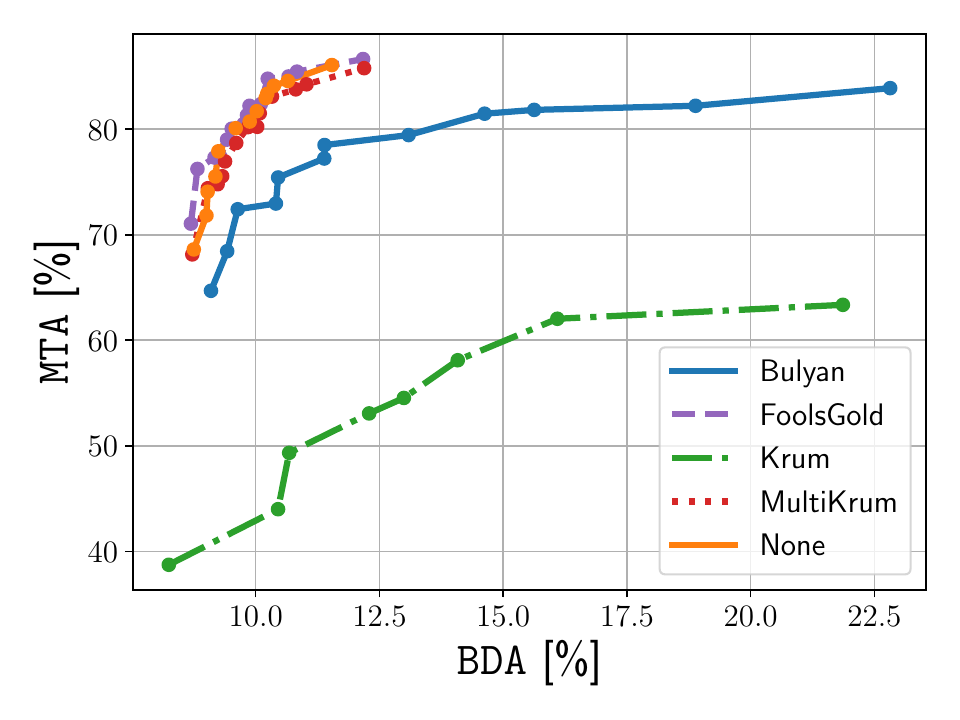}
    }
    \hfill
    \subfloat[DarkFed~\cite{DBLP:conf/ijcai/LiWNHXZW24} \label{fig:pareto_darkfed_accs_during}]{
        \centering
        \includegraphics[width=0.23\linewidth]{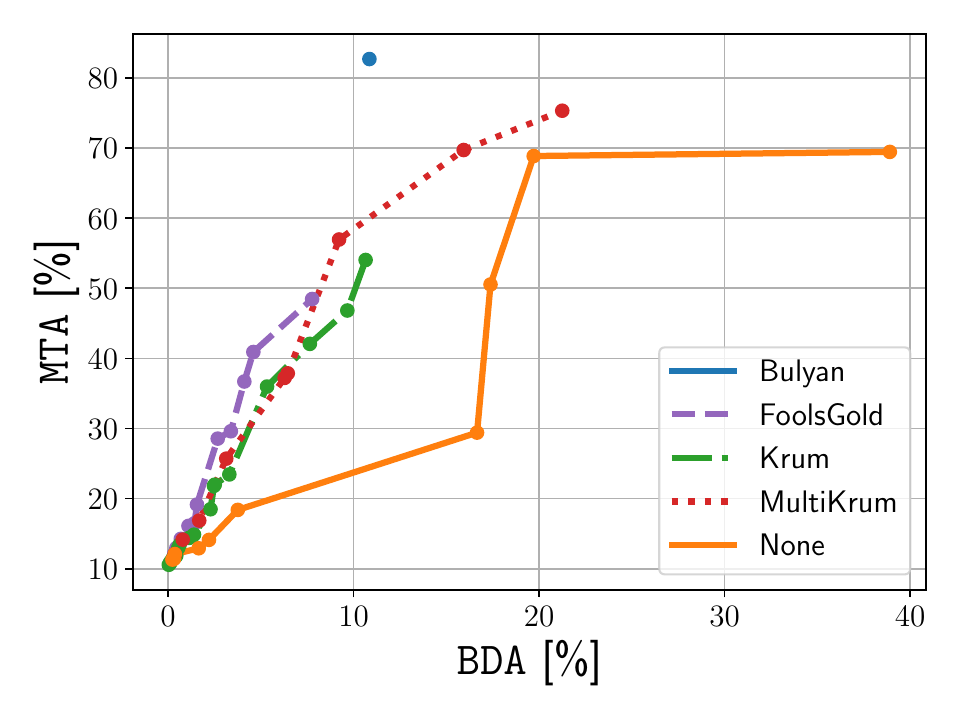}
    }
    \hfill
    \subfloat[FCBA~\cite{DBLP:conf/aaai/LiuZFYXM024} \label{fig:pareto_fcba_accs_during}]{
        \centering
        \includegraphics[width=0.23\linewidth]{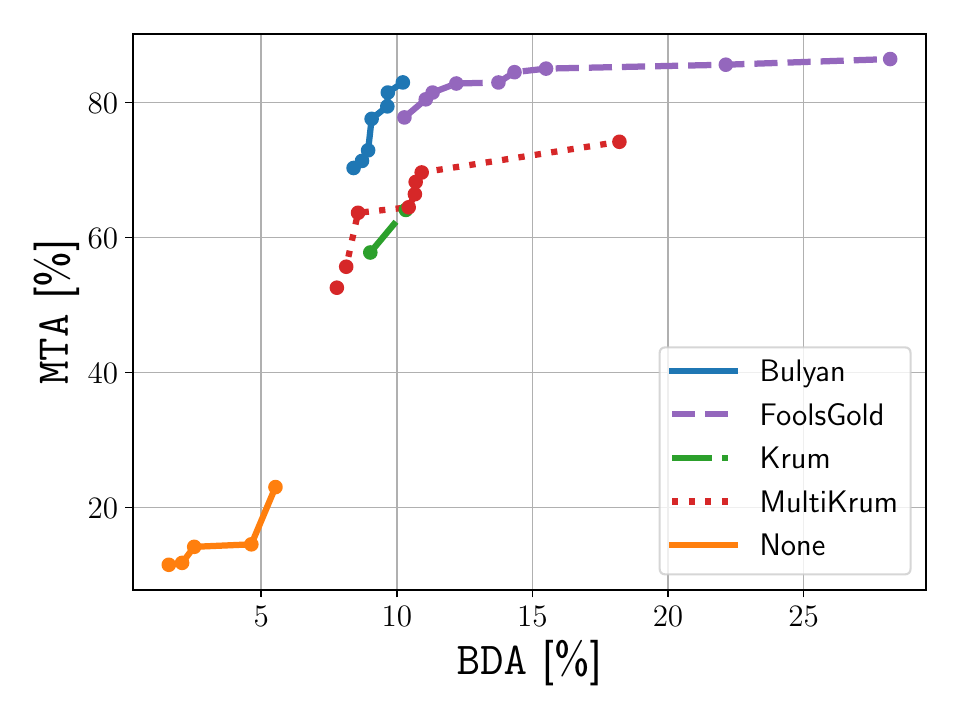}
    }
    \caption{Pareto frontiers of benign hyperparameter configurations for each attack against existing defenses. 
    Here, we assume our greedy adversarial strategy that adaptively adjusts malicious hyperparameters to increase the \bda (i.e., $\bda_\text{m}$).}
    \label{fig:pareto_accs_during}
\end{figure*}

In this section, we examine how to collectively tune benign hyperparameters, using a multi-objective optimization problem, to minimize the \bda and the associated backdoor lifespan while keeping the \mta high.
We achieve this by empirically determining the Pareto Frontier of benign hyperparameter combinations for each considered attack/defense combination, which enables us to pinpoint a recommended set of configurations for the considered hyperparameters.

\subsection{Significant Hyperparameters}
\label{sec:regression}
To guide our evaluation, we first conduct a regression analysis using the Ordinary Least Squares (OLS) method on our results in Sections~\ref{sec:lr_effect} and~\ref{sec:other_params}.
For stability of the regression model, we normalize the independent variables before performing the regression.

Table~\ref{tab:regression} present our regression results, considering the benign hyperparameters $\eta_b$, $\mu_b$, $\lambda_b$, $E_b$, and $B_b$ as independent variables and \bda, \bdaafter, and \lifespan as the dependent variable, respectively.
The fitted models have an $R^2$ of $0.191$, $0.163$, and $0.166$, respectively, indicating a moderate goodness-of-fit~\cite{cohen_statistical_1998}.
\begin{table}
    \centering
    \footnotesize
    \scalebox{0.85}{\begin{tabular}{lc|cccccc}
        \hline
        & & \textbf{coef} & \textbf{std. err} & \textbf{t} & \textbf{P $\mathbf{> |t|}$} & \textbf{[0.025} & \textbf{0.975]} \\
        \hline
        \multirow{6}{*}{\rotatebox[origin=c]{90}{\bda}}
        & $\mathtt{const}$ & 0.6289 & 0.016 & 39.110 & \textbf{$\mathbf{<}$ 0.001} & 0.597 & 0.661 \\
        & $\eta_b$ & -0.0846 & 0.016 & -5.215 & \textbf{$\mathbf{<}$ 0.001} & -0.117 & -0.053 \\
        & $\mu_b$ & -0.0223 & 0.016 & -1.390 & 0.166 & -0.054 & 0.009 \\
        & $\lambda_b$ & -0.0084 & 0.016 & -0.523 & 0.602 & -0.040 & 0.023 \\
        & $B_b$ & 0.0292 & 0.016 & 1.812 & 0.071 & -0.003 & 0.061 \\
        & $E_b$ & -0.0913 & 0.016 & -5.623 & \textbf{$\mathbf{<}$ 0.001} & -0.123 & -0.059 \\
        \hline
        \multirow{6}{*}{\rotatebox[origin=c]{90}{\bdaafter}}
        & $\mathtt{const}$ & 0.1587 & 0.006 & 28.078 & \textbf{$\mathbf{<}$ 0.001} & 0.148 & 0.170 \\
        & $\eta_b$ & -0.0179 & 0.006 & -3.145 & \textbf{0.002} & -0.029 & -0.007 \\
        & $\mu_b$ & -0.0038 & 0.006 & -0.670 & 0.504 & -0.015 & 0.007 \\
        & $\lambda_b$ & -0.0167 & 0.006 & -2.956 & \textbf{0.003} & -0.028 & -0.006 \\
        & $B_b$ & 0.0253 & 0.006 & 4.477 & \textbf{$\mathbf{<}$ 0.001} & 0.014 & 0.036 \\
        & $E_b$ & -0.0149 & 0.006 & -2.607 & \textbf{0.010} & -0.026 & -0.004 \\
        \hline
        \multirow{6}{*}{\rotatebox[origin=c]{90}{\lifespan}}
        & $\mathtt{const}$ & 158.5634 & 22.878 & 6.931 & \textbf{$\mathbf{<}$ 0.001} & 113.515 & 203.612 \\
        & $\eta_b$ & -52.0878 & 23.094 & -2.255 & \textbf{0.025} & -97.562 & -6.614 \\
        & $\mu_b$ & -19.4668 & 22.879 & -0.851 & 0.396 & -64.517 & 25.584 \\
        & $\lambda_b$ & -94.0825 & 22.882 & -4.112 & \textbf{$\mathbf{<}$ 0.001} & -139.139 & -49.026 \\
        & $B_b$ & 131.8755 & 22.910 & 5.756 & \textbf{$\mathbf{<}$ 0.001} & 86.765 & 176.986 \\
        & $E_b$ & -59.2787 & 23.098 & -2.566 & \textbf{0.011} & -104.759 & -13.798 \\
        \hline
        \end{tabular}}
    \caption{Regression analysis of the impact of $\eta_b, \mu_b, \lambda_b, E_b$, and $B_b$ on the \bda, \bdaafter, and \lifespan.}
    \label{tab:regression}
\end{table}
Our results show that $\eta_b$ and $E_b$ significantly impact the \bda and $\eta_b$, $\lambda_b$, $B_b$, and $E_b$ significantly impact the \bdaafter and \lifespan, as their respective $P$ values are below $0.05$.
Additionally, the coefficient signs indicate that increasing $\eta_b$, $\lambda_b$, or $E_b$, or decreasing $B_b$, reduces \bda, \bdaafter, and \lifespan, aligning with our previous empirical findings.
Notably, $\eta_b$ and $E_b$ have a stronger effect on \bda than on \bdaafter---by factors of approximately $\times4.7$ and $\times6.1$, respectively. 
In contrast, $\lambda_b$ and $B_b$ only significantly impact \bdaafter and \lifespan, highlighting the role of implicit (via batch size) and explicit (via weight decay) regularization in attack persistence.
However, we observe no statistically significant effect for the benign momentum $\mu_b$. This is especially surprising, considering that $\mu_b$ is one of the few hyperparameters on which there is broad consensus within the community (see Table~\ref{tab:hyper_params}). 
In contrast, other hyperparameters, such as the learning rate $\eta_b$ and batch size $B_b$, show statistical significance---but are chosen almost ad-hoc in the literature (see Table~\ref{tab:hyper_params}).

\subsection{Adaptive Adversaries} \label{sec:optimal_adversary}
Our measurement results in Section~\ref{sec:lr_effect} and Section~\ref{sec:other_params} reveal that an adversary can adaptively react by choosing appropriate hyperparameters in response to well-chosen benign hyperparameters.
For example, if the benign clients select a learning rate of $0.1$, an adaptive adversary may respond by setting its malicious learning rate to $\eta_m=0.2$ (i.e., $\beta=2$), resulting in an improvement of $\sim 11\%$ percentage points for the \bda in the FCBA~\cite{DBLP:conf/aaai/LiuZFYXM024}  attack compared to setting $\eta_m = \eta_b$.

To better evaluate the impact of adaptive adversarial strategies, we consider two adaptive approaches:
\begin{description}[leftmargin=0.5cm]
    \item[Greedy: ]  This adversary adapts to the benign hyperparameters by choosing malicious hyperparameters that maximize \bda on a per-parameter basis. Table~\ref{tab:optimal_adversary} in Appendix~\ref{app:optimal_adversary} shows the best malicious hyperparameter configurations for each attack and each selected benign hyperparameter, as determined by our results in Sections~\ref{sec:lr_effect} and \ref{sec:other_params}.
    \item[\nsga:] Unlike the greedy adversary, this adversary does not adapt to individual benign choices but instead tunes all hyperparameters jointly to react to the benign choices through a complex stochastic search (e.g., using the NSGA-II~\cite{DBLP:journals/tec/DebAPM02} search algorithm). This makes the stochastic adversary more powerful but also significantly more computationally demanding compared to the aforementioned greedy strategy. 
\end{description}

For each adaptive strategy, we consider an adversary that (i) tries to maximize \bda without a constraint on \mta (denoted by $\bda_\text{m}$), and (ii) seeks to maximize \bda while limiting the drop in \mta by a margin $\epsilon_{adv} = 5\%$ (cf. Equation~\ref{eq:bda_adv}) by selecting malicious hyperparameters that fulfil this constraint (denoted by $\mta_\text{c}$). 

While our analysis indicates that an adversary can indeed adapt to proper benign configurations, we demonstrate in the following section that the effectiveness of such adaptive strategies is ultimately limited---and that careful selection of benign parameters can severely cap their impact. 

\subsection{Recommended Benign Configurations}
\label{sec:optimal_benign}

\setlength\tabcolsep{2.9pt} %
\begin{table*}
    \centering
    \scriptsize
    \scalebox{0.74}{

        \begin{tabular}{|l|cc|c|cc|c|cc|c|cc|c|cc|c|}
        \hline
         & \multicolumn{3}{c|}{\mta} & \multicolumn{3}{c|}{\bda} & \multicolumn{3}{c|}{\mtaafter} & \multicolumn{3}{c|}{\bdaafter} & \multicolumn{3}{c|}{\lifespan} \\\cline{2 - 16}
        \textbf{Benign param.} & \multicolumn{2}{c|}{Recommended} & Orig. & \multicolumn{2}{c|}{Recommended} & Orig. & \multicolumn{2}{c|}{Recommended} & Orig. & \multicolumn{2}{c|}{Recommended} & Orig. & \multicolumn{2}{c|}{Recommended} & Orig. \\\cline{2 - 16}
        \multirow{2}{*}{\textbf{Adapt. malicious param.}} & Greedy & \nsga & \multirow{2}{*}{Orig.} & Greedy & \nsga & \multirow{2}{*}{Orig.} & Greedy & \nsga & \multirow{2}{*}{Orig.} & Greedy & \nsga & \multirow{2}{*}{Orig.} & Greedy & \nsga & \multirow{2}{*}{Orig.} \\
        & $\bda_\text{m}$ / $\mta_c$ & $\bda_\text{m}$ / $\mta_\text{c}$  &  & $\bda_\text{m}$ / $\mta_\text{c}$  & $\bda_\text{m}$ / $\mta_\text{c}$  &  & $\bda_\text{m}$ / $\mta_\text{c}$  & $\bda_\text{m}$ / $\mta_\text{c}$  &  & $\bda_\text{m}$ / $\mta_\text{c}$  & $\bda_\text{m}$ / $\mta_\text{c}$  &  & $\bda_\text{m}$ / $\mta_\text{c}$  & $\bda_\text{m}$ / $\mta_\text{c}$  &  \\
        \hline
        A3FL / FoolsGold & 85.1 \% & 85.5 \% & 85.4 \% & 52.5 \% & 72.8 \% & 95.0 \% & 87.3 \% & 87.3 \% & 85.8 \% & 13.3 \% & 19.9 \% & 65.0 \% & 0 & 8 & 994 \\
        Chameleon / None & 84.4 \% & 84.7 \% & 88.9 \% & 10.6 \% / 10.4 \% & 12.3 \% & 72.3 \% & 86.8 \% & 86.9 \% & 90.5 \% & 10.3 \% / 10.2 \% & 10.3 \% & 40.1 \% & 0 & 0 & 312 \\
        DarkFed / Bulyan & 80.7 \% / 80.6 \% & 81.1 \% & 81.0 \% & 11.2 \% & 11.4 \% & 11.2 \% & 84.6 \% / 84.5 \% & 84.5 \% & 83.0 \% & 10.4 \% / 10.8 \% & 10.6 \% & 10.2 \% & 0 & 0 & 0 \\
        FCBA / FoolsGold & 85.6 \% / 85.0 \% & 85.2 \% & 82.9 \% & 22.1 \% / 10.4 \% & 23.5 \% & 39.4 \% & 87.3 \% & 87.3 \% & 85.4 \% & 10.2 \% / 9.9 \% & 10.3 \% & 11.5 \% & 0 & 0 & 0 \\
        \hline
        \textbf{Avg. improvement} & \textbf{-0.60\% / -0.78\%} & \textbf{-0.43\%} & & \textbf{30.38\% / 33.35\%} & \textbf{24.48\%} & & \textbf{0.33\% / 0.30\%} & \textbf{0.33\%} & & \textbf{20.65\%} & \textbf{18.93\%} & & \textbf{100\%} & \textbf{100 \%/ 99.60\%} &\\
        \hline
        \end{tabular}

    }
    \caption{Selected results for \mta, \bda, \mtaafter, \bdaafter, and \lifespan for attack/defense combination using our  ``recommended'' benign hyperparameters ($\eta_b = 0.15, \mu_b = 0.9, \lambda_b = 0.0005, E_b = 10, B_b = 32$) under our greedy and stochastic adaptive adversaries that aim to maximize BDA ($\bda_\text{m}$), or are MTA-constrained  ($\mta_\text{c}$) within $\epsilon_{adv}=5\%$, and the original malicious/benign parameters (``Orig.''). We include one value in the cell ``$\bda_\text{m}/\mta_\text{c}$'' when these strategies converge on the same result.}
    \label{tab:best_hyperparams_with_defenses}
\end{table*}
\setlength\tabcolsep{6pt} %

Leveraging the aforementioned adaptive strategies, we now empirically determine the Pareto frontier, i.e., the set of optimal configurations for the benign learning rate $\eta_b$, weight decay $\lambda_b$, number of local epochs $E_b$, and batch size $B_b$,\footnote{We excluded the impact of the momentum  $\mu$ when deriving the Pareto frontier, as it had no significant impact on our optimization objectives (see Table~\ref{tab:regression}).} for each considered attack/defense combination (cf. Section~\ref{sec:methodology} for the selection criteria). 
Here, we consider the \mta and \bda as our optimization objectives. 
The Pareto frontier is determined by solving the following optimization problem: $\max_{\omega \in \Omega}{\left( \mta(\omega), -\bda(\omega)  \right)}$, where solutions $\omega = (\eta_b, \mu_b, \lambda_b, E_b, B_b)$ are drawn from the solution space $\Omega = \Omega_{\eta_b} \times \Omega_{\mu_b} \times \Omega_{\lambda_b} \times \Omega_{E_b}, \times \Omega_{B_b}$.

In multi-objective optimization a solution $x \in \Omega$ \textit{dominates} another solution $y \in \Omega$ if it improves upon $y$ in all objectives, i.e., $\mta(x) > \mta(y) \land \bda(x) < \bda(y)$, denoted by $x \succ y$.
A solution $w^*$ is \textit{Pareto optimal} if no other solution in $\Omega$ dominates it.
The \textit{Pareto frontier} is the set of Pareto-optimal solutions: $$PF(\Omega) = \{w^* \in \Omega | \nexists w \in \Omega: w \succ w^* \}$$

To determine the Pareto frontier, we evaluate \mta and \bda for each attack across different benign hyperparameters, using the greedy adversary that seeks to maximize its \bda as a baseline---due to its low computational cost and its close approximation of the stochastic adversary, as shown in Table~\ref{tab:best_hyperparams_no_defense}.
We also included in our evaluation four state-of-the-art defenses, namely Bulyan~\cite{mhamdi_hidden_2018}, FoolsGold~\cite{fung_limitations_2020}, Krum~\cite{blanchard_machine_2017}, and MultiKrum~\cite{blanchard_machine_2017} (see Appendix~\ref{app:defense_selection} for an overview/justification for the defenses and Appendix~\ref{app:attacks_vs_defenses} for a visualization of how the selected attacks perform under the selected defenses using the original parameters).
To reduce the computational load in our experiments, we limit the attack window to $100$ rounds and the post-attack phase to $1000$ rounds, which our prior experiments show is sufficient for all attacks to reach peak backdoor accuracy.\footnote{
Note that we had to adapt the Chameleon learning rate schedule to match our shorter pre-attack window (cf. Appendix~\ref{app:changed_attacks}).}

Note that, in a realistic HFL setting---which our work aims to emulate, the defender's options are fundamentally constrained when compared to the attackers' options. 
Specifically, benign clients must select a fixed set of hyperparameters beforehand, regardless of the specific attack (e.g., A3FL or Chameleon) and adaptive strategy chosen by the adversary. 
In contrast, the adversary operates in a more flexible setting and can dynamically adapt their hyperparameters based on the specific attack they wish to launch. 
The search space for benign hyperparamters includes: $\Omega_{\eta_b} = \{0.1, 0.15, 0.2\}$, $\Omega_{\mu_b} = \{0.9\}$, $\Omega_{\lambda_b} = \{0.0005, 0.001\}$, $\Omega_{E_b} = \{10, 20\}$, and $\Omega_{B_b} = \{16, 32\}$---yielding 24 configurations per attack/defense combination.

To efficiently explore this space, we use two alternative approaches (i) a grid search and (ii) the NSGA-II~\cite{DBLP:journals/tec/DebAPM02} algorithm, a robust multi-objective optimization algorithm that was also used in~\cite{DBLP:conf/ccs/BagdasarianS24} for tuning. 
NSGA-II evolves a population of solutions using genetic operations (selection, crossover, mutation), while maintaining a diverse Pareto front through non-dominated sorting and crowding distance. 
Compared to the brute-force grid search (which gave nearly identical results), NSGA-II significantly reduces computational overhead---by approximately 50\%---while scaling better in multi-dimensional settings.

Our results are shown in Figure~\ref{fig:pareto_accs_during}.
We observe a clear tradeoff between \mta and \bda---decreasing the backdoor accuracy consistently leads to decreases in main task accuracy.
Furthermore, we observe that in most configurations, the DarkFed~\cite{DBLP:conf/ijcai/LiWNHXZW24} and FCBA~\cite{DBLP:conf/aaai/LiuZFYXM024} attacks severely impact \mta and/or do not exceed a \bda of $10\%$ significantly, indicating that these attacks are not as effective as they might initially seem, under well-chosen benign hyperparameters.
Interestingly, we also observe that the absence of any defense can perform comparably to some existing defenses---for example, MultiKrum combined with A3FL and FoolsGold combined with Chameleon.

\vspace{0.5 em}
\noindent 
\fbox{
\parbox{\dimexpr\linewidth-2\fboxsep-2\fboxrule\relax}{
\textit{Analyzing the individual hyperparameter configurations comprising the Pareto frontier yields a recommended configuration for the benign hyperparameters that minimizes \bda \emph{for all considered attacks and defenses}: namely $\eta_b = 0.15$ (round-wise decaying by $\gamma=0.999$)}, $\mu_b = 0.9, \lambda_b = 0.0005, E_b = 10, B_b = 32$.\footnotemark}} 

\addtocounter{footnote}{-1}
\stepcounter{footnote}\footnotetext{Interestingly, in additional experiments we conducted, where benign clients could optimize configurations per attack/defense, the resulting configurations still closely matched our recommended hyperparameters.}

\subsection{Detailed Results}
\label{sub:detailed_results}

Once the benign and malicious hyperparameters are determined using the strategies outlined above, we execute all attacks and defenses with these fixed parameters and report the full results in Table~\ref{tab:best_hyperparams_no_defense} in Appendix~\ref{app:recommended_configs} and Table~\ref{tab:best_hyperparams_with_defenses} (where we show, due to lack of space, those combinations (attacks/defenses) that yield the best tradeoffs). 
Here, we show the \mta, \mtaafter, \bda, \bdaafter, and \lifespan that each attack achieves when no server-side defense is present, or when the considered four defenses are activated as follows: (a) when using the benign and malicious parameters as reported in the original papers (denoted by "{}Orig."{}), (b) when using our recommended benign parameters and the ``Greedy'' adaptive strategy outlined in Section~\ref{sec:optimal_adversary} (with both MTA-constrained ($\mta_\text{c}$) and unconstrained adversaries $\bda_\text{m}$), and (c) when using our recommended benign parameters and the ``\nsga'' adaptive strategy outlined in Section~\ref{sec:optimal_adversary} (with both $\mta_\text{c}$ and $\bda_\text{m}$ adversaries).

We note that the stochastic adversary was able to find malicious hyperparameter configurations that slightly improve the \bda, especially when defenses are in place, and was able to stay better with the \mta constraints (due to its greater flexibility in optimizing all hyperparameters jointly) when compared to the greedy strategy. 
In some cases, however, the stochastic adversary was overly aggressive in sacrificing the main task accuracy (A3FL/None in Table~\ref{tab:best_hyperparams_no_defense}). Interestingly, the $\mta_\text{c}$ variant of the stochastic adversary aligned more consistently with the greedy adaptive strategy (as shown in Table~\ref{tab:best_hyperparams_no_defense}, the difference between those strategies was less than 4\% on average). Moreover, the greedy adversary was able to consistently identify effective configurations, making it a robust alternative to the more computationally expensive stochastic strategy.

Note that some attacks/defense combinations (especially the defense Krum~\cite{blanchard_machine_2017}, and the FCBA~\cite{DBLP:conf/aaai/LiuZFYXM024} attack) significantly lower the main task accuracy in such a way that the \mta constraint could not be satisfied by both adaptive strategies, as the combination already degraded \mta to the extent that no adversarial hyperparameter setting could compensate (we denote those by ``N.A'' in Table~\ref{tab:best_hyperparams_with_defenses} and \ref{tab:best_hyperparams_no_defense}). 
Especially for the stochastic adversary, the configurations that satisfied the \mta constraint also resulted in several cases in the highest \bda, leading both the $\bda_\text{m}$ and $\mta_c$ adversaries to converge on similar parameter sets. 
This outcome stems from two factors: (i) overly aggressive malicious updates (like the ones suggested by the stochastic $\bda_\text{m}$ adversary) tend to be more detectable by defenses, and (ii) when greedily optimizing for \bda, the resulting updates may diverge more strongly from benign ones, pushing the global model into suboptimal regions of the loss landscape for benign clients. 
This causes steeper gradients and accelerates re-training, indirectly benefiting the adversary. 
On the other hand, the greedy $\mta_\text{c}$ adversary did not consistently outperform the \mta compared to the $\bda_\text{m}$ variant, since its greedy per-parameter strategy can sometimes lead to suboptimal choices relative to the \mta. 
Note that in cases where the configurations satisfying the \mta constraint diverged from those selected by $\bda_\text{m}$ (in both the greedy and stochastic adaptive strategies), the resulting BDA was on average lower, reflecting the tighter constraints placed on the adversary.

\emph{Overall, our recommended hyperparameters led to substantial improvements in \bda, \bdaafter, and \lifespan across all adaptive strategies---greedy, stochastic, and \mta-constrained.} In a few cases, our results suggest that the effectiveness of some defenses may have been overstated, as the adversary was able to identify improved malicious hyperparameters---compared to those used in the original papers---that achieve higher \bda. Moreover, our results show that:

\begin{enumerate}[leftmargin=0.5 cm]
    \item \textbf{The backdoor accuracy of existing attacks is exaggerated when benign hyperparameters are poorly chosen: } Appropriately choosing hyperparameters reduces the reported backdoor accuracies of current SOTA attacks considerably. For instance, the \bda, \bdaafter, and \lifespan of the powerful A3FL attack are reduced by $10.3$ percentage points, $51.5$ percentage points, and $98.6\%$ when considering our strongest adaptive adversarial strategy, and even when no defense is active. When FoolsGold is additionally used as a defense, we achieve even higher reductions of $22.2$ percentage points, $45.1$ percentage points, and $99.2\%$, respectively, with a slight increase in \mta of $0.1\%$ percentage points.
    \item \textbf{A proper choice of hyperparameters can also thwart adaptive adversaries: } Against the stochastic $\bda_\text{m}$ adversary, the \bda decreases for all attacks, on average, by $31.3$ percentage points and $\bdaafter$ by $30.1$ percentage points, even when no defenses are employed. %
    \item \textbf{A proper tuning of hyperparameters only negligibly impacts \mta: } Combining our recommended benign hyperparameters with either the FoolsGold~\cite{fung_limitations_2020}, MultiKrum~\cite{blanchard_machine_2017}, or no defense ensures little impact on main task accuracy when no attack is active, with decreases of at most $5$ percentage points. %
    \item \textbf{Finetuning hyperparameters per defense yields even stron\-ger backdoor robustness: } While our recommended configurations minimize \bda across all considered attacks, it is possible to further tailor benign hyperparameters depending on the employed defense. For example, when using no defense and assuming the greedy $\bda_\text{m}$ attacker, the configuration $\eta_b = 0.1$, $\lambda_b=0.0005$, $E_b=10$, $B_b=16$ further decreases the average \bda across all attacks by $5.4$ percentage points and increases \mta by $16.3$ percentage points.%
   \item \textbf{Relying on no defense or FoolsGold~\cite{fung_limitations_2020} may be enough: } Our results show that properly tuning hyperparameters can be used standalone (against the stochastic adversary, the average \bda with our approach without a defense matches that when defenses are used, see Table~\ref{tab:best_hyperparams_no_defense}) and outperforms defenses such as Krum~\cite{blanchard_machine_2017}). Alternatively, FoolsGold emerges as a complementary defense that pairs well with our approach.
\end{enumerate}

\subheading{Impact on computational costs.} 
We emphasize that adjusting the benign hyperparameters as we propose does not introduce additional computational overhead---neither on the coordinating entity (which identifies these well-converging hyperparameters offline, before deployment) nor on the overall HFL training performance.

Most hyperparameter modifications---such as tuning the learning rate—have a negligible impact on computational cost. 
While some changes, like increasing the number of local epochs $E$, may raise per-round runtime, they often lead to faster convergence, resulting in little to no increase in total computation. 
Our experiments support this claim: training ResNet20 with our recommended hyperparameters achieves $\mta = 80\%$ in 164 rounds, a 33\% reduction compared to 247 rounds using baseline settings ($\eta_b=0.1, \mu_b=0.9, \lambda_b=0.0005, E_b=2, B_b=64$; see Table~\ref{tab:hyper_params}).

\subsection{Limitations}
While our results cover a range of model architectures, datasets, and hyperparameter settings, they are inherently limited to image classification tasks. 
Other machine learning domains---such as generative modeling---often involve different threat models and rely on fundamentally different architectures (e.g., transformers or diffusion models), where hyperparameters may behave differently~\cite{DBLP:conf/nips/DAngeloAVF24}.

Although the exact recommended benign hyperparameters may vary across architectures or deployment settings, we believe the general trends are broadly applicable, as they are analytically grounded and consistently observed across different models, datasets, and attack strategies. 
For example, Figure~\ref{fig:diverse_architectures} demonstrates that using MobileNetV2 or VGG17 as the model architecture, or training on Tiny-ImageNet, does not substantially alter the general trend in how benign and malicious learning rates affect the \bda.

%% file: section_07.tex
\section{Concluding Remarks}

In this work, we investigated how training hyperparameters affect the success of backdoor attacks and the effectiveness of defenses in horizontal federated learning (HFL). Specifically, we demonstrated---both analytically and empirically---that hyperparameter choices made by benign clients influence not only model accuracy but also the success rate of backdoor attacks. This finding contrasts with many prior works in HFL security, which often use arbitrary hyperparameter settings for benign clients—frequently resulting in stronger backdoors and weaker defense performance. 

Moreover, we show that there exists a set of Pareto-near-optimal hyperparameter choices that achieve across-the-board improvements in accuracy-robustness tradeoffs against attacks.

\subheading{Open Science.} To ensure the reproducibility of our results, we fixed all PRNG seeds during our experiments and made our code publicly available on GitHub\footnote{\url{https://github.com/RUB-InfSec/federated_learning_hyperparams}}.
Since some FL experiments, e.g., those using NSGA-II, typically rely on stochastic operations, some results are not inherently reproducible with full precision.
To address this, we thoroughly logged all outputs to ensure full reproducibility.

\subheading{Future Work.}  
We believe that our analysis, which covers four state-of-the-art backdoor attacks and includes analytical evaluations, is likely to generalize to future attacks. 
We encourage researchers to adopt our recommended benign hyperparameters when evaluating new attacks, ensuring analysis under realistic and robust conditions. Specifically, attacks should be tested against well-tuned benign configurations to ensure that they generalize well to realistic scenarios. At the same time, defenses should be evaluated against adaptive adversaries---those who optimize their hyperparameters in response to the benign setup—using the strategies we outlined. %

While our analysis provides practical guidance for configuring attacker hyperparameters in defense evaluations,  future research could further expand this analysis to a wider range of attacks, defenses, and ML domains, such as NLP, to better understand the role of hyperparameters in adversarial robustness.

\subheading{Ethical Concerns.} Our primary contribution in this work lies in equipping defenders with insights on how to build resilience against adaptive attackers---regardless of the specific strategy employed. 
In fact, our paper points out that there are hyperparameter choices that the benign clients can choose to mitigate or weaken any attack, regardless of the adversary’s strategy. 
This empowers system designers to proactively enhance robustness, ultimately giving defenders a strategic advantage.

%% file: appendix.tex
\section{Selection of Representative Backdoor Attacks}
\label{app:RW_chain}

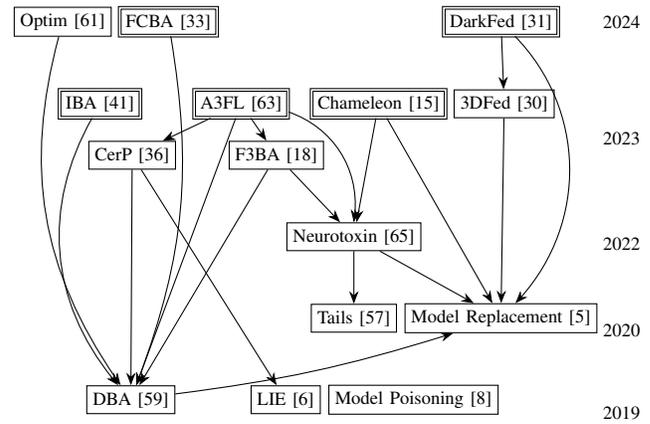
\begin{figure}[tb]
    \scriptsize

    \begin{tikzpicture}[
        node distance=0.1cm,
        every node/.style={rectangle, draw, align=center},
        highlight/.style={draw, dashed},
        best/.style={draw, double},
        every edge/.style={draw, -{Stealth[scale=1.0]}}, %
        arrow/.style={->, >=Stealth} %
    ]

    \node (optim) [] {Optim~\cite{DBLP:conf/cikm/Yang0NHW24}};
    \node (fcba) [right = of optim, best] {FCBA~\cite{DBLP:conf/aaai/LiuZFYXM024}};
    \node (darkfed) [right = 3.0cm of fcba, best] {DarkFed~\cite{DBLP:conf/ijcai/LiWNHXZW24}};

    \node (iba) [below right = 0.7cm and -0.7cm of optim, best] {IBA~\cite{nguyen_iba_2023}};
    \node (a3fl) [right = 0.7cm of iba, best] {A3FL~\cite{zhang_a3fl_2023}};
    \node (chameleon) [right = 0.3cm of a3fl, best] {Chameleon~\cite{dai_chameleon_2023}};
    \node (3dfed) [right = of chameleon] {3DFed~\cite{li_3dfed_2023}};
    
    \node (cerberus) [below right = 0.3cm and -0.7cm of iba] {CerP~\cite{lyu_poisoning_2023}};
    \node (f3ba) [right = 0.7cm of cerberus] {F3BA~\cite{fang_vulnerability_2023}};

    \node (neurotoxin) [below right = 0.7cm and -0.5cm of f3ba] {Neurotoxin~\cite{zhang_neurotoxin_2022}};
    
    \node (tails) [below = 0.7cm of neurotoxin] {Tails~\cite{wang_attack_2020}};
    \node (modelReplacement) [right = of tails] {Model Replacement~\cite{bagdasaryan_how_2020}};
    
    \node (dba) [below left = 0.7cm and 1.8cm of tails] {DBA~\cite{xie_dba_2020}};
    \node (lie) [right = 1.0cm of dba] {LIE~\cite{baruch_little_2019}};
    \node (modelPoisoning) [right = of lie] {Model Poisoning~\cite{bhagoji_analyzing_2019}};

    \node[draw=none] (2024) [right = 0.5cm of darkfed] {2024};
    \node[draw=none] (2023) [below = 1.2cm of 2024]{2023};
    \node[draw=none] (2022) [below = 1.05cm of 2023]{2022};
    \node[draw=none] (2020) [below = 0.8cm of 2022]{2020};
    \node[draw=none] (2019) [below = 0.75cm of 2020] {2019};

    \draw (optim) edge [arrow, bend right = 25] (dba);
    \draw (fcba) edge [arrow, bend left = 15] (dba);
    \draw (darkfed) edge [arrow] (3dfed);
    \draw (darkfed) edge [arrow, bend left = 44] (modelReplacement);
    \draw [arrow] (a3fl) -- (cerberus);
    \draw [arrow] (a3fl) -- (f3ba);
    \draw (a3fl) edge [arrow] (dba);
    \draw (a3fl) edge [arrow, bend left = 40] (neurotoxin);
    \draw (iba) edge [arrow, bend right = 35] (dba);
    \draw (chameleon) edge [arrow] (neurotoxin);
    \draw (chameleon) edge [arrow] (modelReplacement);
    \draw (3dfed) edge [arrow] (modelReplacement);
    \draw (cerberus) edge [arrow] (lie);
    \draw (cerberus) edge [arrow] (dba);
    \draw (f3ba) edge [arrow] (dba);
    \draw [arrow] (f3ba) -- (neurotoxin);
    \draw [arrow] (neurotoxin) -- (tails);
    \draw [arrow] (neurotoxin) -- (modelReplacement);
    \draw [arrow] (dba) edge [arrow, bend right=5] (modelReplacement);
    \end{tikzpicture}
    \centering
    \caption{Overview of backdoor attacks published at A/A* conferences. An arrow from node A to node B indicates that attack A was shown to beat attack B in paper A. Double squares highlight our selected attacks.  Note that \cite{xie_dba_2020} compared against the arXiv version of \cite{bagdasaryan_how_2020} that was already published in 2018.}
    \label{fig:related_work_chain}
\end{figure}

We visually sorted all initially-selected 15 backdoor attacks (cf. Section~\ref{sec:methodology}) in Figure~\ref{fig:related_work_chain}. Here, contributions are shown in nodes / rectangles. An arrow from node $A$ to node $B$ shows that proposal $A$ achieves higher attack success rates than proposal $B$.
To create this graph, we extracted information from the evaluation of each considered paper.
This visualization aided us in selecting the four most recent attacks (shown at the top of the chain) for our evaluation, namely A3FL~\cite{zhang_a3fl_2023}, Chameleon~\cite{dai_chameleon_2023}, DarkFed~\cite{DBLP:conf/ijcai/LiWNHXZW24}, and FCBA~\cite{DBLP:conf/aaai/LiuZFYXM024}.
Note that we excluded the model poisoning attack~\cite{bhagoji_analyzing_2019} from the analysis since it is almost identical to the model replacement attack~\cite{bagdasaryan_how_2020}. 
Further, we excluded the optimization-based attack~\cite{DBLP:conf/cikm/Yang0NHW24} since it requires the adversary to have knowledge of the benign updates, which contradicts our threat model. 
We also excluded IBA~\cite{nguyen_iba_2023} from our primary analysis as it is not a pure backdoor attack but a mixture of a backdoor attack and adversarial example generation. 
For completeness, we, however, show that our results apply to this attack as well in Appendix~\ref{app:IBA}.

\section{Representative Backdoor Defenses}
\label{app:defense_selection}

To obtain an unbiased selection of representative backdoor defenses, we select frequently used defenses against backdoor attacks based on the evaluations of our considered attack papers. 
Precisely, for each paper depicted in Figure~\ref{fig:related_work_chain}, we extract the defenses used for comparison and filter them based on two criteria: namely, publication in an A/A* conference according to the CORE rankings\footnote{\url{https://portal.core.edu.au/conf-ranks/}} and availability of official source code. 
From this, we identify the four most frequently evaluated defenses: Krum~\cite{blanchard_machine_2017} (8), FoolsGold~\cite{fung_limitations_2020} (6), Bulyan~\cite{mhamdi_hidden_2018} (4), and MultiKrum~\cite{blanchard_machine_2017} (4), where the numbers in parentheses indicate the number of attack papers that included the respective defense in their evaluation:

\begin{description}[leftmargin=0.2 cm]
    \item[Krum] 
    The Krum defense~\cite{blanchard_machine_2017} clusters client updates based on their proximity.
    It uses the client's update as the global update that is the center of the cluster with the smallest diameter, assuming that the cluster with the smallest diameter will belong to the set of benign clients as their updates are assumed to be more similar to each other than the updates of malicious clients.

    \item[Multi-Krum] Multi-Krum~\cite{blanchard_machine_2017} extends Krum by using the average of the best $m$ updates as a new global model.
    A common choice for $m$ is the number of assumed benign clients, i.\,e., $m = (M -f)$~\cite{blanchard_machine_2017}.

    \item[Bulyan] Bulyan~\cite{mhamdi_hidden_2018} first selects assumed benign local models using an existing rule like, e.g., Krum. 
    Then, it parameter-wise computes the global model as the average of the $\beta = M - 4f$ models closest to the median of all models in the selected set.

    \item[FoolsGold] FoolsGold~\cite{fung_limitations_2020} detects colluding clients by measuring the cosine similarity between updates, assuming that colluding clients will submit similar updates.
    It keeps track of a per-client global learning rate $\alpha_i$ and reduces it for those with high similarity to limit their influence on the global model.
\end{description}

\section{Attack/Defense Comparison}
\label{app:attacks_vs_defenses}

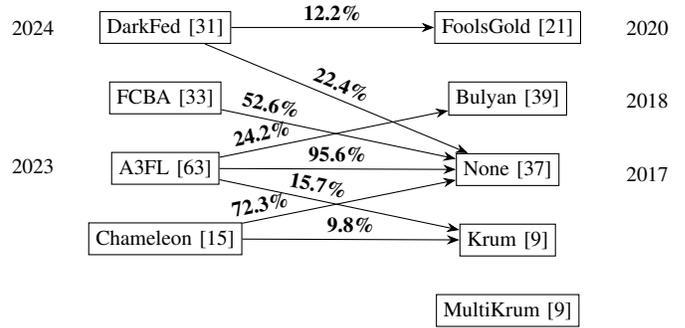
\begin{figure}[tb]
    \footnotesize

    \begin{tikzpicture}[
        node distance=0.5cm,
        every node/.style={rectangle, draw, align=center},
        every edge/.style={draw, -{Stealth[scale=1.0]}}, %
        edgelabel/.style={draw=none, rectangle=none},
        arrow/.style={->, >=Stealth} %
    ]

    \node (darkfed) {DarkFed~\cite{DBLP:conf/ijcai/LiWNHXZW24}};
    \node (fcba) [below = of darkfed] {FCBA~\cite{DBLP:conf/aaai/LiuZFYXM024}};
    \node (a3fl) [below = of fcba] {A3FL~\cite{zhang_a3fl_2023}};
    \node (chameleon) [below = of a3fl] {Chameleon~\cite{dai_chameleon_2023}};

    \node (foolsgold) [right = 2.7cm of darkfed] {FoolsGold~\cite{fung_limitations_2020}};
    \node (bulyan) [below = of foolsgold] {Bulyan~\cite{mhamdi_hidden_2018}};
    \node (none) [below = of bulyan] {None~\cite{mcmahan_communication-efficient_2017}};
    \node (krum) [below = of none] {Krum~\cite{blanchard_machine_2017}};
    \node (multikrum) [below = of krum] {MultiKrum~\cite{blanchard_machine_2017}};
    
    \node[draw=none] (2024l) [left = 0.5cm of darkfed] {2024};
    \node[draw=none] (2023l) [below = 1.45cm of 2024l] {2023};
    
    \node[draw=none] (2020r) [right = 0.5cm of foolsgold] {2020};
    \node[draw=none] (2018r) [below = 0.58cm of 2020r] {2018};
    \node[draw=none] (2017r) [below = 0.58cm of 2018r] {2017};
    
    \newcommand{\drawedge}[5]{%
        \pgfmathsetmacro{\percent}{#3}
        \pgfmathsetmacro{\normval}{(min(1, max(0.1, \percent / 100)) - 0.1) * 10/9}
        \pgfmathsetmacro{\fractionred}{190 /255 * min(1, 2 * \normval)}
        \pgfmathsetmacro{\fractiongreen}{153 /255 * min(1, -2 * \normval + 2)}
        \definecolor{edgecolor}{rgb}{\fractionred, \fractiongreen, 0} %
        \draw [arrow] (#1) to node[pos=#4, sloped, above, edgelabel] {\textbf{#3\%}} (#2);
    }

    \drawedge{a3fl}{krum}{15.7}{0.4}{}
    \drawedge{a3fl}{none}{95.6}{0.5}{}
    \drawedge{a3fl}{bulyan}{24.2}{0.2}{}
    \drawedge{chameleon}{none}{72.3}{0.1}{}
    \drawedge{chameleon}{krum}{9.8}{0.5}{}
    \drawedge{fcba}{none}{52.6}{0.2}{}
    \drawedge{darkfed}{none}{22.4}{0.5}{}
    \drawedge{darkfed}{foolsgold}{12.2}{0.5}{}
    
    \end{tikzpicture}
    \centering
    \caption{Overview of which selected attacks were evaluated against which of the selected defenses in their original papers, alongside the resulting \bda values in our system and threat models.}
    \label{fig:attack_defense_comparison}
\end{figure}

To better understand how the selected attacks (cf. Appendix~\ref{app:RW_chain}) fare with the selected defenses (cf. Appendix~\ref{app:defense_selection}), Figure~\ref{fig:attack_defense_comparison} visualizes which attacks were evaluated under which defenses in the original (attack) papers. 
It also shows the resulting BDA values of this comparison, as determined using our evaluation framework, system, and threat model to ensure comparability.

\section{Notes on the Evaluation of FCBA, DarkFed, and Chameleon}
\label{app:changed_attacks}

\begin{figure}
    \centering
    \subfloat[\mta \label{fig:fcba_changes_MTA}]{
        \centering
        \includegraphics[width=0.45\linewidth]{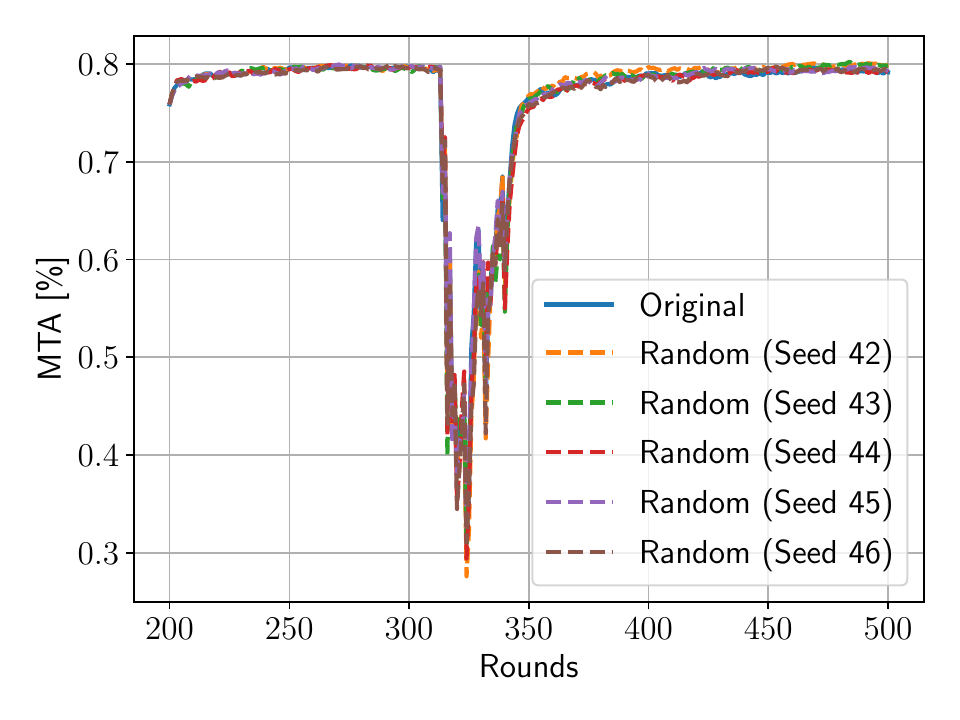}
    }
    \hfill
    \subfloat[\bda \label{fig:fcba_changes_BDA}]{
        \centering
        \includegraphics[width=0.45\linewidth]{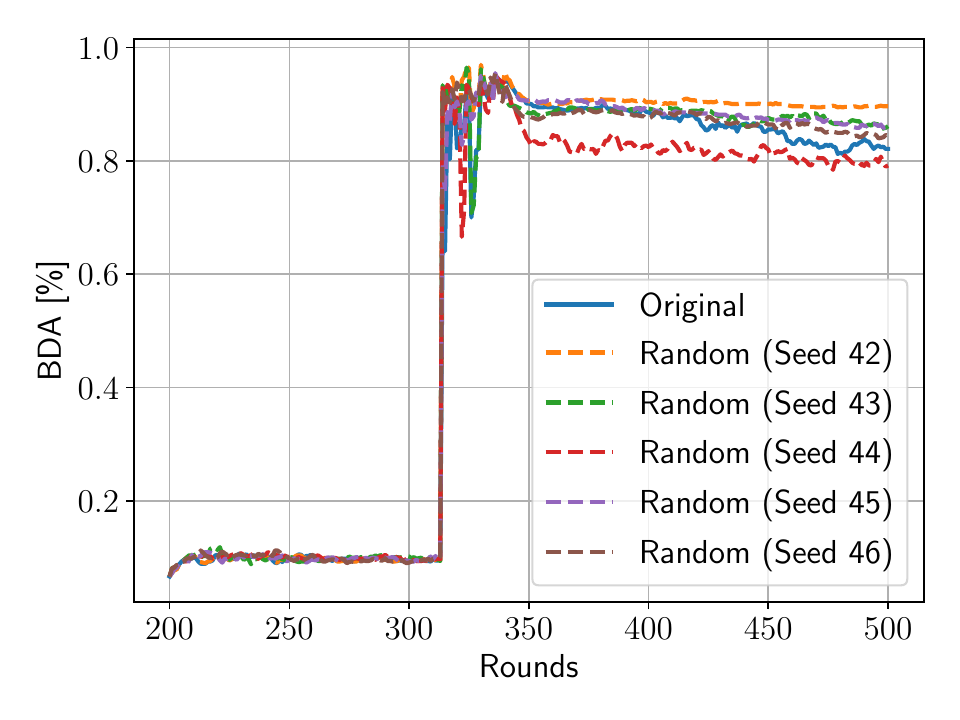}
    }
    \caption{FCBA with malicious clients using randomized (instead of individually assigned)  triggers.}
    \label{fig:fcba_changes}
\end{figure}

\subheading{FCBA.} 
The FCBA attack~\cite{DBLP:conf/aaai/LiuZFYXM024} splits the global trigger into four parts and assigns each malicious client a unique combination of these sub-triggers, resulting in 14 distinct local triggers and, consequently, 14 malicious clients.
As our setup assumes only 10\% malicious clients (i.e., 10 out of 100), assigning a unique trigger to each client is not feasible in this setting. Instead, we opted to allow each active malicious client to randomly select one of the 14 predefined triggers during training.
Figure~\ref{fig:fcba_changes} shows the backdoor accuracy under this randomized trigger selection in the original FCBA setting (with 14 malicious clients) compared to the original setup. 
To account for randomness in trigger assignment, we report results across five different PRNG seeds.
As shown in Figure~\ref{fig:fcba_changes}, this modification does not impact the \mta and only has a slight impact on the \bda, which slightly increases in 4 out of 5 cases.

Further, the FCBA attack considers single-shot adversaries that scale their model updates to perform complete model replacement as introduced in \cite{bagdasaryan_how_2020} when they are active~\cite{DBLP:conf/aaai/LiuZFYXM024}.
In our threat model, more than one malicious client can be active per round. 
If multiple clients perform full model replacement, however, we empirically observed a very high chance of exploding gradients, preventing further training.
Thus, we modified the FCBA attack---as it is common for model replacement-based attacks~\cite{bagdasaryan_how_2020}---to divide the scaling factor by the number of malicious clients that are active in that round.

\subheading{DarkFed.}
The DarkFed attack, in its original version, employs a version of FedAvg that divides the current sum of model updates by the number of total clients $N$ instead of the number of clients per round $M$. 
This version stems from a mistake in the original FedAvg paper~\cite{mcmahan_communication-efficient_2017} and was fixed in a later revision~\cite{mcmahan_communication-efficient_2023}.
We use the corrected version and also apply this to the results of the DarkFed attack under its original hyperparameters to ensure comparability.

\subheading{Chameleon.}
In its original setup, the Chameleon~\cite{dai_chameleon_2023} attack uses a learning rate schedule where the benign learning rate increases from $0.001$ to $0.2$ over the first 500 rounds, then decreases linearly until round 2000 or until reaching a minimum of $0.0001$.
To ensure a fair comparison, we, therefore, adapted this learning rate schedule by compressing it to 1000 rounds, i.e., we assume that $\eta_b$ increases from $0.001$ to $0.2$ in the first 250 rounds and then decreases linearly until round $1000$ (twice as steep as in the original schedule).

\section{Diversifying Hyperparameters across Clients}
\label{app:diversifying_hyperparams}

\begin{figure}
    \centering
    \includegraphics[width=0.7\linewidth]{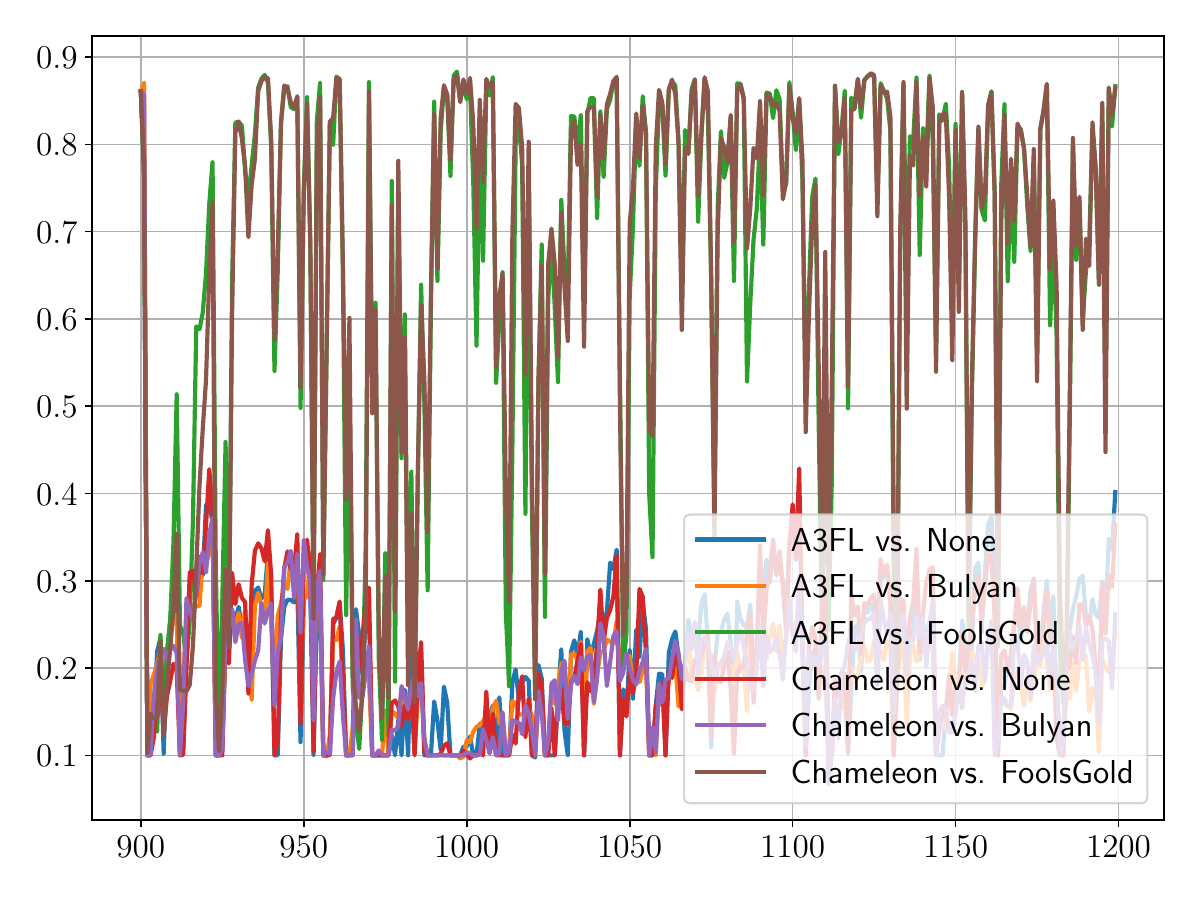}
    \caption{\mta for different attack/defense combinations when diversifying hyperparameters randomly across clients.}
    \label{fig:diversifying_hyperparams}
\end{figure}

Figure~\ref{fig:diversifying_hyperparams} shows the \mta and \bda for the A3FL~\cite{zhang_a3fl_2023} and Chameleon \cite{dai_chameleon_2023} attacks under different aggregation rules and defenses (plain FedAvg~\cite{mcmahan_communication-efficient_2017}, Bulyan \cite{mhamdi_hidden_2018}, and FoolsGold~\cite{fung_limitations_2020}) when diversifying hyperparameters across clients.
We observe that in four out of six cases, convergence is prevented completely with average {\mta}s not exceeding $25\%$.
We attribute this to the excessive heterogeneity between model updates when diversifying local training hyperparameters, leading to unstable training dynamics~\cite{mcmahan_communication-efficient_2017}.

\section{Recommended Benign Configurations}
\label{app:recommended_configs}

In Table~\ref{tab:best_hyperparams_no_defense}, we complement our results from Table~\ref{tab:best_hyperparams_with_defenses} in Section~\ref{sub:detailed_results} by showing the achieved \mta, \mtaafter, \bda, \bdaafter, and \lifespan for all considered attack-defense combinations for three hyperparameter combinations: (a) when using the benign and malicious parameters as reported in the original papers (denoted by "{}Orig."{}), (b) when using our recommended benign parameters and the ``Greedy'' adaptive strategy outlined in Section~\ref{sec:optimal_adversary} (with both MTA-constrained ($\mta_\text{c}$) and unconstrained adversaries $\bda_\text{m}$), and (c) when using our recommended benign parameters and the ``\nsga'' adaptive strategy outlined in Section~\ref{sec:optimal_adversary} (with both MTA-constrained and unconstrained adversaries).

Overall, our measurements suggest that, even in combination with state-of-the-art defenses, our proposed benign hyperparameter setting decreases \bda and \bdaafter compared to the original settings significantly. 
Notably, we observe substantial \lifespan decreases of at least $94\%$ across all settings, underscoring the strong positive effect of benign hyperparameters.

\setlength\tabcolsep{2.9pt} %
\begin{table*}
    \centering
    \scriptsize
    \scalebox{0.71}{
        \begin{tabular}{|l|cc|c|cc|c|cc|c|cc|c|cc|c|}
        \hline
         & \multicolumn{3}{c|}{\mta} & \multicolumn{3}{c|}{\bda} & \multicolumn{3}{c|}{\mtaafter} & \multicolumn{3}{c|}{\bdaafter} & \multicolumn{3}{c|}{\lifespan} \\\cline{2 - 16}
        \textbf{Benign param.} & \multicolumn{2}{c|}{Recommended} & Orig. & \multicolumn{2}{c|}{Recommended} & Orig. & \multicolumn{2}{c|}{Recommended} & Orig. & \multicolumn{2}{c|}{Recommended} & Orig. & \multicolumn{2}{c|}{Recommended} & Orig. \\\cline{2 - 16}
        \textbf{Malicious param.} & Greedy & \nsga & Orig. & Greedy & \nsga & Orig. & Greedy & \nsga & Orig. & Greedy & \nsga & Orig. & Greedy & \nsga & Orig. \\
        \textbf{Attack/Defense} & $\bda_\text{m}$ / $\mta_\text{c}$ & $\bda_\text{m}$ / $\mta_\text{c}$  &  & $\bda_\text{m}$ / $\mta_\text{c}$  & $\bda_\text{m}$ / $\mta_\text{c}$  &  & $\bda_\text{m}$ / $\mta_\text{c}$  & $\bda_\text{m}$ / $\mta_\text{c}$  &  & $\bda_\text{m}$ / $\mta_\text{c}$  & $\bda_\text{m}$ / $\mta_\text{c}$  &  & $\bda_\text{m}$ / $\mta_\text{c}$  & $\bda_\text{m}$ / $\mta_\text{c}$  &  \\
        \hline
        A3FL / Bulyan & 80.9 \% / 81.2 \% & 81.3 \% & 85.2 \% & 29.6 \% / 37.8 \% & 86.6 \% & 45.3 \% & 84.3 \% / 84.5 \% & 84.5 \% & 84.7 \% & 12.8 \% / 12.3 \% & 19.6 \% & 40.6 \% & 0 & 3 & 51 \\
        A3FL / FoolsGold & 85.1 \% & 85.5 \% & 85.4 \% & 52.5 \% & 72.8 \% & 95.0 \% & 87.3 \% & 87.3 \% & 85.8 \% & 13.3 \% & 19.9 \% & 65.0 \% & 0 & 8 & 994 \\
        A3FL / Krum & 60.3 \% / N.A. & 55.2 \% / N.A. & 73.8 \% & 7.5 \% / N.A. & 57.4 \% / N.A. & 15.7 \% & 65.3 \% / N.A. & 65.8 \% / N.A. & 75.9 \% & 7.0 \% / N.A. & 12.9 \% / N.A. & 20.5 \% & 0 / N.A. & 0 / N.A. & 0 \\
        A3FL / MultiKrum & 83.4 \% & 83.7 \% & 87.2 \% & 40.9 \% & 57.4 \% & 83.1 \% & 86.5 \% & 86.5 \% & 87.1 \% & 12.9 \% & 17.6 \% & 47.3 \% & 0 & 0 & 354 \\
        A3FL / None & 83.4 \% & 79.8 \% / 84.0 \% & 86.9 \% & 81.9 \% / 80.2 \% & 85.3 \% / 74.7 \% & 95.6 \% & 86.9 \% & 86.8 \% & 86.9 \% & 16.7 \% / 18.8 \% & 10.8 \% / 23.0 \% & 74.5 \% & 13 & 3 / 14 & 1000 \\
        Chameleon / Bulyan & 82.4 \% / 82.2 \% & 82.1 \% & 88.4 \% & 18.3 \% / 18.9 \% & 27.3 \% & 9.5 \% & 84.5 \% & 84.5 \% & 88.9 \% & 10.3 \% / 10.4 \% & 10.3 \% & 9.5 \% & 0 & 0 & 0 \\
        Chameleon / FoolsGold & 85.5 \% & 85.4 \% & 88.6 \% & 10.6 \% & 11.6 \% & 71.5 \% & 87.3 \% & 87.3 \% & 90.5 \% & 10.4 \% & 10.4 \% & 46.5 \% & 0 & 0 & 383 \\
        Chameleon / Krum & 61.6 \% / N.A. & 61.7 \% / N.A. & 75.4 \% & 18.3 \% / N.A. & 19.6 \% / N.A. & 9.8 \% & 65.3 \% / N.A. & 65.2 \% / N.A. & 78.4 \% & 10.9 \% / N.A. & 11.1 \% / N.A. & 10.0 \% & 0 / N.A. & 0 / N.A. & 0 \\
        Chameleon / MultiKrum & 84.3 \% & 84.4 \% & 89.2 \% & 11.0 \% & 15.1 \% & 29.3 \% & 86.4 \% & 86.5 \% & 90.2 \% & 10.4 \% & 10.4 \% & 18.3 \% & 0 & 0 & 56 \\
        Chameleon / None & 84.4 \% & 84.7 \% & 88.9 \% & 10.6 \% / 10.4 \% & 12.3 \% & 72.3 \% & 86.8 \% & 86.9 \% & 90.5 \% & 10.3 \% / 10.2 \% & 10.3 \% & 40.1 \% & 0 & 0 & 312 \\
        DarkFed / Bulyan & 80.7 \% / 80.6 \% & 81.1 \% & 81.0 \% & 11.2 \% & 11.4 \% & 11.2 \% & 84.6 \% / 84.5 \% & 84.5 \% & 83.0 \% & 10.4 \% / 10.8 \% & 10.6 \% & 10.2 \% & 0 & 0 & 0 \\
        DarkFed / FoolsGold & 16.6 \% / N.A. & 85.4 \% & 84.9 \% & 1.0 \% / N.A. & 12.7 \% & 12.2 \% & 10.0 \% / N.A. & 87.3 \% & 85.4 \% & 0.0 \% / N.A. & 10.1 \% & 10.1 \% & 0 / N.A. & 0 & 0 \\
        DarkFed / Krum & 22.3 \% / N.A. & 60.5 \% / N.A. & 50.0 \% & 2.6 \% / N.A. & 10.7 \% / N.A. & 10.9 \% & 10.0 \% / N.A. & 65.1 \% / N.A. & 55.1 \% & 0.0 \% / N.A. & 11.3 \% / N.A. & 10.8 \% & 0 / N.A. & 0 / N.A. & 0 \\
        DarkFed / MultiKrum & 24.6 \% / N.A. & 83.7 \% & 83.2 \% & 2.6 \% / N.A. & 16.8 \% & 14.1 \% & 10.0 \% / N.A. & 86.4 \% & 84.0 \% & 0.0 \% / N.A. & 10.5 \% & 10.1 \% & 0 / N.A. & 0 & 0 \\
        DarkFed / None & 11.4 \% / N.A. & 84.3 \% & 84.2 \% & 0.2 \% / N.A. & 16.7 \% & 22.4 \% & 10.0 \% / N.A. & 86.9 \% & 85.1 \% & 0.0 \% / N.A. & 10.3 \% & 9.9 \% & 0 / N.A. & 0 & 0 \\
        FCBA / Bulyan & 81.3 \% / N.A. & 81.5 \% & 80.6 \% & 10.2 \% / N.A. & 9.6 \% & 9.6 \% & 84.7 \% / N.A. & 84.5 \% & 82.8 \% & 10.3 \% / N.A. & 10.4 \% & 9.8 \% & 0 / N.A. & 0 & 0 \\
        FCBA / FoolsGold & 85.6 \% / 85.0 \% & 85.2 \% & 82.9 \% & 22.1 \% / 10.4 \% & 23.5 \% & 39.4 \% & 87.3 \% & 87.3 \% & 85.4 \% & 10.2 \% / 9.9 \% & 10.3 \% & 11.5 \% & 0 & 0 & 0 \\
        FCBA / Krum & 61.1 \% / N.A. & 60.6 \% / N.A. & 47.0 \% & 10.5 \% / N.A. & 12.1 \% / N.A. & 9.8 \% & 65.4 \% / N.A. & 65.5 \% / N.A. & 55.2 \% & 11.1 \% / N.A. & 9.7 \% / N.A. & 10.7 \% & 0 / N.A. & 0 / N.A. & 0 \\
        FCBA / MultiKrum & 54.1 \% / N.A. & 61.9 \% / N.A. & 14.2 \% & 10.8 \% / N.A. & 27.2 \% / N.A. & 7.0 \% & 83.2 \% / N.A. & 83.6 \% / N.A. & 63.9 \% & 9.7 \% / N.A. & 10.0 \% / N.A. & 9.9 \% & 0 / N.A. & 0 / N.A. & 0 \\
        FCBA / None & 11.8 \% / N.A. & 11.6 \% / N.A. & 63.3 \% & 5.2 \% / N.A. & 3.4 \% / N.A. & 52.6 \% & 70.1 \% / N.A. & 71.0 \% / N.A. & 83.5 \% & 9.8 \% / N.A. & 8.7 \% / N.A. & 35.8 \% & 0 / N.A. & 0 / N.A. & 237 \\
        \hline
        \textit{Average} & 62.0 \% / 83.5 \% & 74.0 \% / 83.7 \% & 76.0 \% & 17.9 \% / 28.4 \% & 29.5 \% / 32.0 \% & 35.8 \% & 66.8 \% / 86.2 \% & 81.2 \% / 86.2 \% & 81.1 \% & 8.8 \% / 11.9 \% & 11.8 \% / 13.1 \% & 25.1 \% & 0.7 / 1.3 & 0.7 / 1.8 & 169.4 \\
        \textit{Average w/ defenses} & 65.6 \% / 83.4 \% & 76.2 \% / 83.6 \% & 74.8 \% & 16.2 \% / 24.2 \% & 29.5 \% / 31.3 \% & 29.6 \% & 67.6 \% / 86.0 \% & 80.7 \% / 86.1 \% & 79.8 \% & 8.7 \% / 11.3 \% & 12.2 \% / 12.7 \% & 21.3 \% & 0.0 & 0.7 / 1.0 & 114.9 \\
        \textit{Average w/o defenses} & 47.7 \% / 83.9 \% & 65.1 \% / 84.3 \% & 80.8 \% & 24.5 \% / 45.3 \% & 29.4 \% / 34.5 \% & 60.7 \% & 63.5 \% / 86.8 \% & 82.9 \% / 86.9 \% & 86.5 \% & 9.2 \% / 14.5 \% & 10.0 \% / 14.5 \% & 40.1 \% & 3.25 / 6.5 & 0.8 / 4.7 & 387.3 \\
        \hline
        \end{tabular}
    }
    \caption{Full results for \mta, \bda, \mtaafter, \bdaafter, and \lifespan for attack/defense combination using our  ``recommended'' benign hyperparameters ($\eta_b = 0.15, \mu_b = 0.9, \lambda_b = 0.0005, E_b = 10, B_b = 32$) under our greedy and stochastic adaptive adversaries that aim to maximize BDA ($\bda_\text{m}$), or are MTA-constrained  ($\mta_\text{c}$) within $\epsilon_{adv}=5\%$, and the original malicious/benign parameters (``Orig.''). We include one value in the cell ``$\bda_\text{m}/\mta_\text{c}$'' when these strategies converge on the same result.}
    \label{tab:best_hyperparams_no_defense}
\end{table*}

\setlength\tabcolsep{6pt} %

\section{Hyperparameter choices for adaptive adversary}
\label{app:optimal_adversary}

A determined adaptive adversary can leverage our analysis from Section~\ref{sec:lr_effect} and \ref{sec:other_params} to determine malicious hyperparameter choices that maximize the $\bda$ under a given benign hyperparameter choice. 
Table~\ref{tab:optimal_adversary} shows these adaptive parameter choices for the greedy $\bda_\text{max}$ adversary as determined based on our empirical results from Sections~\ref{sec:lr_effect} and \ref{sec:other_params}.
Using this approach, an adversary can tweak the malicious hyperparameters on a per-parameter basis. 

\begin{table}
    \centering
    \footnotesize
    \scalebox{0.87}{\begin{tabular}{cc|cccc}
    \hline
         \multicolumn{2}{c|}{\multirow{2}{*}{\textbf{Benign values}}} & \multicolumn{4}{c}{\textbf{Adaptive malicious values}} \\
         & & A3FL~\cite{zhang_a3fl_2023}  & Chameleon~\cite{dai_chameleon_2023} & DarkFed~\cite{DBLP:conf/ijcai/LiWNHXZW24}  & FCBA~\cite{DBLP:conf/aaai/LiuZFYXM024} \\ \hline
        \multirow[c]{4}{*}{$\eta$} & 0.05 & 0.1 & 0.5 & 0.25 & 0.1 \\
         & 0.1 & 0.1 & 0.5 & 0.2 & 0.1 (0.02)\\
         & 0.2 & 0.4 & 1 & 0.4 & 0.4 \\
         & 0.5 & 1 & 0.25 & 0.5 & 0.5 \\\hline
        \multirow[c]{4}{*}{$E$} & 2 & 20 & 2 & 10 & 10 \\
         & 5 & 20 & 2 & 5 & 10 \\
         & 10 & 5 & 2 & 5 & 10 \\
         & 20 & 10 & 2 & 2 & 10 \\\hline
        \multirow[c]{3}{*}{$B$} & 32 & 32 & 32 & 32 & 64 (64)\\
         & 64 & 32 & 32 & 32 & 128 \\
         & 128 & 32 & 32 & 32 & 128 \\\hline
        \multirow[c]{3}{*}{$\lambda$} & 0.0001 & 0.001 & 0.0005 & 0.0001 & 0.0001 \\
         & 0.0005 & 0.0005 & 0.0005 & 0.0001 & 0.0001 \\
         & 0.001 & 0.0001 & 0.0001 & 0.0001 & 0.0001 \\
        \hline
    \end{tabular}}
    \caption{Example of malicious hyperparameters for the greedy $\bda_\text{max}$ adversary. Here, we only show those significant hyperparameters outlined in Table~\ref{tab:regression}.} 
    \label{tab:optimal_adversary}
\end{table}

\section{Applicability to other Datasets, Models, and Learning rate schedules}

\subsection{Impact of Learning Rate on Attack Success for Uniform Schedule}
\label{app:lr_uniform}

\begin{figure}
    \centering
    \subfloat[A3FL~\cite{zhang_a3fl_2023} \label{fig:lr_effect_a3fl_bda_during_uniform}]{
        \centering
        \includegraphics[width=0.45\linewidth]{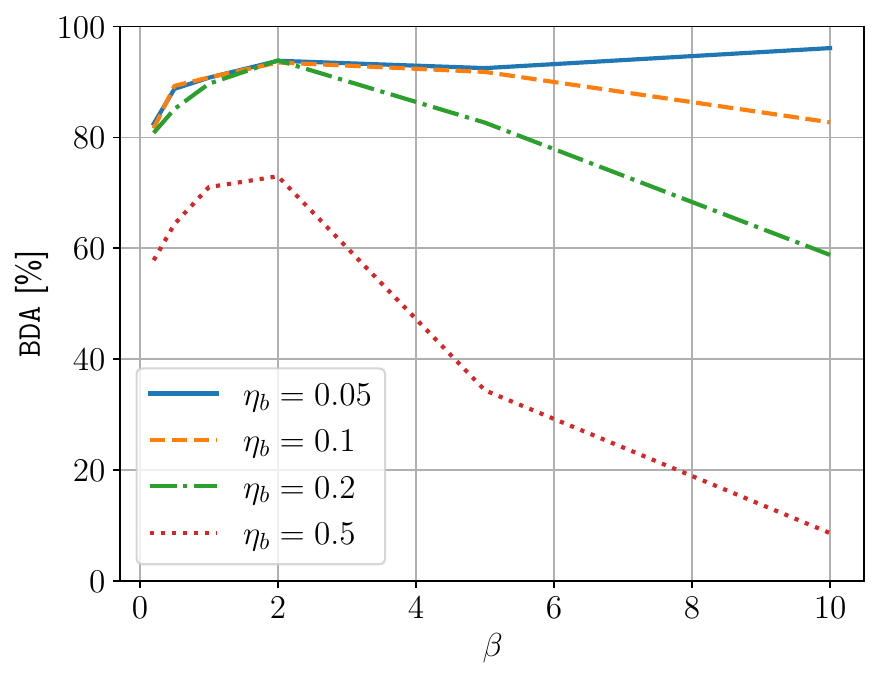}
    }
    \hfill
    \subfloat[Chameleon~\cite{dai_chameleon_2023} \label{fig:lr_effect_chameleon_bda_during_uniform}]{
        \centering
        \includegraphics[width=0.45\linewidth]{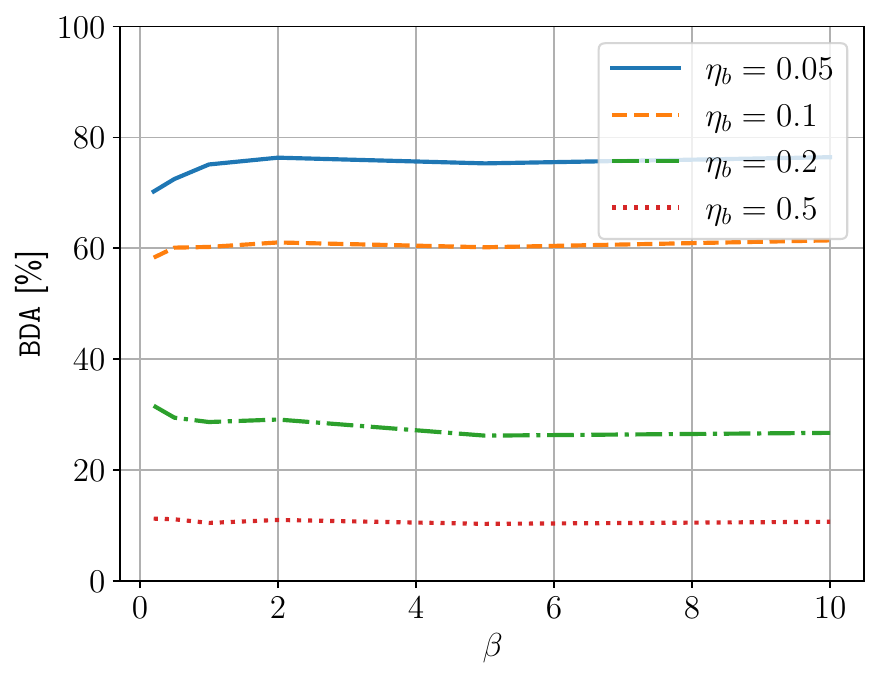}
    }

    \vspace{0.6em}
    
    \subfloat[DarkFed~\cite{DBLP:conf/ijcai/LiWNHXZW24} \label{fig:lr_effect_darkfed_bda_during_uniform}]{
        \centering
        \includegraphics[width=0.45\linewidth]{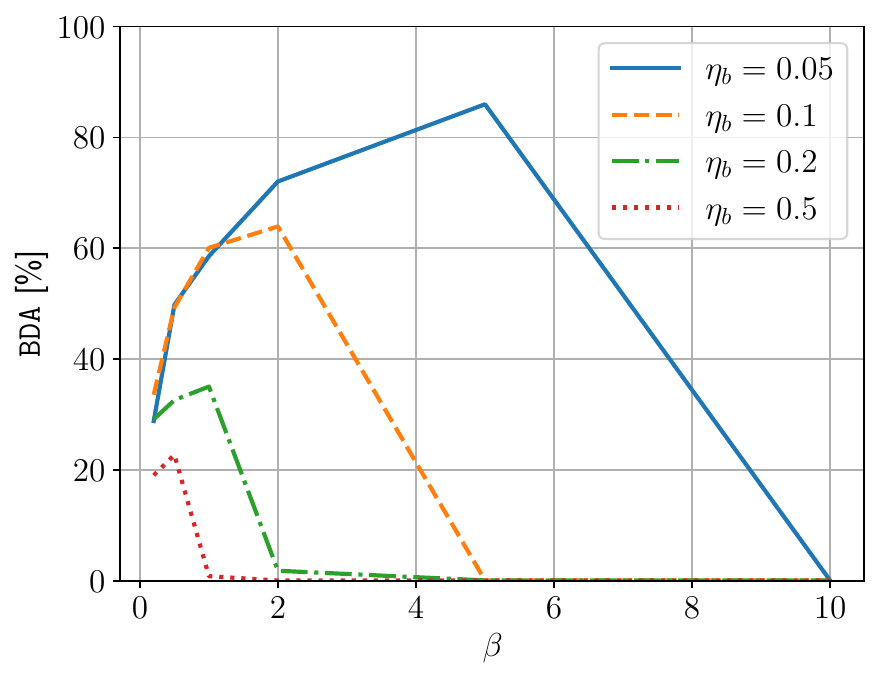}
    }
    \hfill
    \subfloat[FCBA~\cite{DBLP:conf/aaai/LiuZFYXM024} \label{fig:lr_effect_fcba_bda_during_uniform}]{
        \centering
        \includegraphics[width=0.45\linewidth]{figures/results/LR_-_FCBA__uniform__BDA_during.pdf}
    }

    \caption{Impact of $\eta_b$ and $\beta$ on the \bda of SoTA attacks if a uniform learning schedule is applied.}
    \label{fig:lr_effect_bda_during_uniform}
\end{figure}

\begin{figure}
    \centering
    \subfloat[A3FL~\cite{zhang_a3fl_2023} \label{fig:lr_effect_a3fl_bda_after_uniform}]{
        \centering
        \includegraphics[width=0.45\linewidth]{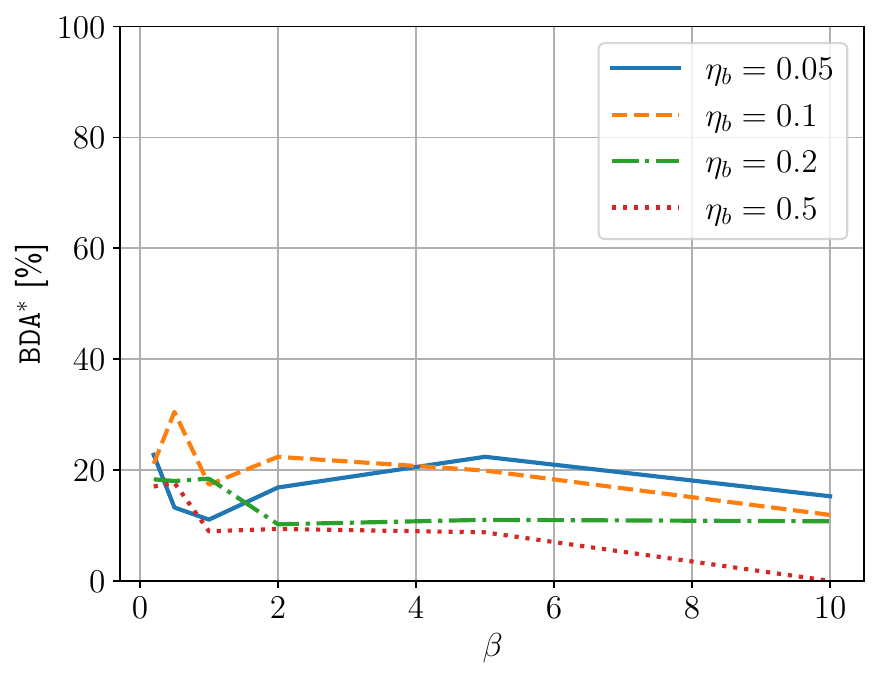}
    }
    \hfill
    \subfloat[Chameleon~\cite{dai_chameleon_2023} \label{fig:lr_effect_chameleon_bda_after_uniform}]{
        \centering
        \includegraphics[width=0.45\linewidth]{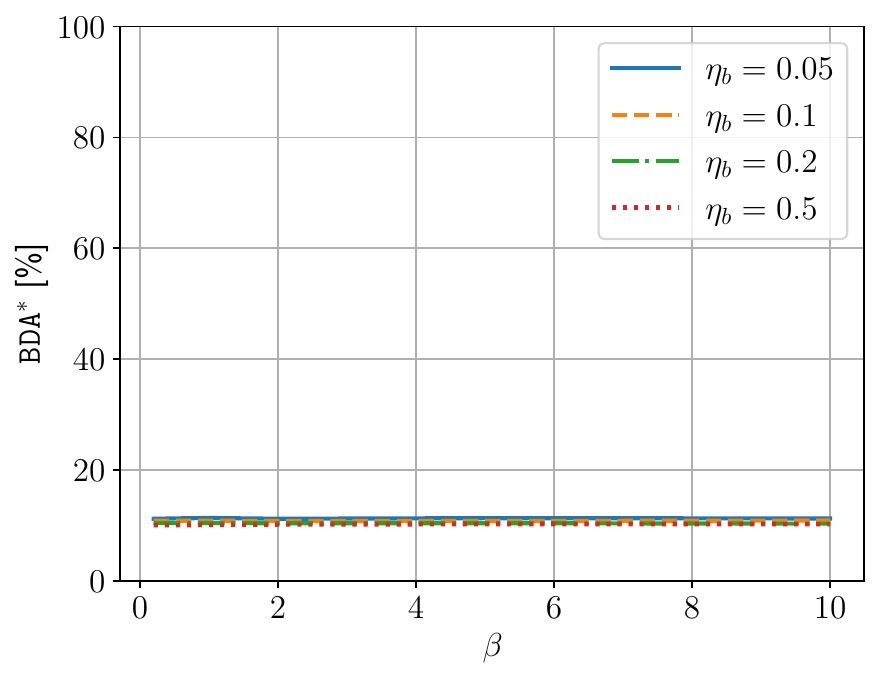}
    }

    \vspace{0.6em}
    
    \subfloat[DarkFed~\cite{DBLP:conf/ijcai/LiWNHXZW24} \label{fig:lr_effect_darkfed_bda_after_uniform}]{
        \centering
        \includegraphics[width=0.45\linewidth]{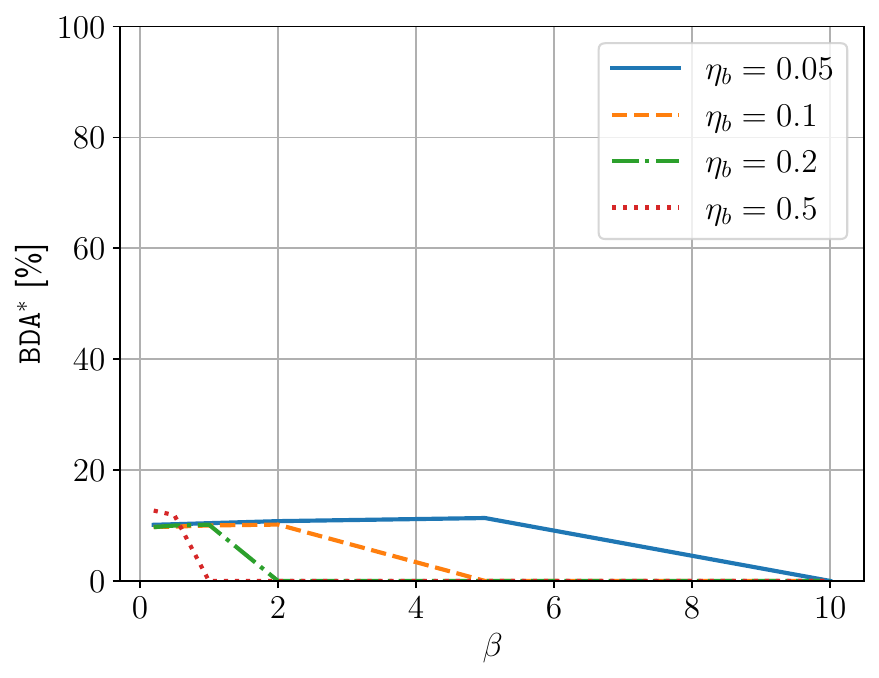}
    }
    \hfill
    \subfloat[FCBA~\cite{DBLP:conf/aaai/LiuZFYXM024} \label{fig:lr_effect_fcba_bda_after_uniform}]{
        \centering
        \includegraphics[width=0.45\linewidth]{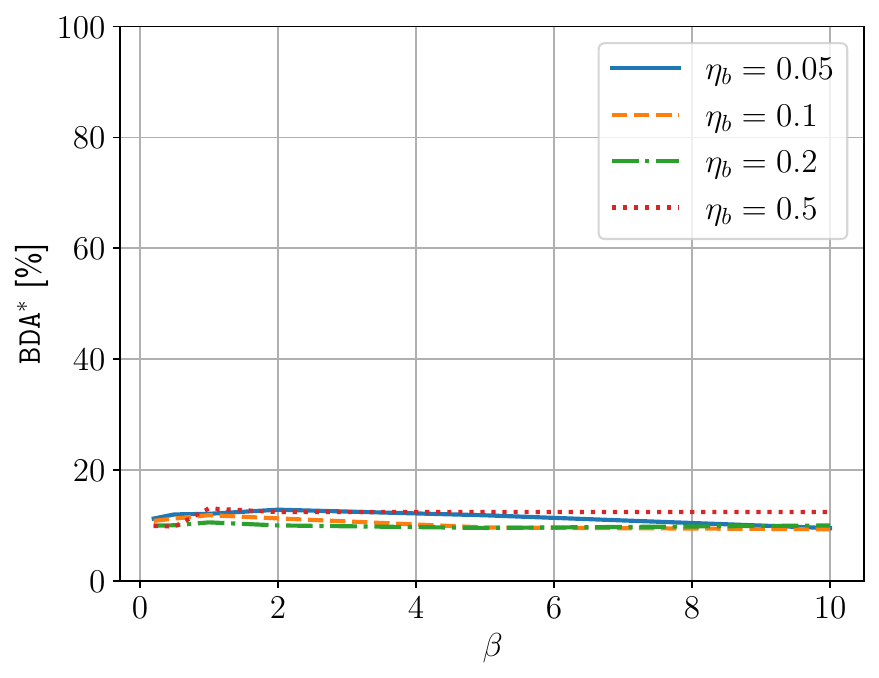}
    }

    \caption{Impact of $\eta_b$ and $\beta$ on the \bdaafter of SoTA attacks if a uniform learning schedule is applied.}
    \label{fig:lr_effect_bda_after_uniform}
\end{figure}

Figure~\ref{fig:lr_effect_bda_during_uniform} shows the \bda, for the A3FL~\cite{zhang_a3fl_2023}, Chameleon~\cite{dai_chameleon_2023}, DarkFed~\cite{DBLP:conf/ijcai/LiWNHXZW24}, and FCBA~\cite{DBLP:conf/aaai/LiuZFYXM024} attacks when using a uniform learning rate schedule.
We notice that the \bda closely mirrors the results for the decaying schedule, with more noticeable effects from increasing the learning rate. Specifically, there are more significant reductions in \bda as both $\beta$ and $\eta_b$ increase. This trend is even more pronounced in the \bdaafter, as shown in Figure~\ref{fig:lr_effect_bda_after_uniform}.
Here, we observe that all attacks do not exceed \bdaafter of $30\%$ while still showing the expected behavior of higher $\eta_b$ reducing the \bdaafter.
This stems from the fact that both the benign and malicious learning rates are much higher in the uniform schedule---namely $\eta_b$ instead of $\eta_b \cdot 0.999^{1000} \approx \eta_b \cdot 0.368$ in the first round of the attack---resulting in the SGD optimization dynamics being much less stable.

\begin{figure}
    \centering
    \subfloat[A3FL~\cite{zhang_a3fl_2023} \label{fig:lr_effect_a3fl_mta_during_uniform}]{
        \centering
        \includegraphics[width=0.45\linewidth]{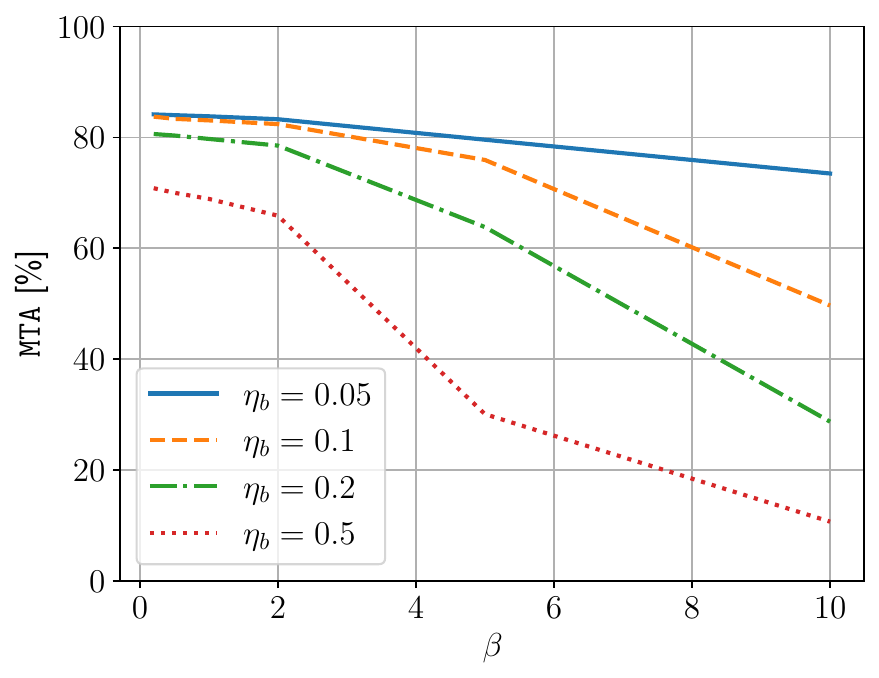}
    }
    \hfill
    \subfloat[Chameleon~\cite{dai_chameleon_2023} \label{fig:lr_effect_chameleon_mta_during_uniform}]{
        \centering
        \includegraphics[width=0.45\linewidth]{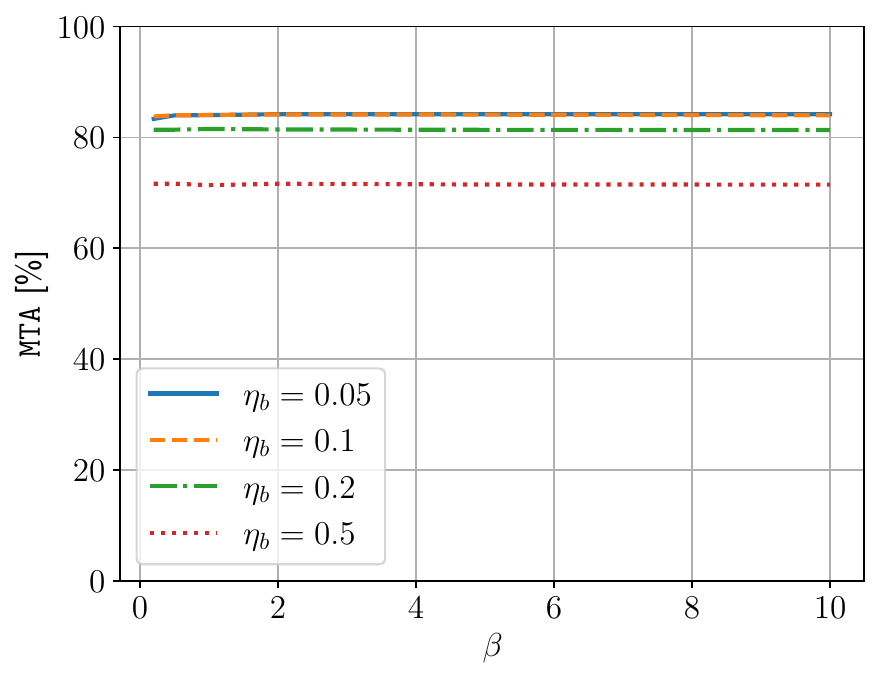}
    }

    \vspace{0.6em}
    
    \subfloat[DarkFed~\cite{DBLP:conf/ijcai/LiWNHXZW24} \label{fig:lr_effect_darkfed_mta_during_uniform}]{
        \centering
        \includegraphics[width=0.45\linewidth]{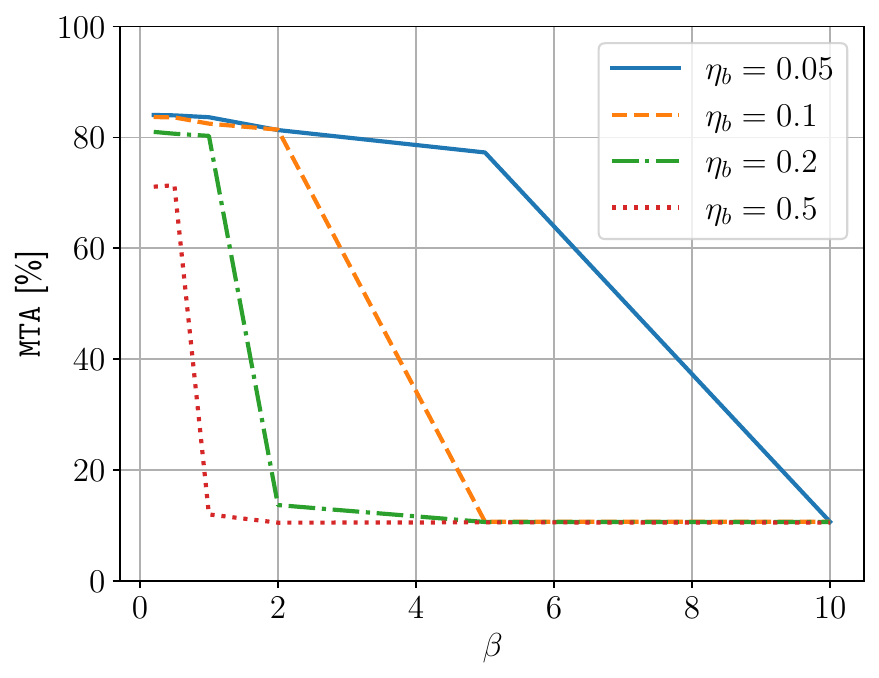}
    }
    \hfill
    \subfloat[FCBA~\cite{DBLP:conf/aaai/LiuZFYXM024} \label{fig:lr_effect_fcba_mta_during_uniform}]{
        \centering
        \includegraphics[width=0.45\linewidth]{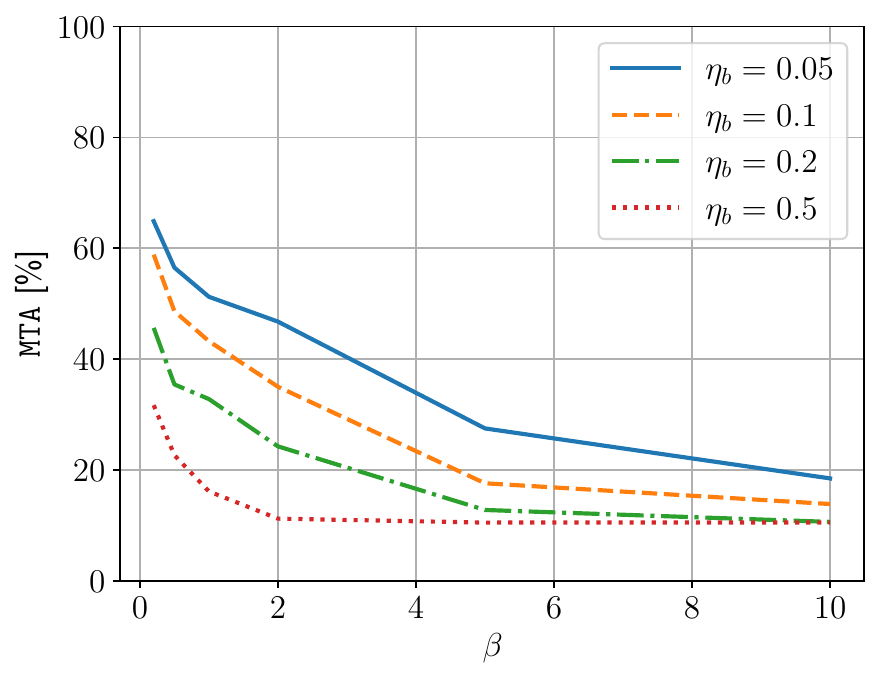}
    }

    \caption{Impact of $\eta_b$ and $\beta$ on the \mta of SoTA attacks if a uniform learning schedule is applied.}
    \label{fig:lr_effect_mta_during_uniform}
\end{figure}

This, however, impacts the \mta negatively, as it is shown in Figure~\ref{fig:lr_effect_mta_during_uniform}.
Again, we observe the same trends as for the decaying learning rate schedule, i.e., a decrease in \mta when increasing $\eta_b$ or $\beta$ but with higher decreases, which is somewhat expected since the learning rate in HFL has to decay over time to ensure proper convergence~\cite{li_convergence_2020}.

\subsection{Ablation on Tiny-Imagenet}
\label{app:lr_imagenet}

\begin{figure}
    \centering
    \subfloat[\bda \label{fig:lr_effect_fcba_imagenet_bda_during}]{
        \centering
        \includegraphics[width=0.45\linewidth]{figures/results/LR_-_FCBA_ImageNet_BDA_during.pdf}
    }
    \hfill
    \subfloat[\bdaafter \label{fig:lr_effect_fcba_imagenet_bda_after}]{
        \centering
        \includegraphics[width=0.45\linewidth]{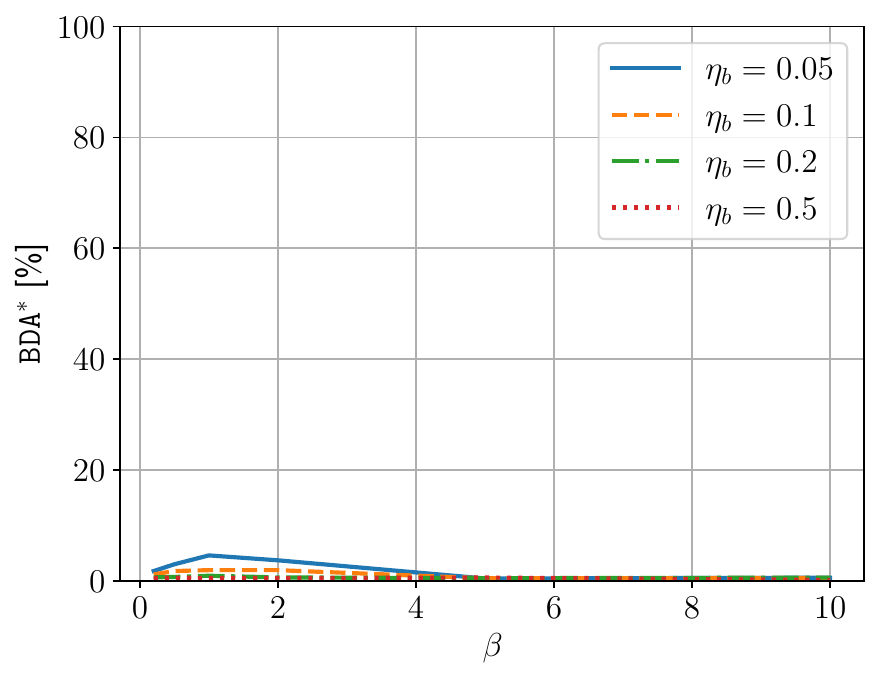}
    }

    \vspace{0.6em}
    
    \subfloat[\lifespan \label{fig:lr_effect_fcba_imagenet_lifespan}]{
        \centering
        \includegraphics[width=0.45\linewidth]{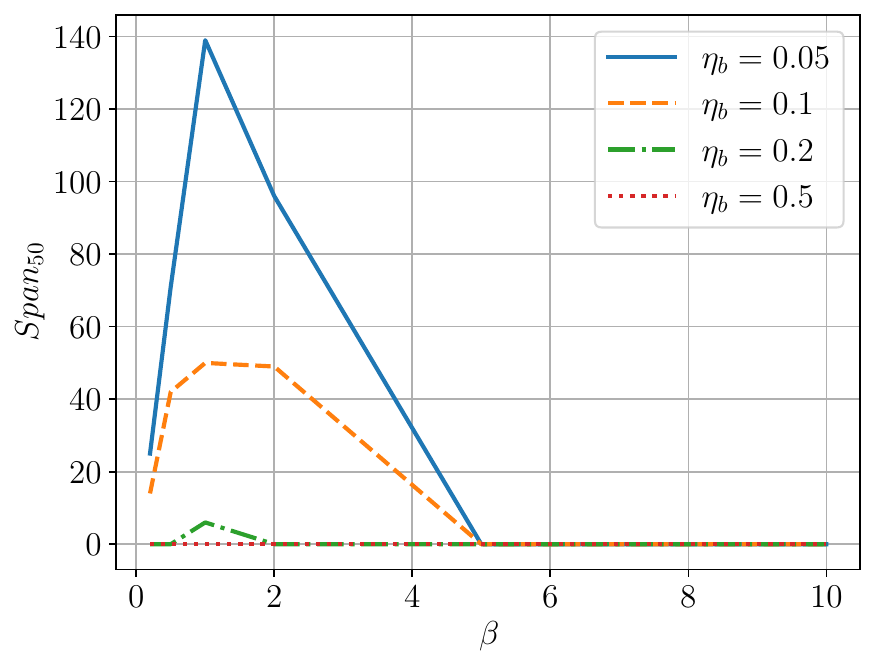}
    }
    \hfill
    \subfloat[\mta \label{fig:lr_effect_fcba_imagenet_mta_during}]{
        \centering
        \includegraphics[width=0.45\linewidth]{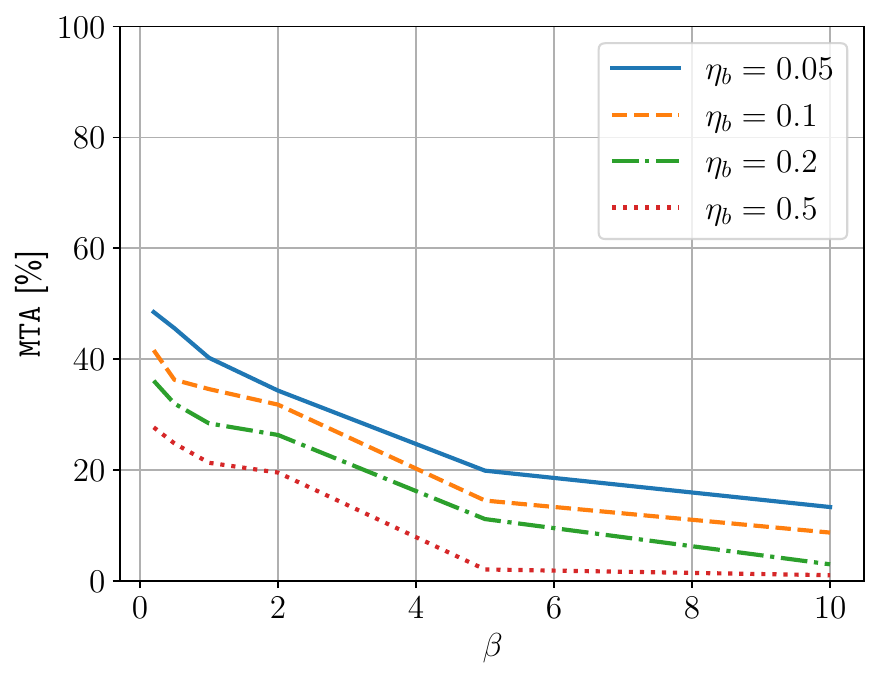}
    }

    \caption{Impact of $\eta_b$ and $\beta$ on the FCBA attack on Tiny-Imagenet.}
    \label{fig:lr_effect_fcba_imagenet}
\end{figure}

Figure~\ref{fig:lr_effect_fcba_imagenet} shows the \bda, \bdaafter, \mta, and \lifespan of the FCBA~\cite{DBLP:conf/aaai/LiuZFYXM024} attack under different benign and malicious learning rates when Tiny-Imagenet~\cite{tinyimagenet} is used as the dataset.
Compared to CIFAR-10~\cite{krizhevsky_learning_2009}, Tiny-Imagenet is a considerably more challenging classification task, comprising $100,000 + 10,000$ $64\times64$ RGB images in $200$ different classes, resulting in $500$ and $50$ images per class in the training and test set, respectively.
Despite the much higher complexity of the classification task, our results in Figure~\ref{fig:lr_effect_fcba_imagenet} indicate that our general findings remain valid. Increasing $\eta_b$ from $0.05$ to $0.5$ decreases the maximum \bda and \bdaafter by $19.5$ percentage points and $4.0$ percentage points, respectively.
Further, we observe an asymptotic behavior with respect to $\beta$ that upper-bounds the attack success, especially in terms of \bdaafter and \lifespan.
As expected, we also observe a decrease of up to $20.8$ percentage points in \mta when increasing $\eta_b$.
This significant decrease is due to Tiny-ImageNet being a more challenging classification task, which generally demands lower learning rates and is more sensitive to changes in the learning rate.

\subsection{Ablation using Different Model Architectures}
\label{app:lr_mobilenet}

To ensure that our findings hold across model architectures of different complexity, we evaluate the effect of modifying $\eta_b$ and $\beta$ for the FCBA~\cite{DBLP:conf/aaai/LiuZFYXM024} attack across two different model architectures: MobileNetV2~\cite{DBLP:conf/cvpr/SandlerHZZC18} (3.5M trainable parameters) and VGG17~\cite{DBLP:journals/corr/SimonyanZ14a} as adapted in \cite{nguyen_iba_2023} (20M trainable parameters).

We note that MobileNetV2 and VGG16 have significantly more trainable parameters than the commonly used ResNet20, demonstrating the applicability of our approach to more complex models.

\begin{figure}
    \centering
    \subfloat[\bda \label{fig:lr_effect_fcba_mobilenet_bda_during}]{
        \centering
        \includegraphics[width=0.45\linewidth]{figures/results/LR_-_FCBA_MobileNetV2_BDA_during.pdf}
    }
    \hfill
    \subfloat[\bdaafter \label{fig:lr_effect_fcba_mobilenet_bda_after}]{
        \centering
        \includegraphics[width=0.45\linewidth]{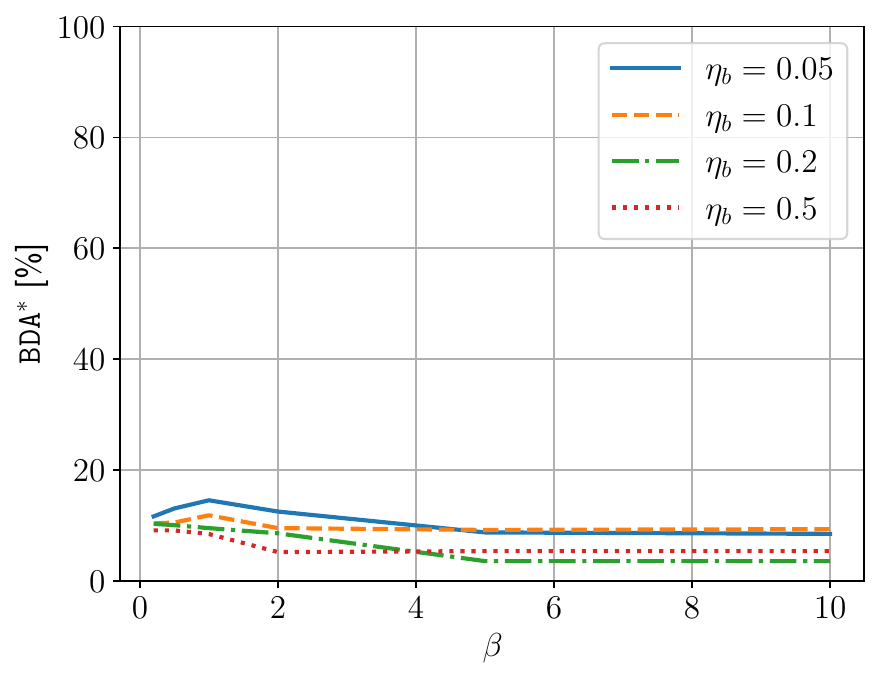}
    }

    \vspace{0.6em}
    
    \subfloat[\lifespan \label{fig:lr_effect_fcba_mobilenet_lifespan}]{
        \centering
        \includegraphics[width=0.45\linewidth]{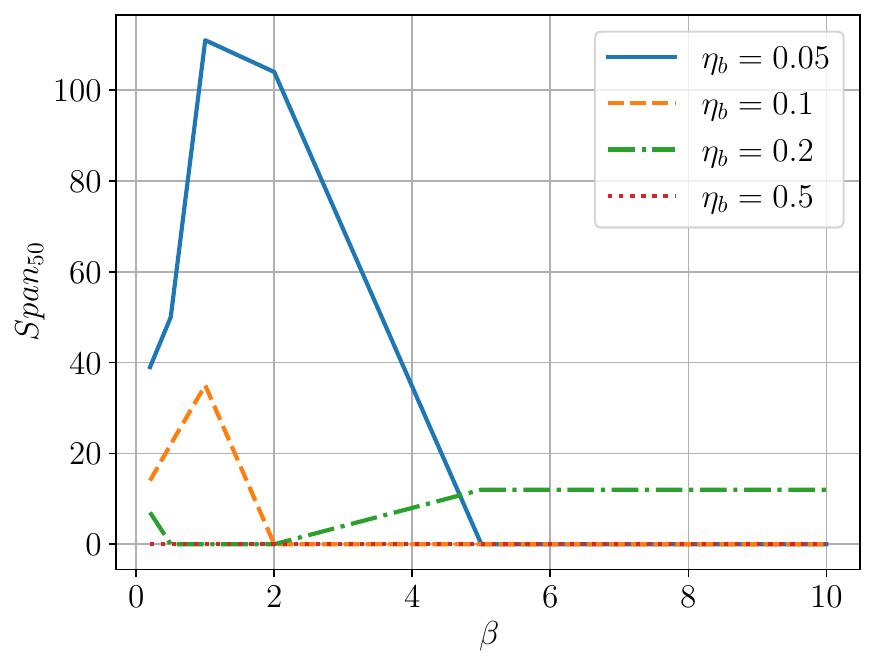}
    }
    \hfill
    \subfloat[\mta \label{fig:lr_effect_fcba_mobilenet_mta_during}]{
        \centering
        \includegraphics[width=0.45\linewidth]{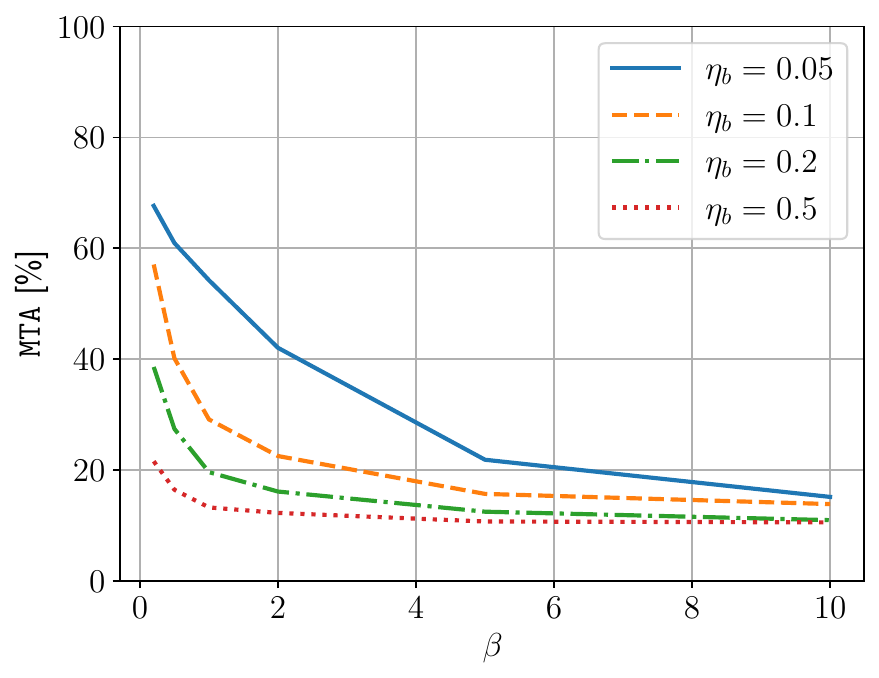}
    }

    \caption{Impact of $\eta_b$ and $\beta$ on the FCBA attack when using MobileNetV2 as the model.}
    \label{fig:lr_effect_fcba_mobilenet}
\end{figure}

\begin{figure}
    \centering
    \subfloat[\bda \label{fig:lr_effect_fcba_vgg17_bda_during}]{
        \centering
        \includegraphics[width=0.45\linewidth]{figures/results/LR_-_FCBA_VGG17_BDA_during.pdf}
    }
    \hfill
    \subfloat[\bdaafter \label{fig:lr_effect_fcba_vgg17_bda_after}]{
        \centering
        \includegraphics[width=0.45\linewidth]{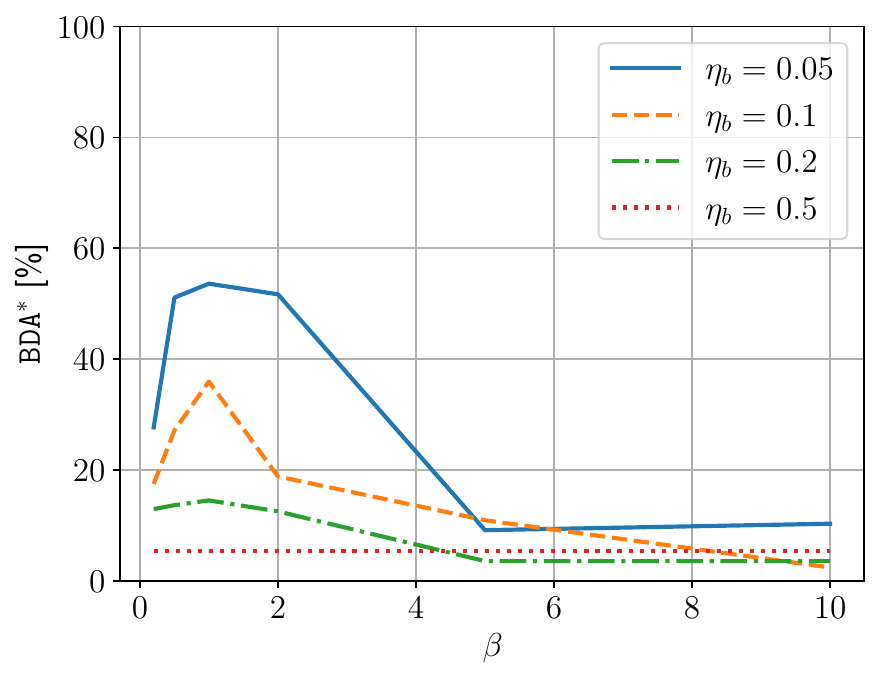}
    }

    \vspace{0.6em}
    
    \subfloat[\lifespan \label{fig:lr_effect_fcba_vgg17_lifespan}]{
        \centering
        \includegraphics[width=0.45\linewidth]{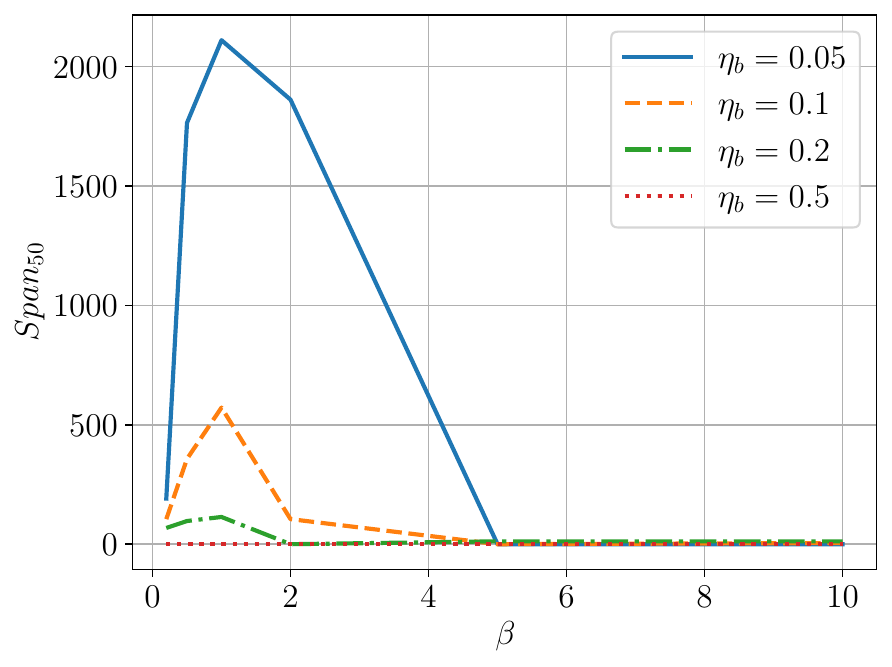}
    }
    \hfill
    \subfloat[\mta \label{fig:lr_effect_fcba_vgg17_mta_during}]{
        \centering
        \includegraphics[width=0.45\linewidth]{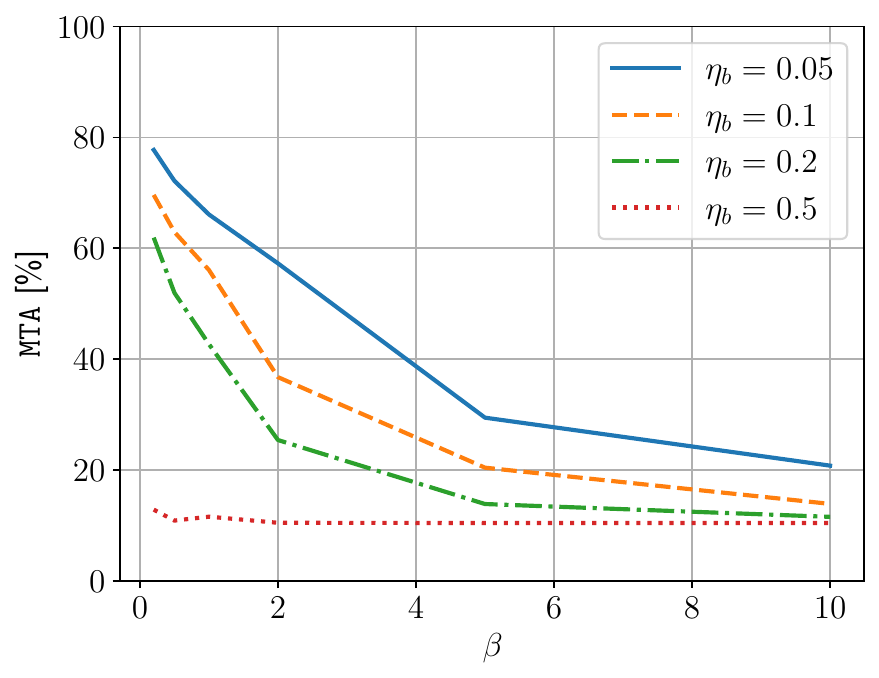}
    }

    \caption{Impact of $\eta_b$ and $\beta$ on the FCBA attack when using VGG17 as the model.}
    \label{fig:lr_effect_fcba_vgg17}
\end{figure}

Our results in terms of \mta, \bda, \bdaafter, and \lifespan for MobileNetV2 and VGG17 are depicted in Figures~\ref{fig:lr_effect_fcba_mobilenet}, \ref{fig:lr_effect_fcba_vgg17}, respectively.
Despite the more complex model architectures, our findings regarding the influence of benign and malicious learning rates on attack success remain consistent.
Decreasing $\eta_b$ from $0.05$ to $0.5$ decreases the maximum \bda by $39.9$ percentage points, and $44.4$ percentage points for MobileNetV2 and VGG17, respectively. 
Here, we also observe an asymptotic behavior with respect to $\beta$, indicating that a proper choice of $\eta_b$ can upper-bound the attack success.
For the maximum \bdaafter, we see similar decreases of $5.4$ percentage points, and $48.2$ percentage points, respectively.
Here, it is interesting to note that for $\eta_b=0.05$, the maximum achievable \bdaafter and \lifespan is significantly higher for the more complex VGG17 model than for MobileNetV2 or the simpler ResNet20 model.
The same pattern appears for \lifespan: setting $\eta_b = 0.5$ reduces it to $0$ regardless of $\beta$.
We attribute this effect to the increased capacity of the VGG17 model compared to the other models, making it easier to learn two tasks (main task and backdoor) simultaneously, thereby slowing down catastrophic forgetting of the backdoor task.
Last, we observe that high values of $\beta$ lead to a sharp drop in \mta to around 10\% on CIFAR-10, indicating complete divergence caused by excessive scaling of model updates.

\section{Analysis of the baseline attack}
\label{app:Baseline}

The most basic backdoor attack is a dirty-label data poisoning attack~\cite{bagdasaryan_how_2020, shejwalkar_back_2022, gu_badnets_2017, bhagoji_analyzing_2019}.
Here, the attacker changes the labels of a fraction of the training data to implant the desired backdoor behavior.
We now present our results for such a baseline data poisoning attack.

\subsection{Learning Rate}

\begin{figure}
    \centering
    \subfloat[\bda \label{fig:lr_effect_data_poisoning_bda_during}]{
        \centering
        \includegraphics[width=0.45\linewidth]{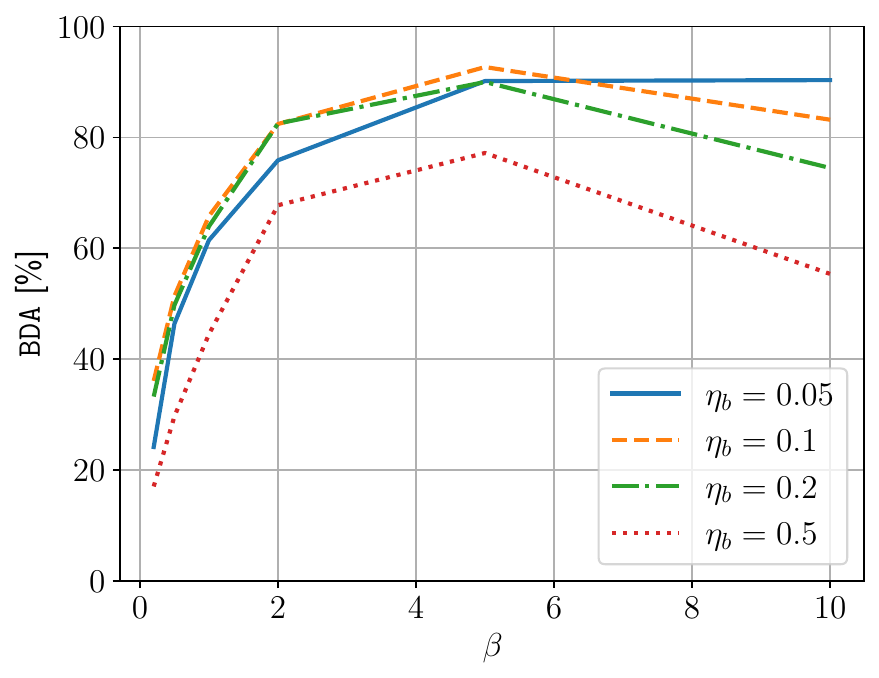}
    }
    \hfill
    \subfloat[\bdaafter \label{fig:lr_effect_data_poisoning_bda_after}]{
        \centering
        \includegraphics[width=0.45\linewidth]{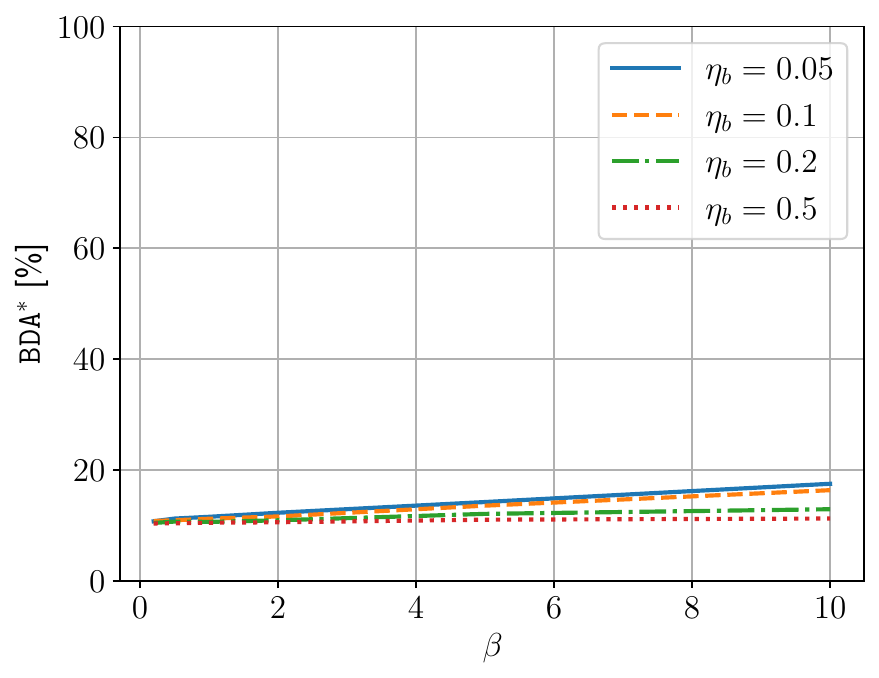}
    }

    \subfloat[\lifespan \label{fig:lr_effect_data_poisoning_lifespan}]{
        \centering
        \includegraphics[width=0.45\linewidth]{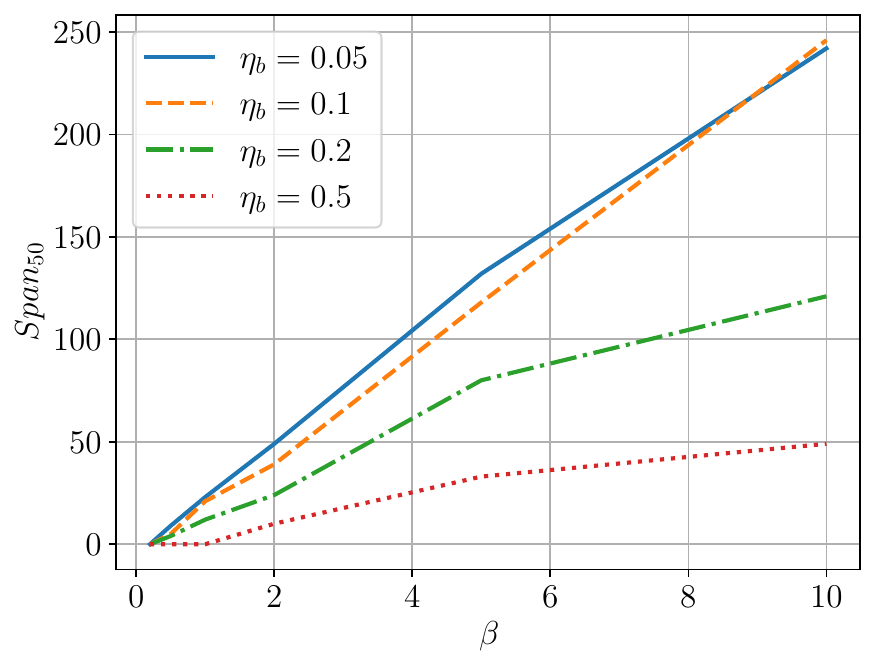}
    }
    \hfill
    \subfloat[\mta \label{fig:lr_effect_data_poisoning_mta_during}]{
        \centering
        \includegraphics[width=0.45\linewidth]{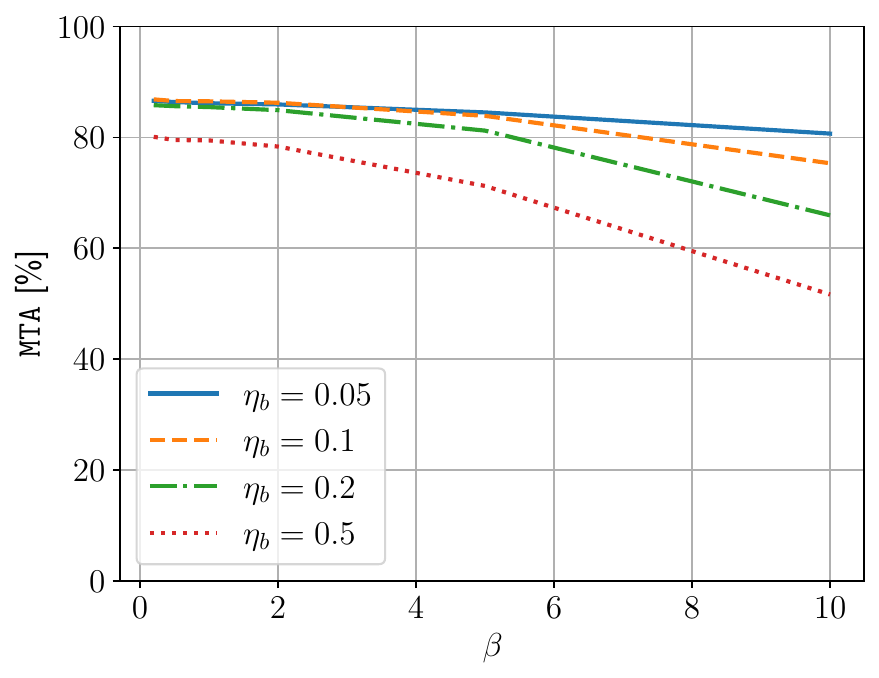}
    }

    \caption{Impact of $\eta_b$ and $\beta$ on the baseline attack.}
    \label{fig:lr_effect_data_poisoning}
\end{figure}

In Figure~\ref{fig:lr_effect_data_poisoning}, we show the \bda and \bdaafter, the \lifespan, and the \mta for the baseline attack.
We observe that the results confirm our analytical findings.
Note that increasing $\eta_b$ from $0.05$ to $0.5$ reduces the maximum \bda by $13.15$ percentage points and the maximum \bdaafter by $6.26$ percentage points.
We also point out that the latter does not exceed $18\%$ for any configuration.
Moreover, in line with the results for the other attacks, we measure a decrease of up to $79.8\%$ in terms of \lifespan.
Choosing a benign learning rate of $0.2$ instead of $0.05$ already achieves significant improvements in terms of \bda and \lifespan while achieving similar {\mta}s to $\eta_b=0.05$.

\subsection{Momentum}

\begin{figure}
    \centering
    \subfloat[\bda \label{fig:mu_effect_data_poisoning_bda_during}]{
        \centering
        \includegraphics[width=0.45\linewidth]{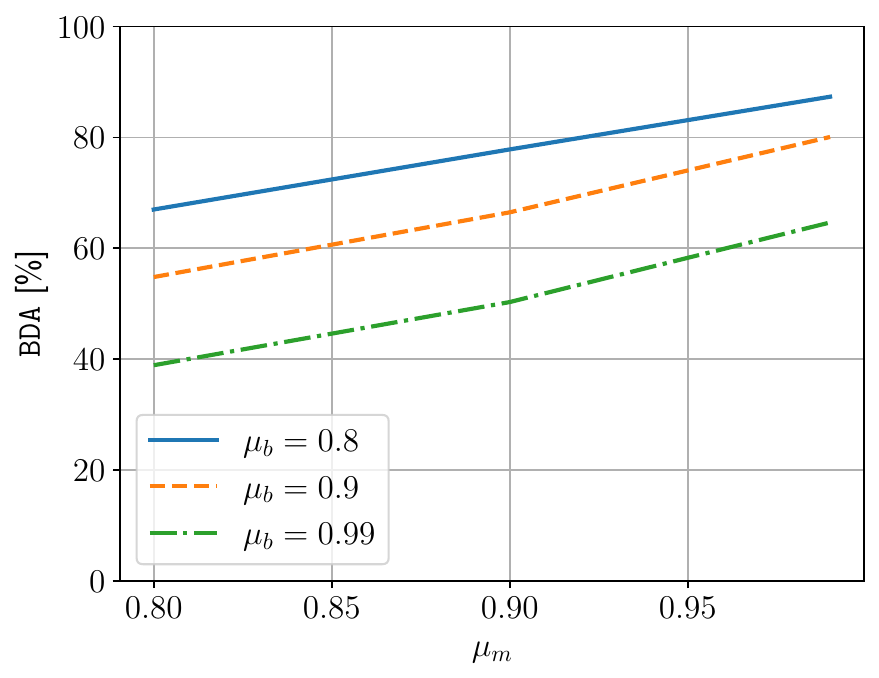}
    }
    \hfill
    \subfloat[\bdaafter \label{fig:mu_effect_data_poisoning_bda_after}]{
        \centering
        \includegraphics[width=0.45\linewidth]{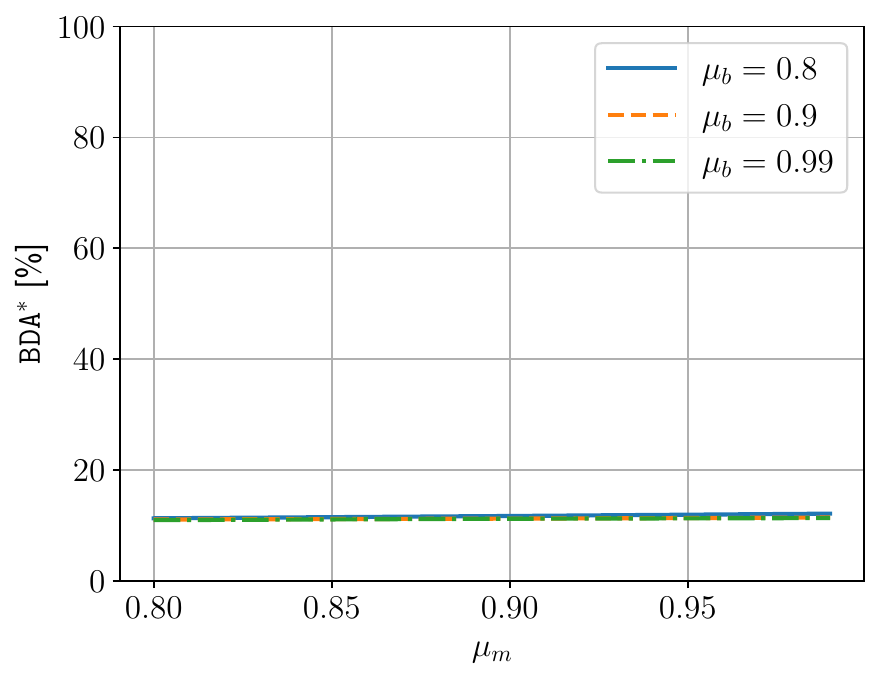}
    }

    \subfloat[\lifespan \label{fig:mu_effect_data_poisoning_lifespan}]{
        \centering
        \includegraphics[width=0.45\linewidth]{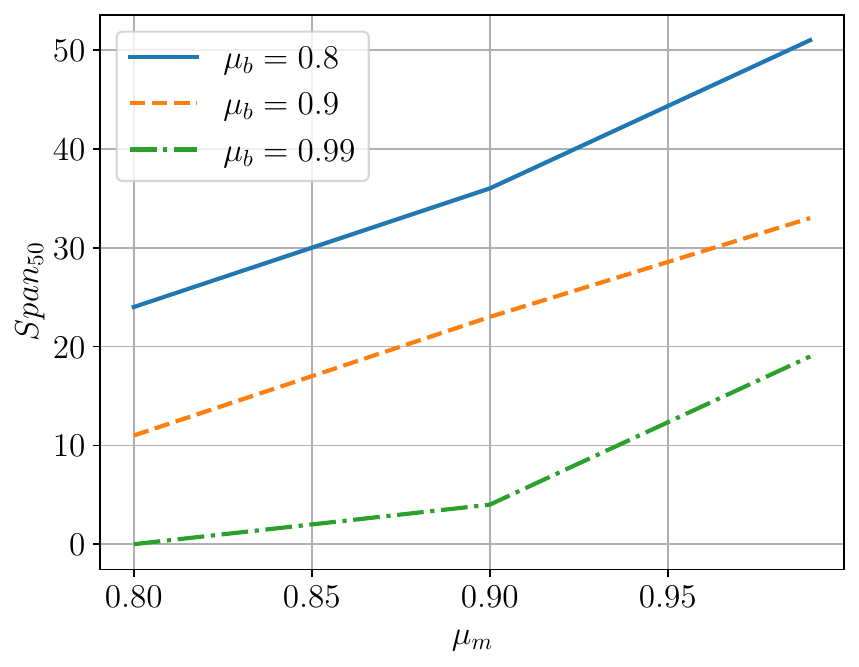}
    }
    \hfill
    \subfloat[\mta \label{fig:mu_effect_data_poisoning_mta_during}]{
        \centering
        \includegraphics[width=0.45\linewidth]{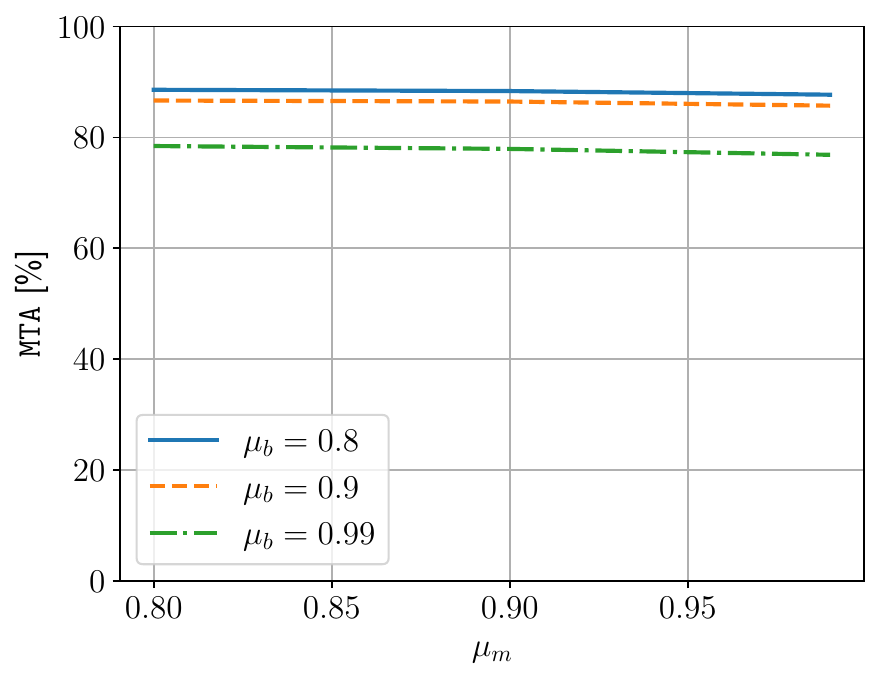}
    }

    \caption{Impact of $\mu_b$ and $\mu_m$ on the baseline attack.}
    \label{fig:mu_effect_data_poisoning}
\end{figure}

In Figure~\ref{fig:mu_effect_data_poisoning}, we evaluate the \bda and \bdaafter, the \lifespan, and the \mta for the baseline attack.
Again here, we observe that our empirical results confirm our analytical findings. Increasing $\mu_b$ from $0.8$ to $0.99$ decreases maximum \bda, \bdaafter and \lifespan by $22.68$ percentage points, $0.79$ percentage points, and $62.75\%$, respectively.
In terms of \mta, choosing higher values for $\mu_b$ decreases \mta by up to $10.84$ percentage points.

\subsection{Batch Size \& Local Epochs}

\begin{figure}
    \centering
    \subfloat[\bda \label{fig:E_effect_data_poisoning_bda_during}]{
        \centering
        \includegraphics[width=0.45\linewidth]{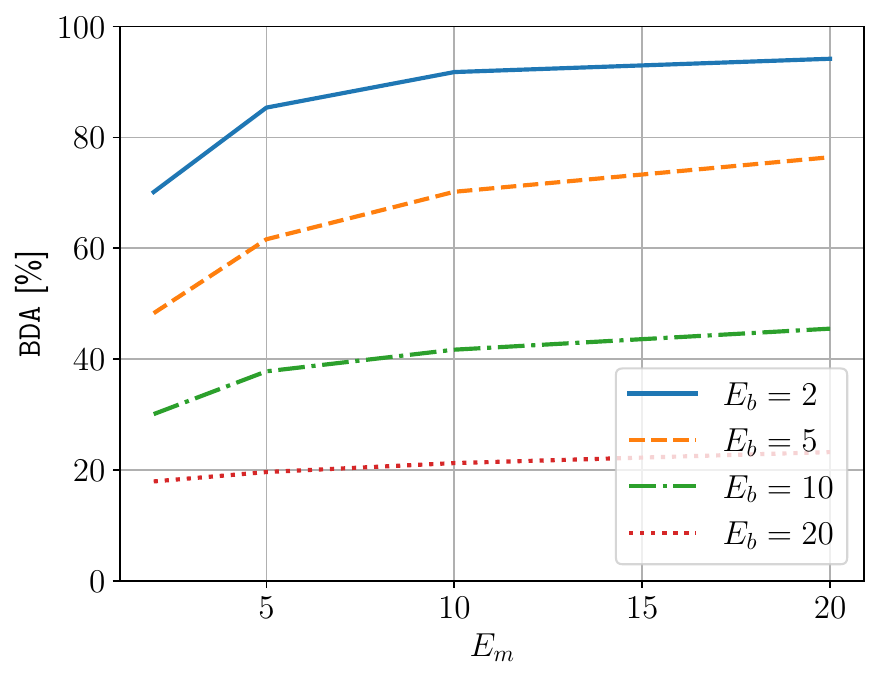}
    }
    \hfill
    \subfloat[\bdaafter \label{fig:E_effect_data_poisoning_bda_after}]{
        \centering
        \includegraphics[width=0.45\linewidth]{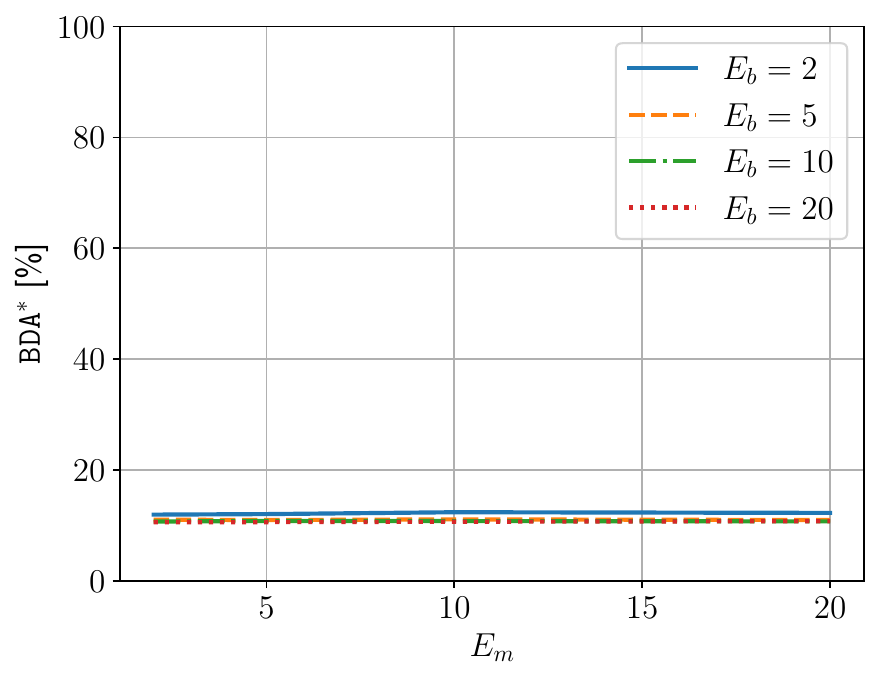}
    }

    \subfloat[\lifespan \label{fig:E_effect_data_poisoning_lifespan}]{
        \centering
        \includegraphics[width=0.45\linewidth]{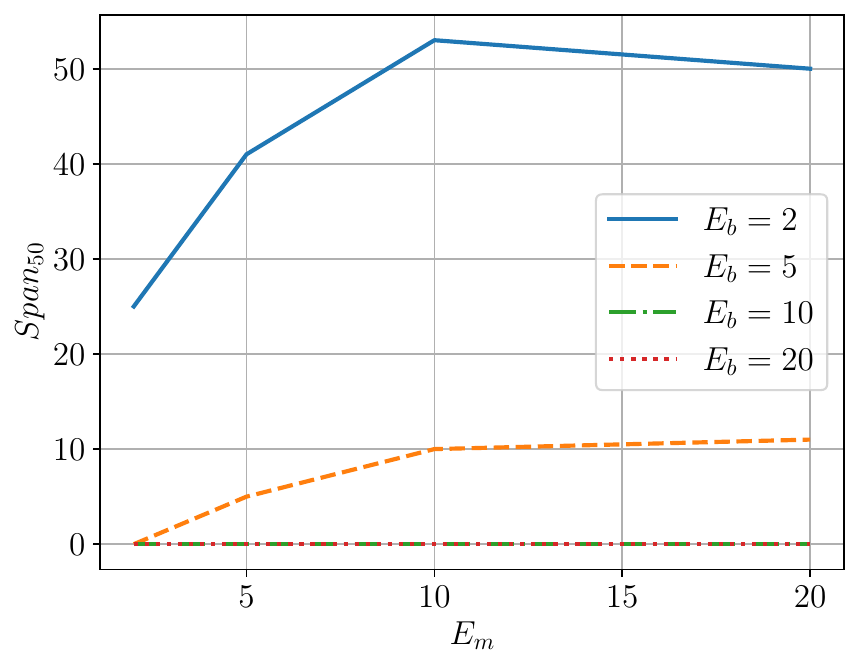}
    }
    \hfill
    \subfloat[\mta \label{fig:E_effect_data_poisoning_mta_during}]{
        \centering
        \includegraphics[width=0.45\linewidth]{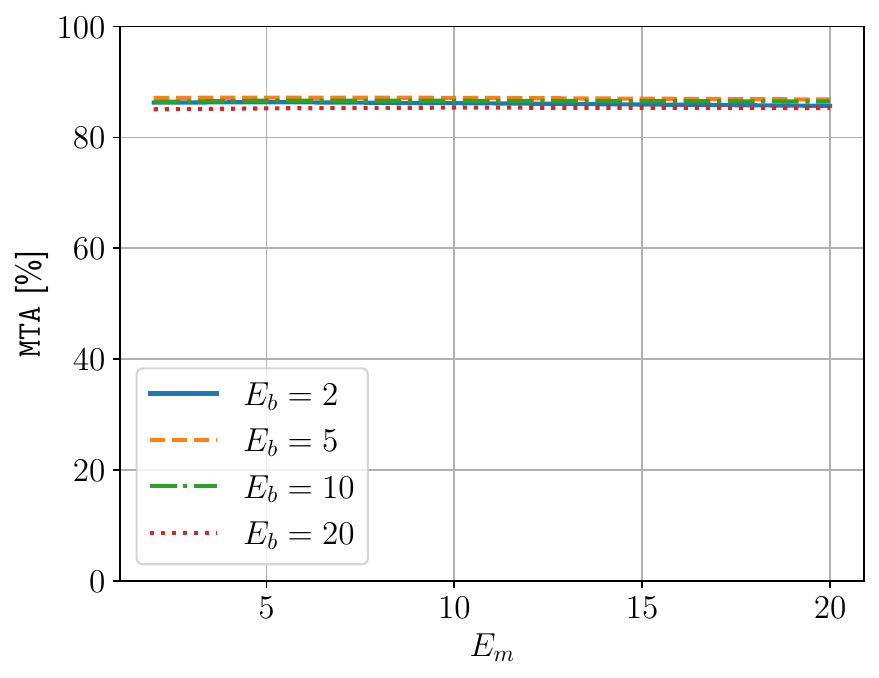}
    }

    \caption{Impact of $E_b$ and $E_m$ on the baseline attack.}
    \label{fig:E_effect_data_poisoning}
\end{figure}

\begin{figure}
    \centering
    \subfloat[\bda \label{fig:B_effect_data_poisoning_bda_during}]{
        \centering
        \includegraphics[width=0.45\linewidth]{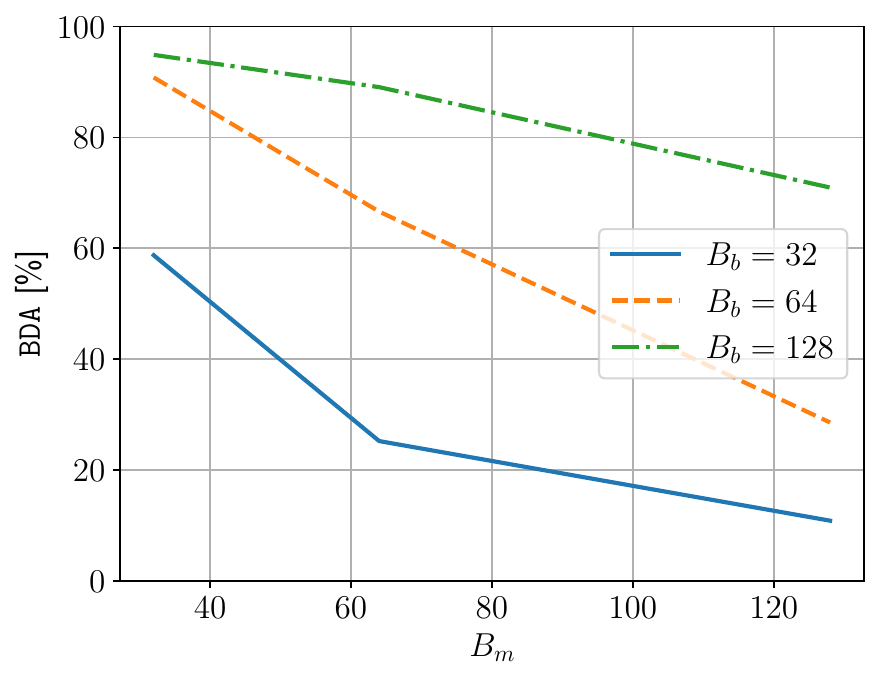}
    }
    \hfill
    \subfloat[\bdaafter \label{fig:B_effect_data_poisoning_bda_after}]{
        \centering
        \includegraphics[width=0.45\linewidth]{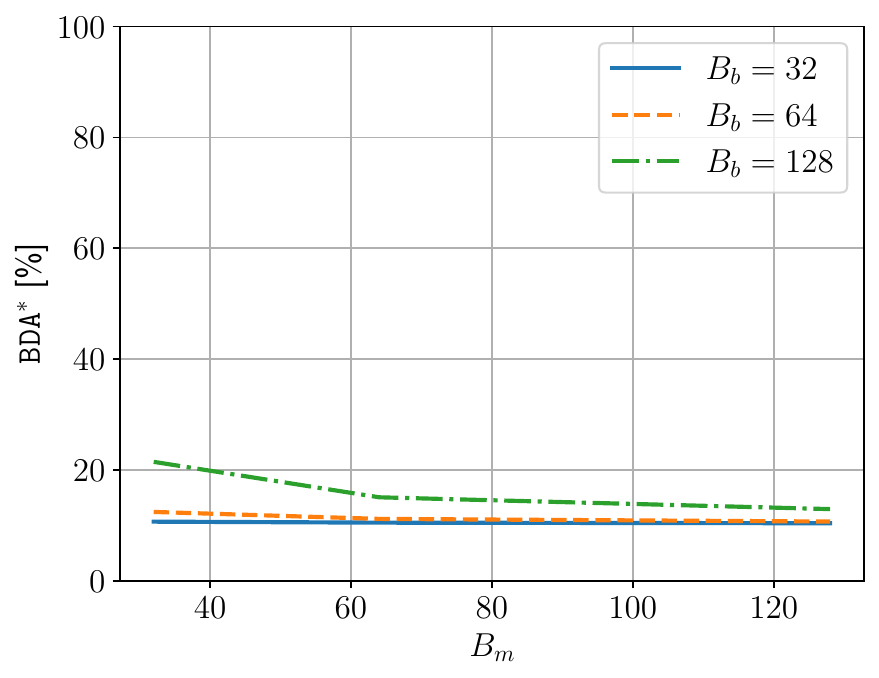}
    }

    \subfloat[\lifespan \label{fig:B_effect_data_poisoning_lifespan}]{
        \centering
        \includegraphics[width=0.45\linewidth]{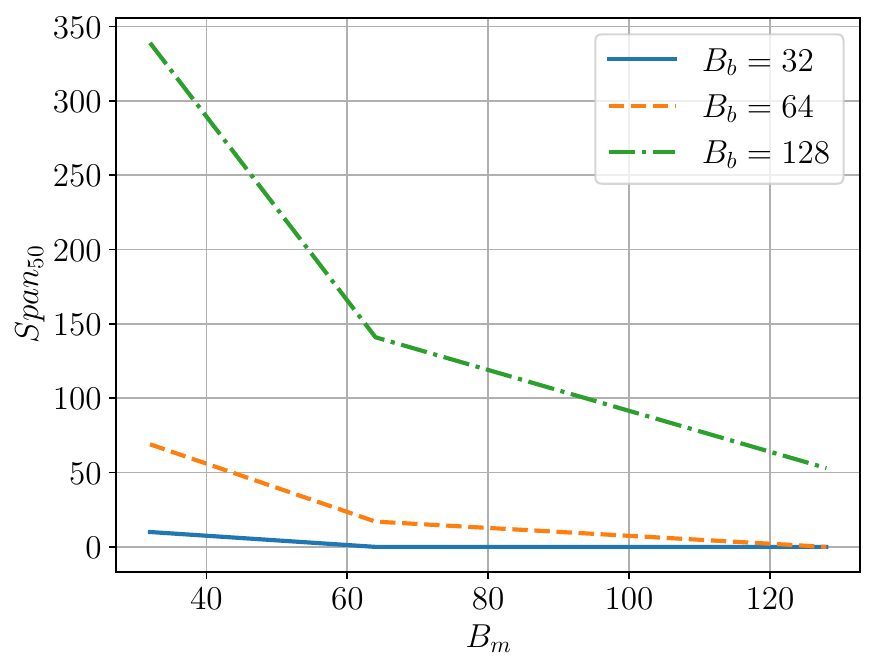}
    }
    \hfill
    \subfloat[\mta \label{fig:B_effect_data_poisoning_mta_during}]{
        \centering
        \includegraphics[width=0.45\linewidth]{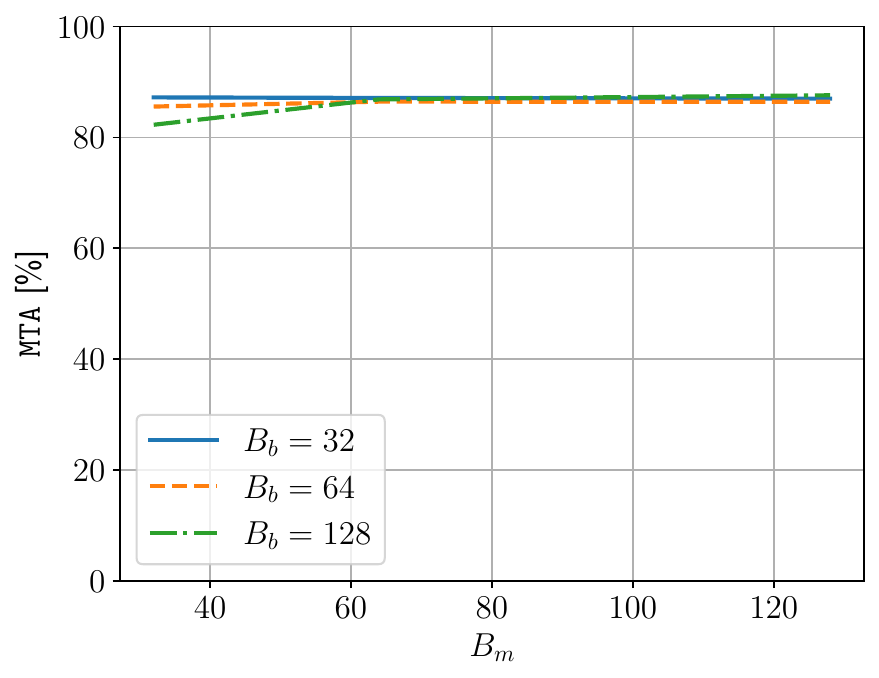}
    }

    \caption{Impact of $B_b$ and $B_m$ on the baseline attack.}
    \label{fig:B_effect_data_poisoning}
\end{figure}

In Figure~\ref{fig:E_effect_data_poisoning} and Figure~\ref{fig:B_effect_data_poisoning}, we evaluate the \bda, \bdaafter, \lifespan, and \mta for the baseline attack for various values of $E_b/E_m$ and $B_b/B_m$, respectively.
Once again, our analytical findings are supported: increasing $E_b$ (from 2 to 20) and decreasing the batch size $B_b$ (from 128 to 32) consistently reduce \bda, \bdaafter, and \lifespan. Specifically, we observe reductions of up to $70.69$ percentage points, $1.66$ percentage points, and $100\%$ for higher $E_b$, and $36.15$ percentage points, $10.79$ percentage points, and $97.05\%$ for smaller $B_b$, respectively.
We also report negligible deviations in \mta and \mtaafter of up to $0.72$ percentage points.

\subsection{Weight Decay}

\begin{figure}
    \centering
    \subfloat[\bda \label{fig:lambda_effect_data_poisoning_bda_during}]{
        \centering
        \includegraphics[width=0.45\linewidth]{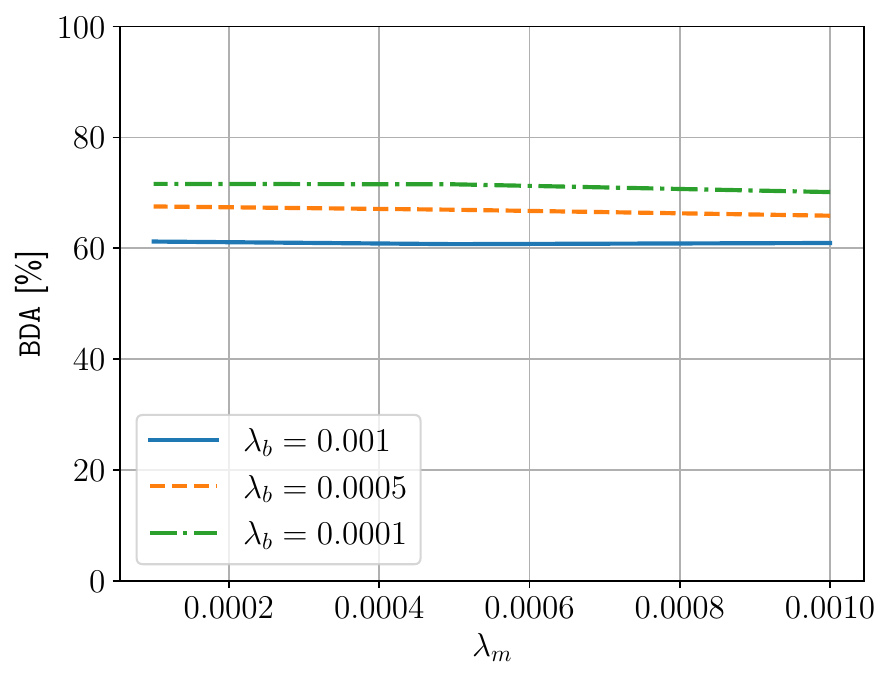}
    }
    \hfill
    \subfloat[\bdaafter \label{fig:lambda_effect_data_poisoning_bda_after}]{
        \centering
        \includegraphics[width=0.45\linewidth]{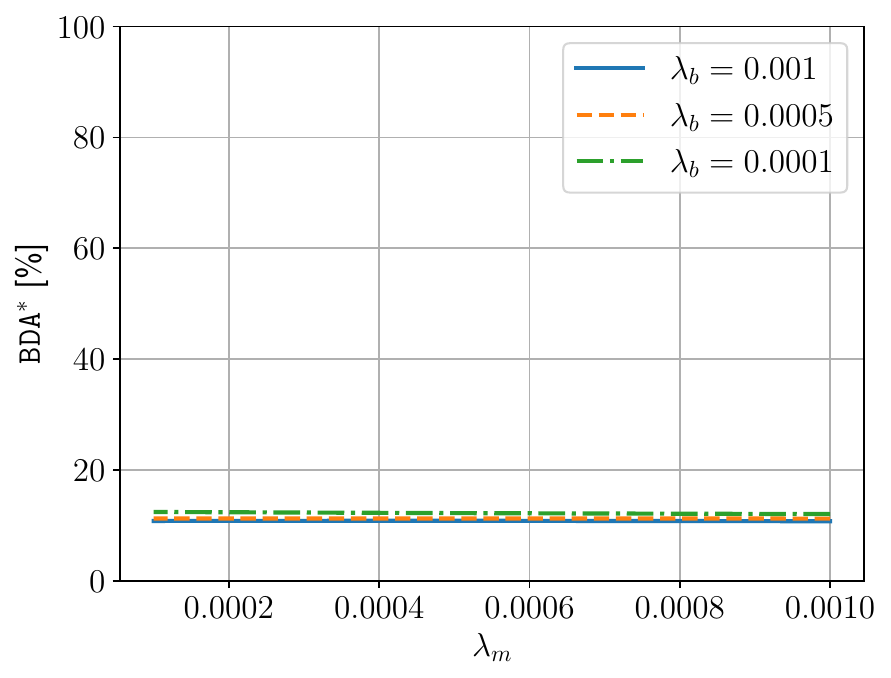}
    }

    \subfloat[\lifespan \label{fig:lambda_effect_data_poisoning_lifespan}]{
        \centering
        \includegraphics[width=0.45\linewidth]{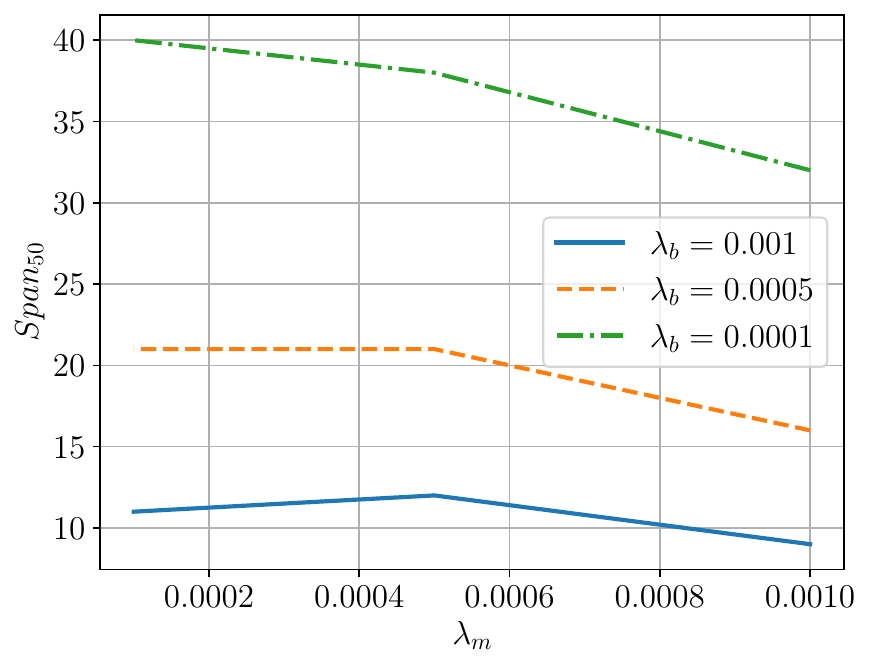}
    }
    \hfill
    \subfloat[\mta \label{fig:lambda_effect_data_poisoning_mta_during}]{
        \centering
        \includegraphics[width=0.45\linewidth]{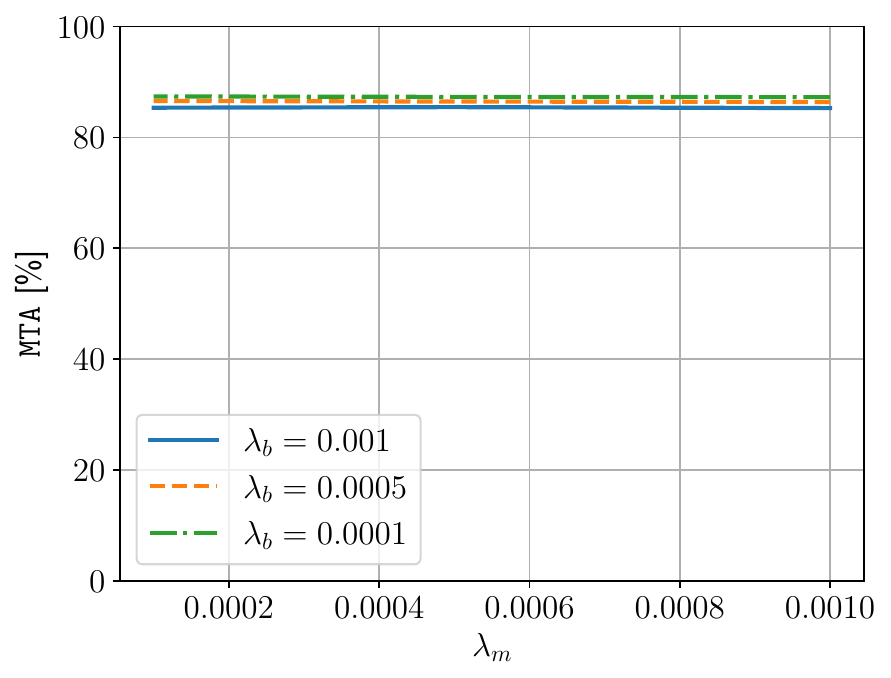}
    }

    \caption{Impact of $\lambda_b$ and $\lambda_m$ on the baseline attack.}
    \label{fig:lambda_effect_data_poisoning}
\end{figure}

In Figure~\ref{fig:lambda_effect_data_poisoning}, we evaluate the \bda and \bdaafter, the \lifespan, and the \mta for the baseline attack.
We find smaller but distinct decreases of maximum \bda, \bdaafter, and \lifespan of $10.40$ percentage points, $1.59$ percentage points, and $70\%$, respectively, when increasing $\lambda_b$ from $0.0001$ to $0.001$, aligning with the results for the other considered attacks.
Further, we observe negligible deviations in $\mta$ of at most $2.03$ percentage points for variations in $\lambda_b$.

\section{Analysis of the IBA attack}
\label{app:IBA}

Similar to A3FL~\cite{zhang_a3fl_2023}, IBA~\cite{nguyen_iba_2023} is a trigger-optimization attack that uses a trigger generation network to generate different triggers for each input---akin to the generation process of adversarial examples~\cite{szegedy_intriguing_2014}.
In the first phase of the attack, IBA optimizes the network that generates the trigger, such that these can already backdoor the system. 
Then, the backdoor behavior is planted into the model using a data poisoning strategy.

We now present the results for the IBA~\cite{nguyen_iba_2023} attack. 
Despite this attack combining backdoor mechanisms with adversarial examples, we show that our findings generally transfer to this attack as well.

\subsection{Learning Rate}

\begin{figure}
    \centering
    \subfloat[\bda \label{fig:lr_effect_iba_bda_during}]{
        \centering
        \includegraphics[width=0.45\linewidth]{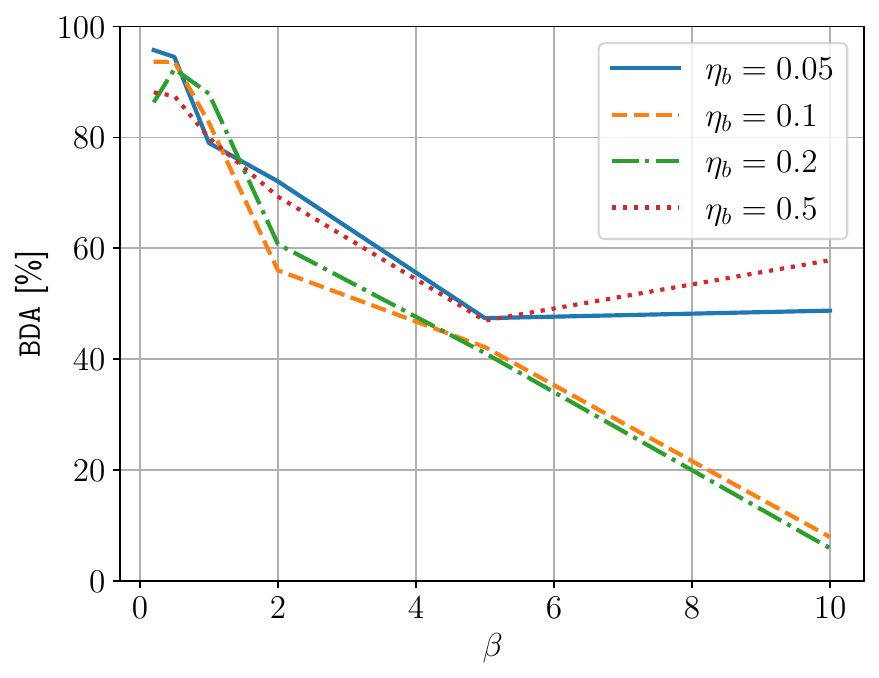}
    }
    \hfill
    \subfloat[\bdaafter \label{fig:lr_effect_iba_bda_after}]{
        \centering
        \includegraphics[width=0.45\linewidth]{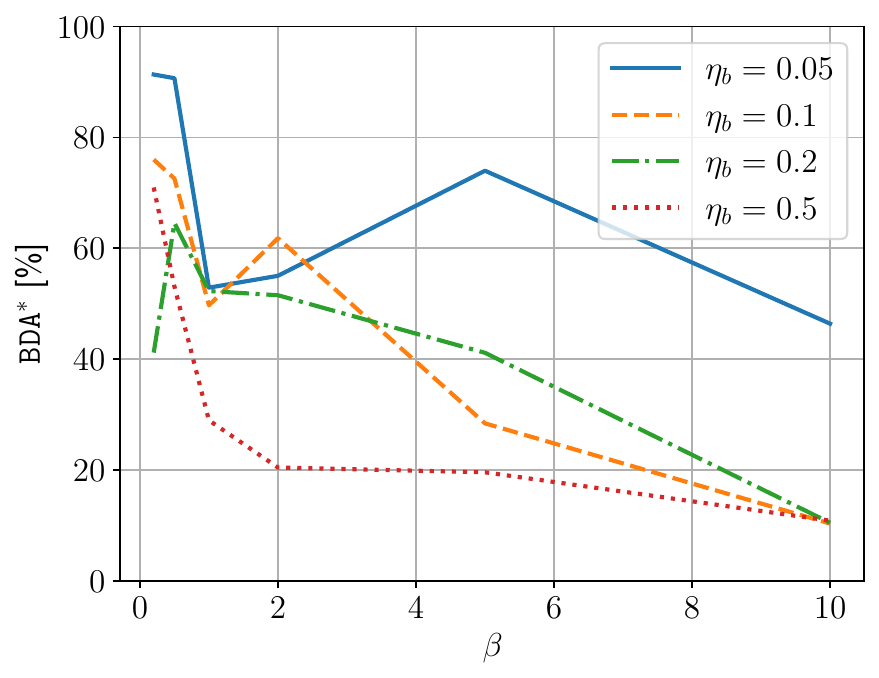}
    }

    \subfloat[\lifespan \label{fig:lr_effect_iba_lifespan}]{
        \centering
        \includegraphics[width=0.45\linewidth]{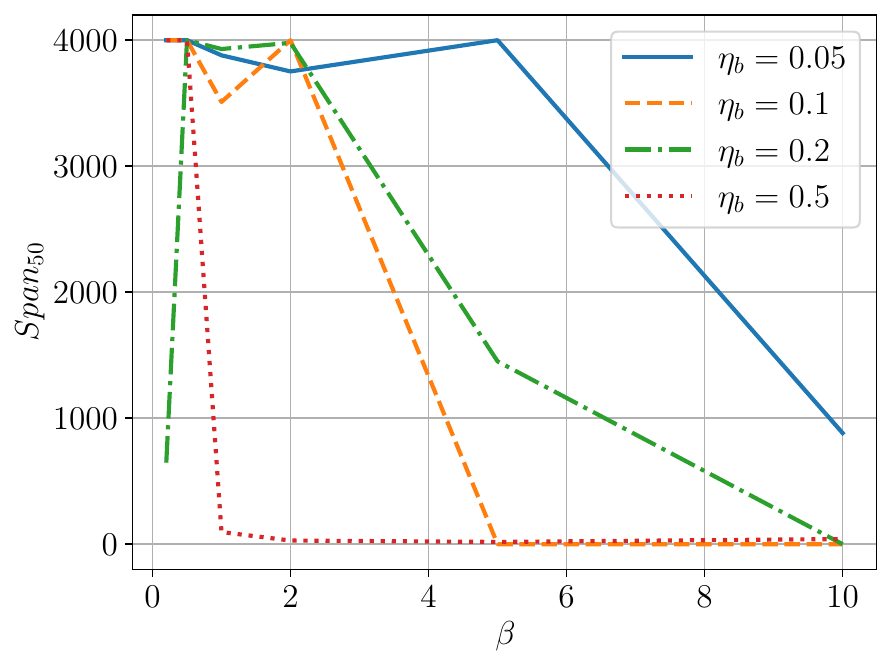}
    }
    \hfill
    \subfloat[\mta \label{fig:lr_effect_iba_mta_during}]{
        \centering
        \includegraphics[width=0.45\linewidth]{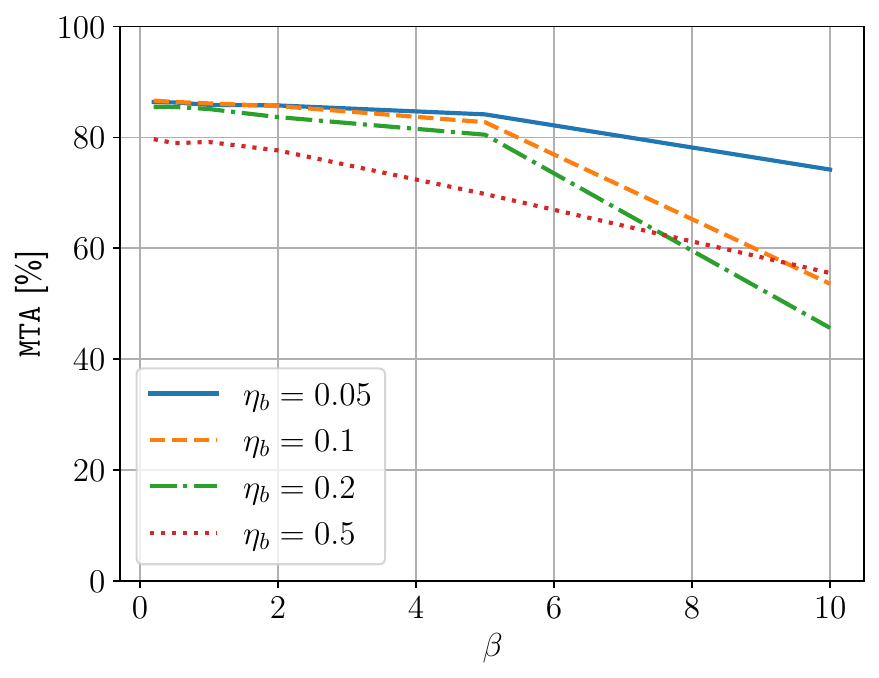}
    }

    \caption{Impact of $\eta_b$ and $\beta$ on the IBA attack.}
    \label{fig:lr_effect_iba}
\end{figure}

In Figure~\ref{fig:lr_effect_iba}, we show the \bda, \bdaafter, \lifespan, and the \mta for the IBA attack.
Overall, the results for the IBA attack align with the results for the other attacks, with decreases of $7.63$ percentage points and $20.38$ percentage points of the maximum achievable \bda and \bdaafter when increasing $\eta_b$ from $0.05$ to $0.5$, respectively. 
There are a few differences, however.
In the IBA attack, we note that a lower malicious learning $\eta_m = \beta\eta_b$ rate causes a higher attack success, for example, for $\eta_b=0.2$, the attacker can achieve the highest \bda by choosing $\beta = 0.5$. 
We argue that this is because this attack generates a per-input trigger in an adversarial example generation fashion (and our analysis does not model the impact of the learning rate on inference time attacks).
Further, for every benign learning rate $\eta_b$, there is a malicious learning rate $\beta \cdot \eta_b$ for which the \lifespan reaches 4000 rounds, i.e., the full evaluation span. 
We also attribute this to the dual nature of the attack that can allow for the trigger itself to achieve a backdoor accuracy of more than $50\%$ without even backdooring the model.
With respect to $\mta$, we again observe that a benign learning rate of $\eta_b=0.2$ can achieve comparable \mta to $\eta_b=0.05$ while reducing the maximum achievable \bda by $3.41$ percentage points, however.

\subsection{Momentum}

\begin{figure}
    \centering
    \subfloat[\bda \label{fig:mu_effect_iba_bda_during}]{
        \centering
        \includegraphics[width=0.45\linewidth]{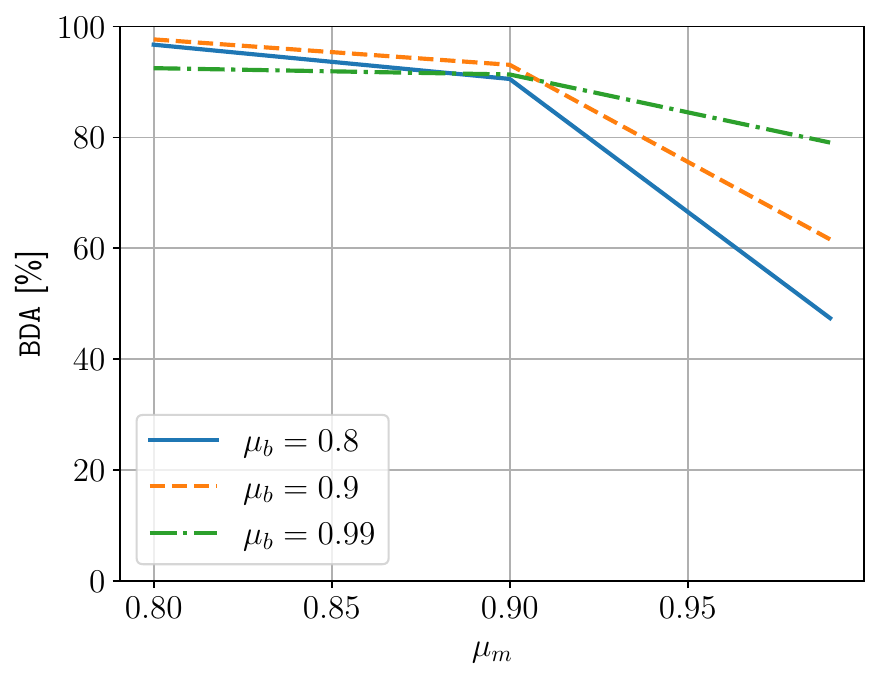}
    }
    \hfill
    \subfloat[\bdaafter \label{fig:mu_effect_iba_bda_after}]{
        \centering
        \includegraphics[width=0.45\linewidth]{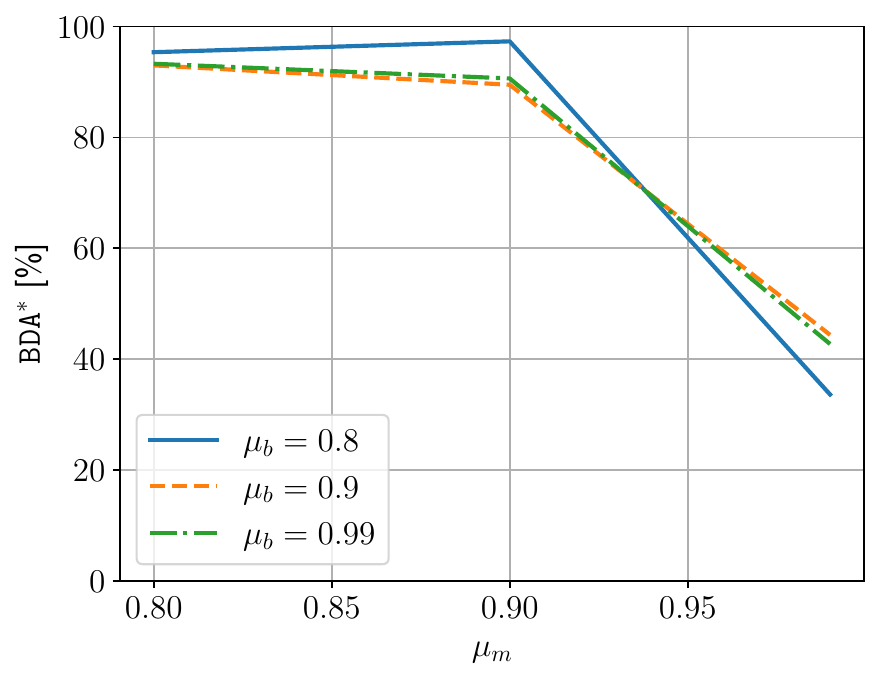}
    }

    \subfloat[\lifespan \label{fig:mu_effect_iba_lifespan}]{
        \centering
        \includegraphics[width=0.45\linewidth]{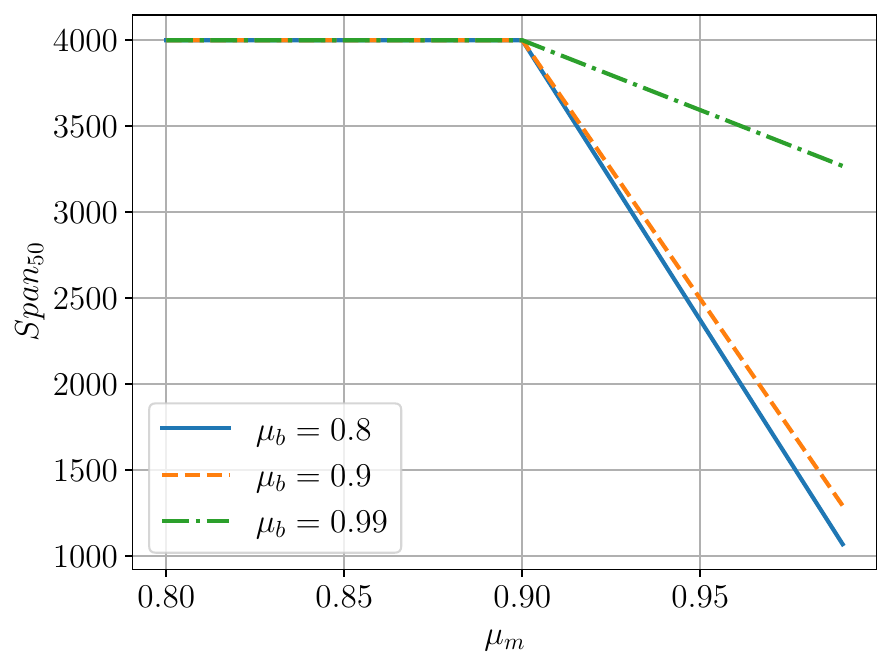}
    }
    \hfill
    \subfloat[\mta \label{fig:mu_effect_iba_mta_during}]{
        \centering
        \includegraphics[width=0.45\linewidth]{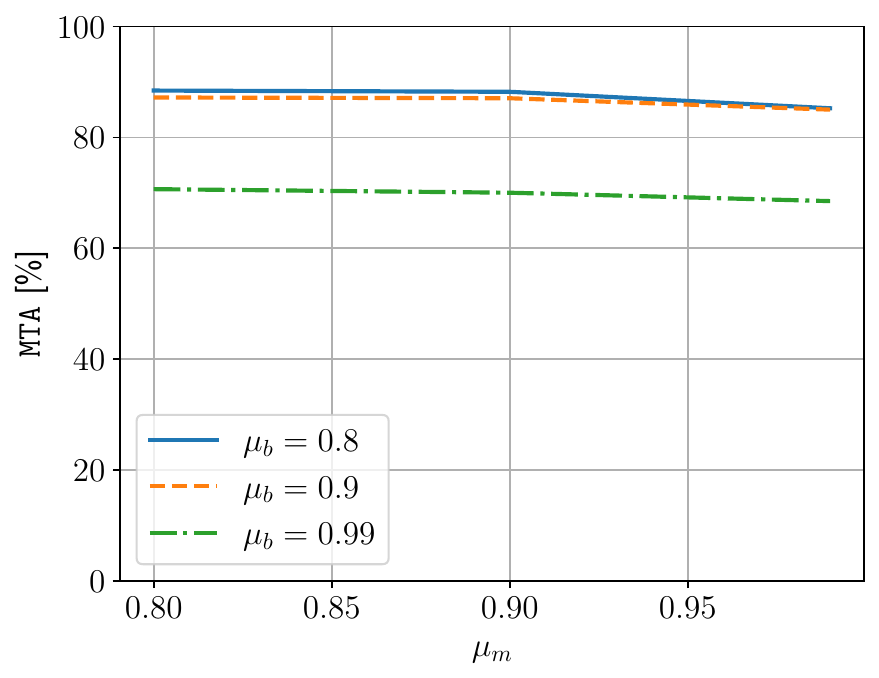}
    }

    \caption{Impact of $\mu_b$ and $\mu_m$ on the IBA attack.}
    \label{fig:mu_effect_iba}
\end{figure}

In Figure~\ref{fig:mu_effect_iba}, we evaluate the \bda, \bdaafter, \lifespan, and \mta for the IBA attack for various benign and malicious momentum factors.
We observe that, overall, the evaluation of the IBA attack aligns with our findings. A higher benign momentum of $\mu_b=0.99$ instead of $\mu_b=0.8$ can decrease maximum \bda and \bdaafter by $4.23$ percentage points and $4.04$ percentage points, respectively.
Similar to the other attacks, these decreases are relatively small, which we attribute to the generally smaller scale of the momentum value.
In terms of \lifespan, we again observe that the momentum cannot upper-bound it, which we again attribute to the dual nature of the attack.
In terms of \mta, we observe quite significant decreases of up to $18.21$ percentage points when choosing $\mu_b=0.99$ instead of $\mu_b=0.8$, which aligns with our main findings.

\subsection{Batch Size \& Local Epochs}

\begin{figure}
    \centering
    \subfloat[\bda \label{fig:E_effect_iba_bda_during}]{
        \centering
        \includegraphics[width=0.45\linewidth]{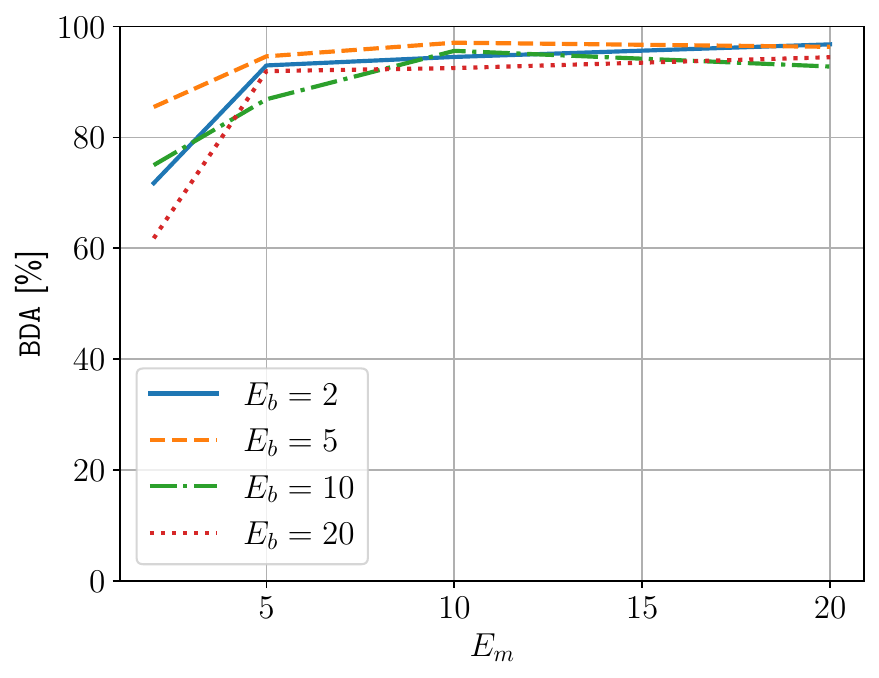}
    }
    \hfill
    \subfloat[\bdaafter \label{fig:E_effect_iba_bda_after}]{
        \centering
        \includegraphics[width=0.45\linewidth]{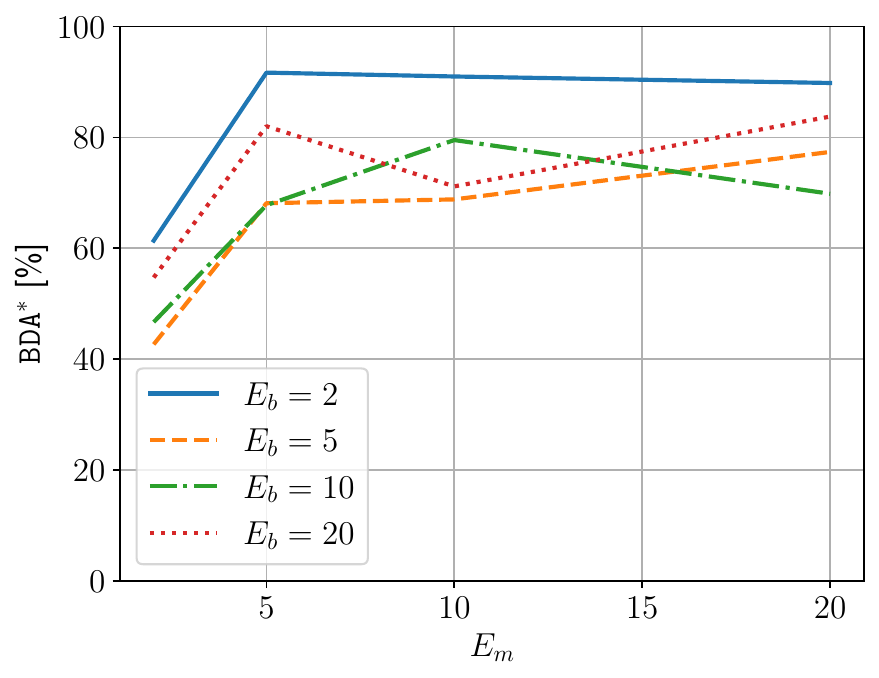}
    }

    \subfloat[\lifespan \label{fig:E_effect_iba_lifespan}]{
        \centering
        \includegraphics[width=0.45\linewidth]{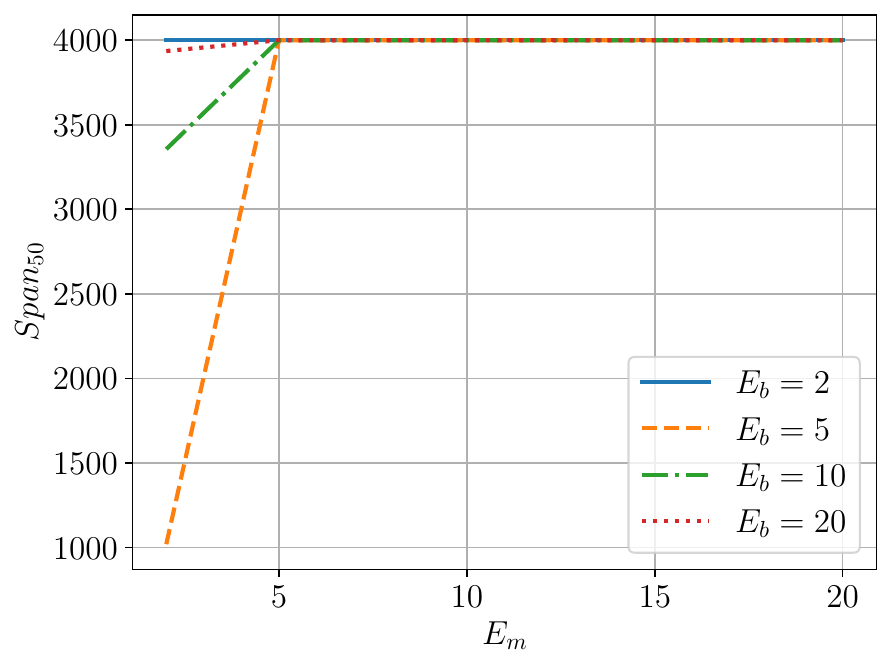}
    }
    \hfill
    \subfloat[\mta \label{fig:E_effect_iba_mta_during}]{
        \centering
        \includegraphics[width=0.45\linewidth]{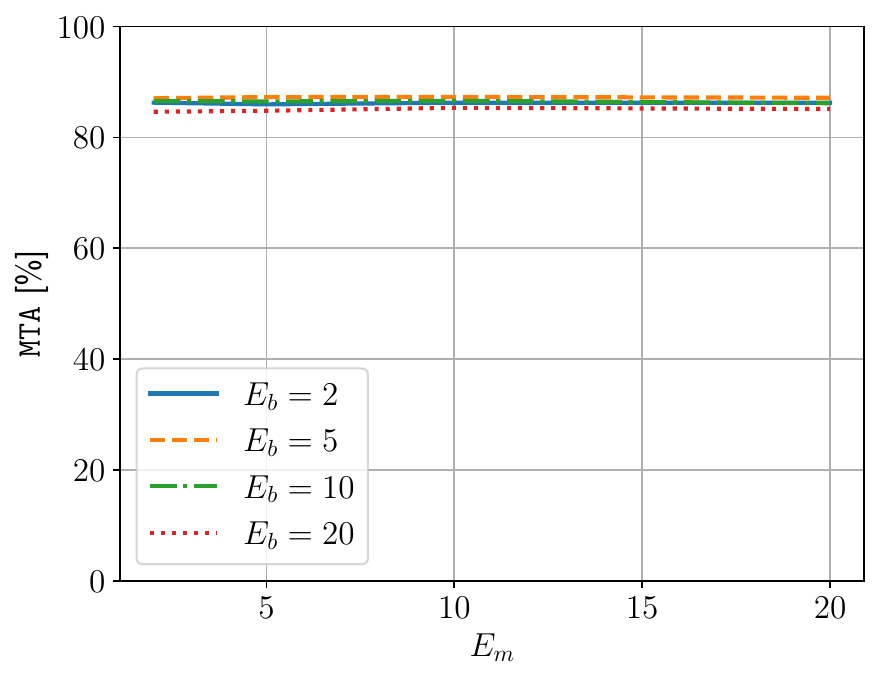}
    }

    \caption{Impact of $E_b$ and $E_m$ on the IBA attack.}
    \label{fig:E_effect_iba}
\end{figure}

\begin{figure}
    \centering
    \subfloat[\bda \label{fig:B_effect_iba_bda_during}]{
        \centering
        \includegraphics[width=0.45\linewidth]{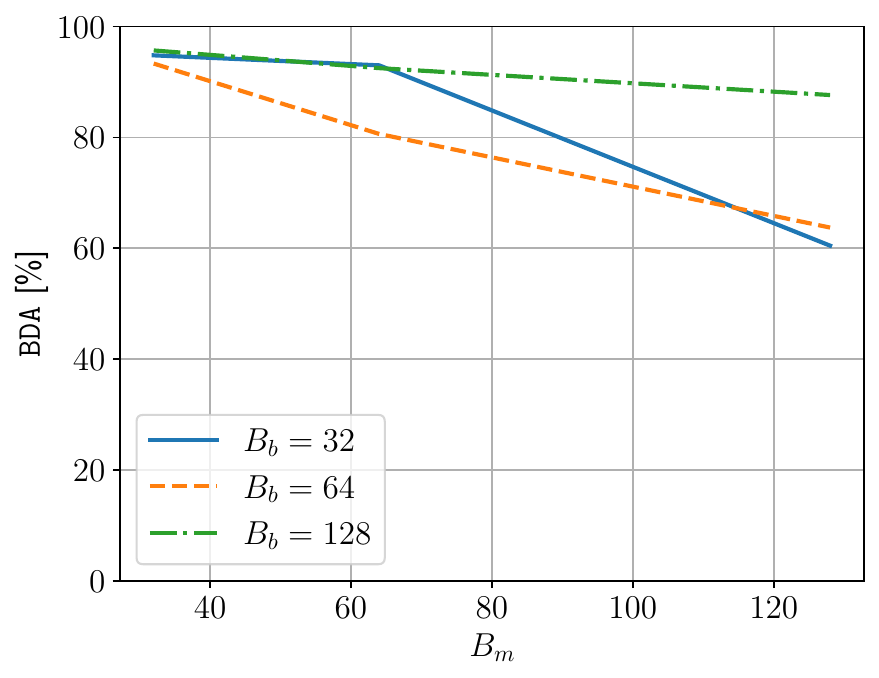}
    }
    \hfill
    \subfloat[\bdaafter \label{fig:B_effect_iba_bda_after}]{
        \centering
        \includegraphics[width=0.45\linewidth]{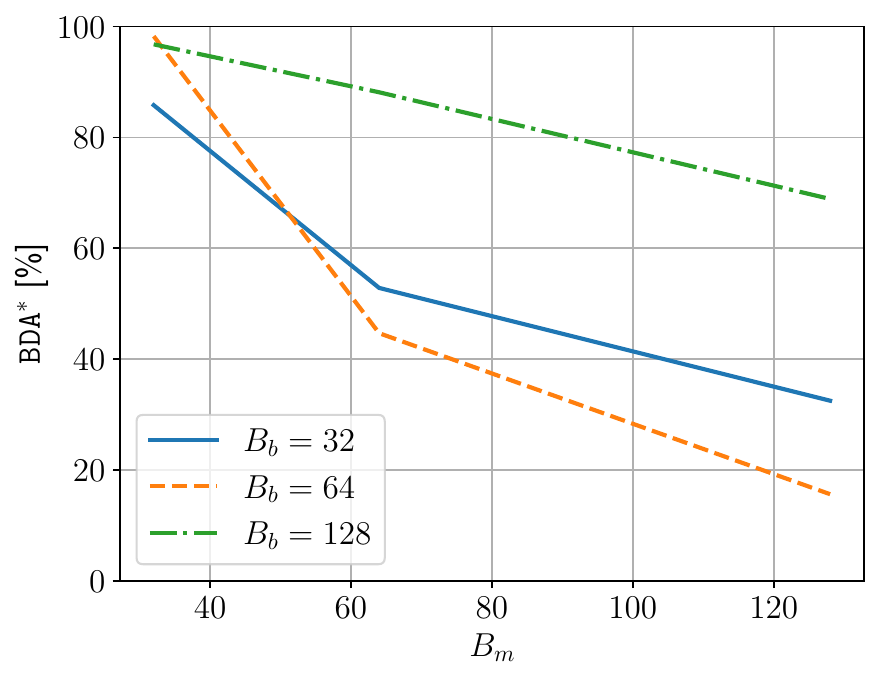}
    }

    \subfloat[\lifespan \label{fig:B_effect_iba_lifespan}]{
        \centering
        \includegraphics[width=0.45\linewidth]{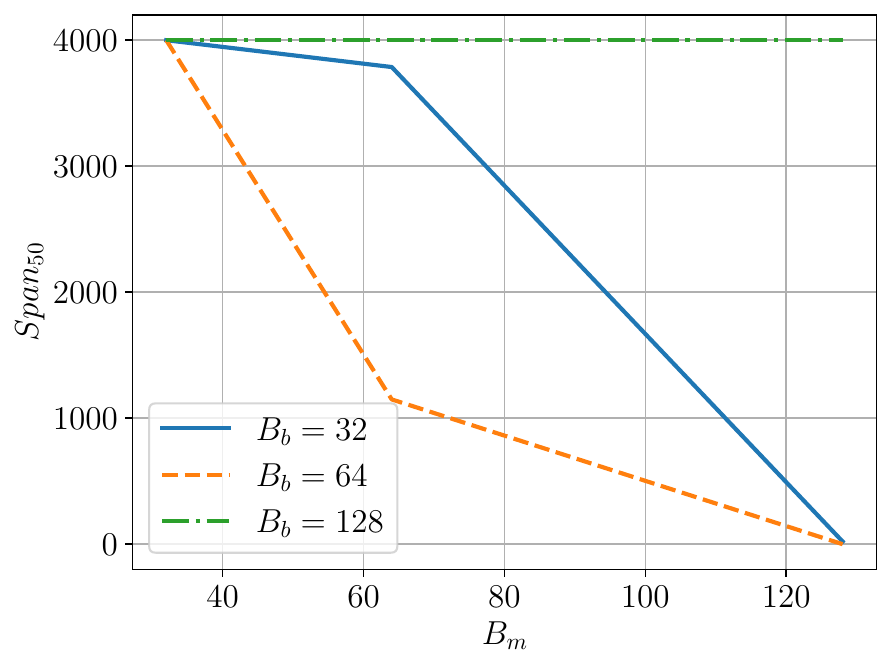}
    }
    \hfill
    \subfloat[\mta \label{fig:B_effect_iba_mta_during}]{
        \centering
        \includegraphics[width=0.45\linewidth]{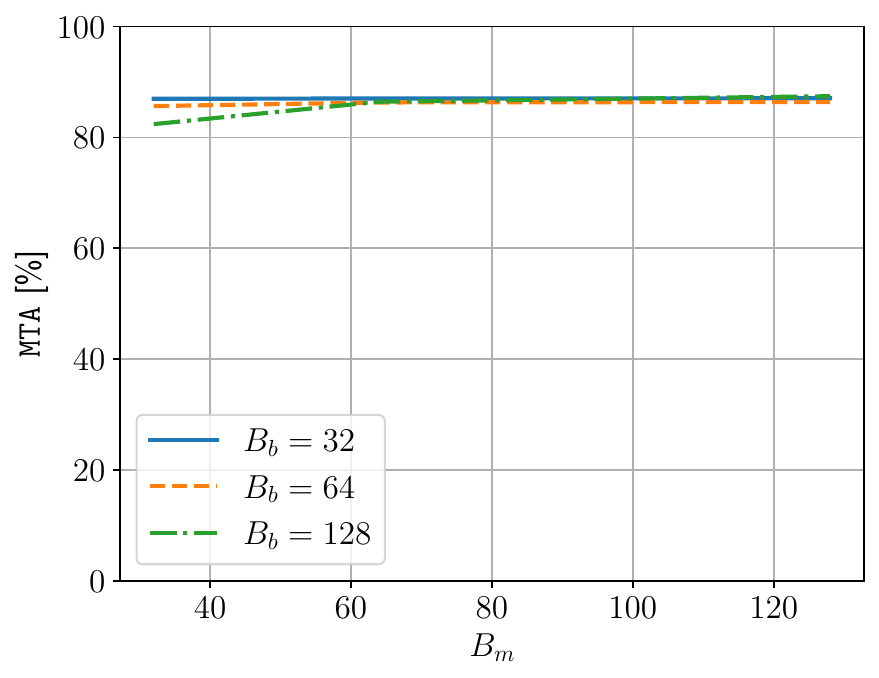}
    }

    \caption{Impact of $B_b$ and $B_m$ on the IBA attack.}
    \label{fig:B_effect_iba}
\end{figure}

In Figure~\ref{fig:E_effect_iba} and Figure~\ref{fig:B_effect_iba}, we evaluate the \bda, \bdaafter, \lifespan, and \mta for the IBA attack for various values of $E_b/E_m$ and $B_b/B_m$, respectively.
Overall, increasing $E_b$ reduces \bda, but only slightly, namely by $4.43$ percentage points when increasing $E_b$ from $2$ to $20$.
However, when considering the \bdaafter, a slightly different pattern emerges than for the other considered attacks:
Here, increasing $E_b$ initially reduces \bda but then leads to an increase beyond $E_b=5$.
A similar trade-off emerges for $B_b$, where the lowest \bda and \bdaafter are reached at $B_b=64$ rather than $32$. 
We attribute this to the attack’s dual nature, combining inference-time and training-time components.
Notice that increasing $E_b$ or decreasing $B_b$ gives the benign clients' updates a more significant influence on the global model due to the increased number of SGD steps performed.
This reduces backdoor persistence while at the same time increasing overfitting, which makes the model more susceptible to adversarial perturbations~\cite{DBLP:conf/nips/YangKYGKGRM20}.
This results in an optimal balance at intermediate values of $E_b=5$ and $B_b=64$, rather than the extremes observed for the other attacks.

In terms of \mta, we report negligible deviations of less than $2\%$ for all considered $E_b$ and $B_b$ values.

\subsection{Weight Decay}

\begin{figure}
    \centering
    \subfloat[\bda \label{fig:lambda_effect_iba_bda_during}]{
        \centering
        \includegraphics[width=0.45\linewidth]{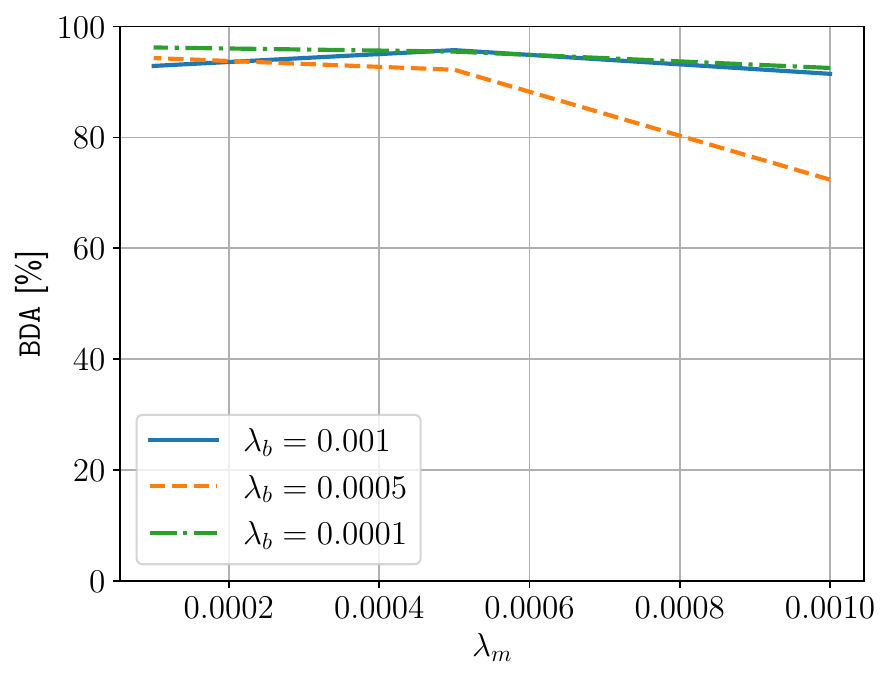}
    }
    \hfill
    \subfloat[\bdaafter \label{fig:lambda_effect_iba_bda_after}]{
        \centering
        \includegraphics[width=0.45\linewidth]{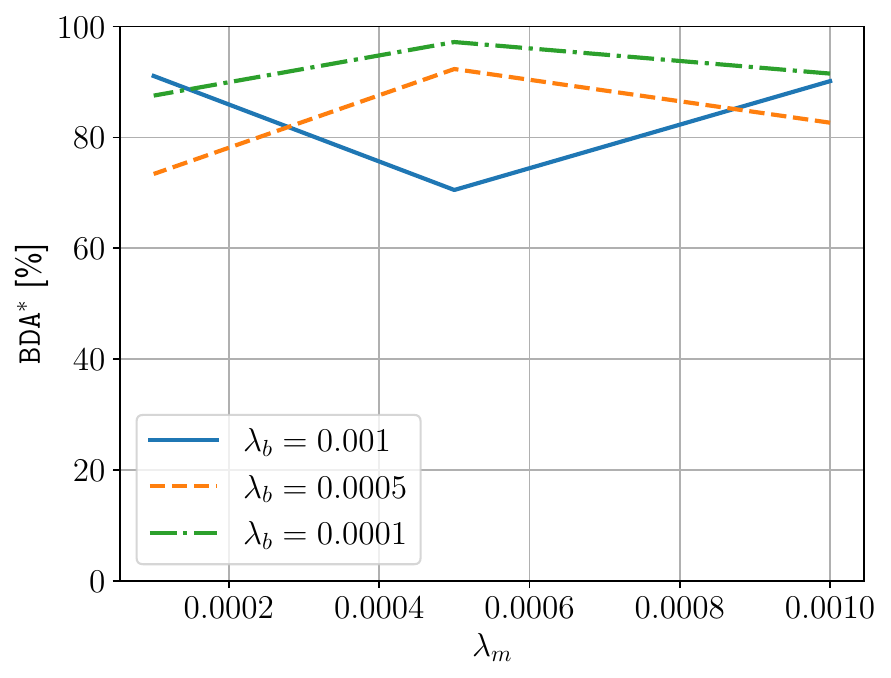}
    }

    \subfloat[\lifespan \label{fig:lambda_effect_iba_lifespan}]{
        \centering
        \includegraphics[width=0.45\linewidth]{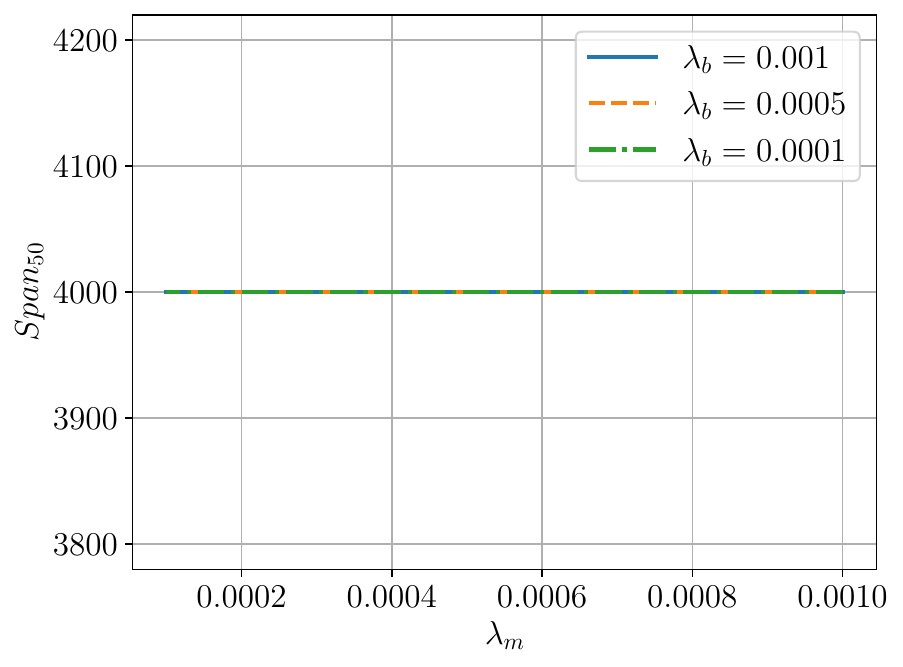}
    }
    \hfill
    \subfloat[\mta \label{fig:lambda_effect_iba_mta_during}]{
        \centering
        \includegraphics[width=0.45\linewidth]{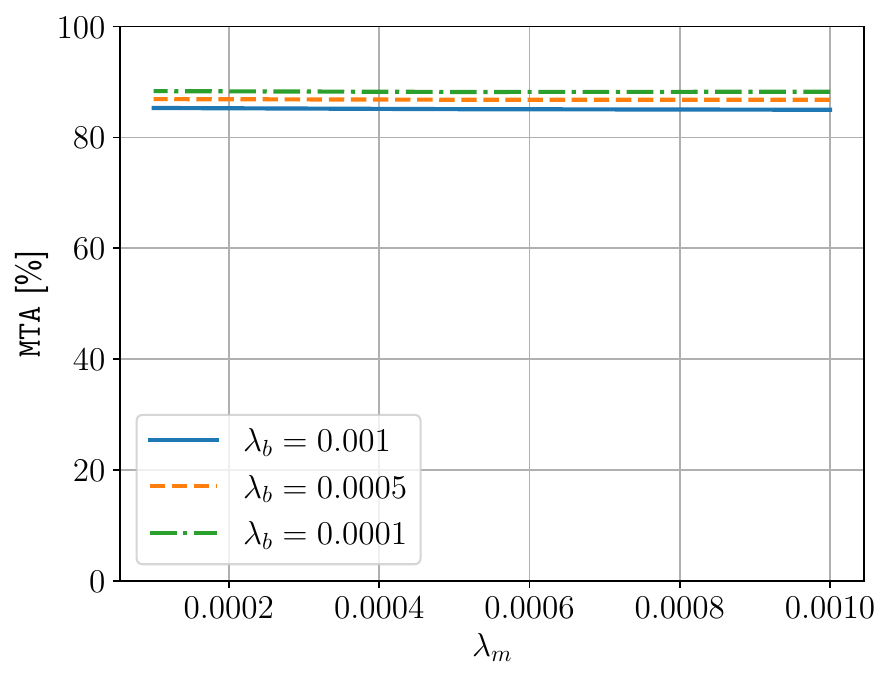}
    }

    \caption{Impact of $\lambda_b$ and $\lambda_m$ on the IBA attack.}
    \label{fig:lambda_effect_iba}
\end{figure}

In Figure~\ref{fig:lambda_effect_iba}, we show the \bda, \bdaafter, \lifespan, and \mta for the IBA attack for various values of $\lambda_b$ and $\lambda_m$, respectively.
We observe no consistent significant impact of setting $\lambda_b$ for this attack, with changes in maximum \bda of not exceeding $1.5\%$, indicating that the attack, due to its special dual nature, is mostly resilient to changes in $\lambda_b$.

\section{Additional Figures}

\subsection{Impact of Learning Rate}
\label{app:lr_effect}

\begin{figure}
    \centering
    \subfloat[A3FL~\cite{zhang_a3fl_2023} \label{fig:lr_effect_a3fl_bda_after}]{
        \centering
        \includegraphics[width=0.45\linewidth]{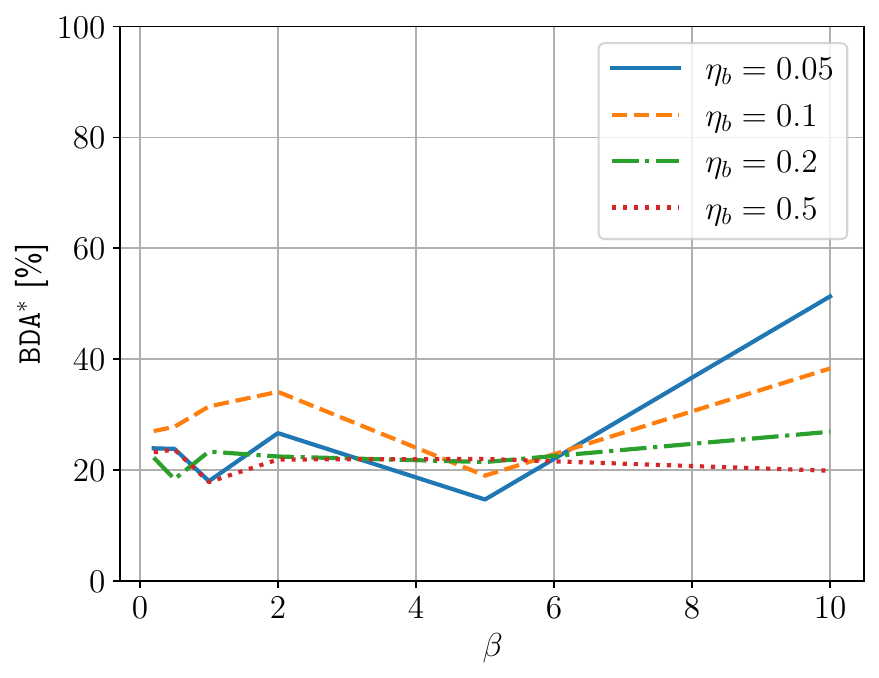}
    }
    \hfill
    \subfloat[Chameleon~\cite{dai_chameleon_2023} \label{fig:lr_effect_chameleon_bda_after}]{
        \centering
        \includegraphics[width=0.45\linewidth]{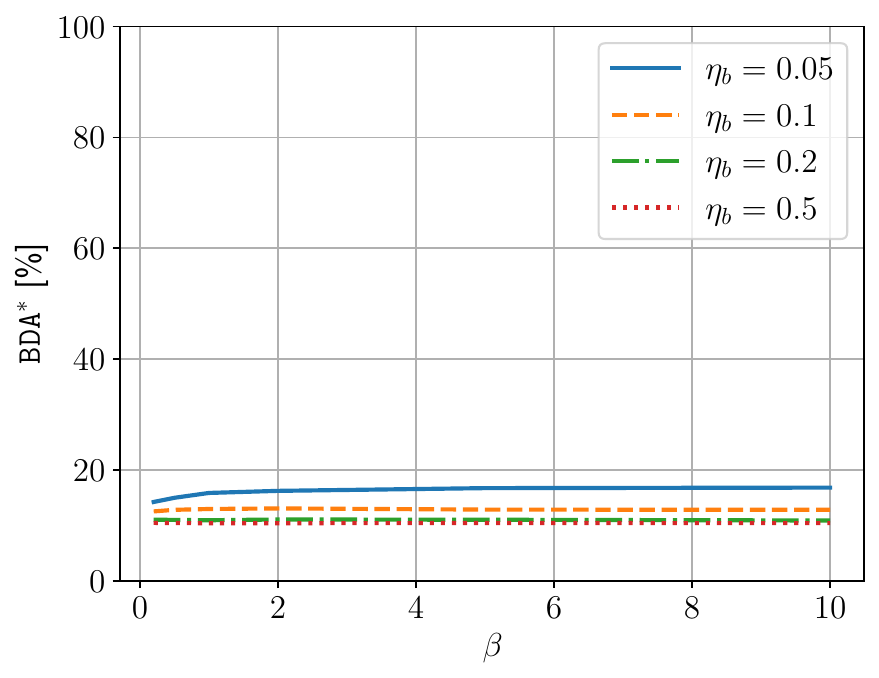}
    }

    \subfloat[DarkFed~\cite{DBLP:conf/ijcai/LiWNHXZW24} \label{fig:lr_effect_darkfed_bda_after}]{
        \centering
        \includegraphics[width=0.45\linewidth]{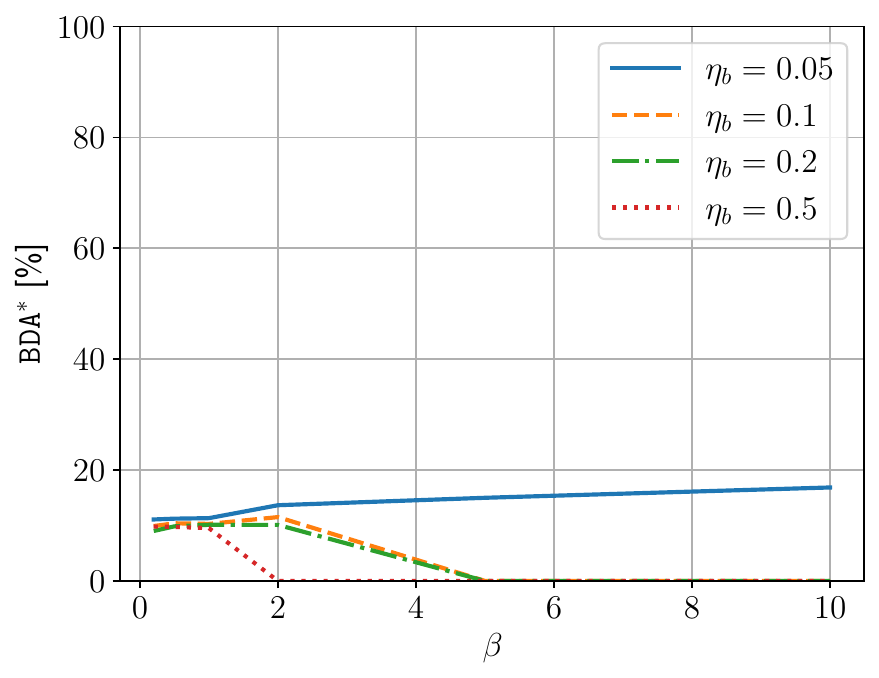}
    }
    \hfill
    \subfloat[FCBA~\cite{DBLP:conf/aaai/LiuZFYXM024} \label{fig:lr_effect_fcba_bda_after}]{
        \centering
        \includegraphics[width=0.45\linewidth]{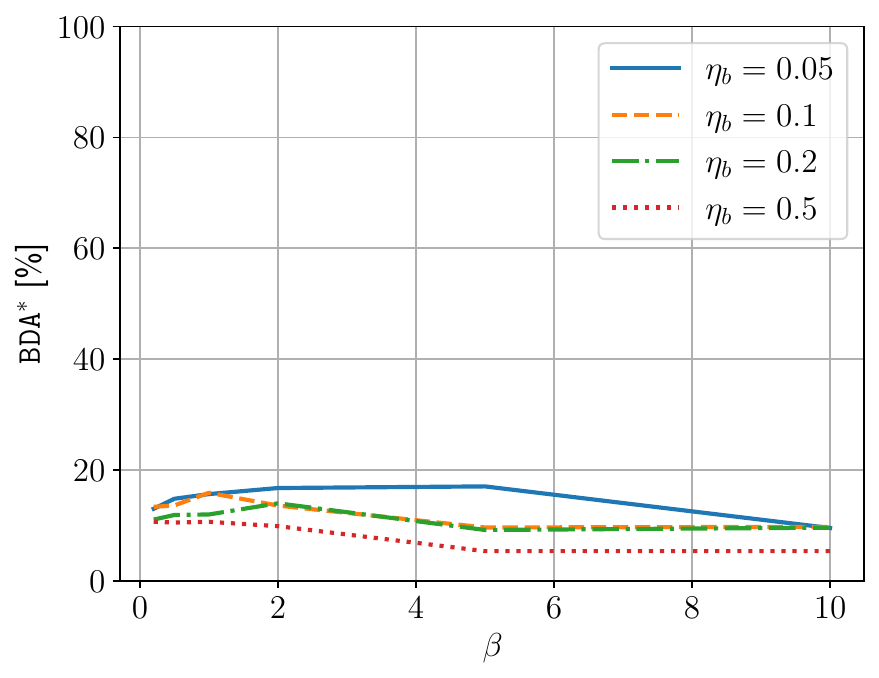}
    }

    \caption{Impact of $\eta_b$ and $\beta$ on the \bdaafter of SoTA attacks.}
    \label{fig:lr_effect_bda_after}
\end{figure}

\begin{figure}
    \centering
    \subfloat[A3FL~\cite{zhang_a3fl_2023} \label{fig:lr_effect_a3fl_lifespan}]{
        \centering
        \includegraphics[width=0.45\linewidth]{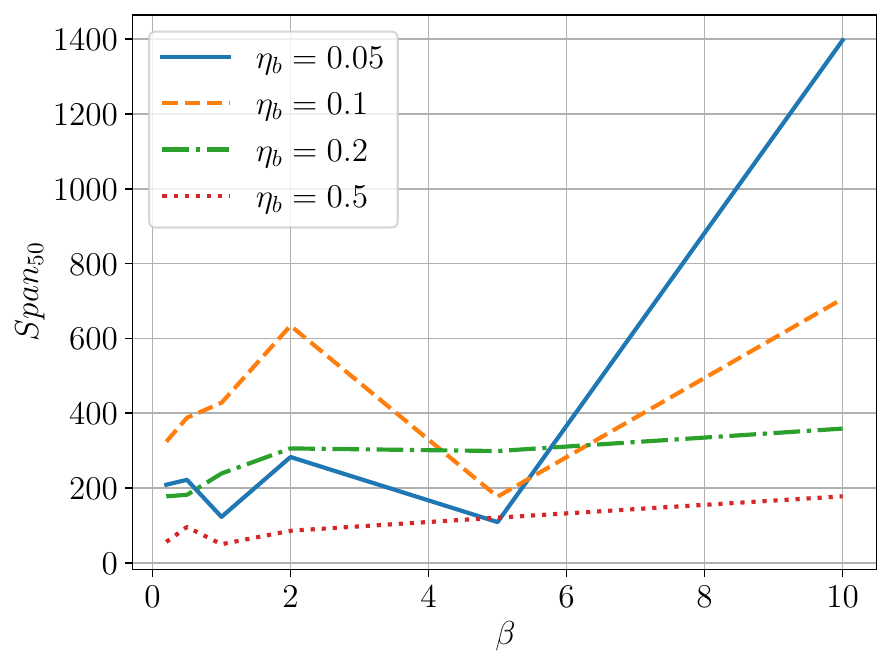}
    }
    \hfill
    \subfloat[Chameleon~\cite{dai_chameleon_2023} \label{fig:lr_effect_chameleon_lifespan}]{
        \centering
        \includegraphics[width=0.45\linewidth]{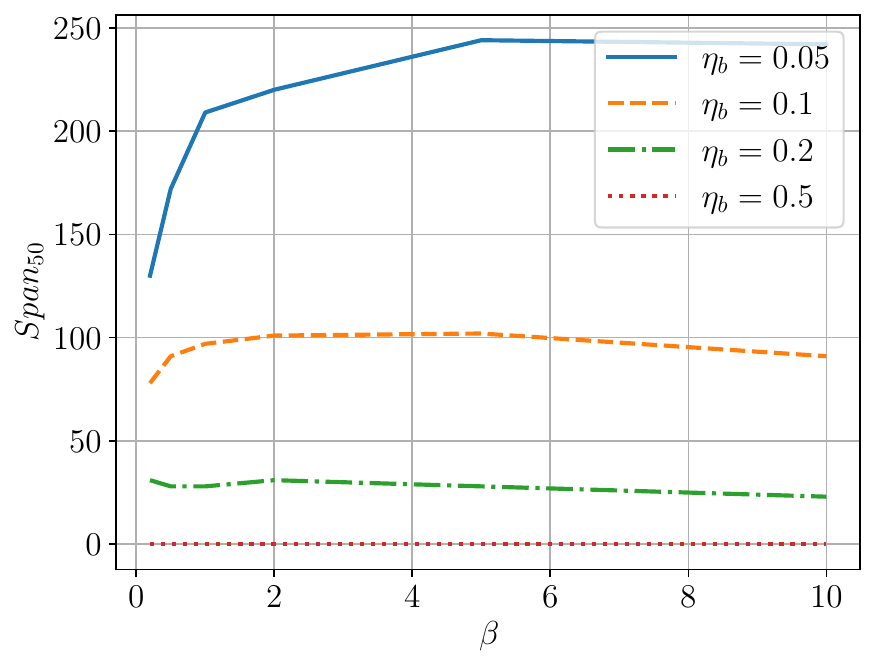}
    }

    \subfloat[DarkFed~\cite{DBLP:conf/ijcai/LiWNHXZW24} \label{fig:lr_effect_darkfed_lifespan}]{
        \centering
        \includegraphics[width=0.45\linewidth]{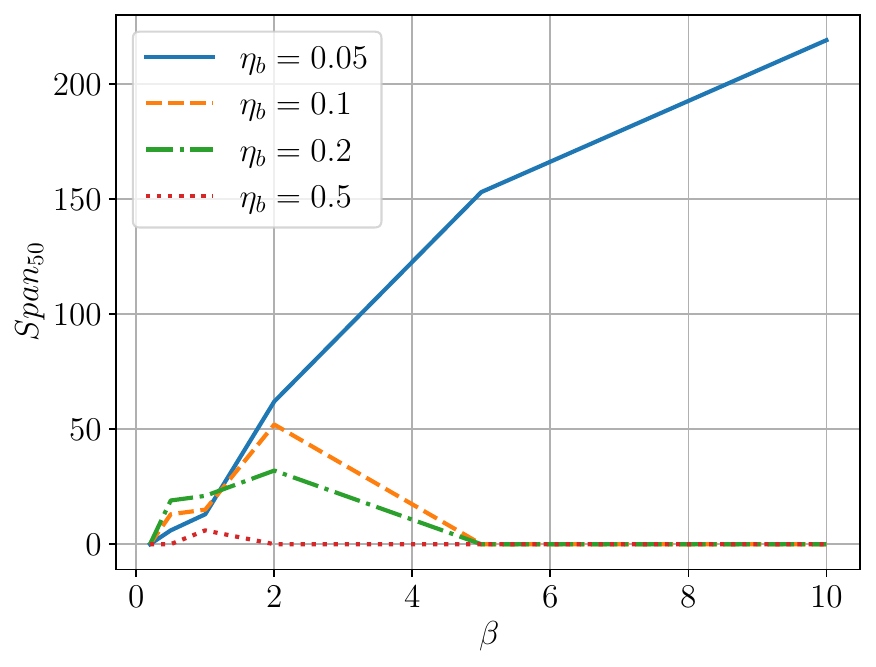}
    }
    \hfill
    \subfloat[FCBA~\cite{DBLP:conf/aaai/LiuZFYXM024} \label{fig:lr_effect_fcba_lifespan}]{
        \centering
        \includegraphics[width=0.45\linewidth]{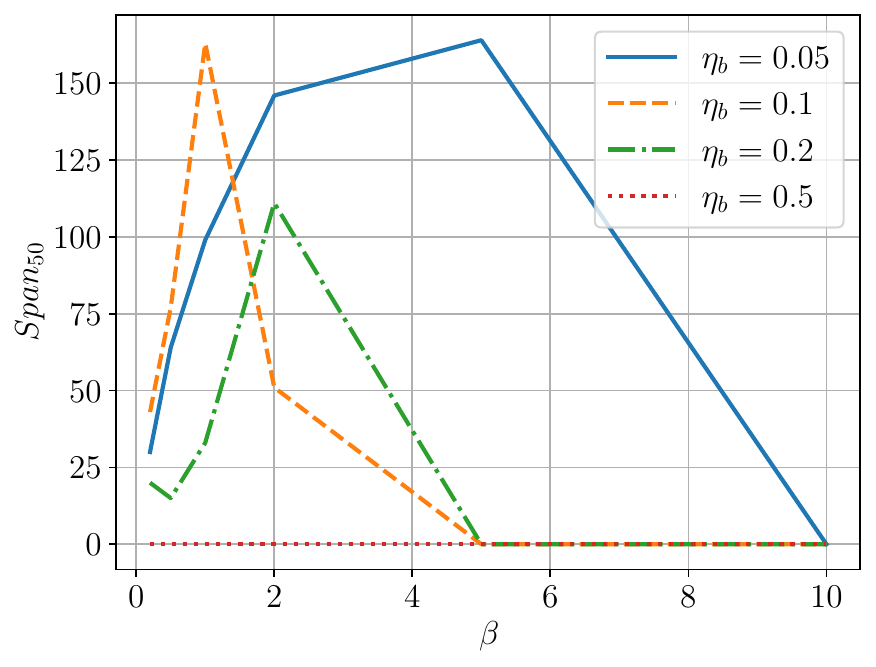}
    }

    \caption{Impact of $\eta_b$ and $\beta$ on the \lifespan of SoTA attacks.}
    \label{fig:lr_effect_lifespan}
\end{figure}

\begin{figure}
    \centering
    \subfloat[A3FL~\cite{zhang_a3fl_2023} \label{fig:lr_effect_a3fl_mta_during}]{
        \centering
        \includegraphics[width=0.45\linewidth]{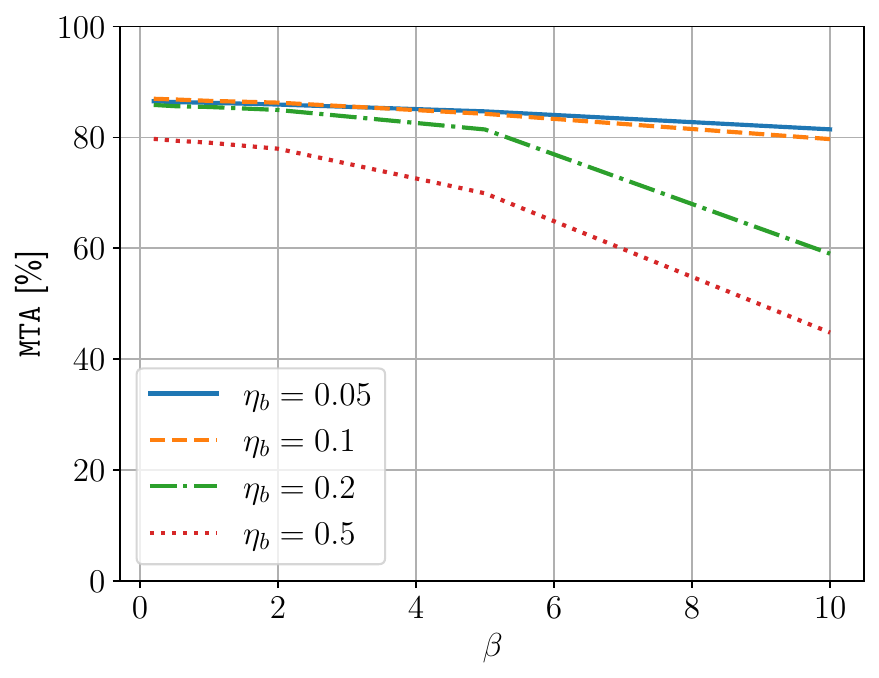}
    }
    \hfill
    \subfloat[Chameleon~\cite{dai_chameleon_2023} \label{fig:lr_effect_chameleon_mta_during}]{
        \centering
        \includegraphics[width=0.45\linewidth]{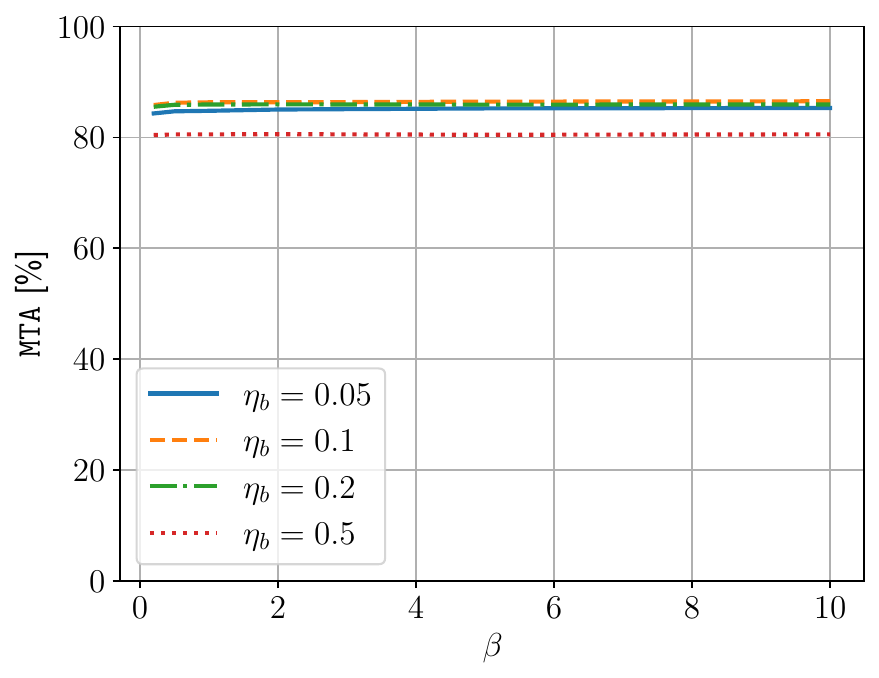}
    }

    \subfloat[DarkFed~\cite{DBLP:conf/ijcai/LiWNHXZW24} \label{fig:lr_effect_darkfed_mta_during}]{
        \centering
        \includegraphics[width=0.45\linewidth]{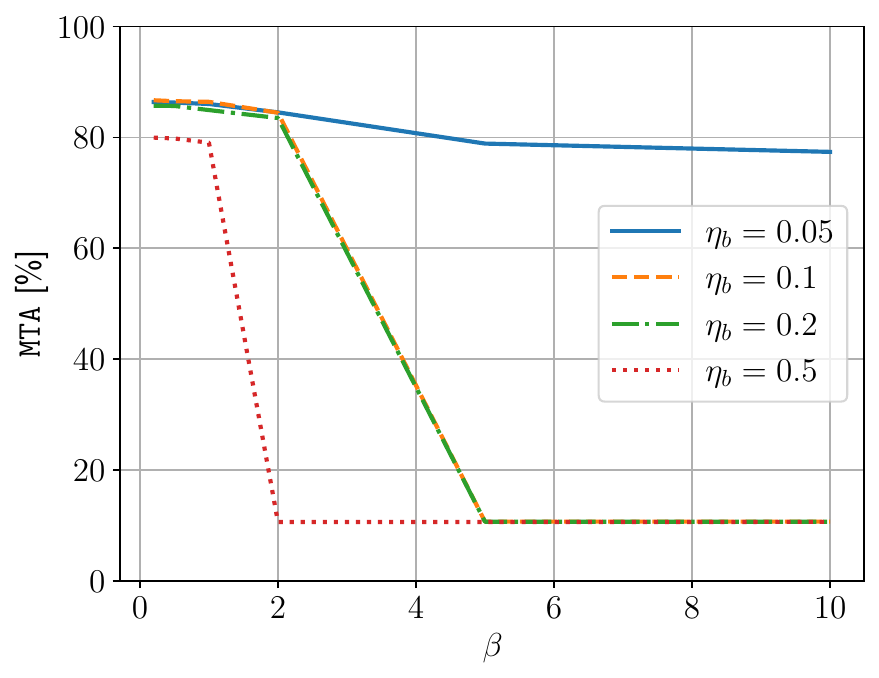}
    }
    \hfill
    \subfloat[FCBA~\cite{DBLP:conf/aaai/LiuZFYXM024} \label{fig:lr_effect_fcba_mta_during}]{
        \centering
        \includegraphics[width=0.45\linewidth]{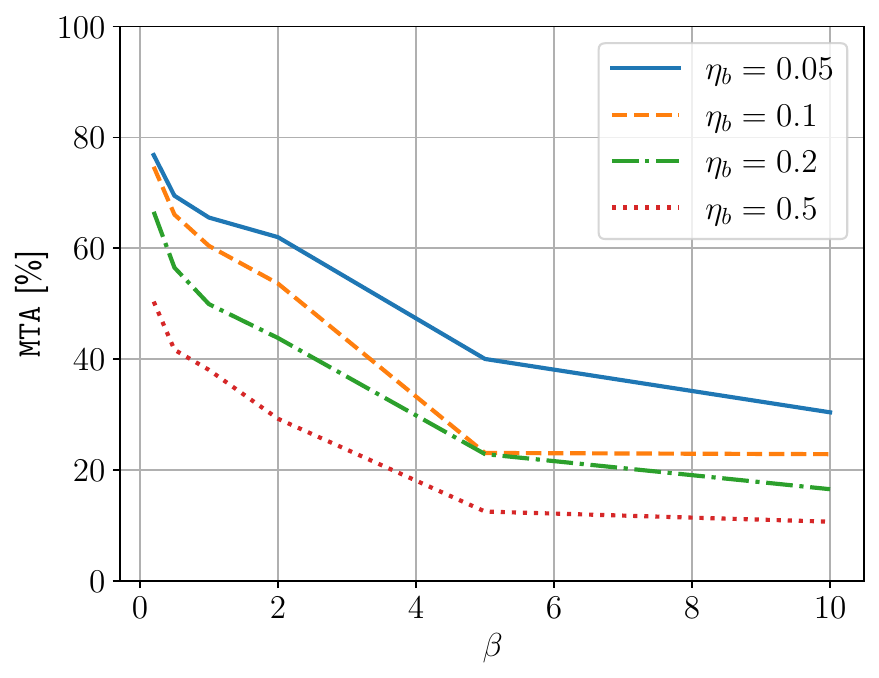}
    }

    \caption{Impact of $\eta_b$ and $\beta$ on the MTA of SoTA attacks.}
    \label{fig:lr_effect_mta_during}
\end{figure}

In Figures~\ref{fig:lr_effect_bda_after} and \ref{fig:lr_effect_lifespan}, we show the \bdaafter and \lifespan for the A3FL~\cite{zhang_a3fl_2023}, Chameleon~\cite{dai_chameleon_2023}, DarkFed~\cite{DBLP:conf/ijcai/LiWNHXZW24}, and FCBA~\cite{DBLP:conf/aaai/LiuZFYXM024} attacks for varying benign and malicious learning rates, complementing our results from Section~\ref{sec:lr_effect}.
In Figure~\ref{fig:lr_effect_mta_during}, for completeness, we also show the corresponding \mta values.

\subsection{Impact of Momentum}
\label{app:mu_effect}

\begin{figure}
    \centering
    \subfloat[A3FL~\cite{zhang_a3fl_2023} \label{fig:mu_effect_a3fl_bda_after}]{
        \centering
        \includegraphics[width=0.45\linewidth]{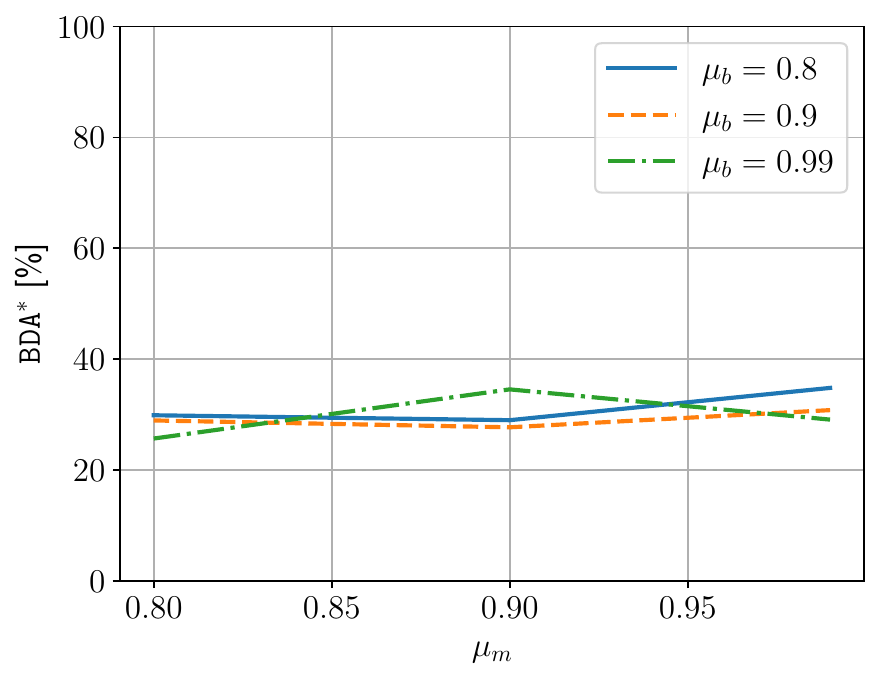}
    }
    \hfill
    \subfloat[Chameleon~\cite{dai_chameleon_2023} \label{fig:mu_effect_chameleon_bda_after}]{
        \centering
        \includegraphics[width=0.45\linewidth]{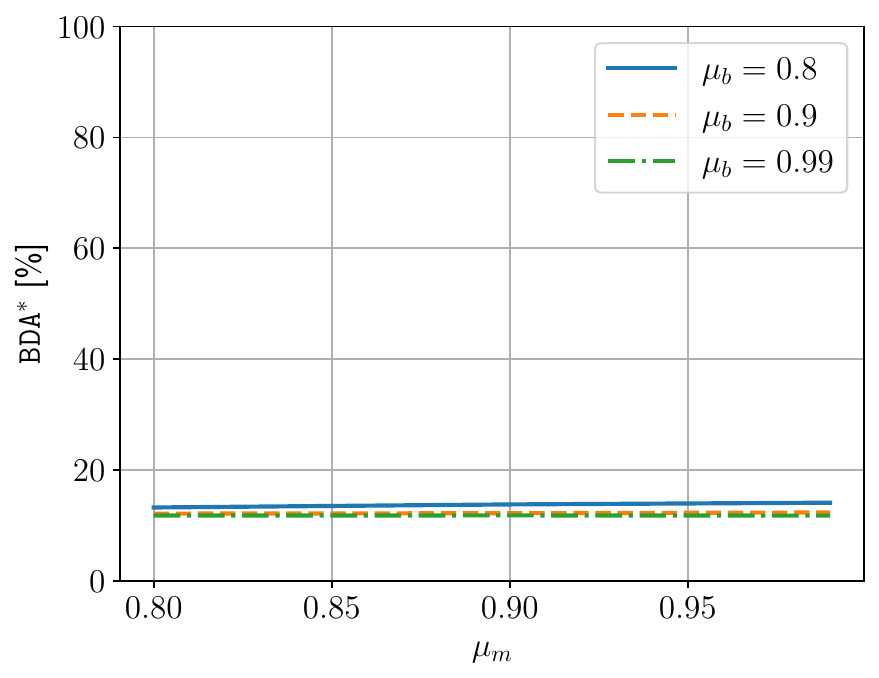}
    }

    \subfloat[DarkFed~\cite{DBLP:conf/ijcai/LiWNHXZW24} \label{fig:mu_effect_darkfed_bda_after}]{
        \centering
        \includegraphics[width=0.45\linewidth]{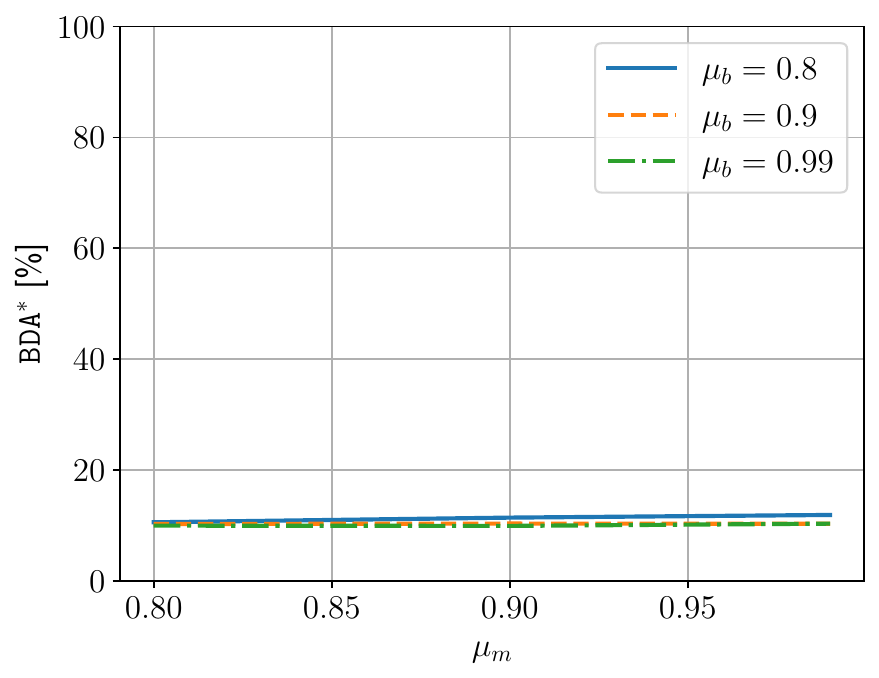}
    }
    \hfill
    \subfloat[FCBA~\cite{DBLP:conf/aaai/LiuZFYXM024} \label{fig:mu_effect_fcba_bda_after}]{
        \centering
        \includegraphics[width=0.45\linewidth]{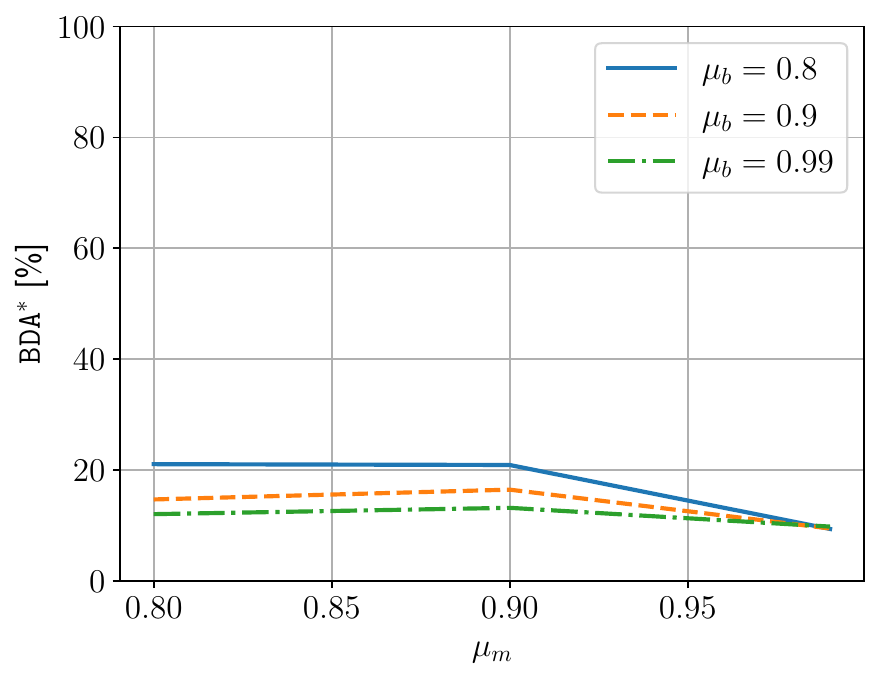}
    }

    \caption{Impact of $\mu_b$ and $\mu_m$ on the \bdaafter of SoTA attacks.}
    \label{fig:mu_effect_bda_after}
\end{figure}

\begin{figure}
    \centering
    \subfloat[A3FL~\cite{zhang_a3fl_2023} \label{fig:mu_effect_a3fl_lifespan}]{
        \centering
        \includegraphics[width=0.45\linewidth]{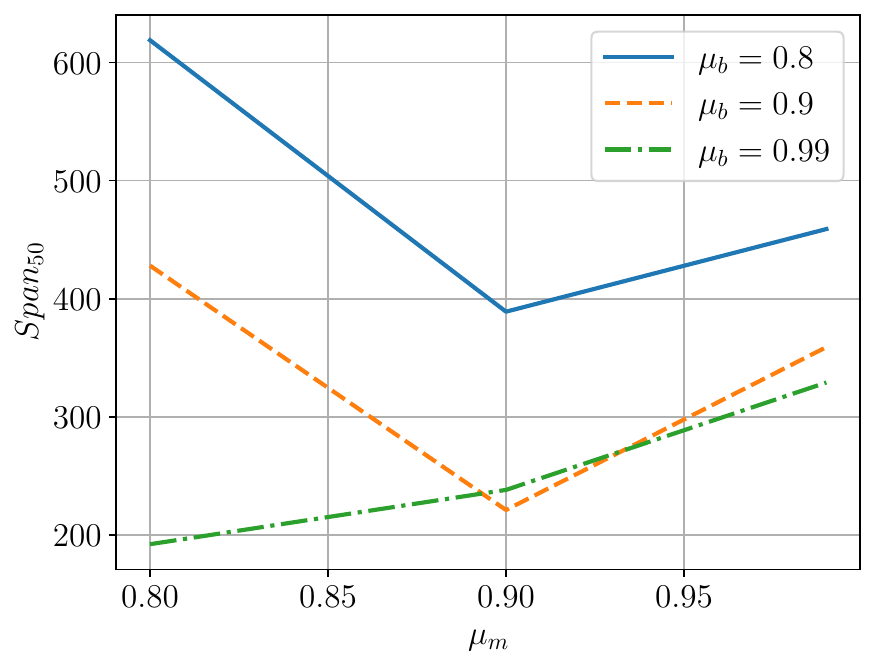}
    }
    \hfill
    \subfloat[Chameleon~\cite{dai_chameleon_2023} \label{fig:mu_effect_chameleon_lifespan}]{
        \centering
        \includegraphics[width=0.45\linewidth]{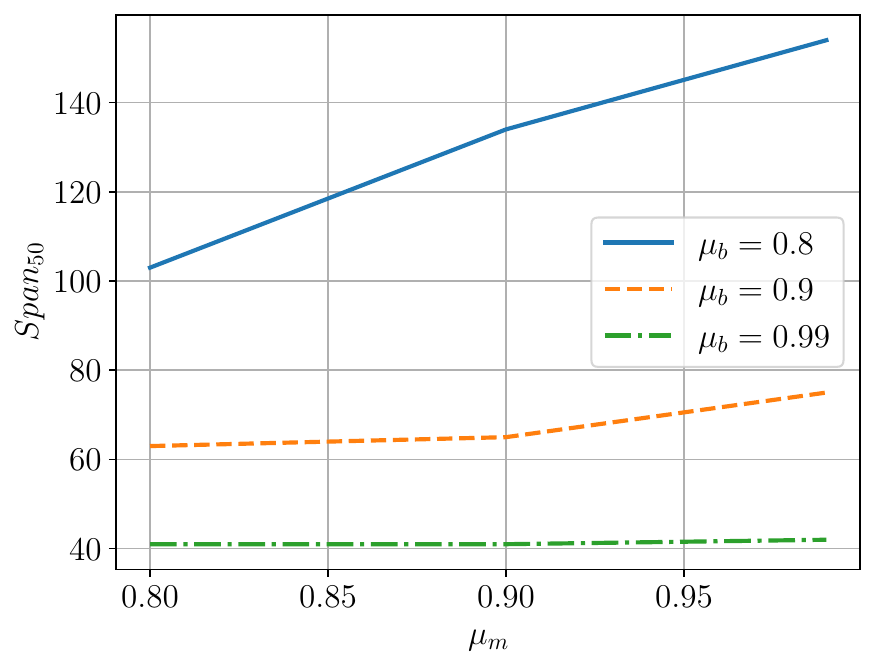}
    }

    \subfloat[DarkFed~\cite{DBLP:conf/ijcai/LiWNHXZW24} \label{fig:mu_effect_darkfed_lifespan}]{
        \centering
        \includegraphics[width=0.45\linewidth]{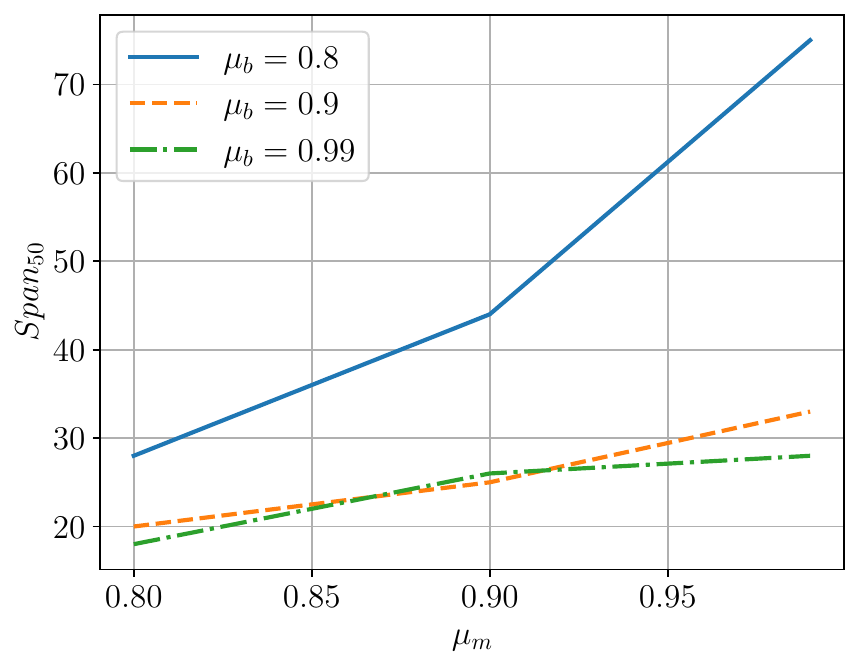}
    }
    \hfill
    \subfloat[FCBA~\cite{DBLP:conf/aaai/LiuZFYXM024} \label{fig:mu_effect_fcba_lifespan}]{
        \centering
        \includegraphics[width=0.45\linewidth]{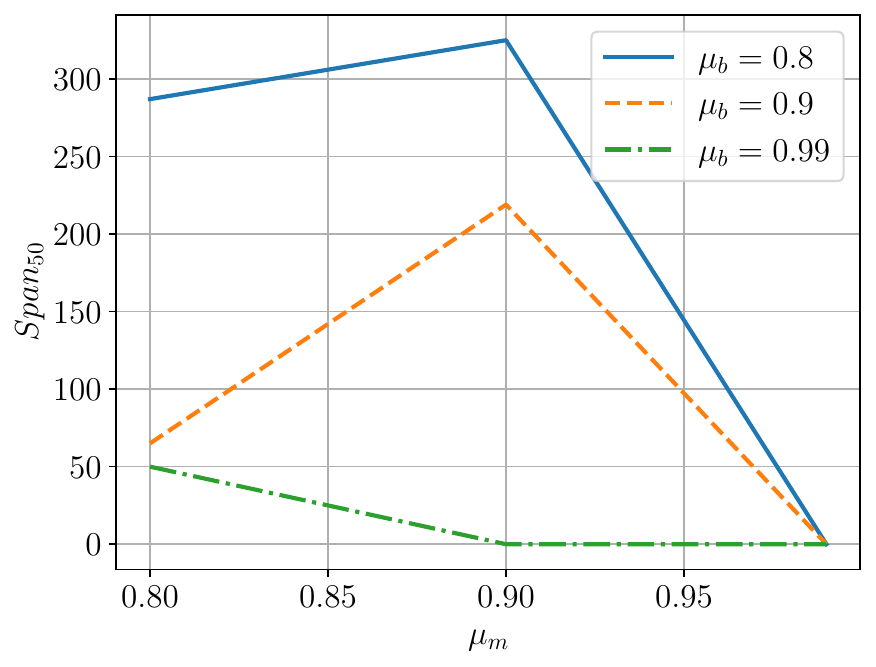}
    }

    \caption{Impact of $\mu_b$ and $\mu_m$ on the \lifespan of SoTA attacks.}
    \label{fig:mu_effect_lifespan}
\end{figure}

\begin{figure}
    \centering
    \subfloat[A3FL~\cite{zhang_a3fl_2023} \label{fig:mu_effect_a3fl_mta_during}]{
        \centering
        \includegraphics[width=0.45\linewidth]{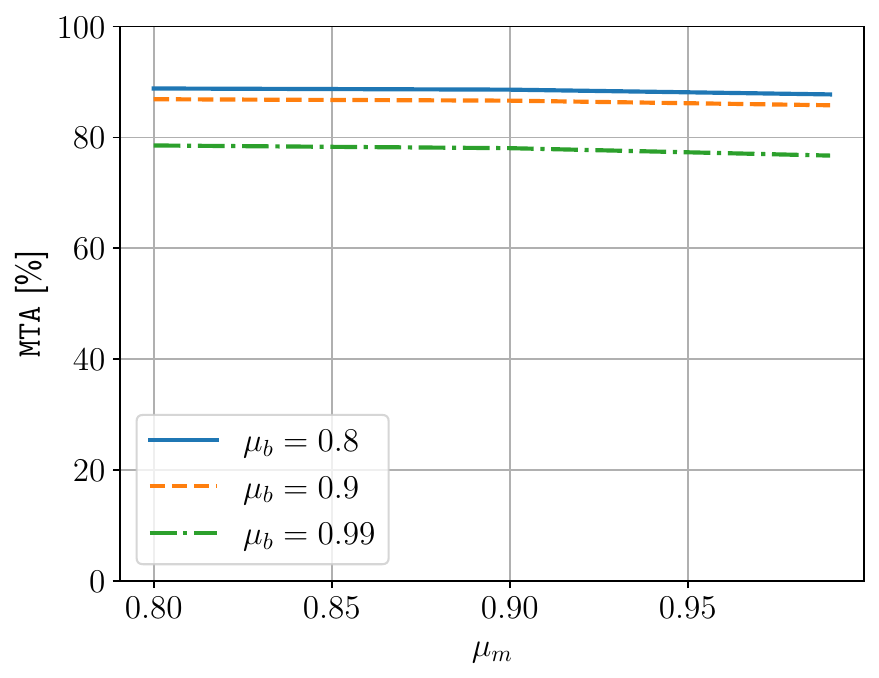}
    }
    \hfill
    \subfloat[Chameleon~\cite{dai_chameleon_2023} \label{fig:mu_effect_chameleon_mta_during}]{
        \centering
        \includegraphics[width=0.45\linewidth]{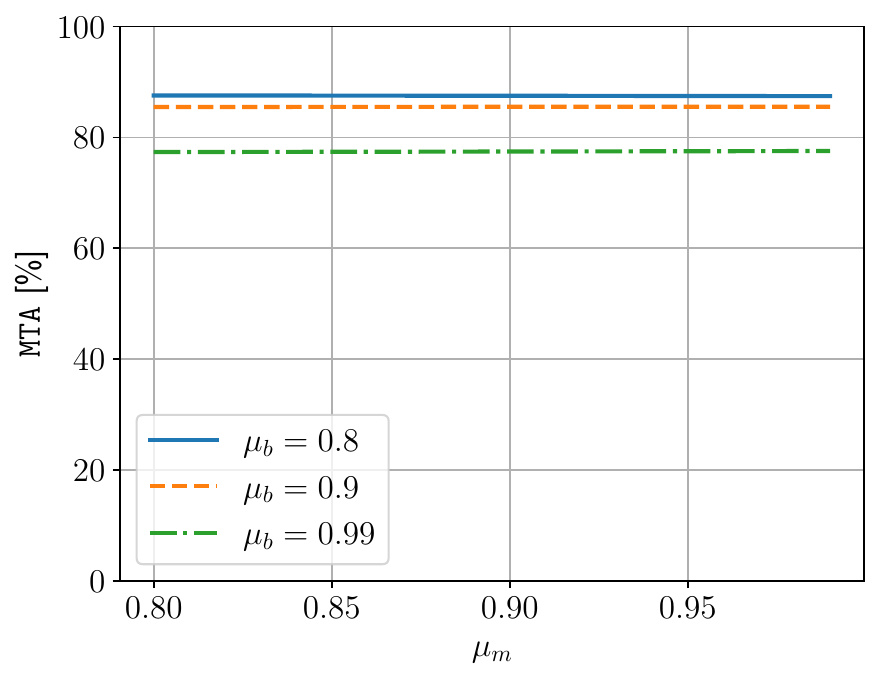}
    }

    \subfloat[DarkFed~\cite{DBLP:conf/ijcai/LiWNHXZW24} \label{fig:mu_effect_darkfed_mta_during}]{
        \centering
        \includegraphics[width=0.45\linewidth]{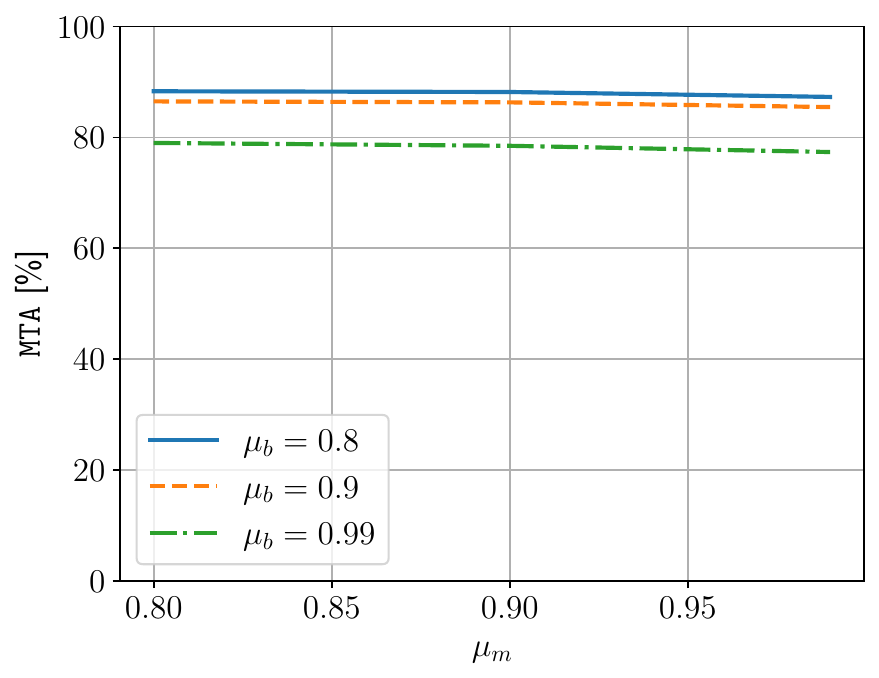}
    }
    \hfill
    \subfloat[FCBA~\cite{DBLP:conf/aaai/LiuZFYXM024} \label{fig:mu_effect_fcba_mta_during}]{
        \centering
        \includegraphics[width=0.45\linewidth]{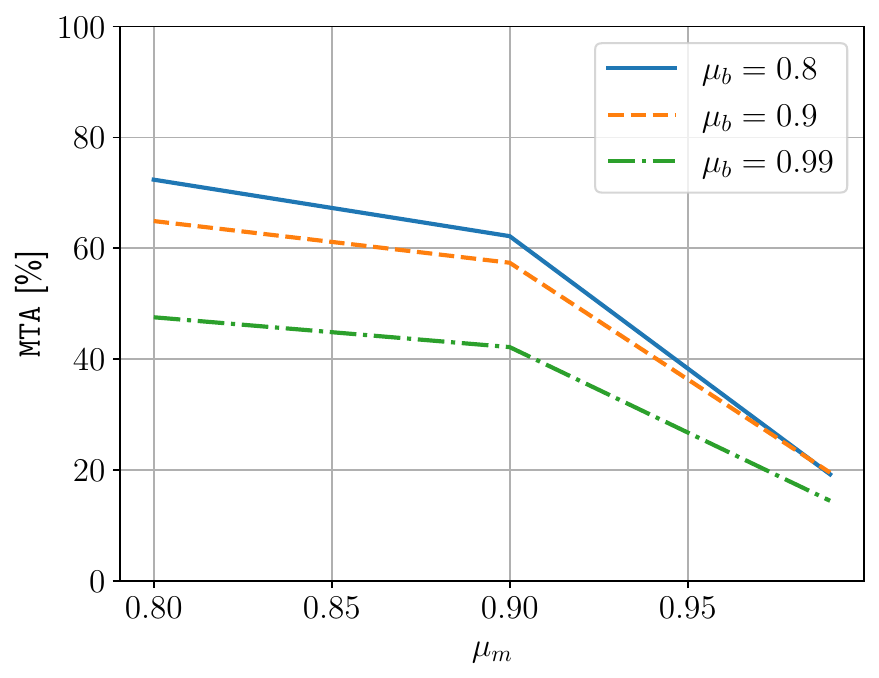}
    }

    \caption{Impact of $\mu_b$ and $\mu_m$ on the \mta of SoTA attacks.}
    \label{fig:mu_effect_mta_during}
\end{figure}

In Figures~\ref{fig:mu_effect_bda_after}, \ref{fig:mu_effect_lifespan}, and \ref{fig:mu_effect_mta_during}, we show the \bdaafter, \lifespan, and \mta for the A3FL~\cite{zhang_a3fl_2023}, Chameleon~\cite{dai_chameleon_2023}, DarkFed~\cite{DBLP:conf/ijcai/LiWNHXZW24}, and FCBA~\cite{DBLP:conf/aaai/LiuZFYXM024} attacks for varying benign and malicious momentum factors, complementing our results from Section~\ref{sec:mu_effect}.

\subsection{Impact of Batch Size \& Local Epochs}
\label{app:E_B_effect}

\begin{figure}[htb!]
    \centering
    \subfloat[A3FL~\cite{zhang_a3fl_2023} \label{fig:E_effect_a3fl_bda_after}]{
        \centering
        \includegraphics[width=0.45\linewidth]{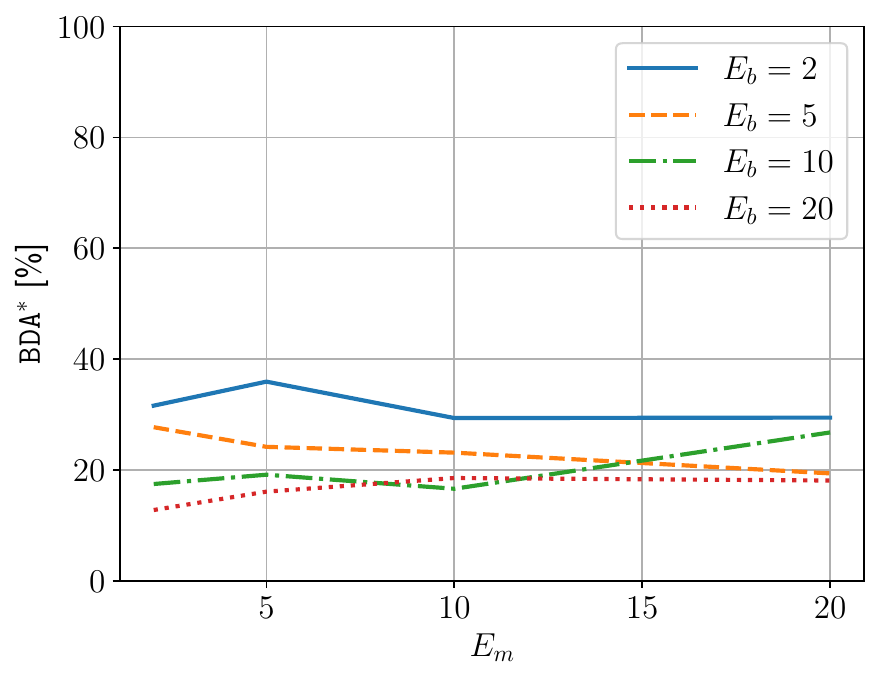}
    }
    \hfill
    \subfloat[Chameleon~\cite{dai_chameleon_2023} \label{fig:E_effect_chameleon_bda_after}]{
        \centering
        \includegraphics[width=0.45\linewidth]{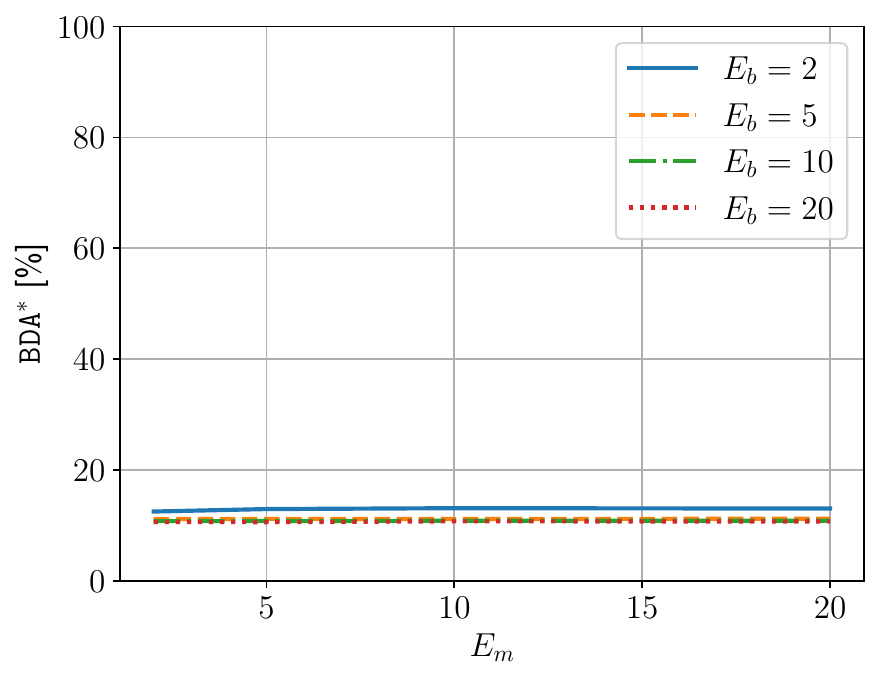}
    }

    \subfloat[DarkFed~\cite{DBLP:conf/ijcai/LiWNHXZW24} \label{fig:E_effect_darkfed_bda_after}]{
        \centering
        \includegraphics[width=0.45\linewidth]{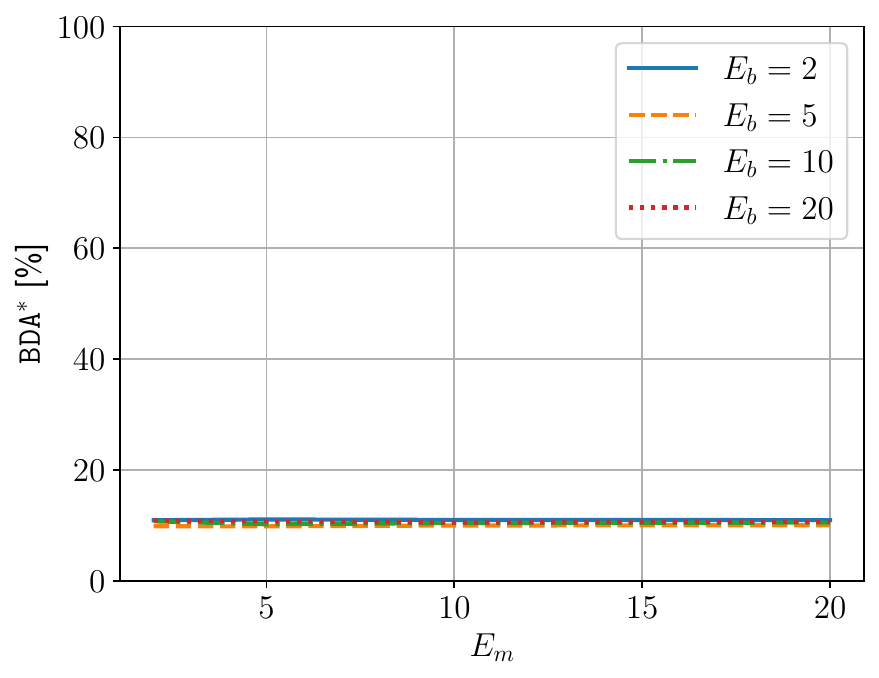}
    }
    \hfill
    \subfloat[FCBA~\cite{DBLP:conf/aaai/LiuZFYXM024} \label{fig:E_effect_fcba_bda_after}]{
        \centering
        \includegraphics[width=0.45\linewidth]{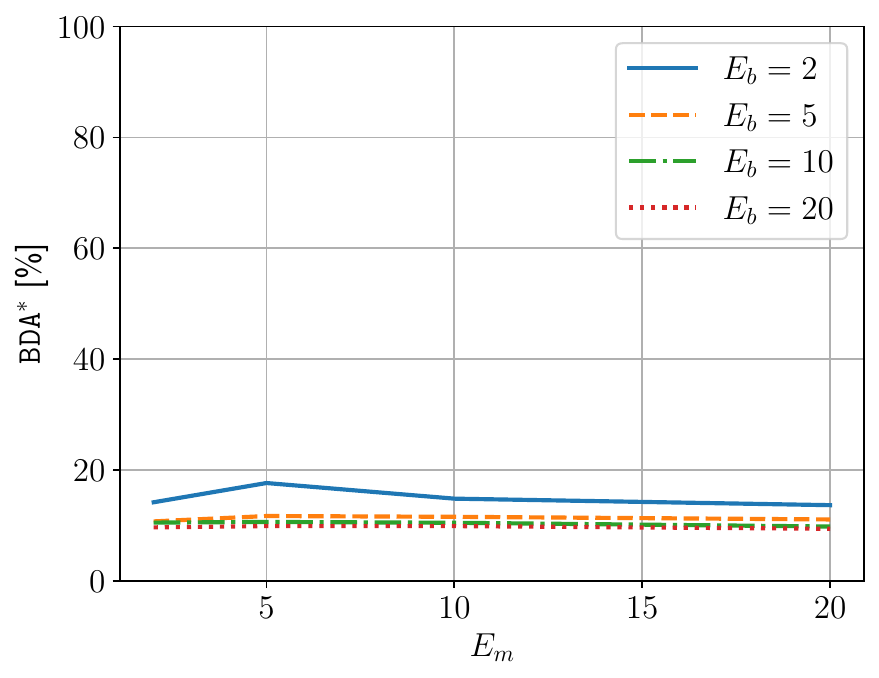}
    }

    \caption{Impact of $E_b$ and $E_m$ on the \bdaafter of SoTA attacks.}
    \label{fig:E_effect_bda_after}
\end{figure}

\begin{figure}[htb!]
    \centering
    \subfloat[A3FL~\cite{zhang_a3fl_2023} \label{fig:B_effect_a3fl_bda_after}]{
        \centering
        \includegraphics[width=0.45\linewidth]{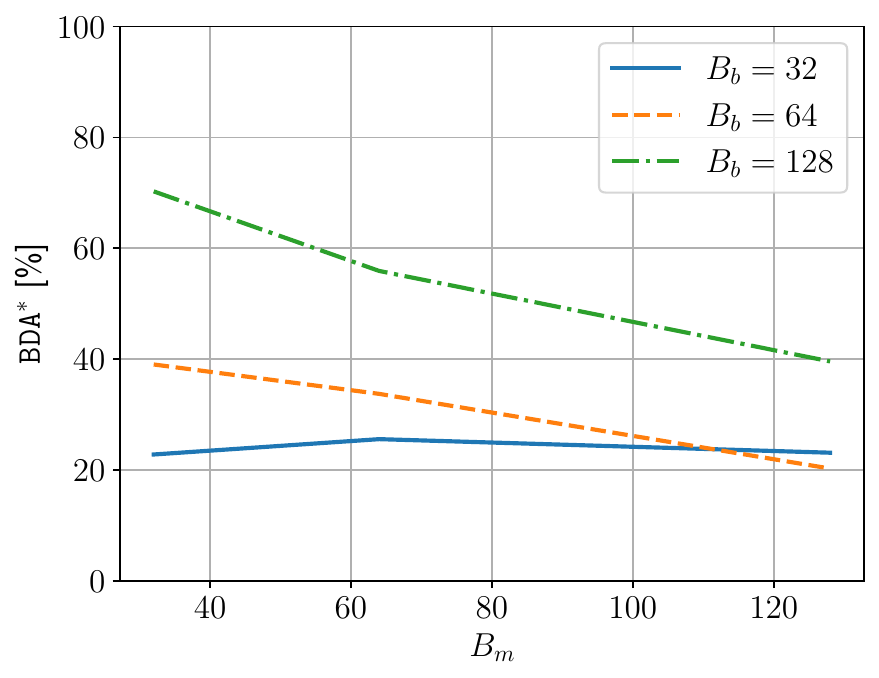}
    }
    \hfill
    \subfloat[Chameleon~\cite{dai_chameleon_2023} \label{fig:B_effect_chameleon_bda_after}]{
        \centering
        \includegraphics[width=0.45\linewidth]{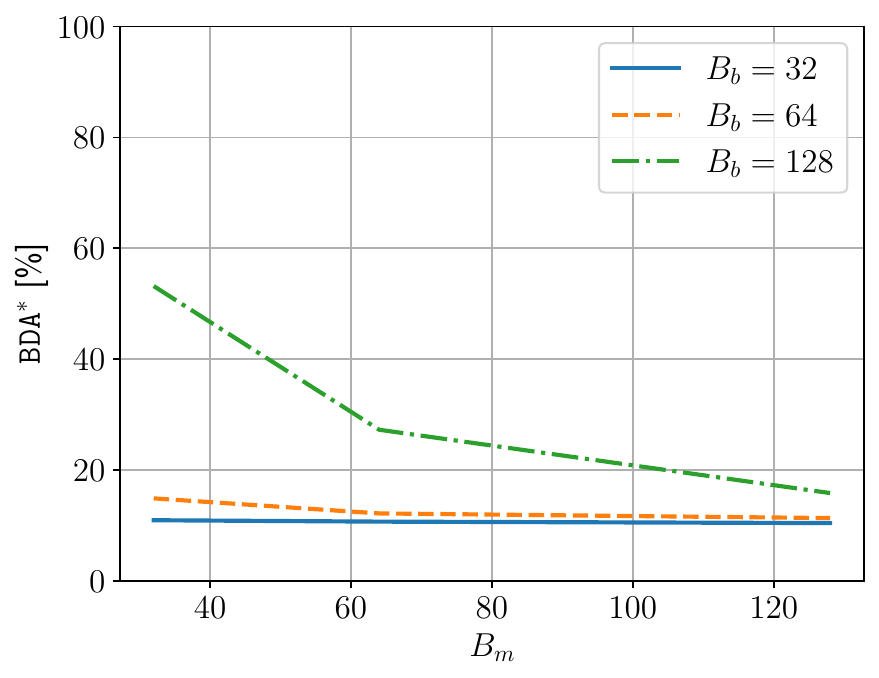}
    }

    \subfloat[DarkFed~\cite{DBLP:conf/ijcai/LiWNHXZW24} \label{fig:B_effect_darkfed_bda_after}]{
        \centering
        \includegraphics[width=0.45\linewidth]{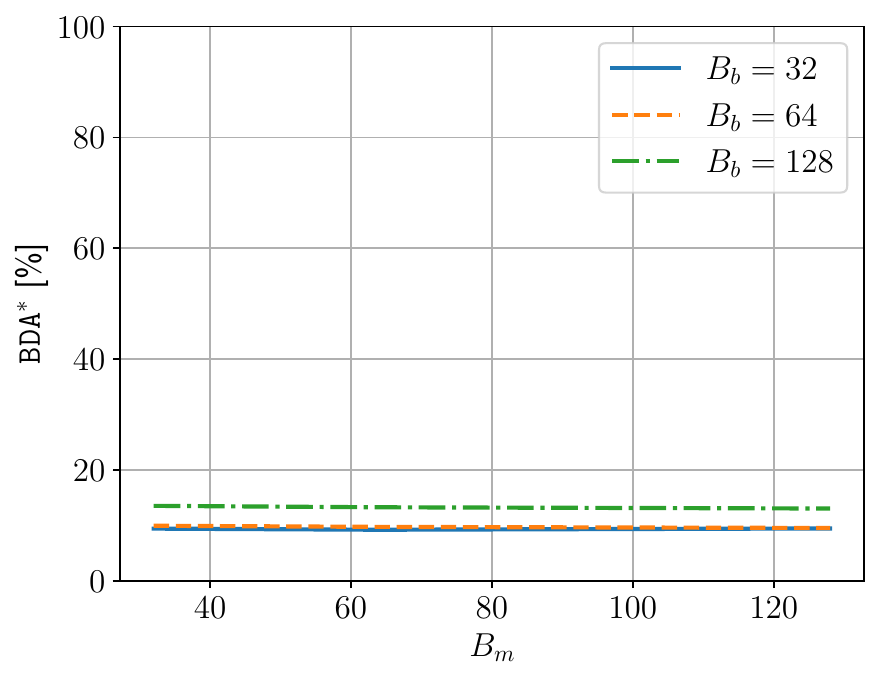}
    }
    \hfill
    \subfloat[FCBA~\cite{DBLP:conf/aaai/LiuZFYXM024} \label{fig:B_effect_fcba_bda_after}]{
        \centering
        \includegraphics[width=0.45\linewidth]{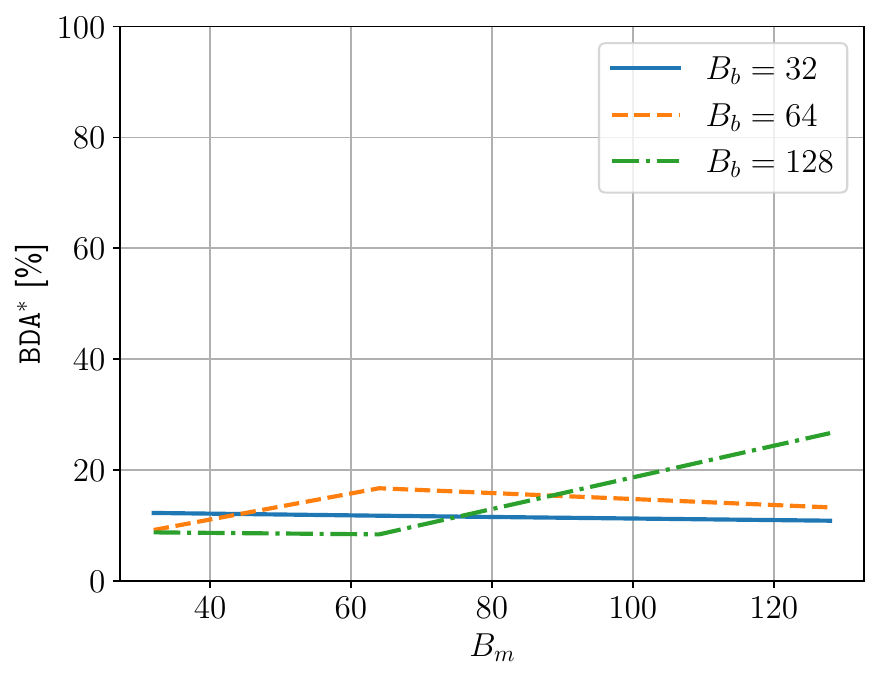}
    }

    \caption{Impact of $B_b$ and $B_m$ on the \bdaafter of SoTA attacks.}
    \label{fig:B_effect_bda_after}
\end{figure}

\begin{figure}[htb!]
    \centering
    \subfloat[A3FL~\cite{zhang_a3fl_2023} \label{fig:E_effect_a3fl_lifespan}]{
        \centering
        \includegraphics[width=0.45\linewidth]{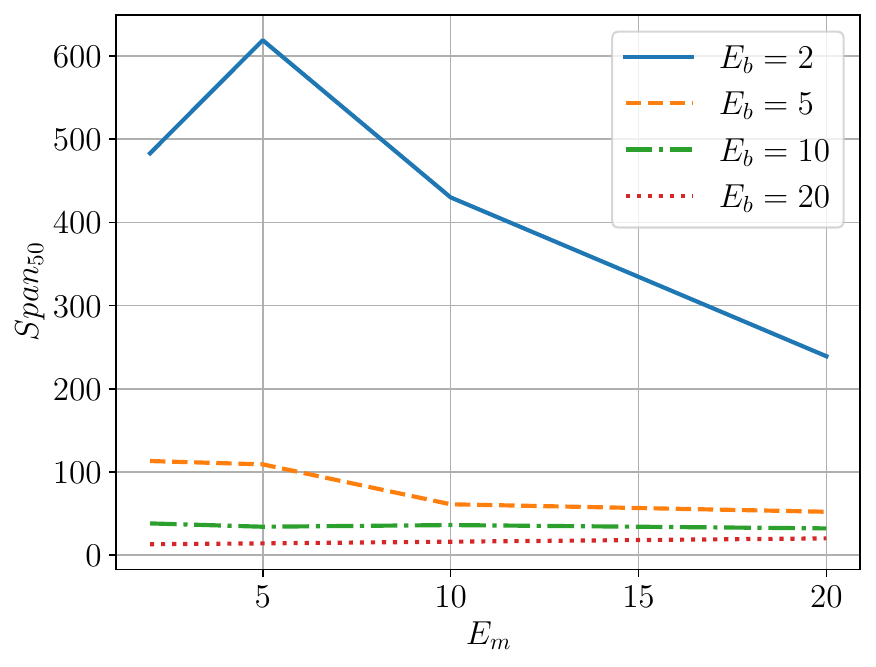}
    }
    \hfill
    \subfloat[Chameleon~\cite{dai_chameleon_2023} \label{fig:E_effect_chameleon_lifespan}]{
        \centering
        \includegraphics[width=0.45\linewidth]{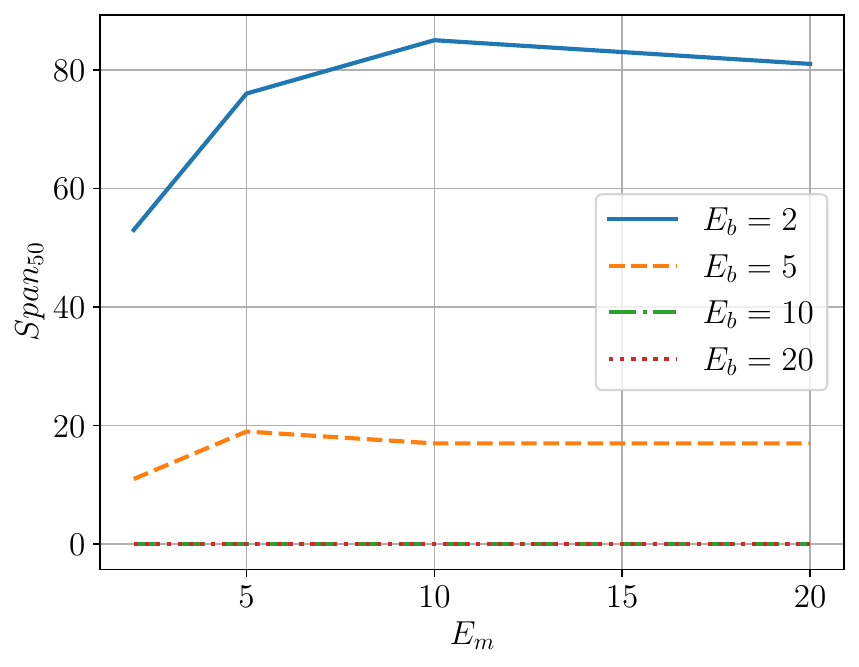}
    }

    \subfloat[DarkFed~\cite{DBLP:conf/ijcai/LiWNHXZW24} \label{fig:E_effect_darkfed_lifespan}]{
        \centering
        \includegraphics[width=0.45\linewidth]{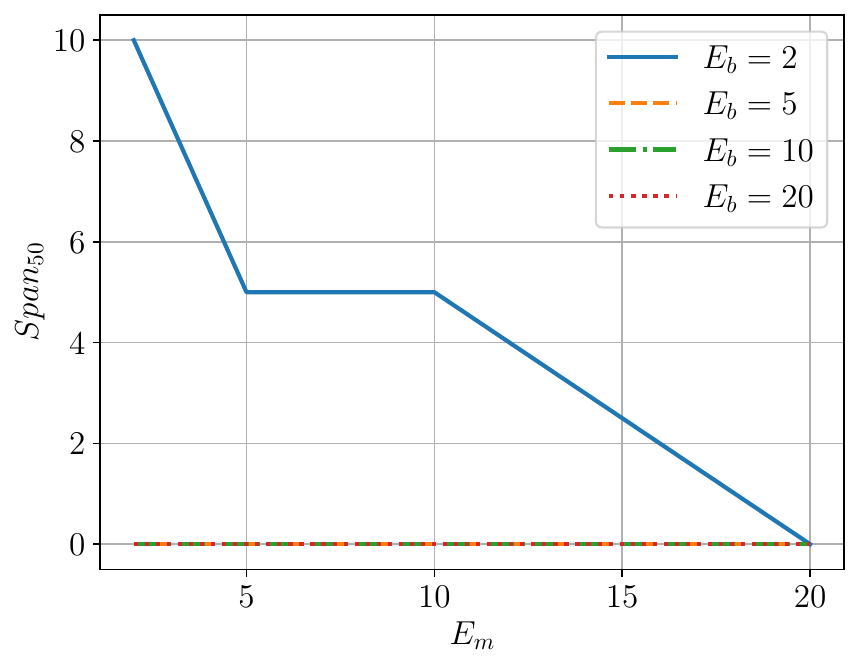}
    }
    \hfill
    \subfloat[FCBA~\cite{DBLP:conf/aaai/LiuZFYXM024} \label{fig:E_effect_fcba_lifespan}]{
        \centering
        \includegraphics[width=0.45\linewidth]{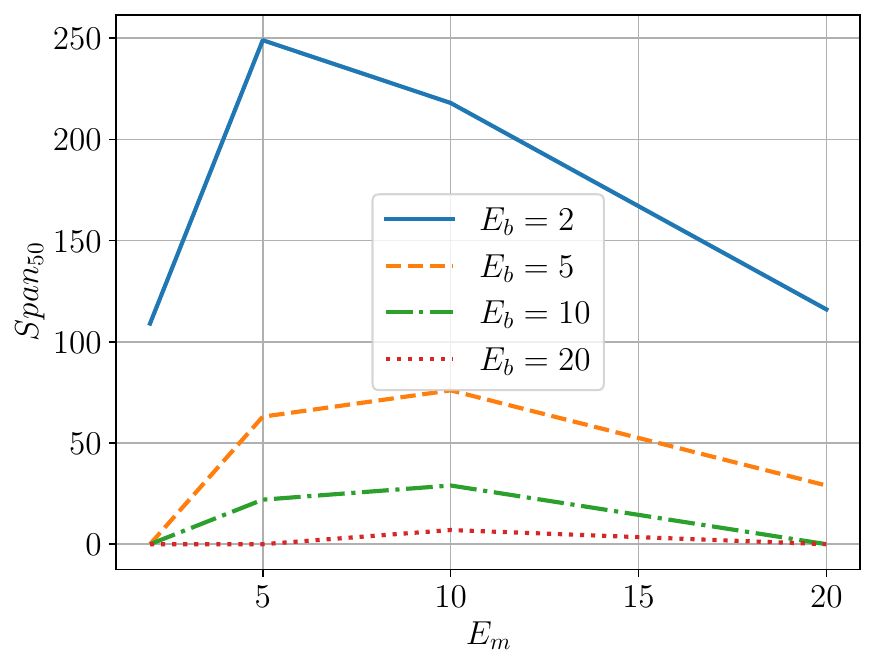}
    }

    \caption{Impact of $E_b$ and $E_m$ on the \lifespan of SoTA attacks.}
    \label{fig:E_effect_lifespan}
\end{figure}

\begin{figure}[htb!]
    \centering
    \subfloat[A3FL~\cite{zhang_a3fl_2023} \label{fig:B_effect_a3fl_lifespan}]{
        \centering
        \includegraphics[width=0.45\linewidth]{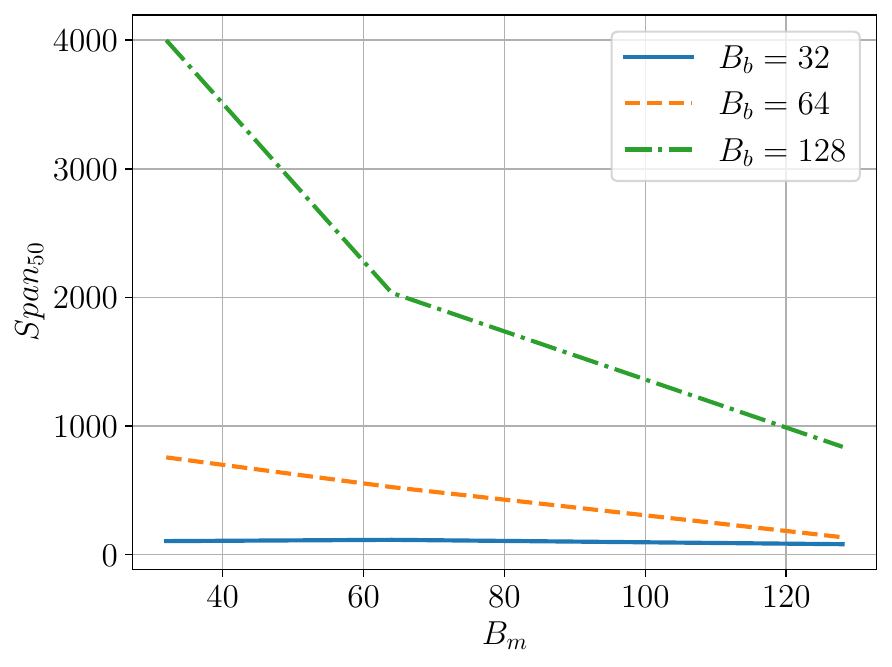}
    }
    \hfill
    \subfloat[Chameleon~\cite{dai_chameleon_2023} \label{fig:B_effect_chameleon_lifespan}]{
        \centering
        \includegraphics[width=0.45\linewidth]{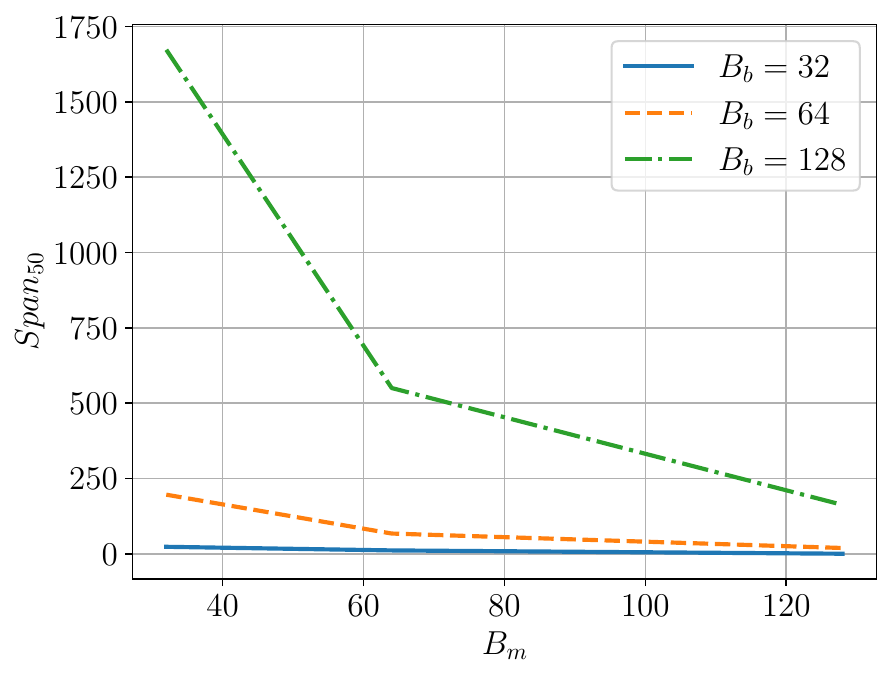}
    }

    \subfloat[DarkFed~\cite{DBLP:conf/ijcai/LiWNHXZW24} \label{fig:B_effect_darkfed_lifespan}]{
        \centering
        \includegraphics[width=0.45\linewidth]{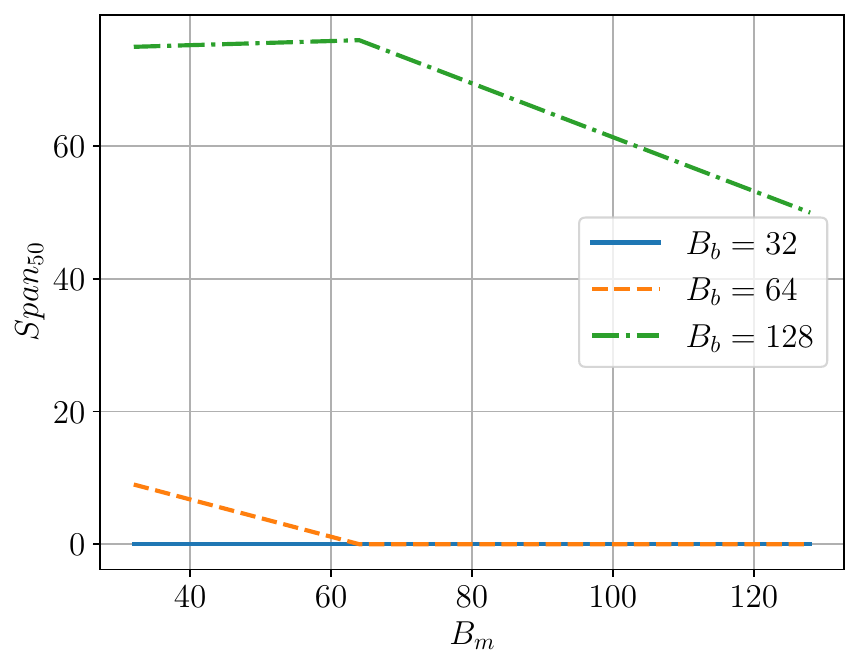}
    }
    \hfill
    \subfloat[FCBA~\cite{DBLP:conf/aaai/LiuZFYXM024} \label{fig:B_effect_fcba_lifespan}]{
        \centering
        \includegraphics[width=0.45\linewidth]{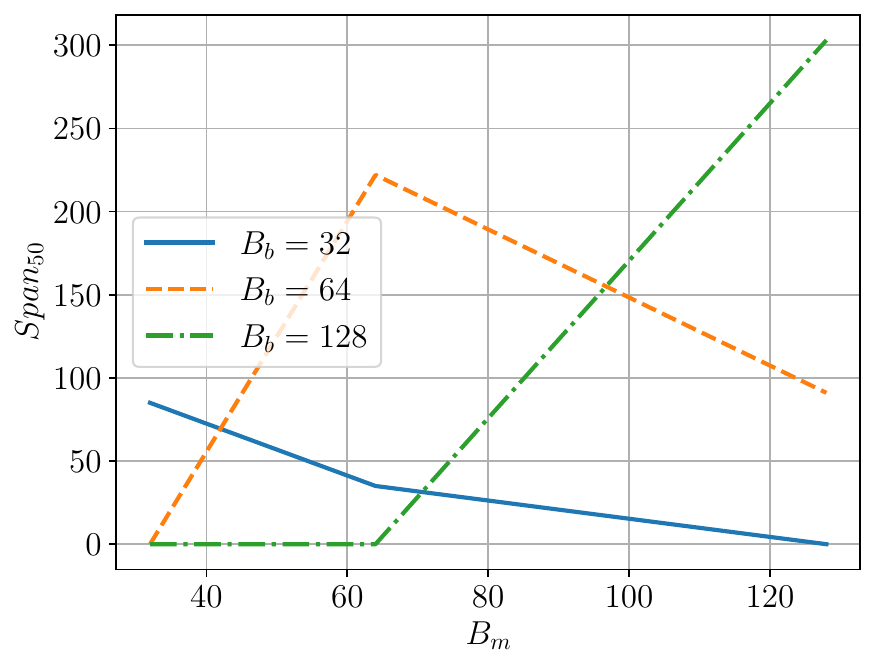}
    }

    \caption{Impact of $B_b$ and $B_m$ on the \lifespan of SoTA attacks.}
    \label{fig:B_effect_lifespan}
\end{figure}

\begin{figure}[htb!]
    \centering
    \subfloat[A3FL~\cite{zhang_a3fl_2023} \label{fig:E_effect_a3fl_mta_during}]{
        \centering
        \includegraphics[width=0.45\linewidth]{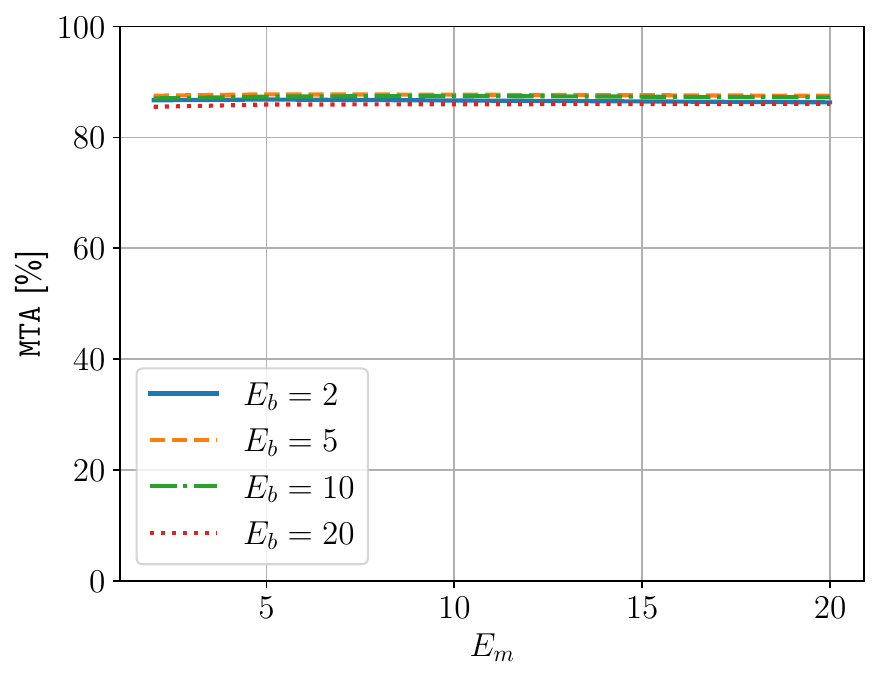}
    }
    \hfill
    \subfloat[Chameleon~\cite{dai_chameleon_2023} \label{fig:E_effect_chameleon_mta_during}]{
        \centering
        \includegraphics[width=0.45\linewidth]{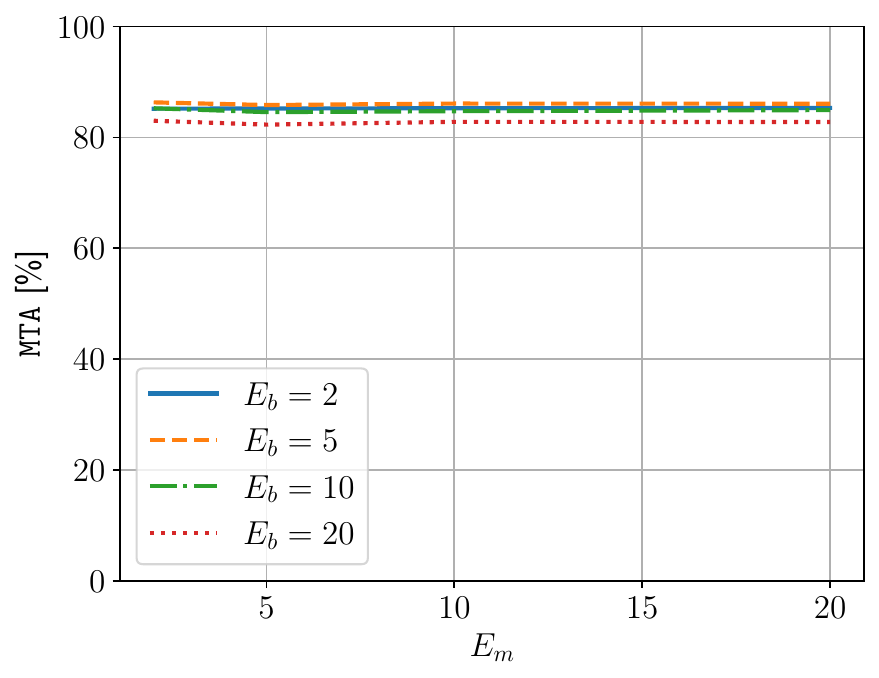}
    }

    \subfloat[DarkFed~\cite{DBLP:conf/ijcai/LiWNHXZW24} \label{fig:E_effect_darkfed_mta_during}]{
        \centering
        \includegraphics[width=0.45\linewidth]{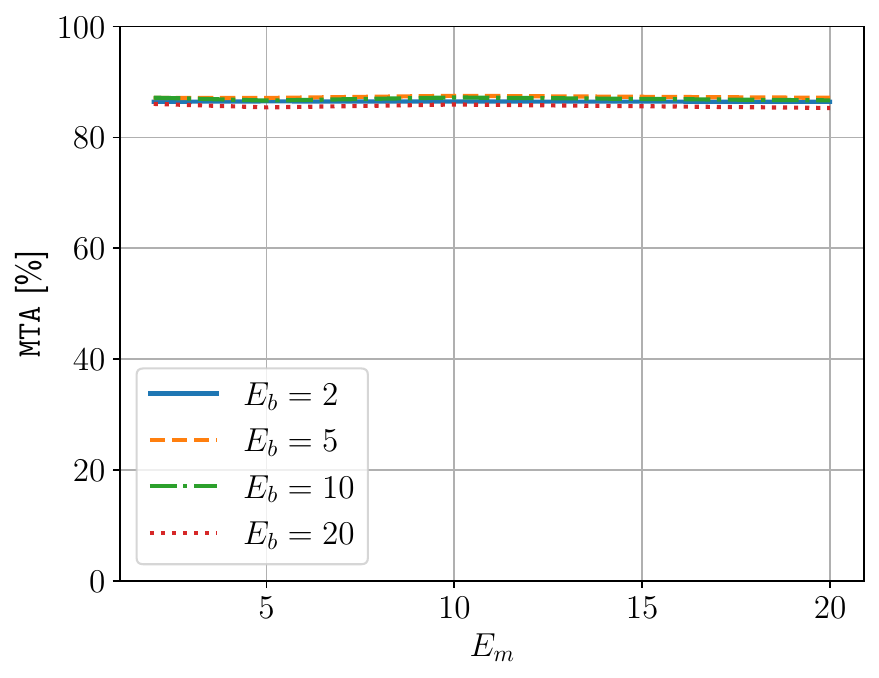}
    }
    \hfill
    \subfloat[FCBA~\cite{DBLP:conf/aaai/LiuZFYXM024} \label{fig:E_effect_fcba_mta_during}]{
        \centering
        \includegraphics[width=0.45\linewidth]{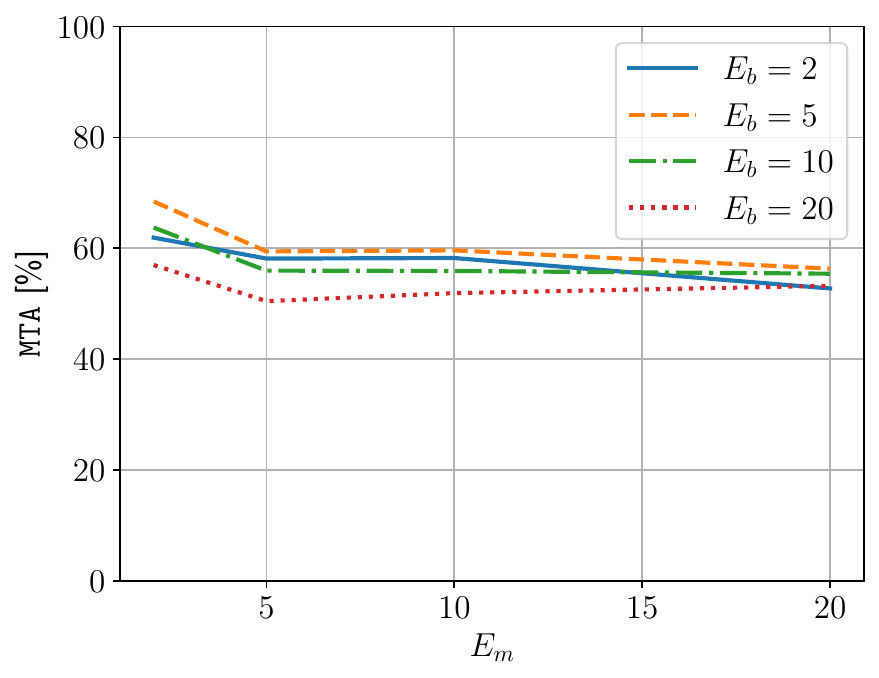}
    }

    \caption{Impact of $E_b$ and $E_m$ on the \mta of SoTA attacks.}
    \label{fig:E_effect_mta_during}
\end{figure}

\begin{figure}[htb!]
    \centering
    \subfloat[A3FL~\cite{zhang_a3fl_2023} \label{fig:B_effect_a3fl_mta_during}]{
        \centering
        \includegraphics[width=0.45\linewidth]{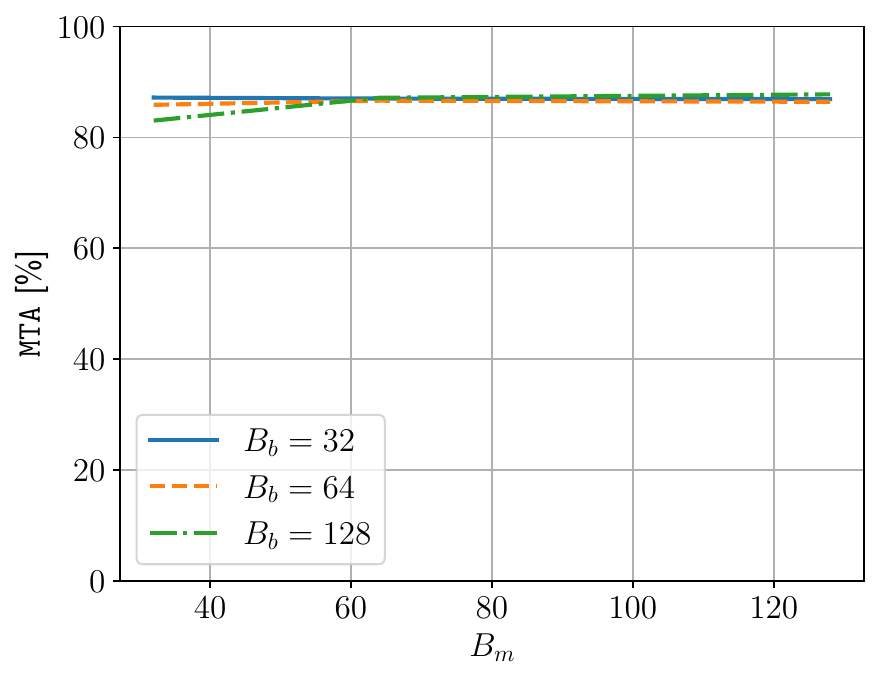}
    }
    \hfill
    \subfloat[Chameleon~\cite{dai_chameleon_2023} \label{fig:B_effect_chameleon_mta_during}]{
        \centering
        \includegraphics[width=0.45\linewidth]{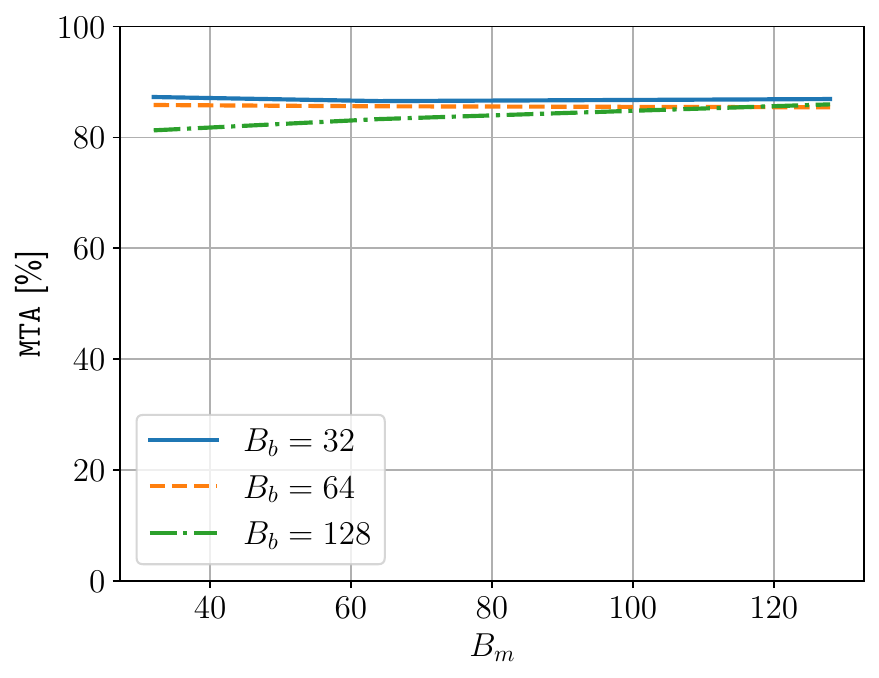}
    }

    \subfloat[DarkFed~\cite{DBLP:conf/ijcai/LiWNHXZW24} \label{fig:B_effect_darkfed_mta_during}]{
        \centering
        \includegraphics[width=0.45\linewidth]{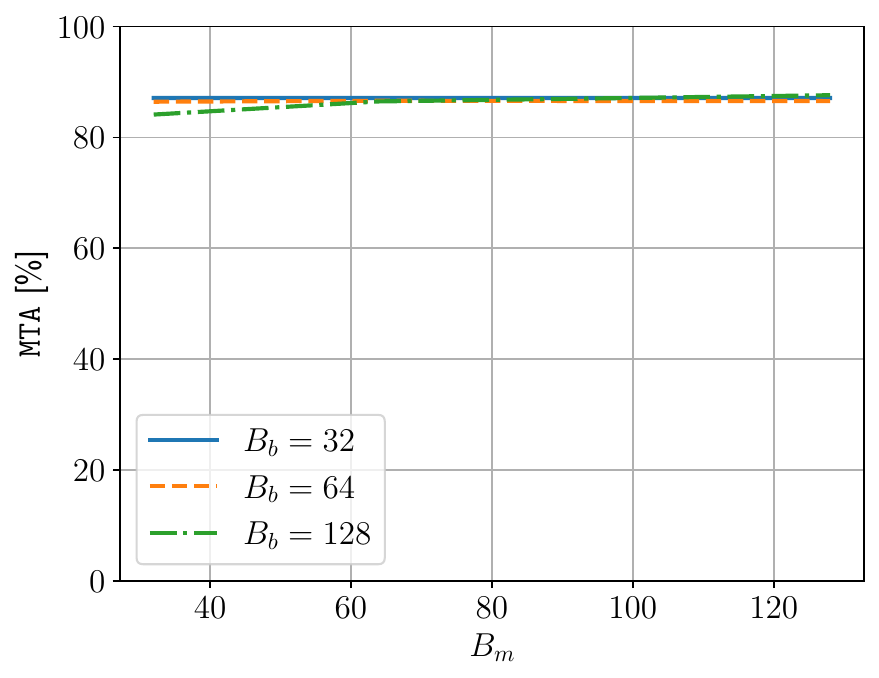}
    }
    \hfill
    \subfloat[FCBA~\cite{DBLP:conf/aaai/LiuZFYXM024} \label{fig:B_effect_fcba_mta_during}]{
        \centering
        \includegraphics[width=0.45\linewidth]{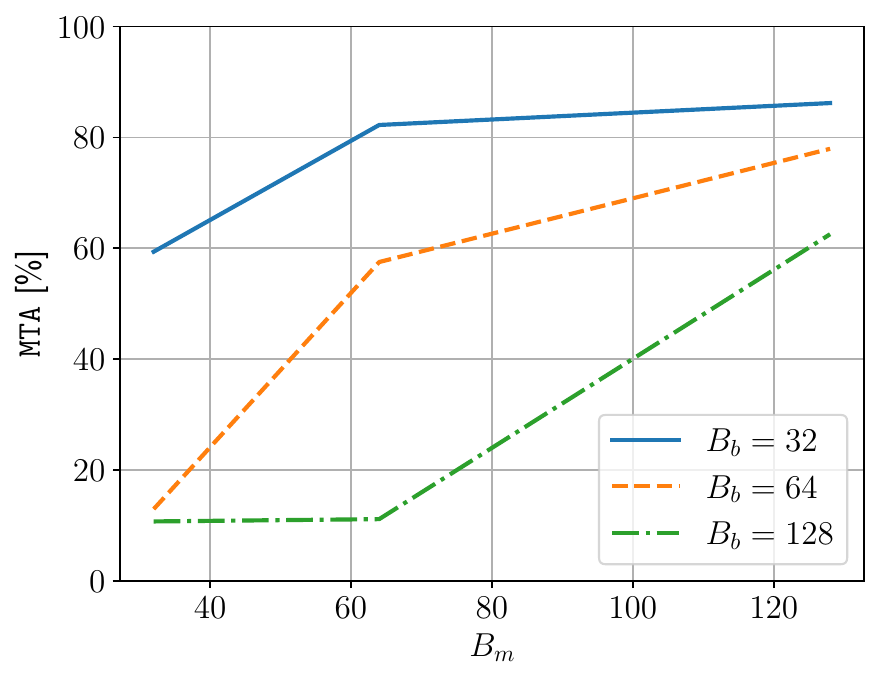}
    }

    \caption{Impact of $B_b$ and $B_m$ on the \mta of SoTA attacks.}
    \label{fig:B_effect_mta_during}
\end{figure}

In Figure~\ref{fig:E_effect_bda_after} and \ref{fig:B_effect_bda_after}, we show the \bdaafter for the A3FL~\cite{zhang_a3fl_2023}, Chameleon \cite{dai_chameleon_2023}, DarkFed~\cite{DBLP:conf/ijcai/LiWNHXZW24}, and FCBA~\cite{DBLP:conf/aaai/LiuZFYXM024} attacks for varying benign and malicious number of local epochs and batch sizes, complementing our results from Section~\ref{sec:E_B_effect}.
Further, Figure~\ref{fig:E_effect_lifespan} and \ref{fig:B_effect_lifespan} depict the corresponding \lifespan and Figure~\ref{fig:E_effect_mta_during} and \ref{fig:B_effect_mta_during} the corresponding \mta. 

\subsection{Impact of Weight Decay}
\label{app:lambda_effect}

\begin{figure}
    \centering
    \subfloat[A3FL~\cite{zhang_a3fl_2023} \label{fig:lambda_effect_a3fl_bda_during}]{
        \centering
        \includegraphics[width=0.45\linewidth]{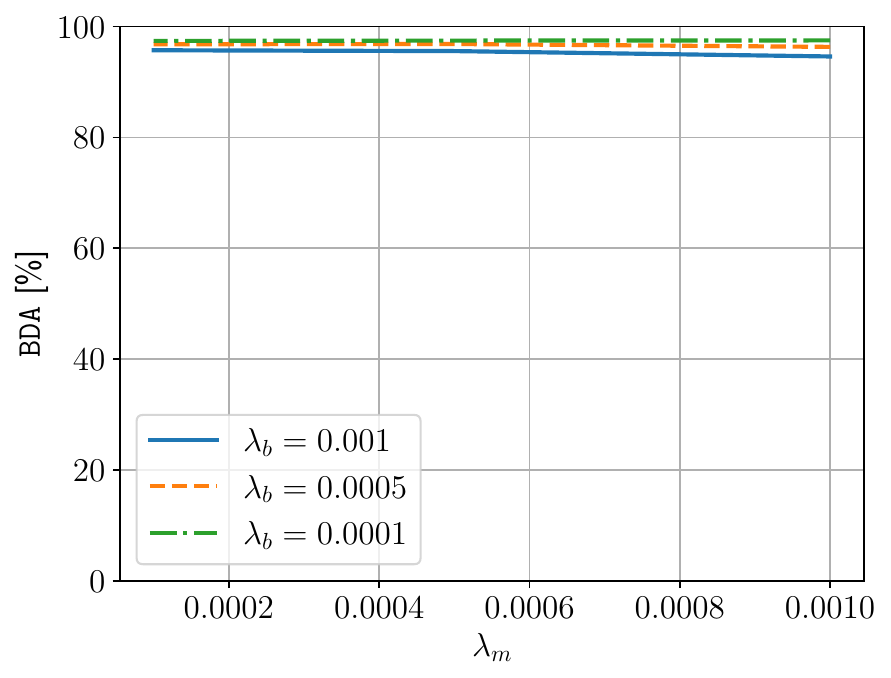}
    }
    \hfil
    \subfloat[Chameleon~\cite{dai_chameleon_2023} \label{fig:lambda_effect_chameleon_bda_during}]{
        \centering
        \includegraphics[width=0.45\linewidth]{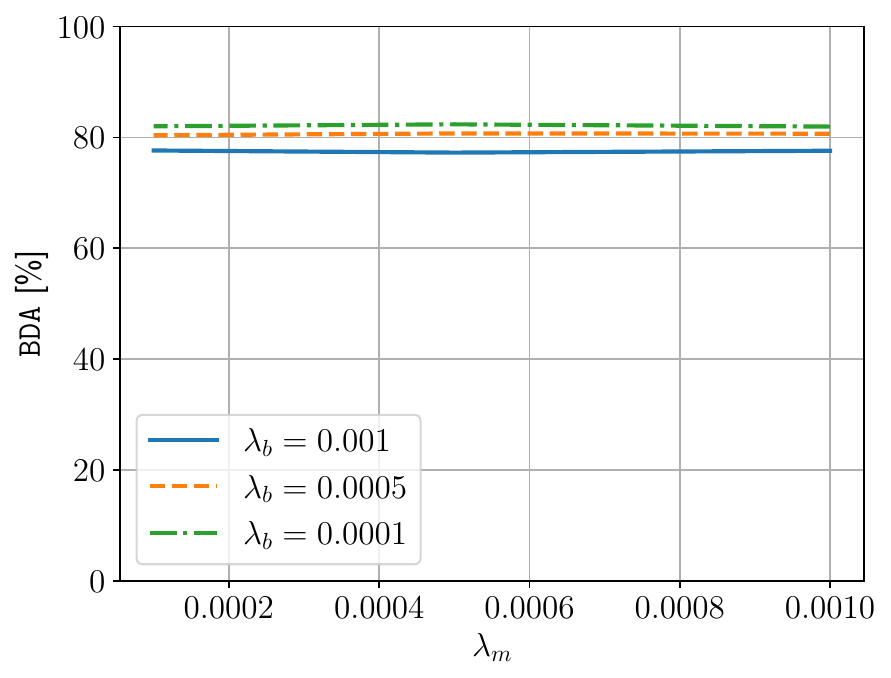}
    }

    \subfloat[DarkFed~\cite{DBLP:conf/ijcai/LiWNHXZW24} \label{fig:lambda_effect_darkfed_auc}]{
        \centering
        \includegraphics[width=0.45\linewidth]{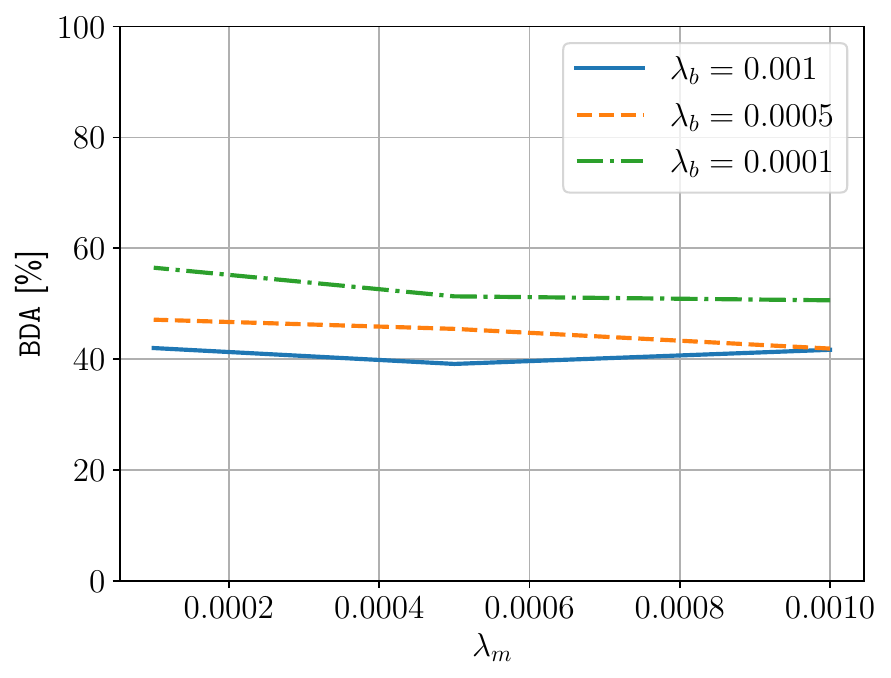}
    }
    \hfill
    \subfloat[FCBA~\cite{DBLP:conf/aaai/LiuZFYXM024} \label{fig:lambda_effect_fcba_bda_during}]{
        \centering
        \includegraphics[width=0.45\linewidth]{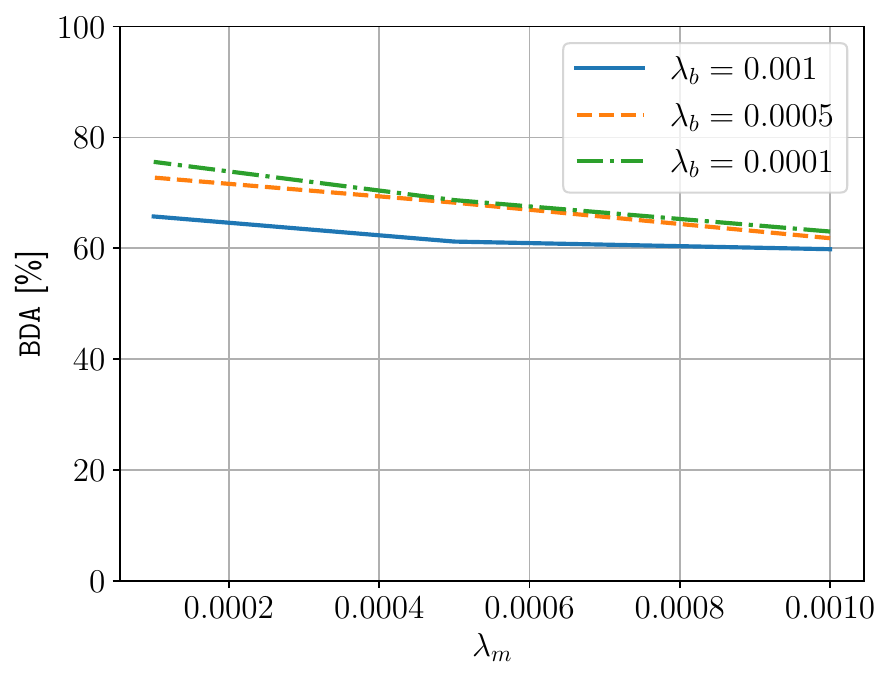}
    }
    \caption{Impact of $\lambda_b$ and $\lambda_m$ on the \bda of SoTA attacks.}
    \label{fig:lambda_effect_bda_during}
\end{figure}

\begin{figure}[htb!]
    \centering
    \subfloat[A3FL~\cite{zhang_a3fl_2023} \label{fig:lambda_effect_a3fl_bda_after}]{
        \centering
        \includegraphics[width=0.45\linewidth]{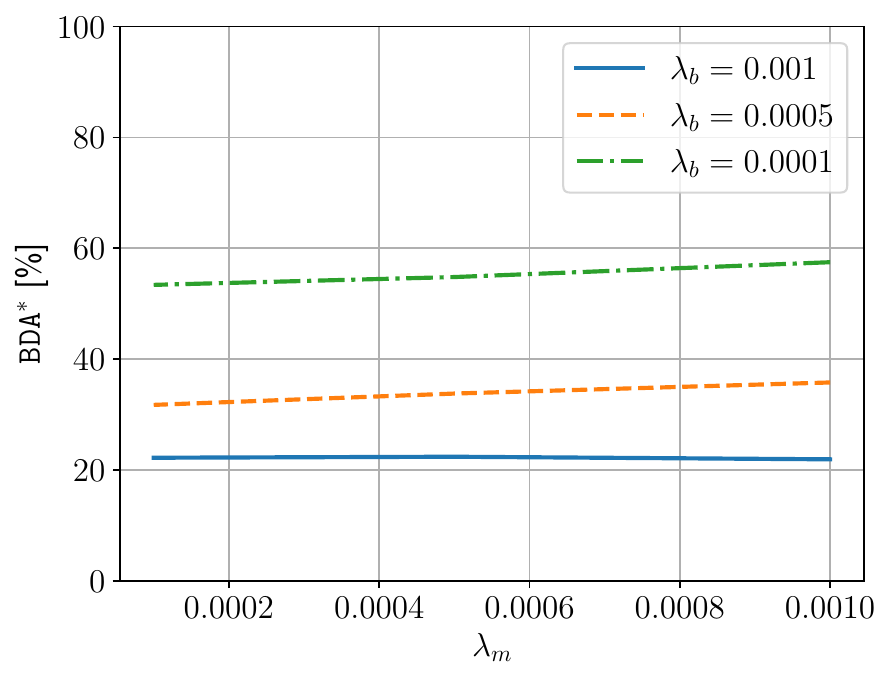}
    }
    \hfill
    \subfloat[Chameleon~\cite{dai_chameleon_2023} \label{fig:lambda_effect_chameleon_bda_after}]{
        \centering
        \includegraphics[width=0.45\linewidth]{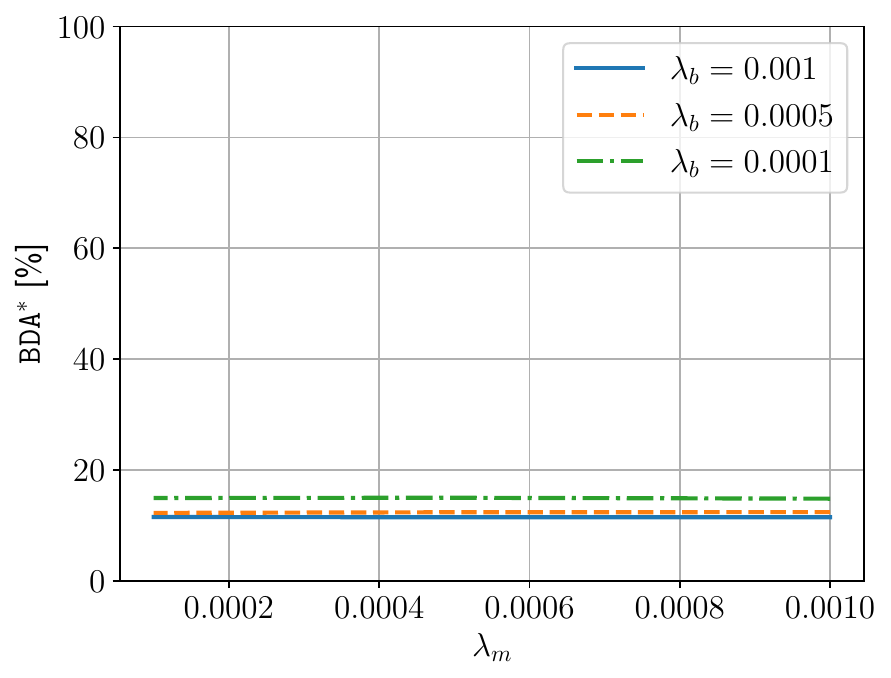}
    }

    \subfloat[DarkFed~\cite{DBLP:conf/ijcai/LiWNHXZW24} \label{fig:lambda_effect_darkfed_bda_after}]{
        \centering
        \includegraphics[width=0.45\linewidth]{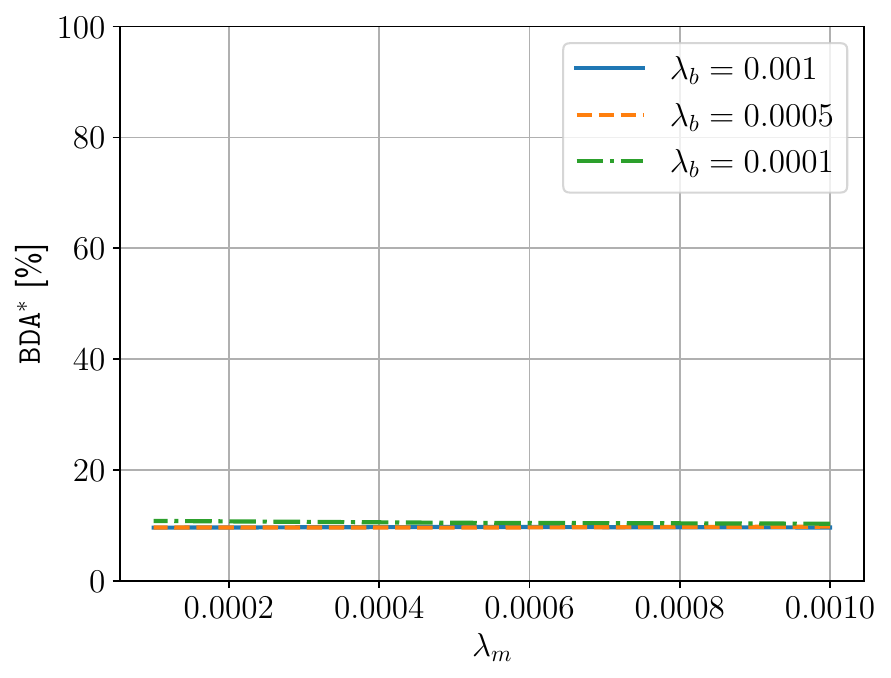}
    }
    \hfill
    \subfloat[FCBA~\cite{DBLP:conf/aaai/LiuZFYXM024} \label{fig:lambda_effect_fcba_bda_after}]{
        \centering
        \includegraphics[width=0.45\linewidth]{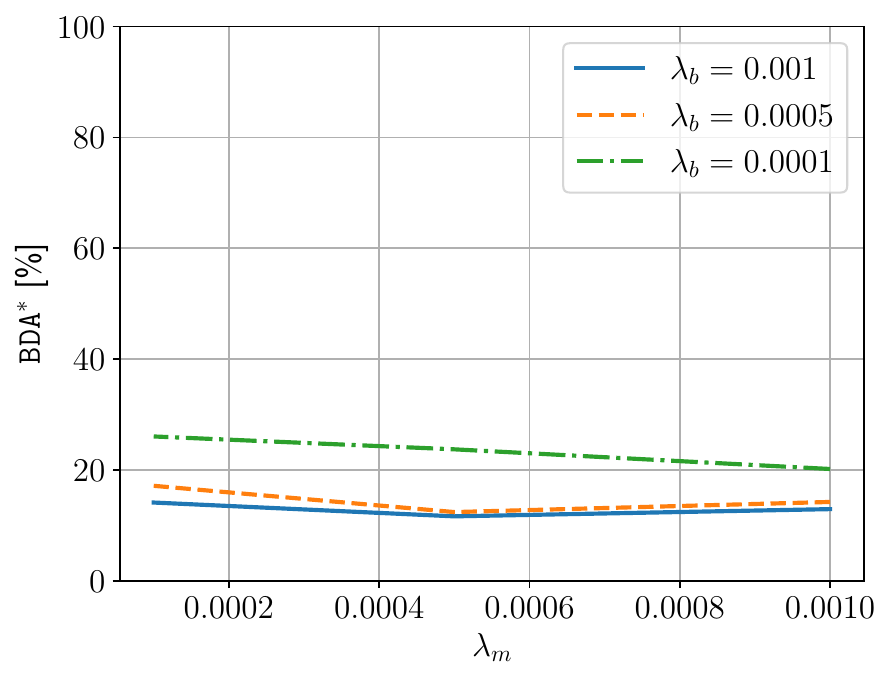}
    }

    \caption{Impact of $\lambda_b$ and $\lambda_m$ on the \bdaafter of SoTA attacks.}
    \label{fig:lambda_effect_bda_after}
\end{figure}

\begin{figure}[htb!]
    \centering
    \subfloat[A3FL~\cite{zhang_a3fl_2023} \label{fig:lambda_effect_a3fl_lifespan}]{
        \centering
        \includegraphics[width=0.45\linewidth]{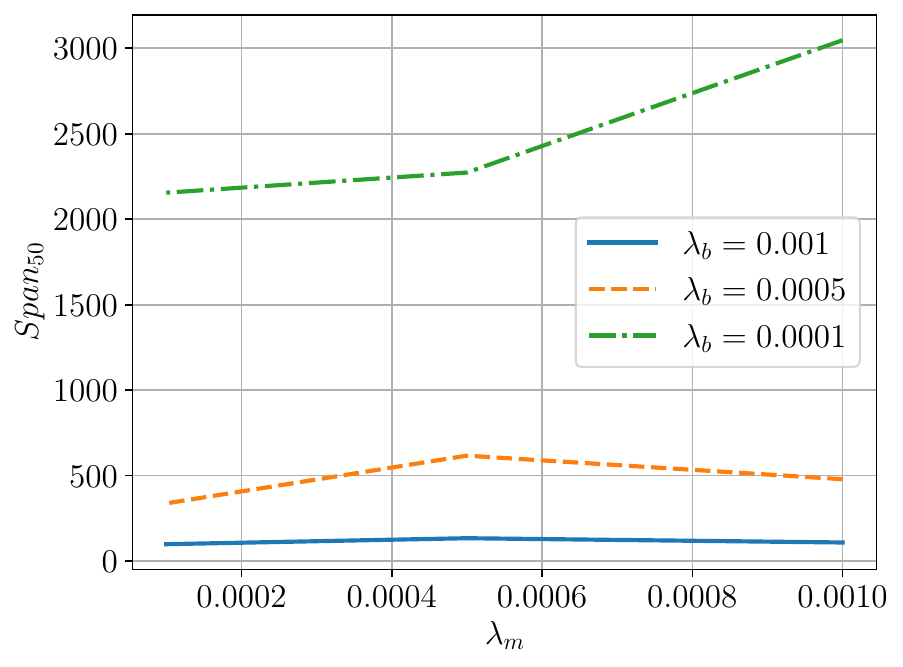}
    }
    \hfill
    \subfloat[Chameleon~\cite{dai_chameleon_2023} \label{fig:lambda_effect_chameleon_lifespan}]{
        \centering
        \includegraphics[width=0.45\linewidth]{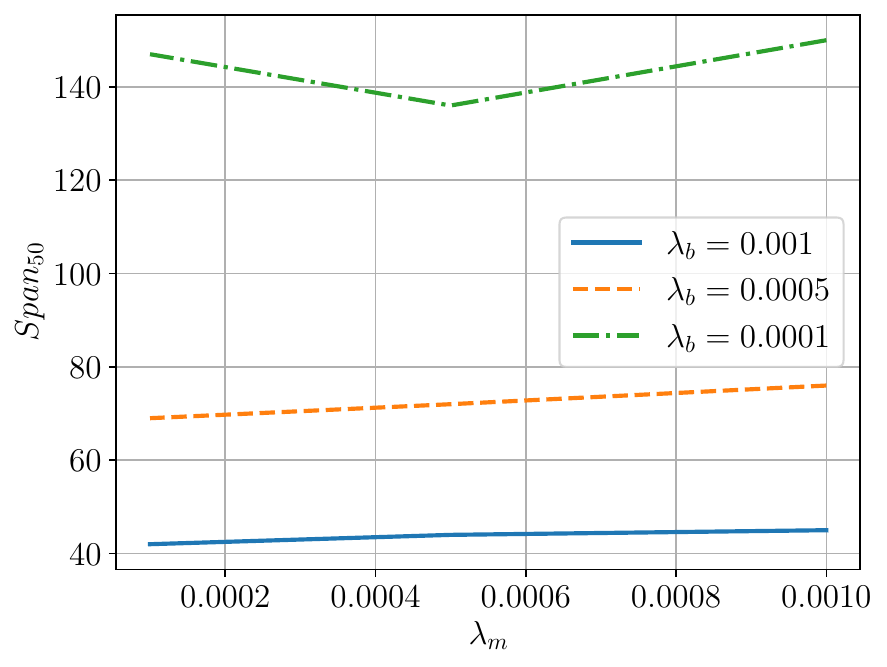}
    }

    \subfloat[DarkFed~\cite{DBLP:conf/ijcai/LiWNHXZW24} \label{fig:lambda_effect_darkfed_lifespan}]{
        \centering
        \includegraphics[width=0.45\linewidth]{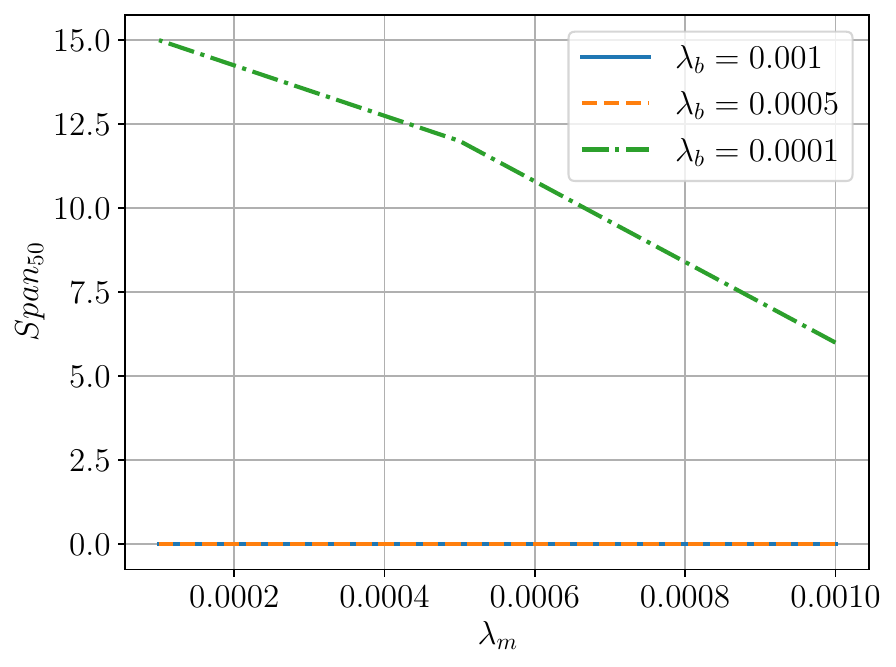}
    }
    \hfill
    \subfloat[FCBA~\cite{DBLP:conf/aaai/LiuZFYXM024} \label{fig:lambda_effect_fcba_lifespan}]{
        \centering
        \includegraphics[width=0.45\linewidth]{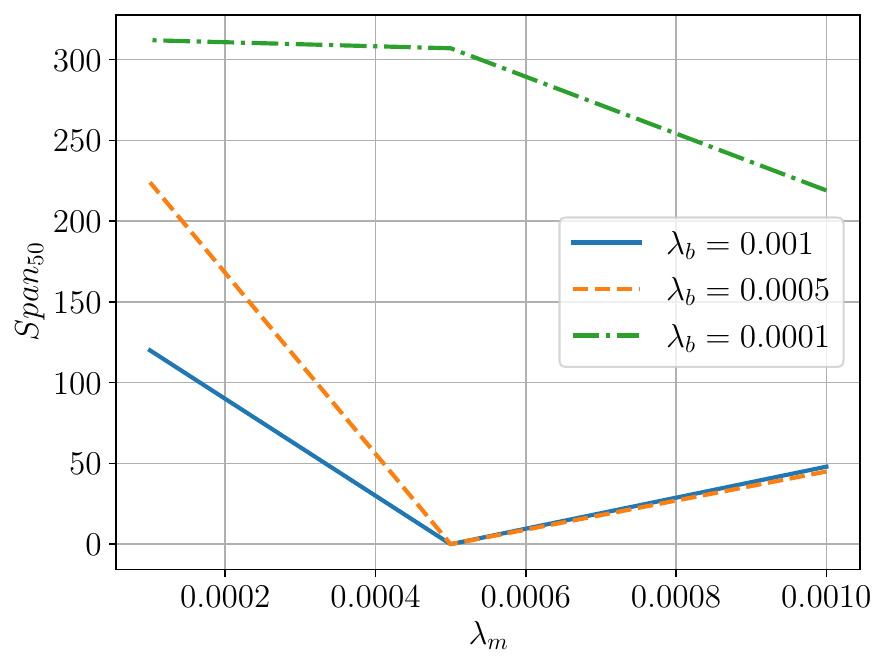}
    }

    \caption{Impact of $\lambda_b$ and $\lambda_m$ on the \lifespan of SoTA attacks.}
    \label{fig:lambda_effect_lifespan}
\end{figure}

\begin{figure}[!]
    \centering
    \subfloat[A3FL~\cite{zhang_a3fl_2023} \label{fig:lambda_effect_a3fl_mta_during}]{
        \centering
        \includegraphics[width=0.45\linewidth]{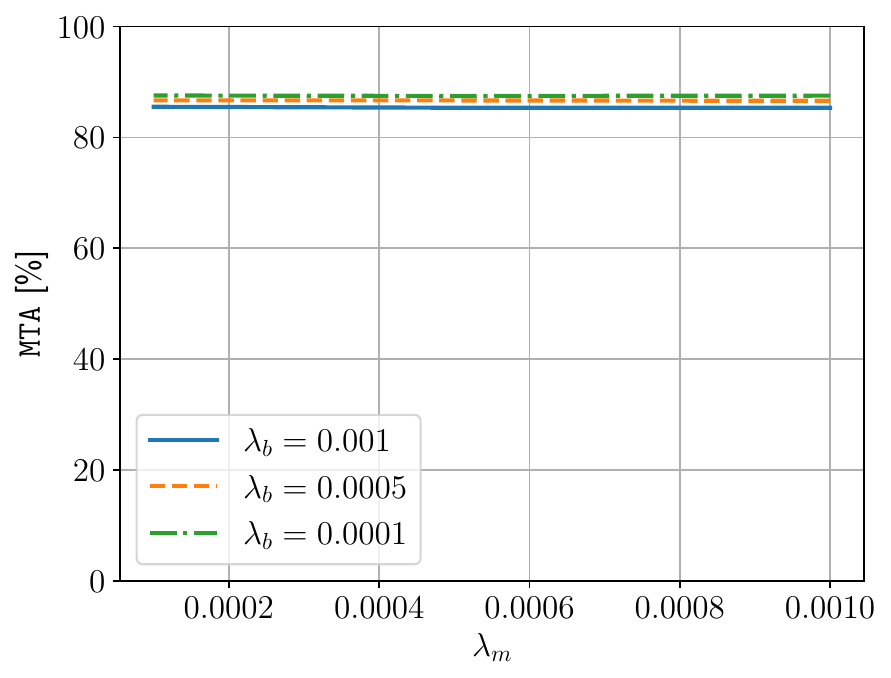}
    }
    \hfill
    \subfloat[Chameleon~\cite{dai_chameleon_2023} \label{fig:lambda_effect_chameleon_mta_during}]{
        \centering
        \includegraphics[width=0.45\linewidth]{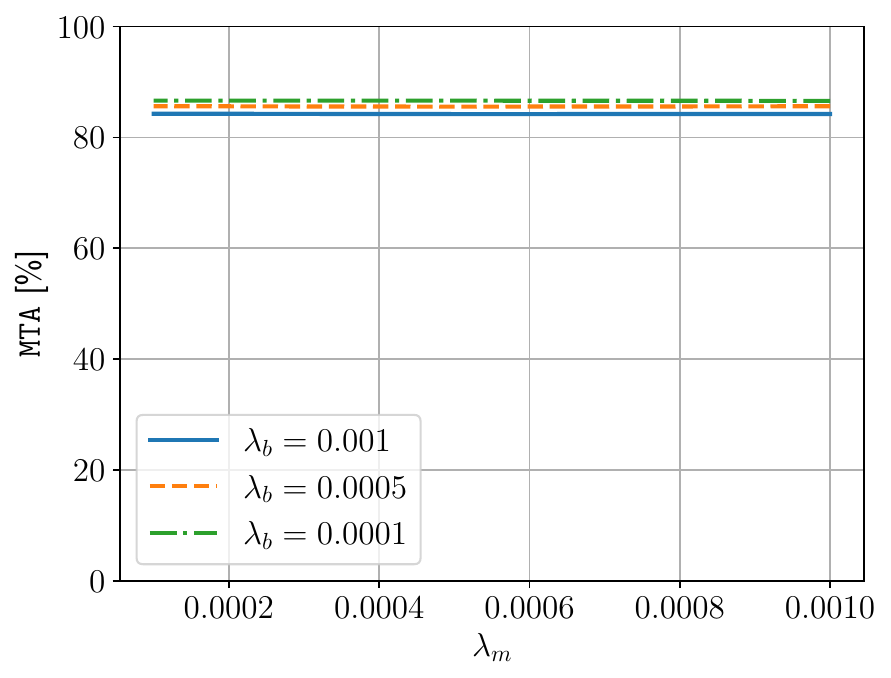}
    }

    \subfloat[DarkFed~\cite{DBLP:conf/ijcai/LiWNHXZW24} \label{fig:lambda_effect_darkfed_mta_during}]{
        \centering
        \includegraphics[width=0.45\linewidth]{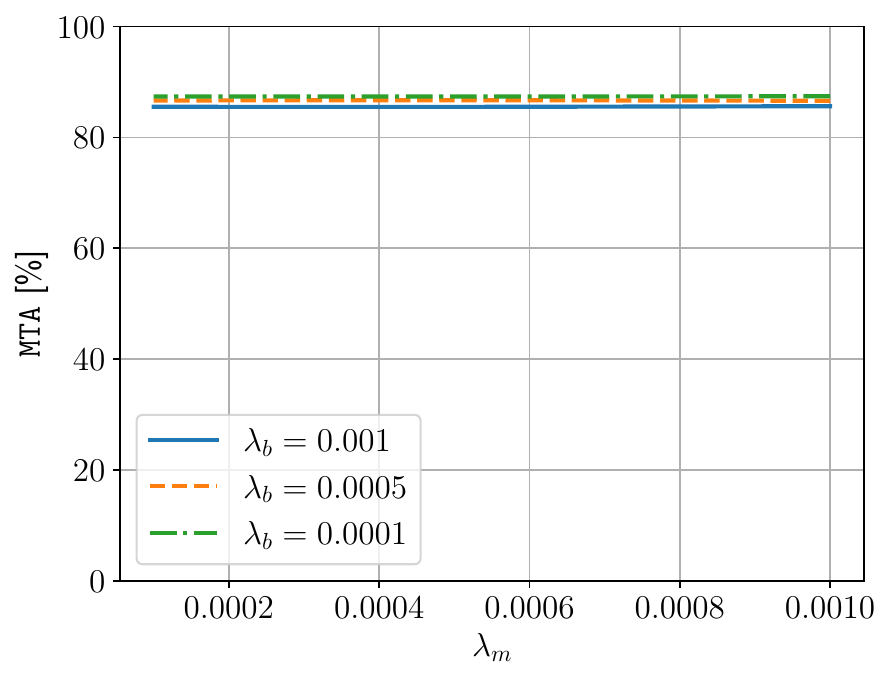}
    }
    \hfill
    \subfloat[FCBA~\cite{DBLP:conf/aaai/LiuZFYXM024} \label{fig:lambda_effect_fcba_mta_during}]{
        \centering
        \includegraphics[width=0.45\linewidth]{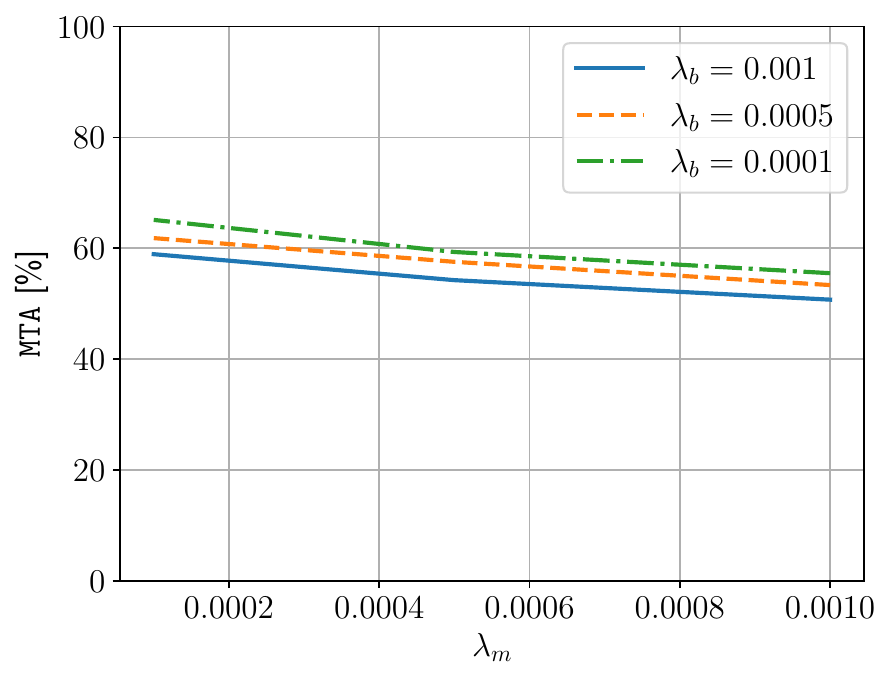}
    }

    \caption{Impact of $\lambda_b$ and $\lambda_m$ on the \mta of SoTA attacks.}
    \label{fig:lambda_effect_mta_during}
\end{figure}

In Figure~\ref{fig:lambda_effect_bda_during}, \ref{fig:lambda_effect_bda_after}, \ref{fig:lambda_effect_lifespan}, and \ref{fig:lambda_effect_mta_during}, we evaluate the \bda, \bdaafter, \lifespan, and \mta for the A3FL~\cite{zhang_a3fl_2023}, Chameleon~\cite{dai_chameleon_2023}, DarkFed~\cite{DBLP:conf/ijcai/LiWNHXZW24}, and FCBA~\cite{DBLP:conf/aaai/LiuZFYXM024} attacks for varying benign and malicious weight decay factors, complementing our results from Section~\ref{sec:lambda_effect}.

